\documentclass{amsart}

\usepackage{srcltx}
\usepackage{deleq}
\usepackage{mhequ}
\usepackage{psfrag}
\usepackage{color}
\usepackage{url}
\usepackage[g]{esvect}
\usepackage{cite}
\usepackage[mathscr]{eucal}
\usepackage{amssymb}
\usepackage{graphicx}

\numberwithin{equation}{section}

\newtheorem{thm}{Theorem}[section]
\newtheorem{prop}[thm]{Proposition}

\newcommand\calV{{\mathcal{V}}}

\newcommand\calT{{\mathcal{T}}}

\newcommand\twog{\gamma}

\newcommand\de{\delta}

\newcommand\bbR{{\mathbb R}}

\newcommand\ol{\overline}

\newcommand\s{\sigma}

\newcommand\f{\phi}
\newcommand\D{\nabla}
\newcommand\e{\epsilon}

\newcommand\ric{{\rm Ric}}

\newcommand\la{\lambda}

\newcommand\ra{\rangle}

\newcommand\<{\langle}

\newcommand\beq{\begin{equation}}
\newcommand\eeq{\end{equation}}
\newcommand\ben{\begin{enumerate}}
\newcommand\een{\end{enumerate}}
\newcommand\bit{\begin{itemize}}
\newcommand\eit{\end{itemize}}

\newcommand\pda{\frac{\d}{\d x^{\a}}}
\newcommand\pdb{\frac{\d}{\d x^{\b}}}

\renewcommand\ss{\smallskip}

\newcommand\chart{$(U,x^{\a})$\,}
\newcommand\X{{\mathfrak X}}

\newcommand\bmp{\begin{minipage}}
\newcommand\emp{\end{minipage}}

\newcommand{\noarxiv}[1]{\ptc{addition for the bams version, to be removed in the arxiv one}}

\newcommand{\bams}[1]{{{\color{blue}\ptc{the stuff in blue should be removed from the bams
version}#1}}}

\newcommand{\bamsgreen}[1]{{\color{green}\ptc{addition for the arxiv version, to be removed in the bams version}#1}}

\newcommand{\bamsf}[1]{{\color{blue}\ptc{the footnote will be removed from the bams version}#1}}

\newcommand{\altreverse}[2]{{\color{red}\ptc{the stuff in red is for the bams version}#1}
 {\color{blue}\ptc{the stuff in blue is old, NOT for the bams version} #2 }}
\newcommand{\zR}{\mathring R}%
\newcommand{\zD}{\mathring D}%
\newcommand{\zA}{\mathring A}%
\newcommand{\mzh}{\mathring h}%
\newcommand{\mzyh}{\mathring \chi}%

\newcommand{\myh}{\chi}
\newcommand{\SI}{\Sigma_{\mbox{\scriptsize inn}}}%
\newcommand{\SO}{\Sigma_{\mbox{\scriptsize out}}}%

\newcommand{\mnote}[1]
{\protect{\stepcounter{mnotecount}}$^{\mbox{\footnotesize $
\bullet$\themnotecount}}$ \marginpar{
\raggedright\tiny\em $\!\!\!\!\!\!\,\bullet$\themnotecount: #1}
}

\newcommand{\Min}[1]{\R^{1,#1}}

\newcommand{\hypwithhat}{\hathyp}
\newcommand{\hathyp}{\,\,\widehat{\!\!\hyp}}

\newcommand{\changedX}{K}

\newcommand{\eean}{\nonumber\end{eqnarray}}

\newcommand{\Mtext}{\Sext}

\newcommand{\val}[1]{\lvert #1\rvert}

\newcommand{\der}[0]{\partial}

\newcommand{\ii}[0]{\val{I}}
\newcommand{\reff}[1]{$(\ref{#1})$}
\newcommand{\ini}[0]{\rvert_{t=0}}

\newcommand{\dive}{\operatorname{div}}

\def\d{\partial}

\def\K0{\phi^{K_0}}

\newcommand{\aregular}{{an {\regular}}}
\newcommand{\regular}{$I^+$--regular}

{\catcode `\@=11 \global\let\AddToReset=\@addtoreset}
\AddToReset{equation}{section}
\renewcommand{\theequation}{\thesection.\arabic{equation}}

{\catcode `\@=11 \global\let\AddToReset=\@addtoreset}
\AddToReset{figure}{section}
\renewcommand{\thefigure}{\thesection.\arabic{figure}}

{\catcode `\@=11 \global\let\AddToReset=\@addtoreset}
\AddToReset{table}{section}
\renewcommand{\thetable}{\thesection.\arabic{table}}

\newcommand{\fourg}{{ g }}

\newcommand{\levoca}[1]{}

\newcommand{\mcN}{{\mycal N}}
\newcommand{\mcNX}{{\mycal N(X)}}

\let\a=\alpha\let\b=\beta \let\g=\gamma \let\d=\delta

\newcommand{\calM}{{\mcM}}

\newcommand{\nopcite}[1]{}

\newcommand{\hypM}{M}

  {\renewcommand{\theenumi}{\roman{enumi}}
   \renewcommand{\labelenumi}{(\theenumi)}}

\newcommand{\const}{\mathrm{const}}

\newcommand{\mcE}{{\mycal E}}

\newcommand{\mcD}{{\mycal D}}
\newcommand{\mcW}{{\mycal W}}

\newcommand{\nablash}{\nabla{\kern -.75 em
     \raise 1.5 true pt\hbox{{\bf/}}}\kern +.1 em}
\newcommand{\Deltash}{\Delta{\kern -.69 em
     \raise .2 true pt\hbox{{\bf/}}}\kern +.1 em}
\newcommand{\Rslash}{R{\kern -.60 em
     \raise 1.5 true pt\hbox{{\bf/}}}\kern +.1 em}

\newcommand{\tthreeg}{\tilde\threeg}
\newcommand{\tcalD}{\widetilde\calD}

\newcommand{\tD}{\tilde D}

\newcommand{\Ric}{\operatorname{Ric}}

\newcommand{\mcO}{{\mycal O}}

\newcommand{\mcU}{{\mycal U}}

\newcommand{\hyp}{M}

\newcommand{\Sp}{\ensuremath{\Sigma_{+}}} 

\newcommand{\threeg}{h}

\newcommand{\mcM}{{\mycal M}}

\newcommand{\mcH}{{\mycal H}}
\newcommand{\mcK}{{\mycal K}}

\newcommand{\bea}{\begin{eqnarray}}
\newcommand{\beaa}{\begin{eqnarray*}}
\newcommand{\bean}{\begin{eqnarray}\nonumber}

\newcommand{\Mext}{\mcM_{\mbox{\scriptsize \rm ext}}}

\newcommand{\mcMext}{\Mext}

\newcommand{\bel}[1]{\begin{equation}\label{#1}}
\newcommand{\beal}[1]{\begin{eqnarray}\label{#1}}
\newcommand{\beadl}[1]{\begin{deqarr}\label{#1}}
\newcommand{\eeadl}[1]{\arrlabel{#1}\end{deqarr}}
\newcommand{\eeal}[1]{\label{#1}\end{eqnarray}}
\newcommand{\eead}[1]{\end{deqarr}}
\newcommand{\eea}{\end{eqnarray}}
\newcommand{\eeaa}{\end{eqnarray*}}

\newcommand{\Hess}{\mathrm{Hess}\,}
\newcommand{\Ricc}{\mathrm{Ric}\,}

\newcommand{\be}{\begin{equation}}
\newcommand{\ee}{\end{equation}}

\newcommand{\tr}{\mbox{\rm tr}\,}

\newcommand{\eq}[1]{\eqref{#1}}
\newcommand{\Eq}[1]{Equation~(\ref{#1})}

\DeclareFontFamily{OT1}{rsfs}{} \DeclareFontShape{OT1}{rsfs}{m}{n}{
<-7> rsfs5 <7-10> rsfs7 <10-> rsfs10}{}
\DeclareMathAlphabet{\mycal}{OT1}{rsfs}{m}{n}

\usepackage{amsmath} 

\newcommand{\ohyp}{\,\,\overline{\!\!\hyp}}
\newcommand{\pohyp}{\partial\ohyp}

\def \R {\mathbb R}

\newcommand{\mcL}{{\mycal L}}
\def \Nat{\mathbb{N}}

\def \N {\Nat}

\newcounter{opp}[section]

\newcommand{\opp}[1]{\label{#1}}

\newcounter{mnotecount}[section]

\renewcommand{\themnotecount}{\thesection.\arabic{mnotecount}}

\newcommand{\ednote}[1]{}

\newcommand{\mcmg}{$(\mcM,\fourg)$}

\newcommand{\Sm}{\ensuremath{\Sigma_{-}}}
\newcommand{\No}{\ensuremath{N_{1}}}
\newcommand{\Nt}{\ensuremath{N_{2}}}
\newcommand{\Nth}{\ensuremath{N_{3}}}

\definecolor{bluem}{rgb}{0,0,0.5}

\definecolor{mycolor}{cmyk}{0.5,0.1,0.5,0}
\definecolor{michel}{rgb}{0.5,0.9,0.9}

\definecolor{turquoise}{rgb}{0.25,0.8,0.7}
\definecolor{bluem}{rgb}{0,0,0.5}

\definecolor{MDB}{rgb}{0,0.08,0.45}
\definecolor{MyDarkBlue}{rgb}{0,0.08,0.45}

\definecolor{MLM}{cmyk}{0.1,0.8,0,0.1}
\definecolor{MyLightMagenta}{cmyk}{0.1,0.8,0,0.1}

\definecolor{HP}{rgb}{1,0.09,0.58}

\newcommand{\ptc}[1]{\mnote{{\bf ptc:} #1}}

\newcommand{\loc}{{\textrm{loc}}}

\newcommand{\Sext}{\hyp_{\mathrm ext}}
\newcommand{\doc}{\langle\langle \mcMext\rangle\rangle}

\newcommand{\mcB}{{\mycal B}}

\def\pa{\partial}

\def\emph#1{{\it #1}}
\def\textbf#1{{\bf #1}}

\def\a{{\alpha}}
\def\b{{\beta}}
\def\ga{\gamma}

\def\Si{\Sigma}

\def\th{\theta}

\def\nab{\nabla}

\def\tr{\mbox{tr}}

\def\AA{{\mycal  A}}

\def\A{{\bf A}}

\def\L{{\bf L}}
\def\O{{\bf O}}

\def\R{{\RR}}

\def\S{{\bf S}}
\def\K{{\bf K}}
\def\g{{\bf g}}

\def\T{{\Bbb T}}
\def\To{\T}

\def\pr{\partial}

\renewcommand{\div}{\mbox{div }}

\def\2{{\overline 2}}

\newcommand{\beqa}{\begin{eqnarray}}
\newcommand{\eeqa}{\end{eqnarray}}

\newcommand{\RR}{\mathbb R}
\newcommand{\calD}{{\mathcal D}}

\renewcommand{\div}{\mathop{\rm div}}

\newcommand{\tsigma}{{\tilde{\sigma}}}    

\newcommand{\laplaciano}[1]{\Delta_{#1}\,}     

\newcounter{shownewstuffflag}

\newcommand{\startnewstuff}{\ifnum\value{shownewstuffflag}>0\color{blue}\fi}
\newcommand{\finishnewstuff}{\ifnum\value{shownewstuffflag}>0\color{black}\fi}

\newcounter{oldeq}

\def\beq{\begin{equation}}
\def\eeq{\end{equation}}

\renewcommand\>{\ra}

\renewcommand\th{\theta}

\renewcommand\a{\alpha}

\renewcommand\div{{\rm div}}
\renewcommand\b{\beta}

\renewcommand\L{\triangle}
\renewcommand\d{\partial}

\renewcommand\S{\Sigma}

\renewcommand\g\gamma

\renewcommand{\bamsgreen}[1]{{\ptc{addition for the arxiv version, to be removed in the bams version}#1}}

\renewcommand{\bamsf}[1]{{\ptc{the footnote will be removed from the bams version}#1}}
\renewcommand{\altreverse}[2]{{\ptc{the stuff in blue is   for the arxiv version}  #2 \ptc{end of the stuff in blue}}}

\renewcommand{\altreverse}[2]{#2}
\renewcommand{\bams}[1]{#1}
\renewcommand{\bamsf}[1]{#1}
\renewcommand{\bamsgreen}[1]{#1}

\newtheorem{theorem}{Theorem}[section]

\newtheorem{Theorem}[theorem]{Theorem}
\newtheorem{Corollary} [theorem] {Corollary}

\newtheorem{Conjecture}[theorem]{Conjecture}

\theoremstyle{definition}

\newtheorem{Definition}[theorem]{Definition}

\theoremstyle{remark}

\newtheorem{Remark}[theorem]{Remark}

\begin{document}

\title[Mathematical General Relativity]{Mathematical general relativity: a sampler}


\author{Piotr T. Chru\'sciel}
\address{LMPT, F\'ed\'eration Denis Poisson, Tours; Mathematical Institute and Hertford College, Oxford}
\curraddr{Hertford College, Oxford OX1 3BW, UK}
\email{chrusciel@maths.ox.ac.uk}
\thanks{}

\author{Gregory J. Galloway}
\address{Department of Mathematics, University of Miami}
\curraddr{Coral Gables, FL 33124, USA}
\email{galloway@math.miami.edu}
\thanks{
Support by the Banff International Research Station (Banff,
Canada), and by Institut Mittag-Leffler (Djursholm, Sweden) is
gratefully acknowledged. The research of GG has been supported
in part by an NSF grant DMS 0708048.}
\author{Daniel Pollack}
\address{Department of Mathematics, University of Washington}
\curraddr{Box 354350, Seattle, WA 98195-4350, USA}
\email{pollack@math.washington.edu}
\thanks{}


\date{}


\begin{abstract}
We provide an introduction to selected recent advances in the
mathematical understanding of Einstein's theory of gravitation.
\end{abstract}

\maketitle \tableofcontents

\section{Introduction}

Mathematical general relativity is, by now, a well-established
vibrant branch of mathematics. It ties fundamental problems of
gravitational physics with beautiful questions in mathematics.
The object is the study of manifolds equipped with a Lorentzian
metric satisfying the Einstein field equations. Some highlights
of its history include the discovery by Choquet-Bruhat of a
well posed Cauchy problem~\cite{ChBActa}, subsequently
globalized by Choquet-Bruhat and
Geroch~\cite{ChoquetBruhatGeroch69}, the singularity theorems
of Penrose and Hawking~\cite{Psing,HawkingPenrose}, the proof
of the positive mass theorem by Schoen and
Yau~\cite{SchoenYauPMT1}, and the proof of stability of
Minkowski space-time by Christodoulou and
Klainerman~\cite{ChristodoulouKlainerman93}.

There has recently been spectacular progress in the field on
many fronts, including the Cauchy problem, stability, cosmic
censorship, construction of initial data, and asymptotic
behaviour, many of which will be described here. Mutual
benefits are drawn, and progress is being made, from the
interaction between general relativity and geometric analysis
and the theory of elliptic and hyperbolic partial differential
equations. The Einstein equation shares issues of convergence,
collapse and stability with other important geometric PDEs,
such as the Ricci flow and the mean curvature flow. Steadily
growing overlap between the relevant scientific communities can
be seen. For all these reasons it appeared timely to provide a
mathematically oriented reader with an introductory survey of
the field. This is the purpose of the current work.

In Section~\ref{Scausal} we survey the Lorentzian causality
theory, the basic language for describing the structure of
space-times. In Section~\ref{Sbh} the reader is introduced to
black holes, perhaps the most fascinating prediction of
Einstein's theory of gravitation, {and the source of many deep
(solved or unsolved) mathematical problems.} In
Section~\ref{SEe} the Cauchy problem for the Einstein equations
is considered, laying down the foundations for a systematic
construction of general space-times. Section~\ref{Sivp}
examines initial data sets, as needed for the Cauchy problem,
and their global properties.  In Section~\ref{SE} we discuss
the dynamics of the Einstein equations, including questions of
stability and predictability; the latter question known under
the baroque name of ``strong cosmic censorship".
Section~\ref{SMts} deals with \emph{trapped} and
\emph{marginally trapped} surfaces, which signal the presence
of black holes, and have tantalizing connections with classical
minimal surface theory. The paper is sprinkled with open
problems, which are collected in Appendix~\ref{ssopp}.

\section{Elements of Lorentzian geometry and causal theory}
 \label{Scausal}
\subsection{Lorentzian manifolds}
In general relativity, and related theories, the space of physical
events is represented by a {\it Lorentzian manifold}. A Lorentzian
manifold is a smooth (Hausdorff, paracompact) manifold $\mcM = \mcM^{n+1}$
of dimension $n+1$, equipped with a Lorentzian metric $\fourg$. A
Lorentzian metric is a smooth assignment to each point $p \in \mcM$
of a symmetric, nondegenerate bilinear form on the tangent space
$T_p\mcM$ of {\it signature} $(- + \cdots +)$.   Hence, if $\{e_0,
e_1, ..., e_n\}$ is an orthonormal basis for $T_p\mcM$ with respect
to $\fourg$, then, perhaps after reordering the basis, the matrix
$[\fourg(e_i, e_j)]$ equals ${\rm diag}\, (-1, +1, ...,+1)$.  A
vector $v =\sum v^{\a} e_{\a}$ then has `square norm',
\beq\label{norm}
\fourg(v,v) =  -(v^0)^2 +\sum (v^i)^2 \,,
\eeq
which can be positive, negative or zero.
This leads to the causal character of vectors, and indeed to the
causal theory of Lorentzian manifolds, which we shall discuss
in Section~\ref{SCT}.

On a coordinate neighborhood \chart $=(U,x^0, x^1, ... , x^n)$
the metric $\fourg$ is completely determined by its metric
component functions on $U$, $\fourg_{\a\b} : =
\fourg(\pda,\pdb)$, $0 \le \a, \b \le n$:  For $v = v^{\a}\pda,
w = w^{\b} \pdb \in T_p\mcM$, $p \in U$, $\fourg(v,w) =
\fourg_{\a\b} v^{\a}w^{\b}$.  (Here we have used the Einstein
summation convention: If, in a coordinate chart, an index
appears repeated, once up and once down, then summation over
that index is implied.)  Classically the metric in coordinates
is displayed via the ``line element", $ds^2 = \fourg_{\a\b}
dx^{\a}dx^{\b}$.

The prototype Lorentzian manifold is Minkowski space $\Min{n}$,
the space-time of special relativity. This is $\bbR^{n+1}$,
equipped with the Minkowski metric, which, with respect to
Cartesian coordinates $(x^0, x^1, ..., x^n)$, is given by
$$
ds^2  = - (dx^0)^2 + (dx^1)^2 + \cdots + (dx^n)^2  \,.
$$
Each tangent space of a Lorentzian manifold is isometric to
Minkowski space, and in this way the local accuracy of special
relativity is built into general relativity.

Every Lorentzian manifold (or, more generally,
pseudo-Riemannian manifold) $(\mcM^{n+1}, \fourg)$ comes
equipped with a {\it Levi-Civita connection} (or covariant
differentiation operator) $\D$ that enables one to compute the
directional derivative of vector fields.
Hence, for smooth vector fields $X, Y \in \X(\mcM)$, $\D_XY \in
\X(\mcM)$ denotes the covariant derivative of $Y$ in the
direction $X$.   The Levi-Civita connection is the unique
connection $\D$ on $(\mcM^{n+1}, \fourg)$ that is (i) symmetric
(or torsion free), i.e., that satisfies $\D_XY - \D_Y X =
[X,Y]$ for all $X,Y \in \X(M)$, and (ii) compatible with the
metric, i.e. that obeys the metric product rule,
$X(\fourg(Y,Z)) = \fourg(\D_XY,Z) + \fourg(Y,\D_XZ)$, for all
$X,Y, Z \in \X(M)$.

In a coordinate chart \chart, one has, \beq\label{covder} \D_XY
= (X(Y^{\mu}) + \Gamma^{\mu}_{\a\b} X^{\a}Y^{\b}) \d_{\mu}  \,,
\eeq where $X^{\a}$, $Y^{\a}$ are the components of $X$ and
$Y$, respectively, with respect to the coordinate basis
$\d_{\a} = \pda$, and where the $\Gamma^{\mu}_{\a\b}$'s are the
classical Christoffel symbols, given in terms of the metric
components by,
\beq\label{chris} \Gamma^{\mu}_{\a\b} =
\frac12g^{\mu\nu}(\d_{\b}g_{\a\nu} + \d_{\a}g_{\b\nu}
-\d_{\nu}g_{\a\b})   \,. \eeq

Note that the coordinate expression \eqref{covder} can also be
written as, \beq \D_XY = X^{\a}\D_{\a}Y^{\mu}\d_{\mu} \,, \eeq
where $\D_{\a}Y^{\mu}$ (often written classically as
$Y^{\mu}{}_{;\a}$)  is given by, \beq \D_{\a}Y^{\mu} =
\d_{\a}Y^{\mu} + \Gamma^{\mu}_{\a\b} Y^{\b}  \,. \eeq

We shall feel free to interchange between coordinate and
coordinate free notations. {The Levi-Civita connection $\D$
extends in a natural way to   a covariant differentiation
operator on all tensor fields.}

The Riemann curvature tensor of $(\mcM^{n+1}, \fourg)$ is the map
$R: \X(M) \times \X(M) \times \X(M) \to \X(M)$, $(X,Y,Z) \to
R(X,Y)Z$, given by \beq\label{riem} R(X,Y)Z = \D_X\D_Y Z- \D_Y\D_X
Z- \D_{[X,Y]}Z  \,. \eeq This expression is linear in $X,Y,Z \in
\X(\mcM)$ with respect to $C^{\infty}(\mcM)$. This implies that $R$
is indeed tensorial,  i.e., that the value of $R(X,Y)Z$ at $p\in M$
depends only on the value of $X,Y,Z$ at $p$.

Equation \eqref{riem} shows that the Riemann curvature tensor
measures the extent to which covariant differentiation
fails to commute. This failure to commute may be seen as an
obstruction to the existence of parallel vector fields.   By
Riemann's theorem, a Lorentzian manifold is locally Minkowskian if
and only if the Riemann curvature tensor vanishes.

The components $R^{\mu}{}_{\g\a\b}$ of the Riemann curvature
tensor $R$ in a coordinate chart \chart are determined by the
equations, $R(\d_{\a},\d_{\b})\d_{\g} = R^{\mu}{}_{\g\a\b}
\d_{\mu}$. Equations \eqref{covder} and \eqref{riem} then yield
the following explicit formula  for the curvature components in
terms of the Christoffel symbols, \beq R^{\mu}{}_{\g\a\b} =
\d_{\a} \Gamma^{\mu}_{\g\b} - \d_{\b} \Gamma^{\mu}_{\g\a} +
\Gamma^{\nu}_{\g\b}\Gamma^{\mu}_{\nu\a}  -
\Gamma^{\nu}_{\g\a}\Gamma^{\mu}_{\nu\b}  \,. \eeq

The Ricci tensor, $\ric$, is a bilinear form obtained by
contraction of the Riemann curvature tensor, i.e., its
components $R_{\mu\nu} = \ric(\d_{\mu},\d_{\nu})$ are
determined by tracing, $R_{\mu\nu} = R^{\a}{}_{\mu\a\nu}$.
Symmetries of the Riemann curvature tensor imply that the Ricci
tensor is symmetric, $R_{\mu\nu} = R_{\nu\mu}$. By tracing the
Ricci tensor, we obtain the scalar curvature, $R =
g^{\mu\nu}R_{\mu\nu}$, where $ g^{\mu\nu} $ denotes the matrix
inverse to $ g_{\mu\nu}$.

\subsection{Einstein equations}
 \label{ssEe}

The Einstein equation  (with cosmological {constant
$\Lambda$}), the field equation of general relativity, is the
tensor equation,
\beq\label{eeqn} \ric -\frac12R\fourg + \Lambda \fourg = 8\pi
\calT \,, \eeq where $\calT$ is the energy-momentum tensor.
(See, e.g., Section~\ref{sLoizelet} for an example of an
energy-momentum tensor.) When expressed in terms of
coordinates, the Einstein equation becomes a system of second
order equations for the metric components $g_{\mu\nu}$ and the
nongravitational field variables introduced through the
energy-momentum tensor. We say that space-time obeys the vacuum
Einstein equation if it obeys the Einstein equation with $\calT
= 0$.

The Riemann curvature tensor has a number of symmetry
 properties, one of which is the so-called first Bianchi
identity:
$$
 R_{\alpha \beta \gamma \delta} +
 R_{\alpha \gamma \delta\beta } +
 R_{\alpha \delta\beta \gamma }
 =0
 \;.
$$
The curvature tensor also obeys a differential identity known
as the second Bianchi identity:
\bel{secBian}
 \nabla_{\sigma}R_{\alpha \beta \gamma \delta} +
 \nabla_{\alpha}R_{\beta \sigma\gamma \delta} +
 \nabla_{\beta}R_{\sigma\alpha  \gamma \delta}
 =0
 \;.
\ee
When twice contracted, \eq{secBian} yields the following
\emph{divergence identity}:
\beq\label{bianchi} \D_{\a}\left(
R^{\a\b} - \frac{R}2 g^{\a\b}\right) = 0\,. \eeq
%
This plays a fundamental role in general
relativity, as, in particular, it implies, in conjunction with
the Einstein equation,  local conservation of energy,
$\D_{\a}T^{\a\b} = 0$. It also plays an important role in the
mathematical analysis of the Einstein equations; see
Section~\ref{SEe} for further discussion.

\subsection{Elements of causal theory}\label{SCT}

\begin{figure}[ht]
\hspace{-2cm}{
 \psfrag{p}{\Large $p$}
 \psfrag{futuret}{\Large future pointing timelike}
 \psfrag{pastt}{\Large past pointing timelike}
 \psfrag{futuren}{\Large future pointing null}
 \psfrag{pastn}{\Large past pointing null}
 \resizebox{4in}{!}{\includegraphics{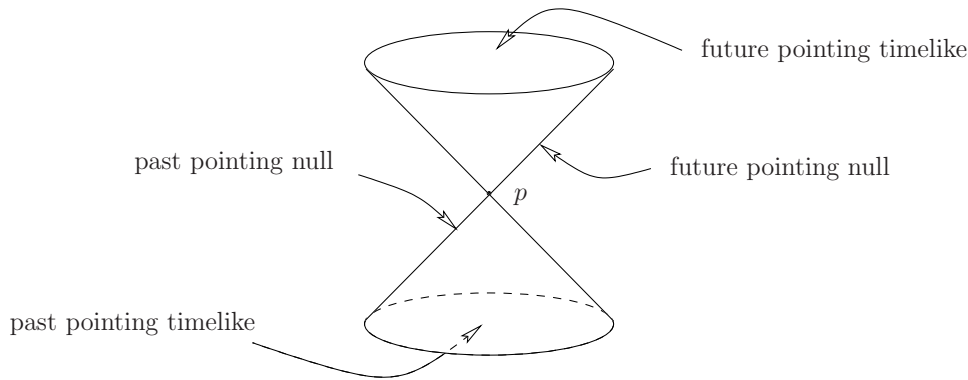}}
}
\caption{The light cone at $p$.
\label{Flightcone}
}
\end{figure}

Many concepts and results in general relativity make use of the
causal theory of Lorentzian manifolds.  The starting point for
causal theory is the causal classification of tangent vectors.
Let $(\mcM^{n+1}, \fourg)$ be a Lorentzian manifold.  A vector
$v \in T_p\mcM$ is timelike (resp., spacelike, null) provided
$g(v,v) < 0$ (resp., $g(v,v) > 0$, $g(v,v) = 0$).  The
collection of null vectors forms a double cone $\calV_p$ in
$T_p\mcM$ (recall \eqref{norm}), called the null cone at $p$;
see Figure~\ref{Flightcone}.

The timelike vectors at $p$ point inside the null cone and the
spacelike vectors point outside. We say that $v \in T_p\mcM$ is
causal if it is timelike or null.  We define the length of
causal vectors as $|v| = \sqrt{-g(v,v)}$.  Causal vectors $v, w
\in T_p\mcM$ that point into the same half-cone of the null
cone $\calV_p$ obey the reverse triangle inequality, $|v+w| \ge
|v| + |w|$.  Geometrically, this is the source of the twin
paradox.

These notions of causality extend to curves. Let $\g: I \to
\mcM$, $t \to \g(t)$, be a smooth curve in $\mcM$, then  $\g$
is said to be timelike (resp., spacelike, null, causal)
provided each of its velocity vectors $\g'(t)$ is timelike
(resp., spacelike, null, causal).  Heuristically, in accordance
with relativity, information flows along causal curves, and so
such curves are the focus of attention in causal theory. The
notion of a causal  curve extends in a natural way to piecewise
smooth curves, and we  will normally work within this class.
 As usual, we define a geodesic to be a curve $t
\to \g(t)$  of zero covariant acceleration, $\D_{\g'}\g'  = 0$.
Since geodesics $\g$ are constant  speed curves ($g(\g',\g') =
const.$),  each geodesic in a Lorentzian manifold is either
timelike, spacelike or null.

The length of a causal curve $\g : [a, b] \to \mcM$, is defined as
$$
L(\g)= \text{Length of } \g = \int_a^b |\g'(t)| dt = \int_a^b
\sqrt{-\fourg(\g'(t),\g'(t))} \, dt  \,.
$$
If $\g$ is timelike one can introduce an arc length parameter
along $\g$.  In general relativity,  a timelike curve
corresponds to the history of an observer, and arc length
parameter, called proper time, corresponds to  time kept by the
observer.  Using the existence and properties of geodesically
convex neighborhoods \cite{ONeill}  one can show that causal
geodesics are locally maximal (i.e., locally longest among
causal curves).

Each null cone $\calV_p$ consists of two half-cones, one of
which may designated as the future cone, and the other as the
past cone at $p$.  If the assignment of a past and future cone
at each point of $\mcM$ can be carried out in a continuous
manner over $\mcM$ then $\mcM$ is said to be time-orientable.
There are various ways to make the phrase ``continuous
assignment" precise, but they all result in the following fact:
A Lorentzian manifold $(\mcM^{n+1},g)$ is time-orientable if
and only if it admits a smooth timelike vector field $Z$. If
$\mcM$ is time-orientable, the choice of a smooth timelike
vector field $Z$ fixes a time orientation on $\mcM$:  For any
$p \in \mcM$,
 a causal vector $v\in T_p\mcM$ is future directed (resp. past directed)
 provided $\fourg(v, Z) < 0$ (resp. $\fourg(v,Z) > 0$).
Thus, $v$ is future
 directed if it points into the same null half cone at $p$ as $Z$.
We note that a  Lorentzian manifold that is not time-orientable
always admits a double cover that is. By  a {\it space-time} we
mean a connected time-oriented Lorentzian manifold
$(\mcM^{n+1}, \fourg)$.  Henceforth, we restrict attention to
space-times.

\subsubsection{Past and futures}
 \label{sspaf}
Let $(\mcM,\fourg)$ be a space-time.    A timelike (resp.
causal) curve $\g: I \to \mcM$ is said to be {\it future
directed} provided each tangent vector $\g'(t)$, $t \in I$,  is
future directed.  ({\it Past-directed} timelike  and causal
curves are defined in a time-dual manner.) $ I^+(p)$, the
timelike future of $p\in \mcM$, is the set consisting of  all
points $q \in \mcM$ for which there exists a future directed
timelike curve from $p$ to $q$. $J^+(p)$, the causal future of
$p\in \mcM$, is the set consisting of  $p$ and all points $q$
for which there exists a future directed causal curve from $p$
to $q$. In Minkowski space $\Min{n}$ these sets have a simple
structure: For each $p \in \Min{n}$, $\d I^+(p) = J^+(p)
\setminus I^+(p)$ is the future cone at $p$ generated by the
future directed null rays emanating from $p$. $I^+(p)$ consists
of the points inside the cone, and  $J^+(p)$ consists of the
points on and inside the cone.  In general, the curvature and
topology of space-time can strongly influence the structure of
these sets.


Since a timelike curve  remains timelike under small smooth
perturbations, it is heuristically  clear that the sets
$I^+(p)$ are in general open; a careful proof makes use of
properties of geodesically convex sets.   On the other hand the
sets $J^+(p)$ need not be closed in general, as can be seen by
considering the space-time obtained by removing a point from
Minkowski space.

 It follows from variational arguments that, for example, if $q \in I^+(p)$ and
 $r \in J^+(q)$ then $r \in I^+(p)$.  This and related claims are in fact a consequence
 of the following fundamental causality result~\cite{ONeill}.

\begin{prop}\label{causality}  If $q \in J^+(p) \setminus I^+(p)$, i.e., if $q$ is in the causal future
 of $p$ but not in the timelike future of $p$ then any future directed causal curve
 from $p$ to $q$ must be a null geodesic.
 \end{prop}

Given a subset $S \subset \mcM$, $I^+(S)$, the timelike future of $S$,
consists of all points $q \in \mcM$ for which there exists a future directed
timelike curve from a point in $S$ to $q$.   $J^+(S)$, the causal future of $S$
consists of the points of $S$ and all points $q \in \mcM$ for which there exists a future directed causal curve from a point in $S$ to $q$.
Note that $I^+(S) = \bigcup_{p \in S} I^+(p)$. Hence, as a union of open sets,
$I^+(S)$ is always open.


The timelike and causal pasts $I^-(p)$, $J^-(p)$, $I^-(S)$,
$J^-(S)$ are defined in a time dual manner in terms of past
directed timelike and causal curves.  It is sometimes
convenient to consider pasts and futures within some open
subset $U$ of $\mcM$. For example,  $I^+(p, U)$ denotes the set
consisting of all points $q \in U$ for which there exists a
future directed timelike from $p$ to $q$ contained in $U$.

Sets of the form $\d I^{\pm}(S)$ are called {\it achronal
boundaries}, and have nice structural properties: They are
achronal Lipschitz hypersurfaces, ruled, in a certain sense, by
null geodesics \cite{ONeill}. (A set is {\it achronal} if no
two of its points can be joined by a timelike curve.)


\subsubsection{Causality conditions}
A number of results in Lorentzian geometry and general
relativity require some sort of causality condition.  It is
perhaps natural on physical grounds to rule out the occurrence
of closed timelike curves. Physically, the existence of such a
curve signifies the existence of an observer who is able to
travel into his/her own past, which leads to variety of
paradoxical situations. A space-time $\mcM$ satisfies the {\it
chronology condition} provided there are no closed timelike
curves in $\mcM$.  It can be shown that all compact space-times
violate the chronology condition, and for this reason compact
space-times have been of limited interest in general
relativity.

A somewhat stronger condition than the chronology condition is
the {\it causality condition}.   A space-time $\mcM$ satisfies
the causality condition provided there are no closed
(nontrivial) causal  curves in $\mcM$. A slight weakness of
this condition is that there are space-times which satisfy the
causality condition, but contain causal curves that are
``almost closed", see e.g.~\cite[p. 193]{HE}.

It is useful to have a  condition that rules out ``almost
closed" causal curves. A space-time $\mcM$ is said to be {\it
strongly  causal} at $p\in \mcM$ provided there are arbitrarily
small neighborhoods $U$ of $p$ such that any causal curve
$\gamma$ which starts in, and leaves, $U$  never returns to
$U$.  $\mcM$ is {\it strongly causal} if it is strongly causal
at each of its points. Thus, heuristically speaking, $\mcM$ is
strongly  causal provided there are no closed or ``almost
closed" causal curves in $\mcM$. Strong causality is  the
``standard" causality  condition of space-time geometry, and
although there are even stronger causality conditions, it is
sufficient for most applications.  A very useful fact about
strongly causal space-times is the following: If $\mcM$ is
strongly causal then any future (or past) inextendible causal
curve $\g$ cannot be ``imprisoned" or ``partially imprisoned"
in a compact set. That is to say, if $\g$ starts in a compact
set $K$, it must eventually leave $K$ for good.

We now come to a fundamental condition in space-time geometry,
that of {\it global hyperbolicity}.  Mathematically, global
hyperbolicity is a basic `niceness' condition that often plays
a role analogous to geodesic completeness in Riemannian
geometry.  Physically, global hyperbolicity is connected to the
notion of strong cosmic censorship, the conjecture that,
generically, space-time solutions to the Einstein equations do
not admit {\it naked} (i.e., observable) singularities; see
Section~\ref{sSScc} for further discussion.

A space-time $\mcM$ is said to be globally hyperbolic provided:
\ben
\item[(1)] $\mcM$ is strongly causal.
\item[(2)] (Internal Compactness) The sets $J^+(p) \cap J^-(q)$
    are compact for all $p,q \in~\mcM$. \een Condition (2) says
    roughly that $\mcM$ has no holes or gaps.  For example
    Minkowski space $\Min{n}$ is globally hyperbolic but the
    space-time obtained by removing one point from it is not.
    Leray~\cite{Leray} was the first to introduce the notion of
    global hyperbolicity (in a somewhat different, but
    equivalent form) in connection with his study of the Cauchy
    problem for hyperbolic PDEs.

We mention a couple of basic consequences of global
hyperbolicity. Firstly, globally hyperbolic space-times are
{\it causally simple}, by which is meant that  the sets
$J^{\pm}(A)$ are closed for all compact $A \subset \mcM$.  This
fact and internal compactness implies that the sets $J^+(A)
\cap J^-(B)$ are compact, for all compact $A,B \subset \mcM$.

Analogously to the case of Riemannian geometry, one can learn
much about the global structure of space-time by studying its
causal geodesics.  Global hyperbolicity is the standard
condition in Lorentzian geometry that guarantees the existence
of maximal timelike geodesic segments joining timelike related
points. More precisely, one has the following.

\begin{prop}
If $\mcM$ is globally hyperbolic and $q \in I^+(p)$, then there
exists a maximal timelike geodesic segment $\g$ from $p$ to $q$
(where by maximal, we mean $L(\g) \ge L(\s)$ for all future
directed causal curves $\sigma$ from $p$ to $q$).
\end{prop}

Contrary to the situation in Riemannian geometry, geodesic
completeness does not guarantee the existence of maximal
segments, as is well illustrated by anti-de Sitter space, see
e.g.~\cite{BeemEhrlichEasley}.

Global hyperbolicity is closely  related to the existence of
certain `ideal initial value hypersurfaces', called  {\it
Cauchy (hyper)surfaces}. There are slight variations in the
literature in the definition of a Cauchy surface.  Here we
adopt the following definition: A Cauchy surface for a
space-time $\mcM$ is a subset $S$ that is met exactly
once by every inextendible causal curve in $\mcM$.  It can be shown
that a Cauchy surface for $\mcM$ is necessarily a $C^0$ (in fact,  Lipschitz)
hypersurface in $\mcM$.  Note also that a Cauchy surface is {\it acausal}, that
is, no two of its points can be joined by a causal curve.
The following result is fundamental.

\begin{prop}[Geroch~\cite{GerochDoD}]  \label{ghcs}
$\mcM$ is globally hyperbolic if and only if $\mcM$ admits a
Cauchy surface. If $S$ is a Cauchy surface for $\mcM$ then
$\mcM$ is homeomorphic to $\Bbb R \times S$.
\end{prop}

With regard to the implication that global hyperbolicity
implies the existence of a Cauchy surface,  Geroch, in fact,
proved something substantially stronger.  (We will make some
comments about the converse in Section~\ref{SDoD}.) A {\it time
function} on $\mcM$ is a $C^0$ function $t$ on $\mcM$ such that
$t$ is strictly increasing along every future directed causal
curve. Geroch established the existence of a time function $t$
all of whose level sets $t = t_0$, $t_0 \in \bbR$, are Cauchy
surfaces.  This result can be strengthened to the smooth
category.  By a {\it smooth time function} we mean a smooth
function $t$ with everywhere past pointing timelike gradient.
This implies that $t$ is strictly increasing along all future
directed causal curves, and that its level sets are smooth
spacelike\footnote{A hypersurface is called \emph{spacelike} if
the induced metric is Riemannian; see Section~\ref{Ssub}.}
hypersurfaces.   It has been shown that a globally hyperbolic space-time
admits a smooth time function all of whose levels sets are Cauchy
surfaces \cite{BernalSanchez, Seifert}.  In fact, one obtains  a diffeomorphism
$\mcM \approx \bbR \times S$, where the $\bbR$-factor
corresponds to a smooth time function, such that each slice
$S_t = \{t\}\times S$, $t \in \bbR$, is a  Cauchy surface.

Given a Cauchy surface $S$ to begin with, to simply show that
$\mcM$ is homeomorphic to $\Bbb R \times S$, consider a
complete timelike vector field $Z$ on $\mcM$ and observe that
each integral curve of $Z$, when maximally extended, meets $S$
in a unique point. This leads to the desired homeomorphism.
 (If $S$
is smooth this will be a diffeomorphism.) In a similar vein,
one can show that any two Cauchy surfaces are homeomorphic.
Thus, the topology of a globally hyperbolic space-time is
completely determined by the common topology of its Cauchy
surfaces.

The following result is often useful.
\begin{prop}\label{compact} Let $\mcM$ be a space-time.
\ben
\item If $S$ is a compact acausal $C^0$ hypersurface and
    $\mcM$ is globally hyperbolic then  $S$  must be a Cauchy
    surface for $\mcM$.
\item  If $t$ is a smooth time function on $\mcM$ all of whose level sets are compact,
then  each level set is a Cauchy surface for $\mcM$, and hence $\mcM$ is globally hyperbolic.
\een
\end{prop}

We will comment on the proof shortly, after
Proposition~\ref{dodcauchy}.

\subsubsection{Domains of dependence}\label{SDoD}
The {\it future domain of dependence} of  an acausal set $S$ is the set
$\mcD^+(S)$ consisting of all points $p\in \mcM$ such that
every past inextendible causal curve\footnote{We note that some
authors use past inextendible {\it timelike} curves to define
the future domain of dependence, which results in some small
differences in certain results.} from $p$ meets $S$.
\begin{figure}[t]
\begin{center} {\psfrag{hyp}{\Huge $\!\hyp$}
\psfrag{remove}{\Huge remove}
\psfrag{dpluss}{\Huge $\!\!\!\!\mcD^+(\hyp)$}
\psfrag{dminus}{\Huge $\mcD^-(\hyp)$}
\resizebox{5in}{!}{\includegraphics{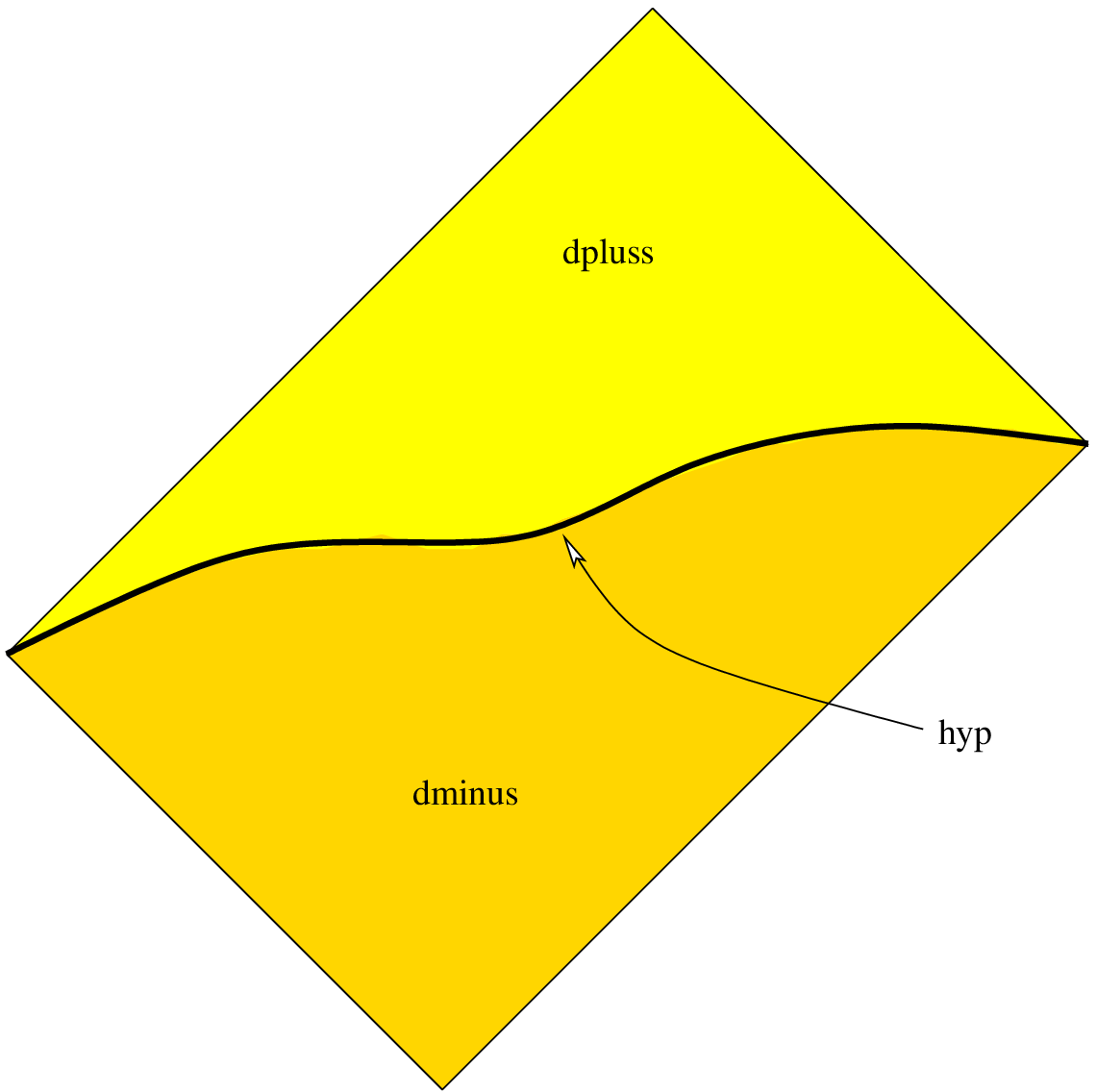}\includegraphics{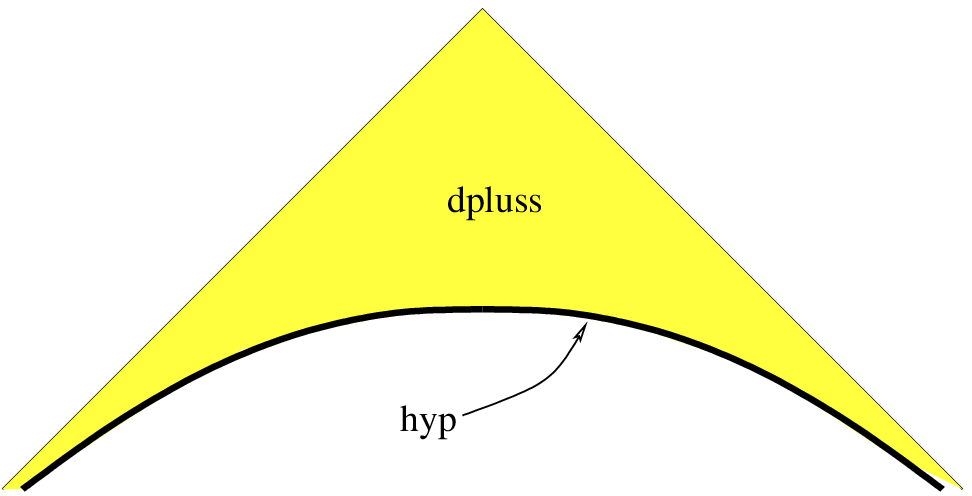}\includegraphics{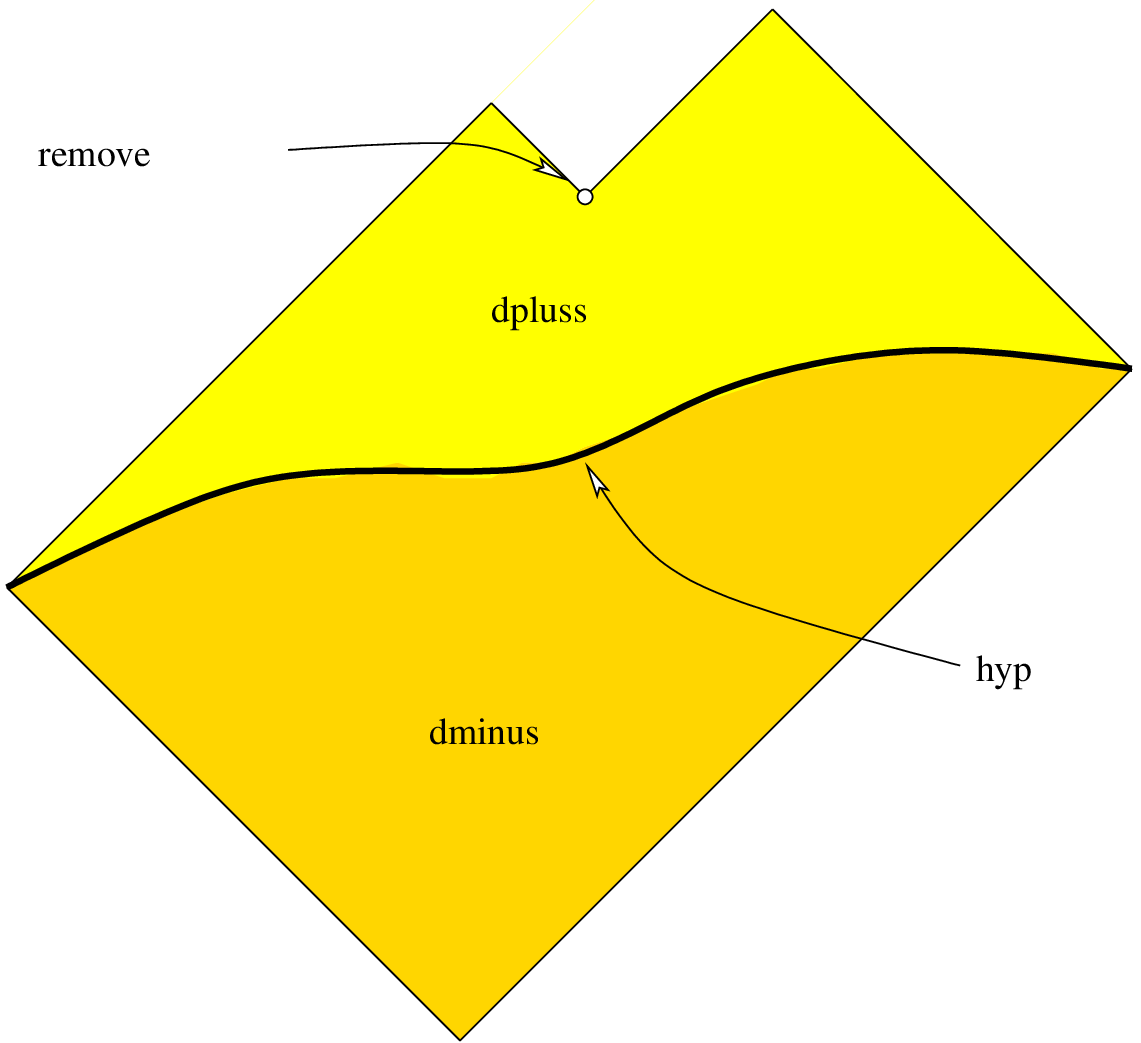}}
}
\caption{Examples of domains of dependence and Cauchy horizons. \label{Fdod}}
\end{center}
\end{figure}%
In physical terms, since information travels  along causal
curves, a point in $\mcD^+(S)$ only receives information  from
$S$.  Thus, in principle, $\mcD^+(S)$ represents the region of
space-time to the future of $S$ that is predictable from $S$.
$\mcH^+(S)$, the {\it future Cauchy horizon} of  $S$,  is
defined to be the future boundary of $\mcD^+(S)$; in precise
terms,  $\mcH^+(S) = \{ p \in \ol{\mcD^+(S)}: I^+(p) \cap
\mcD^+(S) = \emptyset \}$. Physically, $\mcH^+(S)$ is the
future limit of the region of space-time  predictable from $S$.
Some examples of domains of dependence, and Cauchy horizons,
can be found in Figure~\ref{Fdod}.

It follows almost immediately from the definition that $\mcH^+(S)$ is achronal.
In fact,  Cauchy horizons have structural
properties similar to achronal boundaries, as indicated in the following.

\begin{prop}\label{horizon}
Let $S$ be an acausal subset of a space-time $\mcM$. Then
$\mcH^+(S)\setminus \bar{S}$, if nonempty, is an achronal $C^0$
hypersurface of $\mcM$ ruled by null geodesics, called
generators, each of which either is past inextendible in $\mcM$
or has past end point on $\bar{S}$.
\end{prop}

The past domain of  dependence $\mcD^-(S)$ of $S$,  and the
past Cauchy horizon $\mcH^-(S)$ of $S$, are defined in a
time-dual manner. The total domain of dependence $\mcD(S)$ and
the total Cauchy horizon $\mcH(S)$, are defined respectively
as,  $\mcD(S) = \mcD^+(S) \cup \mcD^-(S)$ and $\mcH(S) =
\mcH^+(S) \cup \mcH^-(S)$

Domains of dependence may  be used to characterize Cauchy
surfaces. In fact, it follows easily from the definitions that
an acausal subset $S \subset \mcM$ is a Cauchy surface for
$\mcM$ if and only if $\mcD(S)= \mcM$.  Using the fact that
 $\d \mcD(S) = \mcH(S)$, we obtain the following.

\begin{prop}\label{dodcauchy}
Let $S$ be an acausal subset of a space-time $\mcM$. Then, $S$
is a Cauchy surface for $\mcM$ if and only if $\mcD(S) = \mcM$
if and only if $\mcH(S) = \emptyset$.
\end{prop}

Part 1 of Proposition~\ref{compact} can now be readily proved by
showing, with the aid of Proposition~\ref{horizon}, that
$\mcH(S)=\emptyset$.  Indeed if $\mcH^+(S) \ne \emptyset$ then
there exists a past inextendible null geodesic $\eta \subset
\mcH^+(S)$ with future end point $p$ imprisoned in the compact
set $J^+(S) \cap J^-(p)$ which, as already mentioned, is not
possible in  strongly causal space-times. Part 2 is proved
similarly; compare~\cite{BILY,Galloway:cauchy}.

The following basic result ties domains of dependence to global hyperbolicity.
\begin{prop}\label{dodgh}
Let $S \subset \mcM$ be acausal.
\ben
\item Strong causality holds at each point of ${\rm int}\, \mcD(S)$.
\item Internal compactness holds on ${\rm int}\, \mcD(S)$,
    i.e., for all $p,q \in {\rm int}\, \mcD(S)$, $J^+(p) \cap
    J^-(q)$ is compact. \een
\end{prop}

Propositions \ref{dodcauchy} and \ref{dodgh} immediately imply
that if $S$ is a Cauchy surface for a space-time $\mcM$ then
$\mcM$ is globally hyperbolic, as claimed in  Proposition
\ref{ghcs}.


%



\subsection{Submanifolds}\label{Ssub}
In addition to curves, one may also speak of the causal
character of higher dimensional submanifolds.   Let $V$ be a
smooth submanifold of a space-time $(\mcM,g)$.  For $p \in V$,
we say that the tangent space $T_pV$ is spacelike (resp.
timelike, null) provided $g$ restricted to $T_pV$ is positive
definite (resp., has Lorentzian signature, is degenerate).
Then $V$ is said to be spacelike (resp., timelike, null)
provided each of its tangent spaces is spacelike (resp.,
timelike, null).   Hence if $V$ is spacelike (resp., timelike)
then, with respect to its {\it induced metric}, i.e.,  the
metric $g$ restricted to the tangent spaces of $V$,  $V$ is a
Riemannian  (resp., Lorentzian) manifold.

\section{Stationary black holes}
 \label{Sbh}

{Perhaps the first thing which comes to mind when general
relativity is mentioned are black holes. These} are among  the
most fascinating objects predicted by Einstein's theory of
gravitation.
\bams{%
Although they have been studied for years,\footnote{The
reader is referred to the introduction
  to~\cite{CMSciama} for an excellent concise review of the history of
  the concept of a black hole, and
  to~\cite{Carter:1997im,Israel:bhreview} for a more detailed one.}
they  still attract tremendous attention in the physics and
astrophysics literature. {It is seldom realized that, in
addition to Einstein's gravity, several other} field theories
are known
to possess solutions which exhibit black hole properties,  {amongst which}:%
\footnote{An even longer list of models and submodels can be found
in~\cite{Barcelo:2001ah}, see also~\cite{Novello:2002qg,barcelo-2005-8}.}%
\begin{itemize}
\item The ``dumb holes",  {arising in Euler equations}, which are the sonic counterparts of black
  holes, first discussed by Unruh~\cite{Unruh} {(compare~\cite{Chrusciel:2002mi})}.
 \item The ``optical" ones -- the black-hole-type solutions
     arising in the theory of moving dielectric media, or
     in non-linear
     electrodynamics~\cite{Leonhardt:Piwnicki,Novello:2001fv}.
\end{itemize}
 The
numerical study of black holes has become a science in itself,
see~\cite{Pretorius,Campanelli:2005dd} and references therein.
The evidence for the existence of black holes in our universe
is
growing~\cite{Johnston,Ziolkowski,Miller:2003sc,Peterson,Gillessen:2010ty}.
Reviews of, and further references to, the quantum aspects of
black holes can be found
in~\cite{Brout:1995rd,Wald:LR,Horowitz:1996rn,Padmanabhan:2003gd,AshtekarLewandowski,EmparanReallLR}.

}
In this section we  focus attention on \emph{stationary}
black holes that are solutions of the \emph{vacuum} Einstein
equations with vanishing cosmological constant, with one
exception: the static electro-vacuum  Majumdar--Papapetrou
solutions, an example of physically significant multiple black
holes.  {By definition, a \emph{stationary} space-time is an
asymptotically flat space-time which is invariant under an
action of $\R$ by isometries, such that the associated
generator --- referred to as \emph{Killing
vector} --- is timelike%
\bamsf{%
\footnote{In fact,  in the literature it is always
implicitly assumed that the stationary Killing vector
$\changedX $ is \emph{uniformly timelike} in the asymptotic
region $\Sext$; by this we mean that $\fourg(\changedX
,\changedX )<-\epsilon<0$ for some $\epsilon$ and for all $r$
large enough. This uniformity condition excludes the
possibility of a timelike vector which asymptotes to a null
one. This involves no loss of generality in well behaved
space-times: indeed, uniformity always holds for Killing
vectors which are timelike for all large distances if the
conditions of the positive energy theorem are
met~\cite{ChBeig1,ChMaerten}.}
}
in the asymptotically flat region. These model steady state
solutions. Stationary black holes} are the simplest to
describe, and most mathematical results on black holes, such as
the uniqueness theorems discussed in Section~\ref{sSubh},
concern those. It should, however, be kept in mind that one of
the major open problems in mathematical relativity is the
understanding of the dynamical behavior of black hole
space-times, about which not much is yet known (compare
Section~\ref{sswebhb}).

 \subsection{The Schwarzschild metric}
 \label{sSSm}
  The simplest stationary solutions describing compact
isolated objects are the spherically symmetric ones. According
to Birkhoff's theorem~\cite{Birkhoff23}, any
$(n+1)$-dimensional, $n\ge 3$, spherically symmetric solution
of the vacuum Einstein equations  belongs to the family of
Schwarzschild metrics, parameterized by a mass parameter $m$:
\bea
 \label{Schwarz}
&\fourg=-V^{2}dt^{2}+V^{-2}dr^{2}+r^{2}d\Omega^{2}\;,&
\\ & V^{2}=1-\frac{2m}{r^{n-2}} \;,\quad t\in \R\;,\ r\in
({2m},\infty)\;. & \eeal{schw}
Here $d\Omega^{2}$ denotes the metric of the standard
$(n-1)$-sphere. (This is true \emph{without} assuming stationarity.)

From now on we assume $n=3$, though identical results hold in
higher dimensions.

We will assume
$$m>0\;,
 $$
because $m<0$ leads to metrics which are called ``nakedly
singular"; this deserves a comment. For Schwarzschild metrics
we have
\bel{KretschSchwarz}
 R_{\alpha\beta\gamma\delta}R^{\alpha\beta\gamma\delta} =
 \frac{48m^2}{r^6}
 \;,
\ee
in dimension $3+1$, which shows that the geometry becomes
singular as $r=0$ is approached; this remains true in higher
dimensions. As we shall see shortly, for $m>0$ the singularity
is ``hidden" behind an event horizon, while this is \emph{not}
the case for $m<0$.

One of the first features one notices is that the metric
\eq{Schwarz} is singular as $r=2m
$ is approached. It turns out that this singularity is related
to an unfortunate choice of coordinates (one talks about ``a
coordinate singularity"); the simplest way to see this is to
replace $t$ by a new coordinate $v$ defined as
\bel{vdef} v=t+f(r)\;,\quad f' = \frac 1 {V^2}
 \;,
\ee
leading to
$$
 v=t+r+2m\ln(r-2m)
  \;.
  $$
This brings $\fourg$ to the form
\begin{equation}
\label{Schwarz2} \fourg=-(1-\frac{2m}{r })dv^{2}+2
dvdr+r^{2}d\Omega^{2}\;.
\end{equation} %
We have $\det \fourg = -r^4 \sin^2\theta$, with all
coefficients of $\fourg$  smooth, which shows that $\fourg$ is
a well-defined Lorentzian metric on the set \bel{Schwarz3}v \in
\R\;, \quad r\in (0,\infty)
 \;.
 \ee
More precisely, \eq{Schwarz2}-\eq{Schwarz3} provides an
analytic extension of the original space-time \eq{Schwarz}.

\bams{%
We could have started immediately from the form \eq{Schwarz2}
of $\fourg$, which would have avoided the lengthy discussion of
the coordinate transformation \eq{vdef}. However, the form
\eq{Schwarz} is the more standard one. Furthermore, it makes
the metric $\fourg$ manifestly asymptotically flat (see
Section~\ref{Saf}); this is somewhat less obvious to an
untrained eye in \eq{Schwarz2}.
}

 It is easily seen that the region
$\{r\le 2m\}$ for the metric \eq{Schwarz2} is a \emph{black
hole region}, in the sense that
\bel{bhdefinwords}\mbox{{observers, or signals, can enter this
region, but can never leave it}.}\ee In order to see that,
recall that observers in general relativity always move on
\emph{future directed timelike curves}, that is, curves with
timelike future directed tangent vector. For signals, the
curves are \emph{causal future directed}. Let, then,
$\gamma(s)=(v(s),r(s),\theta(s),\varphi(s))$ be such a timelike
curve; for the metric \eq{Schwarz2} the timelikeness condition
$\fourg(\dot \gamma, \dot \gamma)<0$ reads
$$-(1-\frac{2m}{r})\dot v^{2}+2 \dot v\dot r+r^{2}(\dot \theta^2 +
\sin^2 \theta \dot \varphi^2) <0\;.$$ This implies $$\dot
v\Big(-(1-\frac{2m}{r})\dot v+2 \dot r\Big) <0\;.$$ It follows that
$\dot v$ does not change sign on a timelike curve. The usual choice
of time orientation corresponds to $\dot v>0$ on future directed
curves, leading to $$-(1-\frac{2m}{r})\dot v+2 \dot r <0\;.$$ For
$r\le 2m$ the first term is non-negative, which enforces $\dot r<0$
on all future directed timelike curves in that region.  Thus, $r$ is
a strictly decreasing function along such curves, which implies that
future directed timelike curves can cross the hypersurface
$\{r=2m\}$ only if coming from the region $\{r>2m\}$. This motivates
the name \emph{black hole event horizon} for $\{r=2m, v\in\R\}$. The
same conclusion \eq{bhdefinwords} applies for causal curves: it
suffices to approximate a causal curve by a sequence of timelike
ones.

\begin{figure}[th]
\begin{center}
  \includegraphics[width=.8\textwidth]{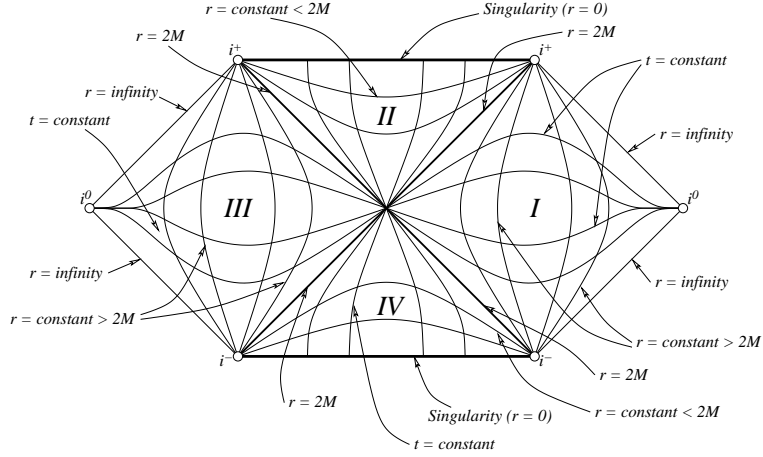}
\end{center}
\caption[FCPd]{The Carter-Penrose
diagram$^{\mbox{\scriptsize\ref{fNi}}}$ for the Kruskal-Szekeres
space-time with mass $M$. There are actually two asymptotically flat
regions, with corresponding event horizons defined with respect to
the second region. Each point in this diagram represents a
two-dimensional sphere, and coordinates are chosen so that
light-cones have slopes plus minus one.} \label{Sfig0}
\end{figure}

The transition from \eq{Schwarz} to \eq{Schwarz2} is not the
end of the story, as further extensions are possible. For the
metric \eq{Schwarz} a maximal analytic extension has been found
independently by Kruskal~\cite{Kruskal},
Szekeres~\cite{Szekeres}, Synge \cite{SyngeSchwarzschild} and
Fronsdal~\cite{Fronsdal};  for some obscure reason neither
Synge nor Fronsdal are almost never mentioned in this context.
This extension is depicted\footnote{\label{fNi}We are grateful
to J.-P.~Nicolas for allowing us to use his figure
from~\cite{NicolasDissMath}.} in Figure~\ref{Sfig0}. The region
$I$ there corresponds to the space-time \eq{Schwarz}, while the
extension just constructed corresponds to the regions $I$ and
$II$.

The Kruskal-Szekeres extension is singled out by being maximal
in the class of vacuum, analytic, simply connected space-times,
with all maximally extended geodesics $\gamma$ either complete,
or with the curvature scalar
$R_{\alpha\beta\gamma\delta}R^{\alpha\beta\gamma\delta}$
diverging along $\gamma$ in finite affine time.

An alternative convenient representation  of the Schwarzschild
metrics, which makes the space-part of $\fourg$ manifestly
conformally flat, is given by
\bel{stschwnd} \fourg= -
\left(\frac{1-m/2|x|^{n-2}}{1+m/2|x|^{n-2}}\right)^2 dt^2+ \left(1 +
\frac{m}{2|x|^{n-2}}\right)^{\frac4{n-2}} \left(\sum_{1=1}^n
(dx^i)^2\right)
 \;.
\ee
%

\subsection{Rotating black holes}
 \label{sSKMP}
Rotating generalizations of the Schwarzschild metrics are given
by the family of \emph{Kerr metrics}, parameterized by a mass
parameter $m$ and an angular momentum parameter $a$. One
explicit coordinate representation of the Kerr metric is
\beal{Kerr2}
 \hat g & =
& -\Big(1-\frac {2mr}{\Sigma }\Big) dv^2 +2 dr dv +\Sigma
d\theta^2 - 2 a \sin^2\theta d\phi dr \\ &&
+\frac{(r^2+a^2)^2-a^2\Delta\sin^2\theta }{\Sigma }
\sin^2\theta d\phi^2 - \frac{4amr\sin^2\theta}{\Sigma } d\phi
dv
 \;,
 \eean
where
\beaa & \Sigma=r^{2}+a^{2}\cos^{2}\theta \;,
%
\qquad  \Delta=r^{2}+a^{2}-2mr
 \;.
& \eeaa
%
Note that \eq{Kerr2}  reduces to the Schwarzschild solution in the representation \eq{Schwarz2} when $a=0$.
The reader is referred to~\cite{BONeillKerr,CarterKerr}
for a thorough analysis. All Kerr metrics satisfying
$$
 m^2\ge  a^2
$$
provide, when appropriately extended,  vacuum space-times
containing a rotating black hole. Higher dimensional analogues
of the Kerr metrics have been constructed by Myers and
Perry~\cite{MyersPerry}.

A fascinating class of black hole solutions of the $4+1$
dimensional stationary vacuum Einstein equations has been found
by Emparan and Reall~\cite{EmparanReall} (see
also~\cite{Emparan:2004wy,EmparanReallReview,PS,CC,CCGP,EmparanReallLR}).
The solutions, called \emph{black rings}, are asymptotically
Minkowskian in spacelike directions, with an event horizon
having $S^1\times S^2$ cross-sections. The ``ring" terminology
refers to the $S^1$ factor in $S^1\times S^2$.

\subsection{Killing horizons}
 \label{sSgn}
 Before continuing
some general notions are in order.  By definition, a \emph{Killing
field} is a vector field the local flow of which preserves the
metric.  Killing vectors are solutions of the over-determined
system of \emph{Killing equations}
\bel{Killingequation}
 \nabla_\a X_\b + \nabla_\b X_\a = 0
 \;.
\ee
One of the features of the metric \eq{Schwarz} is its
\emph{stationarity}, with Killing vector field $X=\partial_t$:
{As already pointed out, a} space-time is called {\em
stationary } if there exists a Killing vector field $X$ which
approaches $\partial_t$ in the asymptotically flat region
(where $r$ goes to $\infty$, see Section~\ref{Saf} for precise
definitions) {\em and} generates a one parameter group of
isometries. A space-time is called {\em static} if it is
stationary and {if the distribution of hyperplanes orthogonal
to the stationary Killing vector $X$ is integrable.}

A space-time is called {\em axisymmetric} if there exists a
Killing vector field $Y$ which generates a one parameter group
of isometries and which behaves like a {\em rotation}: this
property is captured by requiring that all orbits be $2\pi
$--periodic, and that the set $\{Y=0\}$, called the \emph{axis
of rotation}, be non-empty.

Let $X$ be a Killing vector field on $(\mcM,\fourg)$, and
suppose that $\mcM$ contains a null hypersurface (see
Sections~\ref{Ssub} and \ref{sSnullhyp}) $\mcN_0=\mcN_0(X)$
which coincides with a connected component of the set
$$
\mcN(X):= \{p\in \mcM \ |\  g(X_p,X_p)=0\;,\ X_p\ne 0\}
 \;,
$$
with $X$ tangent to $\mcN_0$. Then $\mcN_0$  is called a
\emph{Killing horizon} associated to the Killing vector $X$.
The  simplest example is provided by the ``boost Killing vector
field"
\bel{bLvM} K=z\partial_t+ t\partial_z
 \ee
in four-dimensional Minkowski space-time $\R^{1,3}$:
$\mcN(K)$ has four connected components
$$\mcN(K)_{\epsilon \delta} :=\{ t=\epsilon z\;, \delta t >0\}\;,
\quad \epsilon, \delta \in \{\pm 1\}\;.$$ The closure $\overline{
\mcN(K)}$ of $ \mcN(K)$ is the set $\{|t|=|z|\}$, which is
\emph{not} a manifold, because of the crossing of the null
hyperplanes $\{t=\pm z\}$ at $t=z=0$. Horizons of this type are
referred to as {\em bifurcate Killing horizons}.

A very similar behavior is met in the extended Schwarzschild
space-time: the set $\{r=2m\}$ is a null hypersurface $\mcE$,
the Schwarzschild event horizon. The stationary Killing vector
$X=\partial_t$ extends to a Killing vector $\hat X$ which
becomes tangent to and null on $\mcE$ in the extended
space-time, except at the ``bifurcation sphere" right in the
middle of Figure~\ref{Sfig0}, where $\hat X$ vanishes.

A last noteworthy example in Minkowski space-time $\R^{1,3}$ is
provided by the Killing vector
\bel{nullboost}
 X = y\partial_t + t\partial_y + x \partial_y - y \partial_x = y\partial_t + (t+x)\partial_y   - y \partial_x
 \;.
\ee
Thus, $X$ is the sum of a boost   $y\partial_t + t\partial_y$ and a
rotation  $x \partial_y - y \partial_x$. Note that $X$ vanishes if
and only if
$$
y= t+x=0 \;,
$$
which is a two-dimensional null submanifold of $\R^{1,3}$. The
vanishing set of the Lorentzian length of $X$,
$$
  g(X,X)= (t+x)^2=0
  \;,
$$
is a null hyperplane in $\R^{1,3}$. It follows that, e.g., the
set
$$
 \{t+x=0\;, \ y>0\;, t> 0 \}
$$
is a \emph{Killing horizon} with respect to two different Killing
vectors, the boost Killing vector $x\partial_t+t\partial_x$, and the
Killing vector \eq{nullboost}.

\subsubsection{Surface gravity} \label{ssSg}
The \emph{surface gravity} $\kappa$ of a Killing horizon is
defined by the formula
\begin{equation}
  \label{kdef0}
  d\Big(g(X,X)\Big) =   -2\kappa X^\flat  \ ,
\end{equation}
where $X^\flat$ is the one-form metrically dual to $X$, i.e.
$X^\flat=\fourg_{\mu\nu}\,X^\nu dx^\mu$. Two comments are in
order: First, since $g(X,X)=0$ on $\mcNX$, the differential of
$g(X,X)$ annihilates $T\mcNX$. Now, simple algebra shows that a
one--form annihilating a null hypersurface is proportional to
$g(\ell, \cdot)$, where $\ell$ is any  null vector tangent to
$\mcN$ (those are defined uniquely up to a proportionality
factor, see Section~\ref{sSnullhyp}). We thus obtain that $d
(g(X,X) )$ is proportional to $X^\flat$; whence \eq{kdef0}.
Next, the name ``surface gravity" stems from the following:
using the Killing equations \eq{Killingequation} and \eq{kdef0}
one has
\bel{kapcal}
 X^\mu \nabla_\mu X^\sigma = - X^\mu \nabla^\sigma X_\mu = \kappa X^\sigma
 \;.
\ee
Since the left-hand-side of \eq{kapcal} is the acceleration of
the integral curves of $X$, the equation shows that, in a
certain sense, $\kappa$ measures the gravitational field at the
horizon.

A key property is that {the surface gravity} $\kappa$ is
constant on bifurcate~\cite[p.~59]{KayWald} Killing horizons.
Furthermore, $\kappa$ \cite[Theorem~7.1]{Heusler:book} is
constant for all Killing horizons, whether bifurcate or not, in
space-times satisfying the \emph{dominant energy condition}:
this means that
\bel{DECsts} \mbox{$T_{\mu\nu}X^\mu Y^\nu\ge0$ for causal
 future directed vector fields $X$ and $Y$.}
\ee

As an example, consider the Killing vector $K$ of \eq{bLvM}. We have
$$
d(g(K,K))=d(-z^2+t^2)=2 (-zdz+tdt)
 \;,
  $$
which equals twice $K^\flat$ on $\mcN(K)_{\epsilon \delta}$. On
the other hand, for the Killing vector $X$ of \eq{nullboost} one
obtains
$$
d(g(X,X))=2(t+x) (dt+dx)
 \;,
 $$
which vanishes on each of the Killing horizons $\{t=-x\;, y\ne 0\}$.
This shows  that the same null surface can have zero or non-zero
values of  surface gravity, depending upon which Killing vector has
been chosen to calculate $\kappa$.

The surface gravity of black holes plays an important role in
\emph{black hole thermodynamics}; see~\cite{Brout:1995rd} and
references therein.

A Killing horizon $\mcN_0(X)$ is said to be \emph{degenerate},
or \emph{extreme}, if $\kappa$ vanishes throughout $\mcN_0(X)$;
it is called \emph{non-degenerate} if $\kappa$ has no zeros on
$\mcN_0(X)$. Thus, the Killing horizons
$\mcN(K)_{\epsilon\delta}$ are non-degenerate, while both
Killing horizons of $X$ given by \eq{nullboost} are degenerate.
The Schwarzschild black holes have surface gravity
$$\kappa_m=\frac 1 {2m}
 \;.
$$
So there are no degenerate black holes within the Schwarzschild
family.  Theorem~\ref{Tubhs} below shows that
there are no regular, degenerate, static vacuum black holes at all.

In Kerr space-times  we have $\kappa=0$ if and only if $m=|a|$.
\bams{%
On the other hand, all horizons in the multi-black hole
Majumdar-Papapetrou solutions are degenerate.

\subsection{The orbit-space geometry near Killing horizons}
 \label{sSnhsg}
Consider a space-time $(\mcM,\fourg)$ with a Killing vector
field $X$. On any set $\mcU$ on which $X$ is timelike we can
introduce coordinates in which $X=\partial_t$, and the metric
may be written as
\bel{strep2}
 \fourg = -V^2(dt+ \theta_i dx^i )^2 +  \threeg_{ij}dx^i dx^j
 \;, \quad \partial _t V = \partial_t \theta_i = \partial_t
 \threeg_{ij} = 0
 \;.
\ee
where $\threeg= \threeg_{ij}dx^idx^j$ has Riemannian signature.
The metric $\threeg$ is often referred to as the
\emph{orbit-space metric}.

Let $M$ be a spacelike hypersurface in $\mcM$; then \eq{strep2}
defines a Riemannian metric $\threeg$ on $\hyp\cap \mcU$.
Assume that $X$ is timelike on a one-sided neighborhood $\mcU$
of a Killing horizon $\mcN_0(X)$, and suppose that $\hyp\cap
\mcU$ has a boundary component $S$ which forms a compact
cross-section of $\mcN_0(X)$,  see Figure~\ref{fnearh}. The
vanishing, or not, of the surface gravity has a deep impact on
the geometry of $\threeg$ near $\mcN_0(X)$~\cite{Chstatic}:
\begin{figure}[t]
\begin{center} {
 \psfrag{NOX}{\Huge $\mcN_0(X)$}
\psfrag{S}{\Huge $S$ }
\psfrag{MCAPU}{\Huge $\hyp\cap \mcU$ }
\psfrag{M}{\Huge $\hyp$ }
\resizebox{4in}{!}{\includegraphics{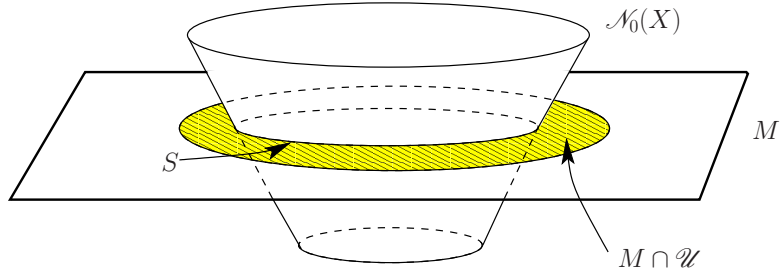}} } \caption{A space-like hypersurface
$\hyp$ intersecting a Killing horizon $\mcN_0(X)$ in a compact cross-section $S$. \label{fnearh}}
\end{center}
\end{figure}

\begin{enumerate}
\item Every differentiable such $S$,  included in a $C^2$
    \emph{degenerate} Killing horizon $\mcN_0(X)$,
    corresponds to a \emph{complete} asymptotic end of
    $(\hyp\cap \mcU,\threeg)$. See Figure~\ref{Fnearh2}.%
\footnote{We are grateful to C.~Williams for providing the
figure.\label{fCW}}
\begin{figure}[t]
\begin{center} {
 \psfrag{(H=0)}{\huge ($\kappa = 0$)}
 \psfrag{(H\\neq0)}{\huge ($\kappa \ne 0$)}
 \psfrag{totally geodesic boundary}{\huge totally geodesic boundary}
\psfrag{infinite cylinder}{\huge Infinite cylinder}
\resizebox{3.5in}{!}{\includegraphics{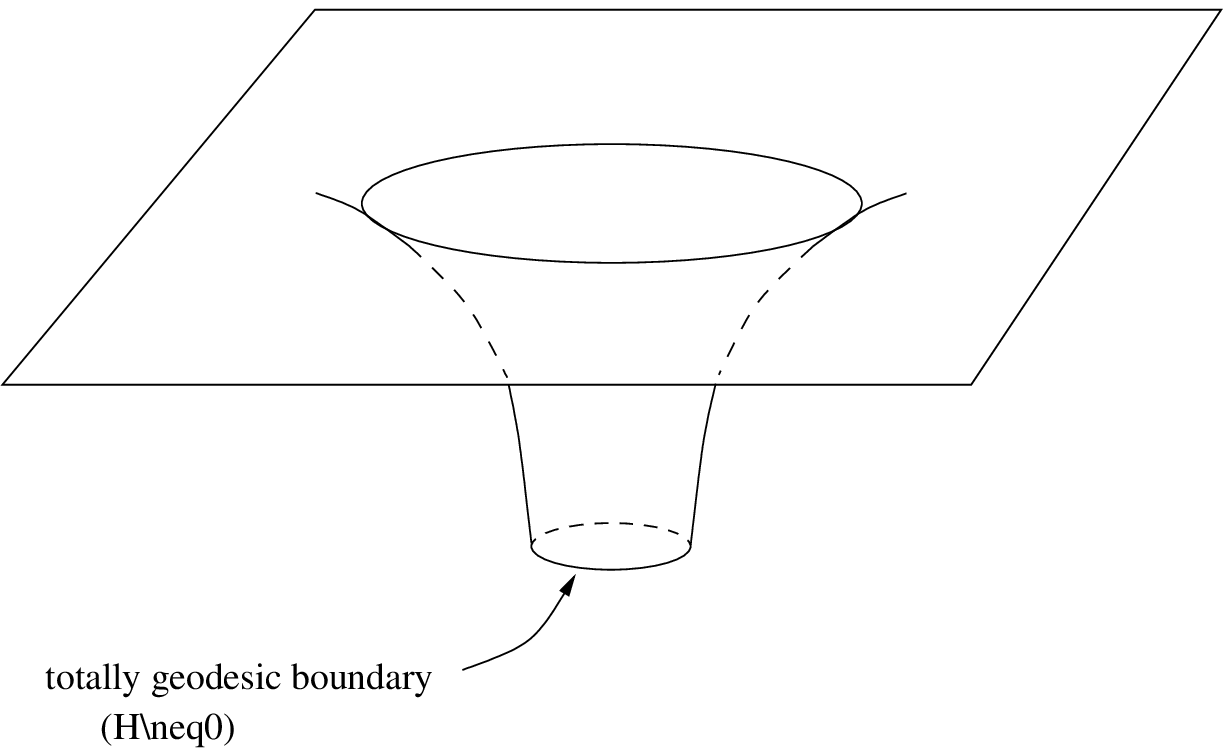}\includegraphics{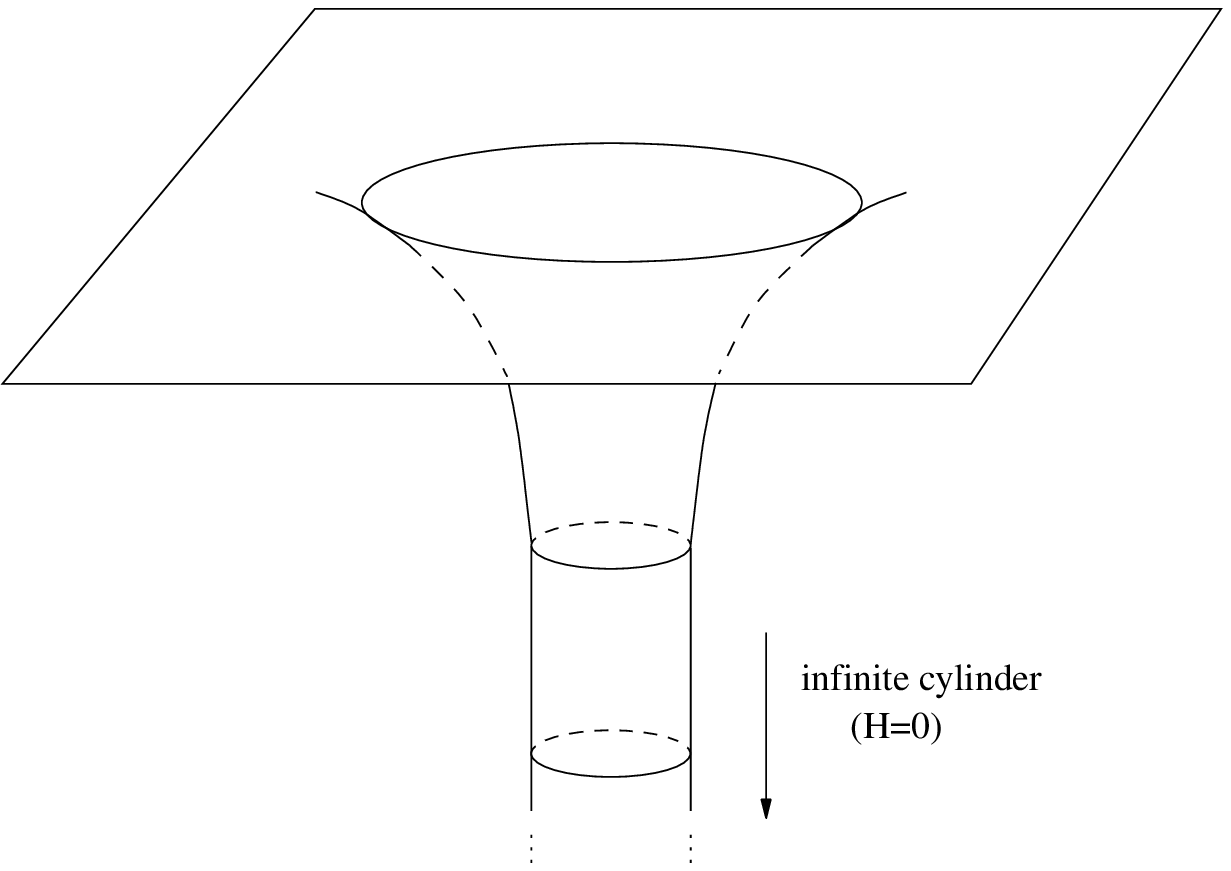}}
}
\caption{The general features of the geometry of the orbit-space metric on a spacelike hypersurface intersecting a non-degenerate (left) and degenerate (right)
Killing horizon, near the intersection, visualized by a co-dimension one embedding in Euclidean space.
\label{Fnearh2}}
\end{center}
\end{figure}

This remains valid for stationary and axi-symmetric
  four-dimensional configurations without the hypothesis
  that $X$ is timelike near the horizon~\cite{ChNguyen}.

  \item Every such $S$  included in  a smooth Killing
      horizon $\mcN_0(X)$ on which
  $$\kappa> 0
   \;,
   $$
  corresponds to a totally geodesic boundary of $(
  {\hyp}\cap\mcU,\threeg)$, with $\threeg$ being smooth
  up--to--boundary at $S$. Moreover
  \begin{enumerate}
  \item a doubling of $(
      \overline{\hyp}\cap\mcU,\threeg)$ across $S$
      leads to a smooth metric on the doubled manifold,
  \item with $\sqrt{-\fourg(X,X)}$ extending smoothly
      to $-\sqrt{-\fourg(X,X)}$ across $S$.
  \end{enumerate}
\end{enumerate}

In the Majumdar-Papapetrou solutions of Section~\ref{sSmbhs},
the orbit-space metric $\threeg$ as in \eq{strep2}  asymptotes
to the usual metric on a round cylinder
as the event horizon is approached.%
\levoca{show this}
One is therefore tempted to think of degenerate event horizons
as corresponding to \emph{asymptotically cylindrical ends} of
$(\hyp,\threeg)$.

}

\subsubsection{Average surface gravity}

Following~\cite{VinceJimcompactCauchyCMP}, near a smooth null
hypersurface one can introduce \emph{Gaussian null
coordinates}, in which the metric takes the form
\bel{GNC1} \fourg=r \varphi dv^2 + 2dv dr + 2r \myh_a dx^a dv +
h_{ab}dx^a dx^b\;. \ee
The hypersurface is given by the equation $\{r=0\}$. Let $S$ be
any smooth compact cross-section of the horizon; then the
\emph{average surface gravity} $\langle \kappa\rangle_S$ is
defined as
\bel{asg}
 \langle \kappa\rangle_S=-\frac 1 {|S|}\int_S \varphi d\mu_h
 \;,
\ee
where $d\mu_h$ is the measure induced by the metric $h$ on $S$,
and $|S|$ is the volume of $S$.  We emphasize that this is
defined regardless of whether or not the stationary Killing
vector is tangent to the null generators of the hypersurface;
on the other hand, $\langle \kappa \rangle_S$ coincides with
$\kappa$ when $\kappa$ is constant and the Killing vector
equals $\partial_v$.

\bams{%
\subsection{Near-horizon geometry}
 \label{ssnhg}
On a degenerate Killing horizon the surface gravity vanishes,
so that the function $\varphi$ in \eq{GNC1} can itself be
written as $rA$, for some smooth function $A$. The vacuum
Einstein equations imply
(see~\cite[eq.~(2.9)]{VinceJimcompactCauchyCMP} in dimension
four and~\cite[eq.~(5.9)]{LP2} in higher dimensions)
\bel{vEe} \zR_{ab} = \frac 12 \mzyh_{a}\mzyh_{b} -  \zD_{(a
}\mzyh_{b)}
 \;,
\ee
where $\zR_{ab}$ is the Ricci tensor of $\mzh_{ab}:=h_{ab}|_{r=0}$,
and $\zD$ is the covariant derivative thereof, while $\mzyh_a:=h_a|_{r=0}$. The Einstein equations also determine
$\zA:=A|_{r=0}$ uniquely in terms of $\mzh_a$ and $\mzh_{ab}$:
\bel{Asol}
\zA = \frac{1}{2} \mzh^{ab} \left( \mzyh_a \mzyh_b -  \zD_a \mzyh_b  \right)
\ee
(this equation
follows again e.g. from~\cite[eq.~(2.9)]{VinceJimcompactCauchyCMP} in
dimension four, and can be checked by a calculation in all higher
dimensions). Equations \eq{vEe} have only been understood under
the supplementary  assumptions of staticity~\cite{CRT},%
\footnote{Some partial results  with a non-zero cosmological
constant have also been proved in~\cite{CRT}.}
or axial symmetry in space-time dimension
four~\cite{Hajicek3Remarks,LP1}:

\begin{Theorem}[~\cite{CRT}]
 \label{TCRT} Let  the space-time dimension be $n+1$, $n\ge 3$, suppose that
 a degenerate Killing horizon $\mcN$ has a compact cross-section,
and   that $\mzyh_a=\partial_a \lambda$ for some function
$\lambda$ (which is necessarily the case in vacuum static
space-times). Then  \eq{vEe} implies $\mzyh_a\equiv0$, so that
$\mzh_{ab}$ is Ricci-flat.
\end{Theorem}

\begin{Theorem}[~\cite{Hajicek3Remarks,LP1}]
 \label{TLP}
In space-time dimension four and in vacuum, suppose that  a
degenerate Killing horizon $\mcN$ has a spherical
cross-section, and that $(\mcM,\fourg)$ admits a second Killing
vector field with periodic orbits. For every connected
component $\mcN_0$ of $\mcN$ there exists a map $\psi$ from a
neighbourhood of $\mcN_0$ into a degenerate Kerr space-time
which preserves $\mzyh_a$, $\mzh_{ab}$ and $\zA$.
\end{Theorem}

It would be of interest to classify solutions of \eq{vEe} in a
useful manner, in all dimensions, without any restrictive
conditions.\opp{nearHorgeom} Note that the proof that all
two-dimensional solutions of \eq{vEe} are axi-symmetric would
remove the ``mean-non-degenerate or rotating" restriction from
Theorem~\ref{Tubh} below.

Theorem~\ref{TLP} plays a  key role in the proof of uniqueness
of stationary axi-symmetric degenerate vacuum black
holes~\cite{ChNguyen}. Higher-dimensional generalisations can
be found
in~\cite{FiguerasLucietti,HollandsDegenerate,KunduriLucietti,KunduriLucietti2}.

In the four-dimensional static case, Theorem~\ref{TCRT}
enforces toroidal topology of cross-sections of $\mcN$, with a
flat $\mzh_{ab}$. On the other hand, in the four-dimensional
axi-symmetric case, Theorem~\ref{TLP} guarantees that the
geometry tends to a Kerr one, up to second order errors, when
the horizon is approached. So, in the degenerate case, the
vacuum equations impose strong restrictions on the near-horizon
geometry. This is not the case any more for non-degenerate
horizons, at least in the analytic setting.
%
%
}

\subsection{Asymptotically flat metrics}
 \label{Saf}
  \renewcommand{\changedX}{X}
In relativity one often needs to consider initial data on
non-compact manifolds, with natural restrictions on the
asymptotic geometry. The most commonly studied such examples
are \emph{asymptotically flat manifolds}, which model isolated
gravitational systems. Now, there exist several ways of
defining asymptotic flatness, all of them roughly equivalent in
vacuum. We will adapt a Cauchy data point of view, as it
appears to be the least restrictive; the discussion here will
also be relevant for Section~\ref{Sivp}.

So, a space-time $(\mcM,\fourg)$ will be said to possess an \emph{asymptotically flat end} if $\mcM$ contains a spacelike hypersurface $\Mtext$  diffeomorphic to $\R^n\setminus B(R)$, where $B(R)$
is a coordinate ball of radius $R$. An end  comes thus equipped with a set of Euclidean
coordinates $\{x^i, i=1,\ldots,n\}$, and one sets $r=|x|:=\left(\sum_{i=1}^n (x^i)^2\right)^{1/2}$. One then assumes that there exists a constant $\alpha>0$ such that, in local coordinates
on $\Mtext$ obtained from $\R^n\setminus B(R)$, the metric $\threeg$ induced by $\fourg$ on $\Mtext$, and the second fundamental form $K$ of $\Mtext$ {(compare \eq{Kdef} below)}, satisfy
the fall-off conditions, for some $k > 1$,
\beal{falloff1}
 & \threeg_{ij}-\delta_{ij}=O_k(r^{-\alpha})\;,  \qquad
  K_{ij}=O_{k-1}(r^{-1-\alpha})\;,
\eea
where we write $f=O_k(r^{\beta})$ if $f$ satisfies
\bel{okdef}
  \partial_{k_1}\ldots\partial_{k_\ell}
f=O(r^{\beta-\ell})\;, \quad 0\le \ell \le k
 \;.
\ee
In applications one needs $(\threeg,K)$ to lie in certain weighted H\"older or
Sobolev space defined on $\hyp$, with the former better suited for the
treatment of the evolution as discussed  in Section~\ref{SE}.%
\footnote{The analysis of elliptic operators
 such as the Laplacian on weighted Sobolev spaces was  initiated by Nirenberg and Walker~\cite{NirenbergWalker}
 (see also~\cite{McOwen79,McOwen80,McOwen80fred,Lockhart,
 CBChristodoulou, LockhartMcOwenActa-cor, LockhartMcOwenActa,
 LockhartMcOwenPisa, Bartnik86} as well as~\cite{YCB:GRbook}).
A readable treatment of analysis on weighted spaces (not  focusing on relativity) can be found
in~\cite{PacardRiviere}.}

\subsection{Asymptotically flat stationary metrics}
 \label{Safsm}

For simplicity we assume that the space-time is vacuum, though similar results hold in general under appropriate
conditions on matter fields, see~\cite{ChMaerten,ChBeigKIDs} and references therein.

Along any spacelike hypersurface $\hyp$, a Killing vector field $\changedX$ of $(\mcM,\fourg)$ can be decomposed as
$$
 \changedX = N n + Y \;,
$$
where $Y$ is tangent to $\hyp$, and $n$ is the unit future-directed normal to $\hyp$. The fields $N$ and $Y$ are called
``Killing initial data", or KID for short. The vacuum field equations, together with the Killing equations, imply the following set of
equations on $\hyp$
\begin{eqnarray}
\label{K.13}
&
 D_i Y_j + D_j Y_i = 2 N K_{ij}\;,
 &
 \\
 &
 R_{ij}(\threeg) +
K^k{_k}
K_{ij} - 2 K_{ik} K^k{}_j    - N^{-1}(\mcL_Y K_{ij} + D_i
D_j N  )=0 \;, & \label{K.15}
\end{eqnarray}
where $
 R_{ij}(\threeg) $ is the Ricci tensor of $\threeg$. These equations play an important role in the gluing constructions described in Section~\ref{ssagt}.

Under the boundary conditions \eq{falloff1}, an analysis of
these equations provides detailed information about the
asymptotic behavior of $(N,Y)$. In particular one can prove
that if the asymptotic region $\Sext$ is contained in a
hypersurface $\hyp$ satisfying the requirements of the positive
energy theorem (see Section~\ref{sSPMT}), and if $\changedX$ is
timelike along $\Sext$, then $(N,Y^i)\to_{r\to\infty}
(A^0,A^i)$, where the $A^\mu$'s are constants satisfying
$(A^0)^2>\sum_i( A^i)^2$~\cite{ChBeig1,ChMaerten}. Further, in
the coordinates of \eq{falloff1},
\beal{foff} 
 & 
  Y^{i } -A^i =O_k(r^{-\alpha})\;, \quad N-A^0= O_k(r^{-\alpha})
 \;.
\eeal{foff2}
As discussed in more detail
in~\cite{ChBeig3}, in $\threeg$-harmonic coordinates, and in e.g. a
maximal (i.e., mean curvature zero) time-slicing, the vacuum equations for $\fourg$ form a
quasi-linear elliptic system with diagonal principal part, with
principal symbol identical to that of the scalar Laplace operator. It can be shown that, in this ``gauge", all metric functions have a full
asymptotic expansion in terms of powers of $\ln r$ and inverse
powers of $r$. In the new coordinates we can in fact take
\bel{decrate}
 \alpha= {n-2}
 \;.
 \ee
By inspection of the equations one can further infer that the
leading order corrections in the metric can be written in the
Schwarzschild form \eq{stschwnd}.


\subsection{Domains of outer communications, event horizons}
 \label{sSdoc}
\begin{figure}[t]
\begin{center} { \psfrag{Mext}{\Large$\,\Mext$}
\psfrag{H}{ } \psfrag{B}{ }
\psfrag{pasthorizon}{\Large $\!\!\!\!\!{\partial I^+(\Mext)}$ }
 \psfrag{pSigma}{$\!\!\pohyp\qquad\phantom{xxxxxx}$}
\psfrag{Sigma}{\Large $\!\Sext$}
 \psfrag{toto}{\Large$\!\!\!\!\!\!\!\!\!\!\!\!\!\!\!\!\! I^-(\Mext)$}
 \psfrag{S}{}
 \psfrag{future}{\Large $\!\!\!\!\!{I^+(\Mext)}$}
\psfrag{H'}{ } \psfrag{W}{$\mathcal{W}$}
\psfrag{scriplus} {} 
\psfrag{scriminus} {} 
 \psfrag{i0}{}
\psfrag{i-}{ } \psfrag{i+}{}
 \psfrag{E+}{\Large $\!\!\!\!\!{\partial I^-(\Mext)}$}
\resizebox{2.3in}{!}{\includegraphics{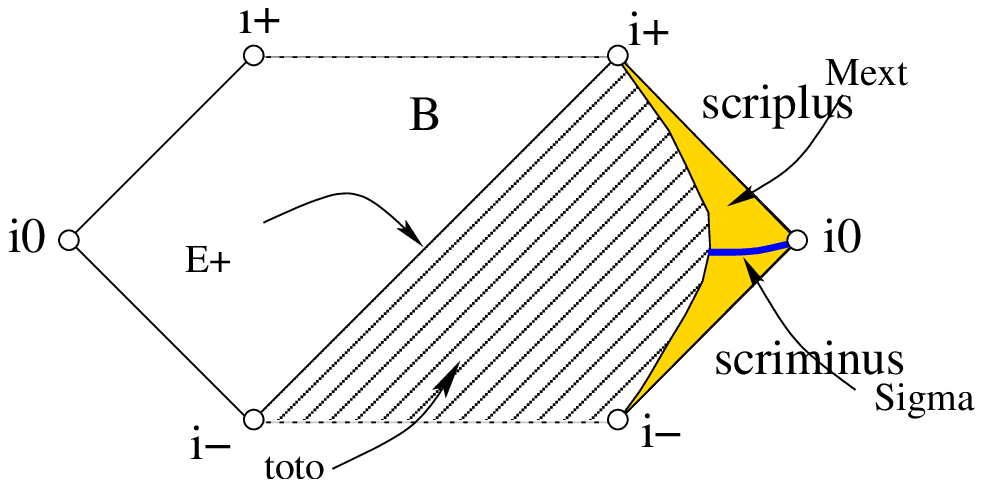}}
\resizebox{2.3in}{!}{\includegraphics{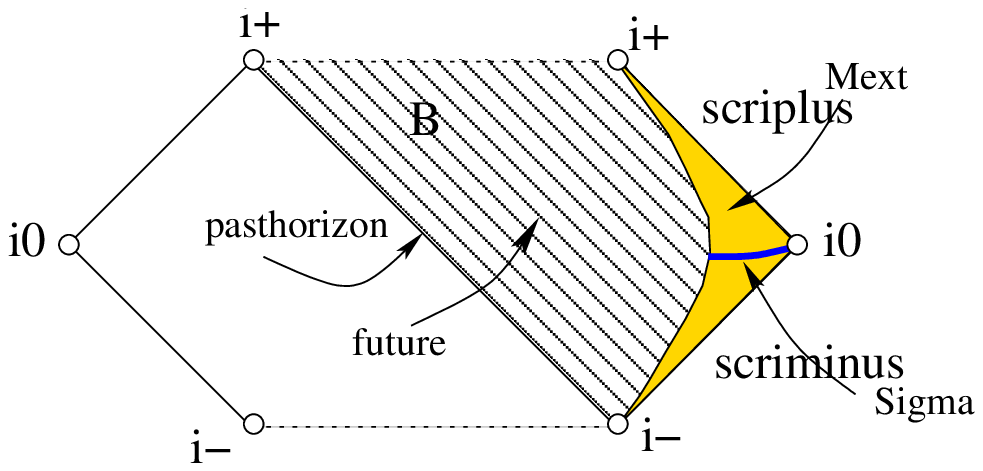}}
}
\caption{$\Sext$, $\Mext$, together with the future and  the past of $\Mext$. One has $\Mext\subset I^\pm(\Mext)$, even
though this is not immediately apparent from the figure.
The domain of outer communications is the intersection $ I^+(\Mext)\cap I^-(\Mext)$, compare Figure~\ref{fregu}.
\label{FPast}}
\end{center}
\end{figure}

A key notion in the theory of asymptotically flat black holes is that of the \emph{domain
of outer communications}, defined for stationary space-times as follows:
For $t\in {\R }$ let $\phi_t[\changedX ]:\mcM\to \mcM$ denote the
one-parameter group of diffeomorphisms generated by $\changedX $; we will
write $\phi_t$ for $\phi_t[\changedX ]$ whenever ambiguities are unlikely to occur.
Let $\Sext$ be as in Section~\ref{Saf}, and assume that $X$ is timelike along $\Sext$.
The exterior region $\Mext$ and the \emph{domain of outer communications} $\doc$ are  then defined as%
\footnote{See Section~\ref{sspaf} for the definition of $I^\pm(\Omega)$.}
\bel{docdef}
{\Mext}:=\cup_t \phi_t(
 \Sext )
 \;, \qquad
 \doc = I^+(\Mext)\cap  I^-(\Mext)
 \;.
\ee
The \emph{black hole region} $\mcB$ and the \emph{black hole event horizon}
$\mcH^+$ are defined  as (see Figures~\ref{FPast} and \ref{fregu})%
\bel{hdefeh}
 \mcB= \mcM\setminus I^-(\Mext)\;,\quad \mcH^+=\partial \mcB
 \;.
\ee
The \emph{white hole region} $\mcW$ and the \emph{white hole
event horizon} $\mcH^-$ are defined  as above after changing
time orientation:
$$
 \mcW= \mcM\setminus I^+(\Mext)\;,\quad \mcH^-=\partial \mcW
 \;.
$$
It follows that the boundaries of  $\doc $ are included in the {event horizons}. We set
\bel{epm}
  \mcE^\pm = \partial \doc \cap I^\pm (\Mext)
 \;, \qquad \mcE=\mcE^+\cup \mcE^- \;.
\ee
The sets $\mcE^\pm$ are achronal boundaries and so, as
mentioned in Section~\ref{Scausal}, they are ruled by null
geodesics, called \emph{generators}.

In general, each asymptotically flat end of $\mcM$ determines a
different domain of outer communications. Although there is
considerable freedom in choosing the asymptotic region $\Sext$
giving rise to a particular end,  it can be shown that $I^\pm
(\Mext)$, and hence $\doc$, $\mcH^\pm$ and $\mcE^\pm$, are
independent of the choice of $\Sext$.

\subsection{Uniqueness theorems}
\label{sSubh}

It is widely expected that the Kerr metrics provide the only
stationary, regular, vacuum, four-dimensional black holes. In
spite of many works on the subject (see~\cite{RobinsonKerr,%
CarterlesHouches,Heusler:book,Weinstein1,%
Neugebauer:2003qe,ChCo,AIK,IonescuKlainerman1,HennigNeugebauer}
and references therein),  the question is far from being
settled. \bams{See~\cite{JP} and references therein for
proposals how to test this idea with astrophysical
observations.}

To describe the current state of affairs, some terminology is
needed. A Killing vector $\changedX $ is said to be complete if
its orbits are complete, i.e., for every $p\in \mcM$ the orbit
$\phi_t[\changedX ](p)$ of $\changedX $ is defined for all
$t\in \R$. $\changedX $ is called \emph{stationary} if it is
timelike at large distances in the asymptotically flat region.

\begin{figure}[t]
\begin{center} { \psfrag{Mext}{$\phantom{x,}\Mext$}
\psfrag{H}{ } \psfrag{B}{ }
\psfrag{H}{ }
 \psfrag{pSigma}{$\!\!\pohyp\qquad\phantom{xxxxxx}$}
\psfrag{Sigma}{ $\hyp$ }
 \psfrag{toto}{$\!\!\!\!\!\!\!\!\!\!\doc$}
 \psfrag{S}{}
\psfrag{H'}{ } \psfrag{W}{$\mathcal{W}$}
\psfrag{scriplus} {} 
\psfrag{scriminus} {} 
 \psfrag{i0}{}
\psfrag{i-}{ } \psfrag{i+}{}
 \psfrag{E+}{ $\phantom{.}{\mycal E}^+$}
\resizebox{3in}{!}
{\includegraphics{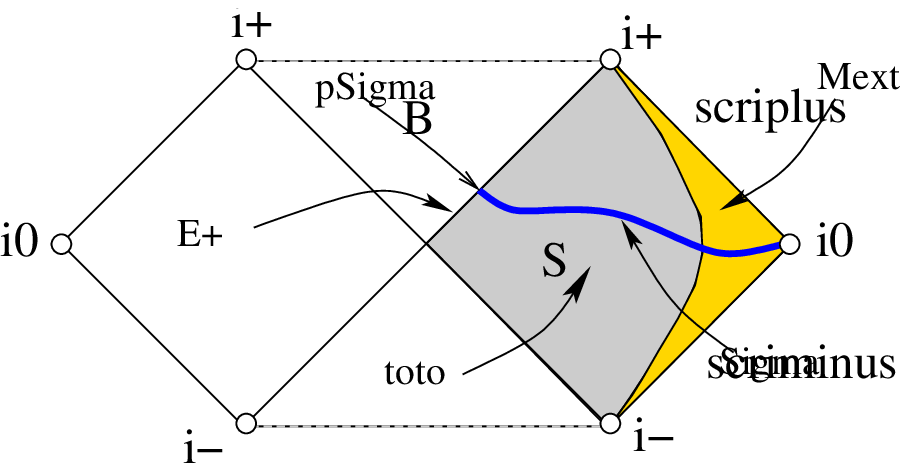}}
}
\caption{The hypersurface $\hyp$ from the definition of \regular ity. To avoid ambiguities, we note that $\Mext$ is a subset of $\doc$.
\label{fregu}}
\end{center}
\end{figure}

A key definition for the uniqueness theory is the following:

\begin{Definition}
 \label{Dmain}
{
Let $(\mcM,\fourg)$ be a space-time containing an asymptotically flat
end $\Sext$, and let  $\changedX $ be a stationary Killing vector field  on $\mcM$.
We will say that $(\mcM,\fourg,\changedX)$ is $\mbox{\rm {\regular}}$%
\index{$\mbox{\rm {\regular}}$}
if $\changedX $ is complete, if the domain of outer communications
$\doc$ is globally hyperbolic, and if $\doc$ contains a spacelike,
connected, acausal hypersurface $\hyp\supset\Sext $,%
\index{$\hyp$}
the  closure $\overline{\hyp} $ of which is a topological manifold with boundary,
consisting of  the union of a compact set and of a finite number of {asymptotically flat
ends, such that the boundary $ \pohyp:= \ohyp \setminus \hyp$ satisfies
\bel{subs}
\pohyp \subset \mcE^+
 \ee
(see \eq{epm})}, with $\pohyp$ meeting every generator of
$\mcE^+$ precisely once; see Figure~\ref{fregu}. }
\end{Definition}

Some comments might be helpful. First one requires completeness
of the orbits of the stationary Killing vector because one
needs an action of $\R$ on $\mcM$ by isometries. Next, one
requires global hyperbolicity of the domain of outer
communications to guarantee its simple connectedness, and to
avoid causality violations. Further, the existence of a
well-behaved spacelike hypersurface gives reasonable control of
the geometry of $\doc$, and is a prerequisite to any elliptic
PDEs analysis, as is extensively needed for the problem at
hand. The existence of compact cross-sections of the future
event horizon $\mcE^+$ prevents singularities on the future
part of the boundary of the domain of outer communications, and
eventually guarantees the smoothness of that boundary.

The event horizon in a stationary space-time will be said to be
\emph{rotating} if the stationary Killing vector is \emph{not}
tangent to the generators of the horizon; it will be said
\emph{mean non-degenerate} if $<\kappa>_{\pohyp}\ne 0$ (compare
\eq{asg}). The proof of the following can be found
in~\cite{ChCo} in the mean non-degenerate case, and
in~\cite{ChNguyen} in the degenerate rotating one:

\begin{theorem}
 \label{Tubh}
Let $(\mcM,\fourg)$ be  \aregular, vacuum,  analytic,
asymptotically flat, four-dimensional stationary space-time. If
$\mcE^+$ is connected and either \emph{mean non-degenerate} or
\emph{rotating}, then $\doc$ is isometric to the domain of
outer communications of a Kerr space-time.
\end{theorem}

Theorem~\ref{Tubh} finds its roots in work by Carter and
Robinson~\cite{RobinsonKerr,CarterlesHouches}, with further key
steps due to Hawking~\cite{Ha1} and Sudarsky and
Wald~\cite{Sudarsky:wald}. It should be emphasized that the
hypotheses of connectedness, analyticity, and non-degeneracy in
the case of non-rotating configurations,  are highly
unsatisfactory, and one believes that they are not needed for
the conclusion.

 Recent progress on the connectedness question has been done
by Hennig and Neugebauer~\cite{HennigNeugebauer}, who excluded
two-component configurations under a non-degeneracy condition
whose meaning remains to be explored; see
also~\cite{Li:Tian2,Weinstein:trans} for previous results.
\bamsgreen{%
The approach in~\cite{HennigNeugebauer} is based on an
observation of Neugebauer and Meinel~\cite{neugebauer:meinel}
that non-singular two-component configurations, if they exist,
must belong to the
Neugebauer-Kramer~\cite{KramerNeugebauerDoubleKerr} family of
solutions. The Neugebauer-Kramer metrics are explicitly known,
but they contain several free parameters, and the metric
functions are too complicated to allow a direct analysis of
global properties. Hennig and Neugebauer exclude most of the
solutions by verifying that they have negative total ADM mass.
For the remaining ones, the authors use a non-degeneracy
condition introduced by Booth and
Fairhurst~\cite{BoothFairhurst}: a black-hole is said to be
\emph{sub-extremal} if any neighborhood of the event horizon
contains trapped surfaces. The key of the analysis is a
beautiful area inequality of Henning, Ansorg, and
Cederbaum~\cite{HennigAnsorgCederbaum}, that on every
sub-extremal component of the horizon  it holds that $8 \pi |
J| < A$, where $J$ is the Komar angular-momentum and $A$ the
area of a section. Hennig and Neugebauer show that all
remaining candidate solutions violate this inequality; this is
their precise non-existence statement. To tie this with the
usual theory one would need to show that non-degeneracy in the
usual sense implies the \emph{sub-extremality} condition. (The
conditions are equivalent for Kerr
solutions~\cite{BoothFairhurst}).
}

\opp{bhanalyticity}%
The analyticity restriction has been removed by Alexakis,
Ionescu and Klainerman in~\cite{AIK} for near-Kerrian
configurations, but the general case remains open.
\bamsgreen{%
The key to the approach in~\cite{AIK} is a unique continuation
theorem near bifurcate Killing horizons proved
in~\cite{IonescuKlainerman1}, which implies the existence of a
second Killing vector field, say $Y$, in a neighbourhood of the
horizon. One then needs to prove that $Y$ extends to the whole
domain of outer communications. This requires another unique
continuation theorem~\cite{IonescuKlainerman2}, which requires
specific convexity conditions, and which has so far only been
shown to apply to near-Kerrian configurations.
}

Partial results concerning uniqueness of higher dimensional
black holes have been obtained by Hollands and
Yazadjiev~\cite{HY,HY2,HY3},
compare~\cite{Harmark,HarmarkOlesen,IdaMorisawa,BMG}.\opp{bhhigher}

\bams{%
The  proof  of Theorem~\ref{Tubh} can be outlined as follows:
First, the event horizon in a smooth or analytic space-time is
a priori only a Lipschitz surface, so the starting point of the
analysis is provided by a result in~\cite{ChDGH}, that event
horizons in regular stationary black hole space-times are as
differentiable as the differentiability of the metric allows.
One then shows~\cite{Ha1}%
\footnote{Compare~\cite{FRW}; the result, proved by Hawking in
space-dimension $n=3$~\cite{Ha1,HE}, has been generalised to
$n\ge 4$  in~\cite{HIW,VinceJimHigh,HICMP}.}
that
  \emph{either}
\begin{itemize}
\item[a)] the stationary Killing vector is tangent to the generators of the event horizon, \emph{or}
    \item[b)] there exists a second Killing vector defined
        near the event horizon.
        \end{itemize}

In  case a) one shows that the domain of outer communications
contains a maximal (mean curvature zero) spacelike
hypersurface~\cite{ChWald1}; to be able to use the result from
that reference it is useful, first, to extend $\doc$ using the
construction in~\cite{RaczWald2}. The construction
in~\cite{ChWald1} assumes a non-degenerate horizon; a proof of
existence of a maximal hypersurface in the degenerate case is
the only element needed to remove the ``mean-non-degenerate or
rotating" restriction of the theorem. (Alternatively, this
restriction could be removed by showing that all solutions of
the near-horizon geometry equations \eq{vEe} are
axi-symmetric.)
 The existence of the maximal surface allows one to establish
staticity~\cite{Sudarsky:wald}, and one concludes using
Theorem~\ref{Tubhs} below.

In case b), analyticity and simple connectedness
imply~\cite{Ch:rigidity} that the isometry group of
$(\mcM,\fourg)$ contains a $U(1)$ factor, with non-empty axis
of rotation. A delicate argument, which finds its roots in the
work of Carter~\cite{CarterlesHouches}, proves that the area
function
$$
 W:=-\det (\fourg(K_a,K_b))\;,\qquad a,b=1,2
$$
where $K_a$ are the stationary and the periodic Killing vector,
is strictly positive on the domain of outer communications.
(This step has been recently simplified in~\cite{ChGstatic}.)
Classical results on group actions on simply connected
manifolds~\cite{Raymond,Orlik} show that the domain of outer
communications is diffeomorphic to $\R\times (\R^3\setminus
\overline{B(1)})$, with the action of the isometry group by
translations in the first factor, and by rotations around an
axis in $\R^3$. The uniformization theorem allows one to
establish that  $\sqrt W$ can be used as the usual polar
coordinate $\rho$ on $\R^3$, leading to a coordinate system in
which the field equations reduce to a harmonic map with values
in two-dimensional hyperbolic space. The map is singular at the
rotation axis (compare~\cite{ChUone}), with rather delicate
singularity structure at points where the event horizon meets
the axis. A uniqueness theorem for such
maps~\cite{RobinsonKerr,Weinstein1} achieves the proof.

The analysis above relies heavily on the fact that the domain
of outer communications is simply connected~\cite{ChWald}
(compare~\cite{Galloway:fitopology}).

\subsubsection{Static case}
 \label{ssscbh}
Assuming  \emph{staticity}, i.e., stationarity and
hypersurface-orthogonality of the stationary Killing vector, a
more satisfactory result is available in space dimensions less
than or equal to seven, and in higher dimensions on manifolds
on which the Riemannian rigid positive energy theorem holds:
non-connected configurations are excluded, without any \emph{a
priori} restrictions on the gradient of the norm of the static
Killing vector at event horizons.

More precisely, we shall say that  a manifold $\hypwithhat$ is
of \emph{positive energy type} if there are no asymptotically
flat complete Riemannian metrics on $\hypwithhat$ with positive
scalar curvature and vanishing mass except perhaps for a flat
one. As made clear in Theorem~\ref{Tposmasman}, this property
has been proved so far for all asymptotically flat
$n$-dimensional manifolds $\hypwithhat$   of dimension  $3\le
n\le 7$~\cite{SchoenCatini}, or under the hypothesis that
$\hypwithhat$ is spin for any $n\ge 3$, and is expected to be
true in general.
 \opp{opppet}

We have the following result, which finds its roots in the work
of Israel~\cite{Israel:vacuum}, with further simplifications by
Robinson~\cite{RobinsonSP}, and with a significant
strengthening by Bunting and
Masood-ul-Alam~\cite{bunting:masood}; the proof of the version
presented here can be found in~\cite{ChCo,Chstatic}, see
also~\cite{ChGstatic}:

\begin{theorem}
 \label{Tubhs}
Let $(\mcM,\fourg)$ be  \aregular, vacuum, static,
$(n+1)$-dimensional space-time, $n\ge 3$. Let
$\,\,\widehat{\!\! \hyp}$ denote the manifold obtained by
doubling the hypersurface $\hyp$ of Definition~\ref{Dmain}
across {all non-degenerate components of its boundary} and
smoothly compactifying,  in the doubled manifold, all
asymptotically {flat} regions but one to a point. If
$\,\,\widehat{\!\! \hyp}$ is of positive energy type, then
$\doc$ is isometric  to the domain of outer communications of a
Schwarzschild space-time.
\end{theorem}

\begin{Remark}
\label{RCRT} {\rm As a corollary of Theorem~\ref{Tubhs} one
obtains non-existence of static, regular, vacuum black holes
with some components of the horizon degenerate. As observed
in~\cite{CRT}, if the space-time dimension is four,
non-existence follows immediately from Theorem~\ref{TCRT} and
from simple connectedness of the domain of outer
communications~\cite{ChWald}, but this does not seem to
generalize to higher dimensions in any obvious way. }
\end{Remark}

\subsubsection{Multi-black hole solutions}
 \label{sSmbhs}

In this section we assume that the space-time dimension is
four. Space-times containing several black holes seem to be of
particular interest, but Theorem~\ref{Tubhs}  implies, {under
the conditions spelled out there,} that no such  \emph{vacuum}
solutions exist in the static class. However, the
\emph{Einstein-Maxwell} equations admit static solutions with
several black holes:  the \emph{Majumdar-Papapetrou} (MP)
solutions. The metric $\fourg $ and the electromagnetic
potential $A$ take the form~\cite{Majumdar,Papapetrou:mp}
\begin{eqnarray}\label{I.0} & \fourg  = -u^{-2}dt^2 + u^2(dx^2+dy^2+dz^2)\,,
\qquad A = \pm u^{-1}dt\,, &
\end{eqnarray}
with $\Delta_\delta u=0$, where $\Delta_\delta$ is the Laplace
operator of the flat metric $\delta$.  \emph{Standard MP black
holes}  are defined by further requiring that
\begin{eqnarray}
  &
  \label{standard}
   \displaystyle
   u=1+\sum_{i=1}^I \frac{\mu_i}{|\vec x - \vec
a_i|} \,, & \end{eqnarray}
for some positive constants $\mu_i$, the  electric charges (up
to the choice of sign for $A$ in \eq{I.0}) carried by the
punctures $\vec x = \vec a_i$. Further, the coordinates $x^\mu$
of (\ref{I.0}) are required to cover the range ${\R }\times({\R
}^3\setminus\{\vec a_i\})$ for a finite set of points $\vec
a_i\in{\R }^3$, $i=1,\ldots,I$. It has been shown by Hartle and
Hawking~\cite{HartleHawking} that standard MP space-times can
be analytically extended to an electro--vacuum space-time with
$I$ black hole regions.

The case $I=1$ is the special case {$m=|q|$} of the so-called
\emph{Reissner-Nordstr\"om}   metrics, which are the charged,
spherically symmetric (connected) generalizations of the
Schwarzschild black holes {with mass $m$ and electric charge
$q$}.

The   static {\regular} electro-vacuum black holes are well
understood: Indeed, the analysis
in~\cite{CarterlesHouches,Masood,Ruback,Simon:elvac}
(compare~\cite{ChstaticelvacarxivErr}) leads to:

\begin{theorem}
 \label{TRN}
Every domain of outer communications in a static, electro-vacuum,
\regular, analytic black hole space-time $\mbox{\rm without degenerate
horizons}$ is isometric  to a domain of outer
communications of a Reissner-Nordstr\"om black hole.
\end{theorem}

The relevance of the standard MP black holes follows now from
the following result~\cite{CT}:

\begin{theorem}
 \label{TMP}
Every domain of outer communications in a static,
electro-vacuum, \regular  black hole space-time $\mbox{\rm
containing degenerate horizons}$ is isometric  to a domain of
outer communications of a standard MP space-time.
\end{theorem}

It thus follows that the MP family provides the only static,
electro-vacuum, analytic, regular black holes with
\emph{non-connected} horizons.

\subsection{Non-zero cosmological constant}
 \label{ssnzcc}
A family of black hole solutions with non-zero cosmological constant has been
discovered by Kottler~\cite{Kottler} (compare~\cite{CadeauWoolgar,Birmingham}).
These metrics are also known as
the Schwarzschild-anti de Sitter metrics (when $\Lambda<0$) or
the Schwarzschild-de Sitter metrics (when $\Lambda>0$), and take the form
\bel{basemet}
 \fourg=-e^{2\mathring\lambda(r)}dt^2 +  e^{-2\mathring\lambda(r)}dr^2 +   r^2
\mathring k\;,
\ee
where $\mathring k$ is an Einstein metric on a compact
$(n-1)$-dimensional manifold $N$, $n\ge 3$.  Here
$$
 e^{2\mathring \lambda(r)}  = \alpha r^2 + \beta + \frac {2m}r\;,
$$
with  $\alpha= -2\Lambda/n(n-1)$,   $\beta = R(\mathring
k)/(n-1)(n-2)$, where $R(\mathring k)$ is the scalar curvature
of the metric $\mathring k$, while $m\in \R$ is a constant,
called the \emph{mass} of $g$. The global structure of
(suitably extended) Kottler space-times has been analyzed
in~\cite{GibbonsHawkingCEH,BLP} in dimension $3+1$; the results
extend to $n\ge 3$ dimension.
}

\section{The Cauchy problem}
 \label{SEe}

The component version of the {\em vacuum Einstein equations with
cosmological constant $\Lambda$}  \eq{eeqn} reads
\begin{equation}\label{vE.5}
G_{\alpha\beta} +  \Lambda \fourg_{\alpha\beta}=0\;,
\end{equation}
where $G_{\alpha\beta}$ is the Einstein tensor defined as
\bel{Einstein}
 G_{\a\b}:= R_{\a\b} - \frac 12 R\fourg_{\a\b}
 \;,
\ee
while $R_{\a\b}$ is the Ricci tensor and $R$  the scalar
curvature. We will refer to those equations as {\em the vacuum
Einstein equations}, regardless of whether or not the
cosmological constant vanishes, and in this work we will mostly
assume $\Lambda=0$. Taking the trace of \eq{vE.5} one obtains
\be\label{vE.4} R= \frac{2(n+1)}{n-1} \Lambda\;, \ee where, as
elsewhere, $n+1$ is the dimension of space-time. This leads to
the following equivalent version of \eq{vE.5}:
\be \label{vE.3} \Ric
= \frac{2\Lambda}{n-1} \fourg\;. \ee
Thus the Ricci tensor of the metric is proportional to the
metric. Pseudo-Riemannian manifolds with metrics satisfying
Equation~\eq{vE.3} are called {\em Einstein manifolds} in the
mathematical literature, see e.g.~\cite{Besse}.

Given a manifold $\mcM $, Equation~\eq{vE.5} or, equivalently,
Equation~\eq{vE.3} forms a system of second order partial
differential equations for the metric, linear in the second
derivatives of the metric, with coefficients which are rational
functions of the $\fourg_{\alpha\beta}$'s, quadratic in the
first derivatives of $\fourg$, again with coefficients rational
in $\fourg$. Equations linear in the highest order derivatives
are called {\em quasi-linear}, hence the vacuum Einstein
equations constitute a second order system of quasi-linear
partial differential equations for the metric $\fourg$.

 {In the
discussion above we assumed that the manifold $\mcM $ has been
given. In the evolutionary point of view, which we adapt in
most of this work,   all space-times of main interest have
topology $\R\times\hyp$, where $\hyp$ is an $n$-dimensional
manifold {carrying initial data}. Thus, solutions of the Cauchy
problem (as defined precisely by Theorem~\ref{TCBG} below) have
topology and differential structure which are determined by the
initial data. As will be discussed in more detail in
Section~\ref{sSScc}, the space-times obtained by evolution of
the data are sometimes extendible; there is then a lot of
freedom in the topology of the extended space-time,  and we are
not aware of conditions which would guarantee \emph{uniqueness}
of the extensions. So in the evolutionary approach the manifold
is best thought of as being given {\em a priori}
--- namely $\mcM =\R\times\hyp$, but it should be kept in mind that
there is no {\em a priori} known natural time coordinate which can
be constructed by evolutionary methods, and which leads to the
decomposition $\mcM =\R\times\hyp$. }

Now, there exist standard classes of partial differential
equations which are known to have good properties. They are
determined by looking at the algebraic properties of those
terms in the equations which contain derivatives of highest
order, in our case of order two. Inspection of \eq{vE.5} shows
that this equation does not fall in any of the standard
classes, such as hyperbolic, parabolic, or elliptic. In
retrospect this is not surprising, because equations in those
classes typically lead to unique solutions. On the other hand,
given any solution $g$ of the Einstein equations \eq{vE.3} and
any diffeomorphism $\Phi$, the pull-back metric $\Phi^*g$ is
also a solution of \eq{vE.3}, so whatever uniqueness there
might be will hold only \emph{up to diffeomorphisms}. An
alternative way of describing this, often found in the physics
literature, is the following: suppose that we have a matrix
$g_{\mu\nu}(x)$ of functions satisfying \eq{vE.5} in some
coordinate system $x^\mu$. If we perform a coordinate change
$x^\mu \to y^\alpha(x^\mu)$, then the matrix of functions $\bar
\fourg_{\alpha \beta}(y)$ defined as
\bel{coorfree}
 \fourg_{\mu\nu}(x) \to \bar \fourg_{\alpha\beta}(y)= \fourg_{\mu\nu}(x(y))
 \frac{\partial x^\mu}{\partial y^\alpha} \frac {\partial x^\nu}
 {\partial y^\beta}
\ee
will also solve \eq{vE.5}, if the $x$-derivatives there are
replaced by $y$-derivatives. This property is known under the
name of \emph{diffeomorphism invariance}, or \emph{coordinate
invariance}, of the Einstein equations. Physicists say that
``the diffeomorphism group is the gauge group of Einstein's
theory of gravitation".

Somewhat surprisingly, Choquet-Bruhat~\cite{ChBActa} proved in
1952 that there exists a set of \emph{hyperbolic} equations
underlying \eq{Einstein}. This proceeds by the introduction of
so-called ``harmonic coordinates", to which we turn our
attention in the next section.

\subsection{The local evolution problem}
 \label{sSHr}
\subsubsection{Wave coordinates}
 \label{ssShc}
 A set of coordinates $\{y^\mu\}$ is called \emph{harmonic} if each of the
 functions $y^\mu$ satisfies
\bel{harmonic}
 \Box_\fourg y^\mu =0
 \;,
\ee
where $\Box_\fourg$ is 
the d'Alembertian associated with $\fourg$ acting on scalars:
\bel{wave}
 \Box_\fourg f := \tr_\fourg \Hess f=
 \frac 1 {\sqrt {|\det \fourg|}} \partial_\mu\left(\sqrt {|\det \fourg|} \fourg^{\mu\nu}\partial_\nu f\right)
 \;.
\ee
One also refers to these as \emph{``wave coordinates"}.
Assuming that \eq{harmonic} holds, \eq{vE.3} can be written as
 \newcommand{\hE}{\hat E}
\beqa 0  =  \hE^{\a\b} & :=& \Box_g \fourg^{\alpha\beta} -
\fourg^{\epsilon\phi}\Big(2
\fourg^{\gamma\delta}\Gamma^\alpha_{\gamma\epsilon}\Gamma^\beta_{\delta\phi}
 +    ( \fourg^{\alpha\gamma}
 \Gamma^\b _{\gamma\delta} + \fourg^{\b\gamma}
 \Gamma^\a _{\gamma\delta}) \Gamma^\delta_{\epsilon\phi}\Big)
\label{wm.14}
 \\
 & &
    - \frac{4\Lambda}{n-1} \fourg^{\alpha\beta}\;.
 \nonumber
 \eeqa
Here the $\Gamma^\alpha_{\beta\gamma}$'s should be calculated
in terms of the $\fourg_{\alpha\beta}$'s and their derivatives
as in \eq{chris}, and the wave operator $\Box_\fourg$ is as in
\eq{wave}. So, {\em in  wave coordinates, the Einstein equation
forms a second-order quasi-linear wave-type system of
equations} \eq{wm.14} for the metric functions
$\fourg^{\alpha\beta}$. (This can of course be rewritten as a
set of quasi-linear equations for the $\fourg_{\alpha\beta}$'s
by algebraic manipulations.)

Standard theory of hyperbolic PDEs~\cite{Evans:book} gives:%
\footnote{If $k$ is an integer, then the Sobolev spaces
$H_\loc^k$ are defined as spaces of functions which are in
$L^2(K)$ for any compact set $K$,with their distributional
derivatives up to order $k$ also in $L^2(K)$. In the results
presented here one can actually allow non-integer $k$'s, the
spaces $H_\loc^k$ are then defined rather similarly using the
Fourier transformation.}

\begin{Theorem}\label{Tlcp} For any initial data
\bel{wm.19}\fourg^{\a\b}(0,y^i)\in H_\loc^{k+1}\;,\quad
\partial_0\fourg^{\a\b}(0,y^i)\in H_\loc^k\;,\qquad k>n/2
 \;,
\ee
prescribed on an open subset $\mcO \subset\{0\}\times
\R^n\subset\R\times\R^{n}$ there exists a unique solution
$\fourg^{\a\b}$ of \eq{wm.14} defined on an open neighborhood
$\mcU\subset \R\times\R^{n}$ of $\mcO$. The set $\mcU$ can be
chosen so that $(\mcU,\fourg)$ is globally hyperbolic with
Cauchy surface $\mcO$.
\end{Theorem}

\begin{Remark}
 \label{Rlcp}
The  results
 \opp{oppdiff}
in~\cite{KlainermanRodnianski:r1,KlainermanRodnianski:r2,KlainermanRodnianski:r3,SmithTataru:sharp}
and references therein allow one to reduce the differentiability
threshold above.
\end{Remark}

\Eq{wm.14} would establish the {\em hyperbolic}, evolutionary
character of the Einstein equations, if not for the following
problem: Given initial data for an equation as in \eq{wm.14}
there exists a unique solution, at least for some short time.
But there is \emph{a priori} no reason to expect that the
solution will satisfy \eq{harmonic}; if it does not, then a
solution of \eq{wm.14} will \emph{not} solve the Einstein
equation. In fact, if we set
\bel{lambdel}
 \lambda^\mu:= \Box_\fourg y^\mu\;,
\ee
then
\bel{wm.18} R^{\alpha\beta}= \frac 12 (\hE^{\alpha\beta} -
\nabla^\alpha \lambda^\beta- \nabla^\beta \lambda^\alpha) +
\frac{2\Lambda}{n-1}\fourg^{\a\b}
 \;,
 \ee
so that it is precisely the vanishing -- or not -- of $\lambda$
which decides whether or not a solution of \eq{wm.14} is a solution
of the vacuum Einstein equations.

This problem has been solved by Choquet-Bruhat~\cite{ChBActa}.
The key observation is that \eq{wm.18} and the Bianchi identity
imply a wave equation for the $\lambda^\a $'s. In order to see
that, recall the twice-contracted Bianchi identity
\eq{bianchi}:
$$
\nabla_\a \Big(R^{\a\b}- \frac R2 \fourg^{\a\b}\Big)=0\;.$$
 Assuming that
\eq{wm.14} holds, one finds
\beaa0 & =& -\nabla_\a
\Big(\nabla^\a  \lambda^\b  + \nabla^\b  \lambda^\a  - \nabla_\gamma
\lambda^\gamma \fourg^{\a\b}\Big)
\\
& =& -\Big(\Box_\fourg \lambda^\b  + R^\b {}_\a \lambda^\a  \Big)
 \;.
 \eeaa
This shows that $\lambda^\a $ necessarily satisfies the second
order hyperbolic system of equations
$$\Box _\fourg
\lambda^\b  + R^\b {}_\a \lambda^\a =0\;.$$
Now, it is a standard
fact in the theory of hyperbolic equations that we will have
$$\lambda^\a \equiv 0$$ on the domain of dependence
$\mcD(\mcO)$, provided that both $\lambda^\a $ and its
derivatives vanish at $\mcO$. To see how these initial
conditions on $\lambda^\alpha$ can be ensured, it is convenient
to assume that $y^0$ is the coordinate along the $\R$ factor of
$\R\times \R^n$, so that the initial data surface $\{0\}\times
\mcO$ is given by the equation $y^0=0$. We have
\beaa \Box_\fourg y^\a  &=& \frac 1 {\sqrt{|\det \fourg|}}\partial
_\b\Big(\sqrt{|\det \fourg|} \fourg^{\b\gamma}\partial_\gamma  y^\a \Big)\\
&=& \frac 1 {\sqrt{|\det \fourg|}}\partial _\b \Big(\sqrt{|\det
\fourg|} \fourg^{\b\a}\Big)
 \;.
\eeaa
Clearly a necessary condition for the vanishing of $\Box_\fourg y^\a
$ is that it vanishes at $y^0=0$, and this allows us to calculate
some time derivatives of the metric in terms of space ones:
\bel{wm.22} \partial _0\Big(\sqrt{|\det \fourg|}
\fourg^{0\a}\Big)=-\partial _i\Big(\sqrt{|\det \fourg|}
\fourg^{i\a}\Big)\;.\ee
This implies that the initial data \eq{wm.19} for the equation
\eq{wm.14} cannot be chosen arbitrarily if we want both
\eq{wm.14} and the Einstein equation to be simultaneously
satisfied.

Now, there is still freedom left in choosing the  wave
coordinates. Using this freedom, one can show that there is
\emph{no loss of generality} in assuming that on the initial
hypersurface $\{y^0=0\}$ we have
\bel{stupidchoice}
 \fourg^{00}=-1\;,\qquad \fourg^{0i}=0
 \;,
\ee
and this choice simplifies the algebra considerably. \Eq{wm.22}
determines the  time derivatives
$\partial_0\fourg^{0\mu}|_{\{y^0=0\}}$ needed in
Theorem~\ref{Tlcp}, once   $\fourg_{ij}|_{\{y^0=0\}}$ and
$\partial_0\fourg_{ij}|_{\{y^0=0\}}$ are given. So, from this
point of view, the essential initial data for the evolution
problem are the space metric
$$
 \threeg:=\fourg_{ij}dy^i dy^j
 \;,
$$
together with its time derivatives.

It turns out that further constraints arise from the requirement of
the vanishing of the derivatives of $\lambda$. Supposing that
\eq{wm.22} holds at $y^0=0$ --- equivalently, supposing  that
$\lambda$ vanishes on $\{y^0=0\}$, we then have
$$\partial_i\lambda^\a =0$$
on $\{y^0=0\}$. To obtain the vanishing of all derivatives
initially it remains to ensure
that some transverse derivative does. A convenient transverse
direction is provided by the field $n$ of unit timelike normals
to $\{y^0=0\}$, and the vanishing of $\nabla_n\lambda^\alpha$
is guaranteed by requiring that
\bel{wm.24}\Big(G_{\mu\nu}+\Lambda
g_{\mu\nu}\Big)n^\mu=0\;.\ee
This follows by simple algebra from the equation $\hE_{\a\b}=0$
and \eq{wm.18},
$$G_{\mu\nu}+\Lambda \fourg_{\mu\nu}=-\Big(\nabla_\mu \lambda_\nu +
\nabla_\nu \lambda_\mu - \nabla^\alpha \lambda_\alpha
g_{\mu\nu}\Big)\;,
$$
using that  $\lambda_\mu|_{y^0=0}=
\partial_i \lambda_\mu|_{y^0=0} =0$.

Equations \eq{wm.24} are called the \emph{Einstein constraint
equations}, and will be discussed in detail in
Section~\ref{Sivp}.

Summarizing, we have proved:

\begin{Theorem}
 \label{Tlocex} Under the hypotheses of Theorem~\ref{Tlcp}, suppose
that the initial data \eq{wm.19} satisfy \eq{wm.22},
\eq{stupidchoice} as well as the constraint equations
\eq{wm.24}. Then the metric given by Theorem~\ref{Tlcp} on a
globally hyperbolic set $\mcU$ satisfies the vacuum Einstein
equations.
\end{Theorem}

%
%
%
%
%
%
%

\subsection{Cauchy data}
 \label{sSCd}
In Theorem~\ref{Tlcp} we consider initial data given in a
single coordinate patch $\mcO\subset \R^n$. This suffices for
applications such as the Lindblad-Rodnianski stability theorem
discussed in Section~\ref{sSsmst} below, where $\mcO=\R^n$. But
a correct geometric picture is to start with an $n$-dimensional
manifold $\hyp$, and prescribe initial data there; the case
where $\hyp$ is $\mcO$ is thus a special case of this
construction. At this stage there are two attitudes one may
wish to adopt: the first is that $\hyp$ is  a subset of the
space-time $\mcM$
--- this is essentially what we assumed in Section~\ref{sSHr}.
The alternative  is to consider $\hyp$ as a manifold of its own,
equipped with an embedding
$$i:\hyp\to\mcM\;.$$
The most  convenient approach is to go back and forth between those
points of view, and this is the strategy that we will follow.

A \emph{vacuum initial data set} $(\hyp,\threeg,K)$ is a triple
where $\hyp$ is an $n$-dimensional manifold, $\threeg$ is a
Riemannian metric on $\hyp$, and $K$ is a symmetric
two-covariant tensor field on $\hyp$. Further $(\threeg,K)$ are
supposed to satisfy the vacuum constraint equations that result
from \eq{wm.24}, and which are written explicitly in terms of
$K$ and $\threeg$ in Section~\ref{CE}. Here the tensor field
$K$ will eventually become the second fundamental form of $M$
in the resulting space-time $\mcM$, obtained by evolving the
initial data. Recall that the second fundamental form of a
spacelike hypersurface $M$ is defined as
\bel{Kdef}
\forall X\in T\hyp\qquad
 K(X,Y) = \fourg(\nabla_Xn, Y)
 \;,
\ee
where $n$ is the future pointing unit normal to $M$. The tensor
$K$ is often referred to as the \emph{extrinsic curvature
tensor of $M$} in the relativity literature. Specifying $K$ is
equivalent to prescribing the time-derivatives of the
space-part $\fourg _{ij}$ of the resulting space-time metric
$\fourg $; this can be seen as follows: Suppose, indeed, that a
space-time $(\mcM,\fourg)$    has been constructed (not
necessarily vacuum) such that $K$ is the extrinsic curvature
tensor of $\hyp$ in $(\mcM,g)$. Consider any domain of
coordinates $\mcO\subset \hyp$, and construct coordinates
$y^\mu$ in a space-time neighborhood  $\mcU$ such that
$\hyp\cap\mcU=\mcO$; those coordinates could be wave
coordinates, obtained by solving the wave equations
\eq{harmonic},  but this is not necessary at this stage. Since
$y^0$ is constant on $\hyp$ the one-form $dy^0$ annihilates
$T\hyp\subset T\mcM$, as does the 1--form $\fourg (n,\cdot)$.
Since $\hyp$ has codimension one, it follows that $dy^0$ must
be proportional to $\fourg (n,\cdot)$:
$$n_\a dy^\a =n_0 dy^0$$
on $\mcO $. The normalization $-1=g(n,n)=g^{\mu\nu}n_\mu n_\nu
= g^{00}(n_0)^2$ gives
$$ n_\a  dy^\a  = \frac{1}{\sqrt{| \fourg^{00}|}} dy^0\;.$$
We then have, by \eq{Kdef},
\bea
 K_{ij}
& = & -\frac 12 \fourg^{0\sigma}\Big(\partial_j \fourg_{\sigma i} +
\partial_i \fourg_{\sigma j} - \partial_\sigma
 \fourg_{ij}\Big)n_0
 \;.
\eeal{wm.30}
This shows that the knowledge of $\fourg _{\mu\nu}$ \emph{and}
$\partial_0\fourg_{ij}$ at $\{y^0=0\}$ allows one to calculate
$K_{ij}$. Reciprocally, \eq{wm.30} can be rewritten as
$$\partial_0\fourg_{ij} = \frac {2}{\fourg^{00}n_0} K_{ij} + \mbox{ terms determined by the
$\fourg _{\mu\nu}$'s and their space--derivatives} \;,$$
so that the knowledge of the $\fourg _{\mu\nu}$'s and of the
$K_{ij}$'s at $y^0=0$ allows one to calculate
$\partial_0\fourg_{ij}$. Thus, $K_{ij}$ is the geometric counterpart
of the $\partial_0\fourg_{ij}$'s.

\subsection{Solutions global in space}
 \label{sSgsp}
In order to globalize the existence Theorem~\ref{Tlcp} \emph{in
space}, the key point is to show that two solutions differing
only by the values $\fourg _{0\a}|_{\{ y^0=0\}}$ are (locally)
isometric: so suppose that $\fourg$ and $\tilde\fourg$ both
solve the vacuum Einstein equations in a globally hyperbolic
region $\mcU$, with the same Cauchy data $(\threeg,K)$ on
$\mcO:=\mcU\cap \hyp$. One can then introduce wave coordinates
in a globally hyperbolic neighborhood of $\mcO$ both for
$\fourg$ and $\tilde\fourg$, satisfying \eq{stupidchoice}, by
solving
\bel{twowave}
 \Box_\fourg y^\mu =0\;,\qquad \Box_{\tilde \fourg} \tilde y^\mu = 0
  \;,
\ee
with the same initial data for $y^\mu$ and $\tilde y^\mu$.
Transforming both metrics to their respective wave-coordinates,
one obtains two solutions of the reduced equation \eq{wm.14}
with the same initial data.

{The question then arises whether the resulting metrics will be
sufficiently differentiable to  apply the uniqueness part of
Theorem~\ref{Tlcp}. Now,   the metrics obtained so far are in a
space $C^1([0,T],H^s)$, where the Sobolev space $H^s$ involves
the space-derivatives of the metric. The initial data for the
solutions $y^\mu$ or $\tilde y^\mu$ of \eq{twowave}  may be
chosen to be in $H^{s+1}\times H^s$. However, a rough
inspection of \eq{twowave} shows that the resulting solutions
will be only in $C^1([0,T],H^s)$, because of the low regularity
of the metric. But then \eq{coorfree} implies that the
transformed metrics will be in $C^1([0,T],H^{s-1})$, and
uniqueness can only be invoked \emph{provided that
$s-1>n/2+1$}, which is  one degree of differentiability more
than what was required for existence. This was the state of
affairs for some fifty-five years until the following  simple
argument of Planchon and Rodnianski~\cite{PlanchonRodnianski}:
To make it clear that the functions $y^\mu$ are considered to
be scalars in \eq{twowave}, we shall write $y$ for $y^\mu$.
Commuting derivatives with $\Box_\fourg$ one finds, for metrics
satisfying the vacuum Einstein equations,
\beaa%
 &
 \Box_\fourg \nabla_\alpha y  =  \nabla_\mu
 \nabla^\mu \nabla_\alpha y = [\nabla_\mu
 \nabla^\mu,\nabla_\alpha] y = \underbrace{R^{\sigma\mu}{}_{\alpha\mu}}_{=R^{\sigma }{}_{\alpha }=0} \nabla_\sigma y = 0
\;.
 \eeaa
Commuting once more one obtains an evolution equation for the field
$\psi_{\alpha\beta}:=\nabla_\alpha \nabla_\beta y$:
\beaa
 &
 \Box_\fourg  \psi_{\alpha\beta} + \underbrace{\nabla_\sigma
 R_{\beta}{}^\lambda{}_\alpha{}^\sigma}_{=0} \nabla_\lambda  y + 2R_\beta {} ^
 \lambda {} _ { \alpha}{}^\sigma    \psi_{\sigma \lambda}
  =0
 \;,
 \eeaa
where the underbraced term vanishes, for vacuum metrics, by a
contracted Bianchi identity. So the most offending term in this
equation for $\psi_{\alpha\beta}$, involving three derivatives
of the metric, disappears when the metric is vacuum. Standard
theory of hyperbolic PDEs  shows now that the functions
$\nabla_\alpha \nabla_\beta y$ are in $C^1([0,T],H^{s-1})$,
hence $y \in C^1([0,T],H^{s+1})$, and the transformed metrics
are regular enough to invoke uniqueness without having to
increase $s$. }

Suppose, now, that an initial data set $(\hyp,\threeg,K)$ as in
Theorem~\ref{Tlcp} is given. Covering $\hyp$ by coordinate
neighborhoods $\mcO_p$, $p\in\hyp$, one can use Theorem~\ref{Tlcp}
to construct globally hyperbolic developments $(\mcU_p, \fourg_p)$
of $(\mcU_p,\threeg,K)$. By the argument just given the metrics so
obtained will coincide, after performing a suitable coordinate
transformation, wherever simultaneously defined. This allows one to
patch the $(\mcU_p, \fourg_p)$'s together to a globally hyperbolic
Lorentzian manifold, with Cauchy surface $\hyp$. Thus:

{
\begin{Theorem}
 \label{Tglobinspace}
Any vacuum initial data set $(\hyp,\threeg,K)$ of differentiability
class $H^{s+1}\times H^s$, $s>n/2$, admits a globally hyperbolic
development.
\end{Theorem}

The solutions are locally unique, in a sense made clear by the
proof. The important question of \emph{uniqueness in the large}
will be addressed in Section~\ref{sSScc}.}

\bams{%
\subsection{Other hyperbolic reductions}
 \label{ssSoa}
The wave-coordinates approach of Choquet-Bruhat, presented
above, is the first hyperbolic reduction discovered for the
Einstein equations. It has been given new life by the
Lindblad-Rodnianski stability theorem, presented in
Section~\ref{sLoizelet} below.  However, one should keep in
mind the existence of several other such reductions.

An example is given by the symmetric-hyperbolic first order
system of Baumgarte, Shapiro, Shibata and
Nakamura~\cite{BaumgarteShapiro,ShibataNakamura,Sarbachetalt},
known as the BSSN system, widely used in numerical general
relativity. Another noteworthy example is the
elliptic-hyperbolic system of~\cite{AnderssonMoncriefAIHP}, in
which  the elliptic character of some of the equations provides
increased control of the solution.  A notorious problem in
numerical simulations is the lack of constraint preservation,
see~\cite{Paschalidis,HolstScheelLindblom} and references
therein for attempts to improve the situation. The reader is
referred
to~\cite{FriedrichRendall00,Friedrich:hyperbolicreview} for a
review of many other possibilities.

\subsection{The characteristic Cauchy problem}
 \label{sScCp}
Another important systematic construction of solutions of the
vacuum Einstein equations proceeds via a \emph{characteristic
Cauchy problem}. In this case the initial data are prescribed
on Cauchy hypersurfaces which are allowed to be piecewise null.
This problem has been considerably less studied than the
spacelike one described above. We will not go into any details
here;
see~\cite{RendallCIVP,DossaAHP,BishopWinicour,CaciottaNicolo,%
CaciottaNicolo2,ChristodoulouBHFormation,CCM2,CCG} for further
information.
 \opp{oppccp}

\subsection{Initial-boundary value problems}
 \label{sSIbvp}
Numerical simulations necessarily take place on a finite grid,
which leads to the need of considering  initial-boundary value
problems. In general relativity those are considerably more
complicated than the Cauchy problem, and much remains to be
understood. In pioneering work, Friedrich and
Nagy~\cite{FriedrichNagy} constructed  a system of equations,
equivalent to Einstein's, for a set of fields that includes
some components of the Weyl tensor, and proved well-posedness
of an initial-boundary value problem for those equations. It
would seem that the recent work by Kreiss \emph{et
al.}~\cite{KRSW} might lead to a simpler formulation of the
problem at hand. \opp{oppibvp}
}

\section{Initial data sets}
\label{Sivp}

We now turn our attention to  an analysis of the constraint
equations, returning to the evolution problem in Section~\ref{SE}.

An essential part of the mathematical analysis of the Einstein
field equations of general relativity is the rigorous
formulation of the Cauchy problem, which is  a means to
describe solutions of a dynamical theory via the specification
of initial data and the evolution of that data. In this section
we will be mainly concerned with the initial data sets for the
Cauchy problem. As already explained in Section~\ref{ssShc},
those initial data sets have to satisfy the relativistic
constraint equations (\ref{wm.24}). This leads to the following
questions: What are the sets of allowable initial data? Is it
possible to parameterize them in a useful way? What global
properties of the space-time can be seen in the initial data
sets? How does one engineer  initial data so that the
associated space-time has some specific properties?

\subsection{The constraint equations}
\label{CE} As explained in Section~\ref{sSCd}, an initial data
set for a vacuum space-time consists of  an $n$-dimensional
manifold $\hyp$ together with a Riemannian metric $\threeg$ and
a symmetric tensor $K$. In the non-vacuum case we also have a
collection of non-gravitational fields which we collectively
label ${\mathcal F}$ (usually these are sections of a bundle
over $\hyp$). We have already seen the relativistic vacuum
constraint equations expressed as the vanishing of the normal
components of the Einstein equations \eq{wm.24}. Now, if
$\threeg$ is the metric induced on a spacelike hypersurface in
a Lorentzian manifold, it has its own curvature tensor $R
^i{}_{jk\ell}$. If we denote by $K_{ij}$ the second fundamental
form of $\hyp$ in $\mcM$, and by $ {\mathscr R} ^i{}_{jk\ell}$
the space-time curvature tensor, the Gauss-Codazzi equations
provide the following relationships:
\beal{firsemb} &  R ^i{}_{jk\ell} = {\mathscr R}  ^i{}_{jk\ell}
+ K^i{}_\ell K_{jk} -K^i{}_k K_{j\ell}
 \;,
 &
 \\
 & D_i K_{jk} -  D_j K_{ik}=   {\mathscr R} _{ i j k\mu}n^\mu
 \;.
 &
\eeal{secemb}
Here $n$ is the timelike normal to the hypersurface, and we are
using a coordinate system in which the $\partial_i$'s are
tangent to the hypersurface $\hyp$.

Contractions of \eq{firsemb}-\eq{secemb} and simple algebra
allow one to reexpress \eq{wm.24} in the following form, where
we have now allowed for the additional presence of
non-gravitational fields:
\begin{eqnarray}
\div K - d (\tr K) & = & 8 \pi J \;,\label{eq:c1}\\
R(\threeg) -2\Lambda - |K|^2_\threeg + (\tr K)^2 & = & 16 \pi \rho \;,\label{eq:c2}\\
\mathcal C({\mathcal F}, \threeg) & = & 0 \;,\label{eq:c3}
\end{eqnarray}
where $R(\threeg)$ is the scalar curvature of the metric
$\threeg$, $J$ is the momentum density of the non-gravitational
fields, $\rho$ is the energy density,%
\footnote{If $\mathcal T$ is the stress-energy tensor of the
non-gravitational fields, and $n$ denotes the unit timelike
normal to a hypersurface $\hyp$ embedded in a  space-time, with
induced data $(\hyp, \threeg, K, {\mathcal F})$, then $J=
-{\mathcal T}(n,\cdot)$ and $\rho={\mathcal T}(n, n)$. However,
in terms of the initial data set itself we shall regard
(\ref{eq:c1})-(\ref{eq:c2}) as the definitions of the
quantities $J$ and $\rho$.} and $\mathcal C({\mathcal F},
\threeg)$ denotes  the set of additional constraints that might
come from the non-gravitational part of the theory. The first
of these  equations is known as the \emph{momentum constraint}
and is a vector field equation on $\hyp$. The second, a scalar
equation, is referred to as the \emph{scalar}, or
\emph{Hamiltonian, constraint}, while the last are collectively
labeled the non-gravitational constraints.
  These are what we shall
henceforth call the \emph{Einstein constraint equations}, or
simply the \emph{constraint equations} if ambiguities are
unlikely to occur.

As an example, for the Einstein-Maxwell theory in 3+1
dimensions, the non-gravitational fields consist of the
electric and magnetic vector fields $E$ and $B$. In this case
we have $\rho=\frac{1}{2}(|E|^2_\threeg +|B|^2_\threeg$),
$J=(E\times B)_\threeg$, and we have the extra
(non-gravitational) constraints $\dive_{\threeg} E=0$ and
$\dive_{\threeg} B=0$.

Equations (\ref{eq:c1})-(\ref{eq:c3}) form an underdetermined
system of partial differential equations. In the classical
vacuum setting of $n=3$ dimensions, these are locally four
equations for the twelve unknowns given by the components of
the symmetric tensors $\threeg$ and $K$.  This section will
focus primarily on the vacuum case with a zero cosmological
constant. However, we will allow arbitrary values of $\Lambda $
in Section \ref{ssSnonzeroLambda}.

The most successful approach so far for studying the existence
and uniqueness of solutions to   (\ref{eq:c1})-(\ref{eq:c3}) is
through the conformal method of Lichnerowicz~\cite{Lich44},
Choquet-Bruhat and York~\cite{CBY}. The idea is to introduce a
set of unconstrained ``conformal data", which are freely
chosen, and find  $(\threeg,K)$ by solving a system of
determined partial differential equations.  In the vacuum case
with vanishing cosmological constant~\cite{CBY}, the free
conformal data consist of a manifold $\hyp$, a Riemannian
metric $\tthreeg$ on $\hyp$, a trace-free symmetric tensor
$\tsigma$, and the \emph{mean curvature function} $\tau$. The
initial data $(\threeg,K)$ defined as
\begin{eqnarray}
\threeg & = & \phi^q \tthreeg \;, \quad
q=\frac{4}{n-2}\;,\label{threegeq}\\
K & = & \phi^{-2} (\tsigma + \tcalD W) + \frac{\tau}{n}
 \phi^q \tthreeg \;, \label{Keq}
\end{eqnarray}
where $\phi$ is positive, will then solve \eq{eq:c1}-\eq{eq:c2}
if and only if the   function $\phi$ and the vector field $W$
solve the equations
 \beq \div_{\tthreeg}(\tcalD
W + \tsigma)=\frac{n-1}{n}\phi^{q+2} \tD \tau
 \;,
 \label{vacmom:noncmc}
\eeq
\beq \label{vacham:noncmc} \laplaciano{\tthreeg} \phi -
\frac{1}{q(n-1)} R(\tthreeg)\phi + \frac{1}{q(n-1)}|\tsigma + \tcalD
W |^2_{\tthreeg} \phi^{-q-3}-\frac{1}{qn}\tau^2\phi^{q+1}=0\;. \eeq
We use the symbol $\tD$ to denote the covariant derivative of
$\tthreeg$; $\tcalD $ is the \emph{conformal Killing operator}:
\beq \tcalD W_{ab} = \tD_a W_b + \tD_b W_a -
\frac{2}{n}\tthreeg_{ab} \tD_c W^c \;.
\label{LWeq}
 \eeq
 Vector fields $W$  annihilated by
$\calD$ are called \emph{conformal Killing vector fields}, and are
characterized by the fact that they generate (perhaps local)
conformal diffeomorphisms of $(\hyp, \threeg)$.
 The semi-linear scalar
equation (\ref{vacham:noncmc}) is often referred to as the
\emph{Lichnerowicz equation}.

Equations (\ref{vacmom:noncmc})-(\ref{vacham:noncmc}) form a
{determined} system of equations for the  $(n+1)$ functions $(\phi,
W)$. The operator $\div_{\tthreeg}(\tcalD \;\cdot)$ is a linear,
formally self-adjoint, elliptic operator on vector fields. What
makes the study of the system
(\ref{vacmom:noncmc})-(\ref{vacham:noncmc}) difficult in general is
the nonlinear coupling between the two equations.

The explicit choice of \eq{threegeq}-\eq{Keq} is motivated by the
two identities (for $\tilde \threeg=\phi^q \threeg$)
\beq R({\tilde \threeg}) = -\phi^{-q-1} (q(n-1)\Delta_\threeg \phi -
R({\threeg})\phi)
 \;,
  \label{ConfReq} \eeq
{where $q=\frac{4}{n-2}$, which is the unique exponent that
does not lead to supplementary $|D \phi|^2$ terms in
(\ref{ConfReq}),} and
\beq D_{\tilde
\threeg}^a (\phi^{-2} B_{ab}) = \phi^{-q-2} D_{\threeg}^a
B_{ab} \label{confdiveq} \eeq
which holds for any trace-free
tensor $B$. Equation (\ref{ConfReq}) is the well known identity
relating the scalar curvatures of two conformally related
metrics.

In the space-time evolution $(\mcM, \fourg)$ of the initial
data set $(\hyp, \threeg,  K)$,
the function $\tau= \tr_\threeg K  $ 
is the mean curvature of the hypersurface $\hyp\subset\mcM$.  The
assumption that the mean curvature function $\tau$ is
 constant on $\hyp$ significantly simplifies the analysis of the vacuum constraint
equations because it decouples equations  (\ref{vacmom:noncmc}) and
(\ref{vacham:noncmc}). One can then attempt to solve
(\ref{vacmom:noncmc}) for  $W$, and then solve the Lichnerowicz
equation (\ref{vacham:noncmc}).

Existence and uniqueness of solutions of this problem for constant
mean curvature (``CMC") data  has been studied extensively. For
compact manifolds this was exhaustively analysed by
Isenberg~\cite{Jimconstraints}, building upon a large amount of
previous work~\cite{Lich44,OMYork73,York74,CBY}; the proof was
simplified  by Maxwell in~\cite{Maxwell:compact}. If we let
${\mathcal Y}([\threeg])$ denote the Yamabe invariant of the
conformal class $[\threeg]$ of metrics determined by $\threeg$
(see~\cite{LeeParker}), the result reads as follows:

\begin{Theorem}[\cite{Jimconstraints}]
\label{TJimC}
 Consider a smooth conformal initial data set
$(\tthreeg, \tsigma, \tau)$ on a compact manifold $\hyp$, with
constant $\tau$. Then there always exists a solution $W$ of
\eq{vacmom:noncmc}. Setting $\sigma= \tcalD W + \tsigma$,  the
existence, or not, of a positive solution $\phi$ of the
Lichnerowicz equation is shown in Table~\ref{tIsenberg}.
\bigskip
\begin{table}[ht]
\label{tIsenberg}
\begin{tabular}{|c||c|c|c|c|}
\hline
&$\sigma \equiv 0, \tau=0$& $\sigma \equiv 0, \tau\not=0$& $\sigma \not\equiv 0, \tau=0$& $\sigma \not\equiv 0, \tau\not=0$\\
\hline \hline
${\mathcal Y}([\threeg])<0$&No&Yes&No&Yes\\
\hline
${\mathcal Y}([\threeg])=0$&Yes&No&No&Yes\\
\hline
${\mathcal Y}([\threeg])>0$&No&No&Yes&Yes\\
\hline
\multicolumn{5}{c}{}\\
\end{tabular}
 \caption{Existence of solutions in the conformal method for CMC data on compact manifolds.}
\end{table}
%
%
%
%
%
%
\end{Theorem}
More recently, work has been done on analyzing these equations for
metrics of low
differentiability~\cite{Choquet-Bruhat:safari,Maxwell:compact}; this
was motivated in part by recent work  on the evolution problem for
``rough initial
data"~\cite{KlainermanRodnianski:r1,KlainermanRodnianski:r2,KlainermanRodnianski:r3,SmithTataru:sharp}.
Exterior boundary value problems for the constraint equations, with
nonlinear boundary conditions motivated by  black holes, were
considered in~\cite{Maxwell:AH,Dain:AH}.

The conformal method easily extends to  CMC constraint
equations for some non-vacuum initial data, e.g. the
Einstein-Maxwell system~\cite{Jimconstraints} where one obtains
results very similar to those of Theorem~\ref{TJimC}. However,
other important examples, such as the Einstein-scalar field
system~\cite{CBIP1,CBIP2,CBIP-Leray,HPP}, require more effort
and are not as fully understood.

Conformal data close to being CMC  (e.g. via a smallness
assumption on $|\nabla\tau|$) are usually referred to as
``near-CMC". Classes of near-CMC conformal data solutions have
been constructed~\cite{IM,CBIM91,ICA07,VinceJim:noncmc} and
there is at least one example of a non-existence theorem
\cite{IsenbergOMurchadha} for a class of near-CMC conformal
data. However, due to the non-linear coupling in the system
(\ref{vacmom:noncmc})-(\ref{vacham:noncmc}),  the question of
existence for unrestricted choices of the mean curvature $\tau$
appears to be significantly more difficult, and until recently
all results assumed strong restrictions on the gradient of
$\tau$. The first general result in this context  is due to
Holst, Nagy, and Tsogtgerel~\cite{HNT07,HNT08}, who construct
solutions with freely specified mean curvature in the presence
of matter. In~\cite{Maxwell08}, Maxwell provides a sufficient
condition, with no restrictions on the mean curvature, for the
conformal method to generate solutions to the vacuum constraint
equations on compact manifolds.  As an application, Maxwell
demonstrates the existence of a large class of solutions to the
\emph{vacuum} constraint equations with freely specified mean
curvature. These results together represent a significant
advance in our understanding of how the conformal method may be
used to generate solutions of the vacuum constraint equations.
However the existence question for generic classes of large
conformal data remains wide open; compare~\cite{MaxwellNonCMC}
for some new results. \opp{nonCMC}

The analysis of the conformal constraint equations
(\ref{vacmom:noncmc})-(\ref{vacham:noncmc})  discussed above
proceeds either via the method of sub- and super-solutions (which is
a barrier argument exploiting the maximum principle), or a
perturbation or fixed point method. In~\cite{HPP} Hebey, Pacard and
Pollack used   the mountain pass lemma to analyse Lichnerowicz-type
equations   arising in certain cases of the Einstein-scalar field
system. Such arguments may conceivably prove useful in studying
(\ref{vacmom:noncmc})-(\ref{vacham:noncmc}) for general $\tau$'s.

A natural question  is whether the set of  solutions to the
constraint equations   forms a manifold. This was first
considered by Fisher and Marsden~\cite{FischerMarsdenHI}, who
provided a Fr\'echet manifold structure; Banach manifold
structures have been obtained in~\cite{ChDelayHilbert}, and a
Hilbert manifold structure (for asymptotically flat initial
data sets) in~\cite{bartnik:phase}.

In~\cite{BartnikIsenberg} the reader will find a presentation
of alternative approaches to constructing solutions of the
constraints, covering work done up to 2003.

\bams{%
\subsubsection{The constraint equations on asymptotically
flat manifolds}
 \label{ssceafm}
Recall that asymptotically flat initial data sets have been
defined in Section~\ref{Saf}. There is a large number of
well-established results concerning the existence of CMC and
``near CMC" solutions of the Einstein constraint equations on
asymptotically flat manifolds \cite{Cantor77, ChIY, CBY,
CBChristodoulou, YCB:GRbook, Maxwell:AH, Maxwell:rough}. We
mention here only that there is a precise conformally invariant
criterion   sufficient  to prove the existence of a positive
solution to the Lichnerowicz
equation relative to a given set of asymptotically flat conformal data~\cite{Cantor77,Maxwell:AH}%
\footnote{Maxwell~\cite{Maxwell:AH} uncovered an error in
Cantor's definition of the invariant~\cite{Cantor77} and
provided the correct definition.}.
}

\subsection{Mass inequalities}
 \label{smineq}
 Among the deepest results in mathematical general relativity
 are the global mass inequalities for asymptotically flat
 manifolds. Those have been discussed extensively in the
 existing
 literature~\cite{LeeParker,BrayNotices,ChBray,BraySchoen,SchoenCatini,SchoenClay},
 and therefore will only be given the minimum amount of
 attention, as needed for the remaining purposes of this work.

\subsubsection{The Positive Mass Theorem} \label{sSPMT}
Using the coordinate system of \eq{falloff1}, one defines the
{\em Arnowitt-Deser-Misner~\cite{ADM} mass}  of $(\hyp,
\threeg)$ of an asymptotically flat end  as
\begin{equation}
\label{mass}
m=\frac{1}{16\pi}\lim_{r\rightarrow\infty}\int_{S_r}\sum_{i,j}\left(\frac{\partial
\threeg_{ij}}{\partial x^i} - \frac{\partial \threeg_{ii}}{\partial
x^j}\right)\,dS_i
 \; .
\end{equation}
Here $S_r$ is the coordinate sphere at radius $r$ and
$dS_i=\partial_i\rfloor d\mu$, and $d\mu$ is the Riemannian
volume form of $\threeg$.  The factor $16\pi$ is  a matter of
convention and is natural in space-dimension three. The
integral converges to a finite, coordinate-independent limit
if, for some $\alpha>\frac{n-2}{2}$,
\begin{equation}
\label{massdecay}
|\threeg_{ij}-\delta_{ij}|\leq cr^{-\alpha},\qquad|\partial \threeg|\leq cr^{-\alpha-1}
\qquad\mbox{and}\qquad R(\threeg)\in L^1(\Sext),
\end{equation}
with those conditions being essentially optimal~\cite{Bartnik86,ChErice}.

For time-symmetric initial data  the vacuum constraint
equations (\ref{eq:c1})-(\ref{eq:c2}) reduce to the condition
that the metric $\threeg$ be scalar-flat, i.e.\ $R(\threeg)=0$.
On the other hand, if one considers time-symmetric data for a
non-vacuum space-time, then from (\ref{eq:c2}) we see that
the scalar curvature is twice the energy density of  the matter  fields. The
non-negativity of $R$, assuming a vanishing cosmological constant
$\Lambda$, is then a consequence of the dominant energy condition
for initial data (which follows from \eq{DECsts}),
\bel{DEC}
 \rho \ge |J|_\threeg
 \;,
\ee
where $\rho$ and $J$ are defined in \eq{eq:c1}-\eq{eq:c2}; see
also Section~\ref{ssEMOTSs}.
One checks that the dominant energy condition  \eq{DECsts} holds on $\mcM$ if and only if
\eq{DEC} holds
relative to each spacelike hypersurface in $\mcM$.

Now, the ADM mass is thought to represent the total mass of the
system as viewed on $\hyp$, which contains contributions from
the matter fields, the gravitational field, as well as their
binding energy. The long-standing question of its positivity
was resolved by Schoen and Yau~\cite{SchoenYauPMT1} in
dimension three, and is now known as the Positive Mass Theorem:

\begin{Theorem}
 \label{Tposmasman}
Let $(\hyp,\threeg)$ be an asymptotically flat Riemannian
manifold  with nonnegative scalar curvature. Suppose that
either $\hyp$ is spin, or the dimension $n\le 7$, or that
$\threeg$ is conformally flat. Then the total ADM mass $m$
satisfies $m\geq0$, with equality if and only if
$(\hyp,\threeg)$ is isometric to Euclidean space $(\R^n,
\delta)$.
\end{Theorem}

As remarked in the introduction, this theorem, and its
generalizations, stands as one of the cornerstones of
mathematical relativity.  Accessible introductions to the
positive mass theorem may be found in~\cite{LeeParker,
ChBray,ChBeijing,SchoenCatini}.  The restriction on the dimension arises
from the use of area minimizing
hypersurfaces~\cite{SchoenCatini}, which are known to sometimes
possess singularities in higher dimensions, {and it is expected
that positivity is true in all dimensions}.
  \opp{opppet2}
The Positive Mass Theorem was proven in all dimensions for
conformally flat manifolds by Schoen and
Yau~\cite{SchoenYauKlienian} by a different argument, and in
all dimensions for spin manifolds by Witten~\cite{WittenPMT}
(see also~\cite{ParkerTaubes,Bartnik86,ChBlesHouches}). The
result was generalized in~\cite{SchoenYauELMGR, SchoenYauPMT2}
(compare~\cite{SchoenClay}) to asymptotically flat initial data
sets $(\hyp,\threeg, K, \mathcal F)$ satisfying the dominant
energy condition (\ref{DEC}).

\subsubsection{Riemannian Penrose Inequality}
\label{sSpenroseinequality}
An important generalization of the Positive Mass Theorem is given by the Riemannian Penrose Inequality.

\begin{Theorem}
\label{RPenroseIneq} Let $(\hyp,\threeg)$ be a complete,
smooth, asymptotically flat $3$-manifold with nonnegative
scalar curvature with total mass $m $ and which has an
outermost minimal surface $\Sigma_0$ of area $A_0$. Then
\begin{equation}
\label{penroseineq}
m\geq\sqrt{\frac{A_0}{16\pi}}
 \;,
\end{equation}
with equality if and only if $(\hyp,\threeg)$ is isometric to
the Schwarzschild metric $(\R^3\setminus\{0\}, (1 +
\frac{m}{2|x|})^4 \delta)$ outside their respective outermost
minimal surfaces.
\end{Theorem}

Theorem~\ref{RPenroseIneq} was first proved by  Huisken and
Ilmanen~\cite{HuiskenIlmanen} under the restriction that $A_0$
is connected, or assuming instead that $A_0$ is   the area of
the largest connected component of $\Sigma_0$. The version
above, with a proof that uses completely different methods, is
due to Bray~\cite{BrayPenroseIneq}. The proofs are  beautiful
applications of geometric flows to a fundamental problem in
relativity. A number of accessible reviews has been written on
these important results, to which we refer the interested
reader~\cite{BrayNotices,BrayICM,ChBray,BraySchoen,MalecPenrose}.
{A generalization of Theorem~\ref{RPenroseIneq} to dimensions
$n\le 7$ has been established in~\cite{BrayLee}.}

One expects that some form of \eq{penroseineq} holds for
general relativistic initial data sets $(\threeg,K)$ satisfying
the dominant energy condition. A suggestion how one could prove
this has been put forward by  Bray and Khuri
in~\cite{BrayKhuri,BrayKhuri2},
compare~\cite{Frauendiener:Penrose,MMS,BHMS,CarrascoMars}. An
inequality in the spirit of \eqref{penroseineq}, but involving
some further geometric constants, has been proved by
Herzlich~\cite{mh-inegalite-penrose} in the Riemannian case.

\subsubsection{Quasi-local mass} \label{ssql}
In the context of asymptotically flat space-times, there are
well-defined global notions of mass and energy, and these are
central to the celebrated positive mass theorem discussed in
Section \ref{sSPMT}.  One would, however, like to have a
well-defined useful \emph{local} notion of mass or energy, with
natural properties -- e.g.,  monotonicity -- that one has in
other physical theories. Such a definition has been elusive
despite a great deal of effort by many people and this remains
an important open problem.
 \opp{oppqlm}
We refer the reader to the Living Reviews article by
Szabados~\cite{SzabadosLR} for a survey, and note that there
have been interesting recent mathematical developments in the
area~\cite{WangYau,WangYau5,WangYau6,ShiTam3,BrayMiao,MST}, not
described in the currently available version
of~\cite{SzabadosLR}.

\subsection{Applications of gluing techniques}
\label{ssagt}
 Over  the past 25 years, ``gluing techniques" have become a
standard tool in geometric analysis. Since the construction by
Taubes of self-dual Yang-Mills connections on
four-manifolds~\cite{Taubes82}, which played a crucial role in
Donaldson's construction of exotic smooth structures in
four-dimensions~\cite{DonaldsonJDG83}, gluing has been applied
in important ways across a very wide range of areas. What
gluing typically refers to is a construction in which solutions
of a nonlinear partial differential equation or system, which
correspond to some geometric quantity of interest, e.g.\
self-dual connections, are fused together to create new
solutions.  This is done by a mix of geometry and analysis in
which one ultimately studies the linearization of the relevant
PDEs, and in most cases one has to overcome analytic degeneracy
introduced in the gluing procedure.  Thus, from an analytic
point of view, gluing should be regarded as a singular
perturbation method. Part of the usefulness of the technique
lies in the fact that, away from the small set about which one
fuses the two solutions, the new solution is very close to the
original ones.  The fact that the original solutions  are {\em
usually} not exactly preserved is a reflection of the fact that
the relevant equations satisfy a unique continuation property:
any two solutions which agree on an open set must agree
everywhere.  This is a well known property for, say, a scalar
semi-linear elliptic equation. The constraint equations,
however, are {\it underdetermined}.  As explained below,
Corvino and Schoen have introduced gluing techniques which
exploit this and show the failure (to a high degree) of unique
continuation. This leads to localized gluings of initial data
sets, which has proven very valuable.

\subsubsection{The linearized constraint equations and KIDs}
\label{sslcekids}
The starting point of gluing constructions
for the constraint equations is the  linearization of these
equations about a given solution $(\hyp, \threeg, K)$. We let
$\mathcal{P}^*_{(\threeg, K)}$ denote the $L^2$ adjoint of the
linearization of the constraint equations at this solution.
Viewed as an operator acting on a scalar function $N$ and a
vector field $Y$, $\mathcal{P}^*_{(\threeg, K)}$ takes the
explicit form~\cite{ChDelay}
 \be
\label{adlince} \mathcal{P}^*_{(\threeg, K)}(N,Y)=\left(
\begin{array}{l}
2(\D_{(i}Y_{j)}-\D^lY_l g_{ij}-K_{ij}N+\tr K\; N g_{ij})\\
 \\
\D^lY_l K_{ij}-2K^l{}_{(i}\D_{j)}Y_l+
K^q{}_l\D_qY^lg_{ij}\\
\;-\Delta N g_{ij}+\D_i\D_j N
+(\D^{p}K_{lp}g_{ij}-\D_lK_{ij})Y^l\\
\;-N \Ricc(g)_{ij} +2NK^l{}_iK_{jl}-2N (\tr \;K) K_{ij}
\end{array}
\right)\;.\ee Now this operator does not, on first inspection, appear to be very ``user friendly".
However, our immediate concern is solely with its kernel, and the pairs  $(N,Y)$ which lie in its
kernel have a very straightforward geometric and physical characterization.  In particular, let
$\Omega$ be an open subset of $\hyp$. By definition, the set of ``KIDs" on $\Omega$, denoted
$\mcK(\Omega)$, is the set of all solutions of the equation
 \begin{equation}
 \mathcal{P}^*_{(\gamma, K)|_{\Omega}} (N,Y)=0\;.
 \label{NoKIDs}
 \end{equation}
Such a solution $(N,Y)$, if nontrivial, generates a space-time
Killing vector field in the domain of dependence of $(\Omega,
\threeg|_{\Omega}, K|_{\Omega})$~\cite{Moncrief75}; compare
Section~\ref{Saf}.

From a geometric point of view one expects that solutions with
symmetries should be rare.  This was made rigorous
in~\cite{CHBeignokids},  where it is  shown that the generic
behaviour among solutions of the constraint equations is the
absence of KIDs on any open set. On the other hand, one should
note that  essentially every explicit solution has symmetries.
In particular, both the flat initial data for Minkowski space,
and the initial data representing the constant time slices of
Schwarzschild have KIDs.

\subsubsection{Corvino's result}
As we have already pointed out, the Einstein constraint
equations form an underdetermined system of equations, and as
such, it is unreasonable to expect that they (or their
linearizations) should satisfy the unique continuation
property. In 2000, Corvino established a gluing result for
asymptotically flat metrics with zero scalar curvature which
dramatically illustrated this point~\cite{Corvino}.  In the
special case when one considers initial data with vanishing
second fundamental form $K\equiv 0$, the momentum constraint
equation (\ref{eq:c1}) becomes trivial and the Hamiltonian
constraint equation (\ref{eq:c1}) reduces to simply
$R(\threeg)=0$, i.e. a scalar flat metric.  Such initial data
sets are referred to as ``time-symmetric" because the
space-time obtained by evolving them possesses a time-reversing
isometry which leaves the initial data surface fixed. Beyond
Euclidean space itself, the constant time slices of the
Schwarzschild space-time form the most basic examples of
asymptotically flat, scalar flat manifolds.  One long-standing
open problem~\cite{SchoenYauLectures, Bartnikopen} in the field
had been whether there exist scalar flat metrics on $\R^n$
which are not globally spherically symmetric but which are
spherically symmetric in a neighborhood of infinity and hence,
by Birkhoff's theorem, Schwarzschild there.

Corvino resolved this by showing that he could deform any asymptotically flat, scalar flat metric to
one which is exactly Schwarzschild  outside of a compact set.

\begin{theorem}[\cite{Corvino}]
\label{Corvino} Let $(\hyp,\threeg)$  be  a smooth Riemannian
 manifold with zero scalar
curvature containing an asymptotically flat end
$\Sext=\{|x|>r>0\}$. Then  there is a $R>r$ and a smooth metric
$\bar\threeg$ on $\hyp$ with zero scalar curvature such that
$\bar\threeg$ is equal to $\threeg$ in $\hyp\setminus\Sext$ and
$\bar\threeg$  coincides  on $\{|x|>R\}$ with the metric
induced on a standard time-symmetric slice in the Schwarzschild
solution. Moreover the mass  of $\bar \threeg$ can be made
arbitrarily close to that of  $\threeg$ by choosing $R$
sufficiently large.
\end{theorem}

Underlying this result is a gluing construction where the
deformation has compact support.  The ability to do this is a
reflection of the underdetermined nature of the constraint
equations.  In this setting, since $K\equiv 0$, the operator
takes a much simpler form, as a two-covariant tensor valued
operator acting on a scalar function $u$ by
\[
\mathcal{P}^*u = -(\Delta_\threeg u)\threeg + \Hess_\threeg u - u\Ric(\threeg)\;.
\]
An elementary illustration of how an underdetermined system can
lead to compactly supported solutions is given by the
construction of compactly supported transverse-traceless
tensors on $\R^3$ in Appendix B of~\cite{CorvinoAHP}  (see
also~\cite{BeigTT, DainFriedrich}).

An additional challenge in proving Theorem \ref{Corvino} is the
presence of KIDs on the standard slice of the Schwarzschild
solution.  If the original metric had ADM mass $m(\threeg)$, a
naive guess could be that the best fitting Schwarzschild
solution would be the one with precisely the same mass. However
the mass, and the coordinates of the center of mass, are in
one-to-one correspondence with obstructions arising from KIDs.
To compensate for this co-kernel in the linearized problem,
Corvino uses these ($n+1$ in dimension $n$) degrees of freedom
as effective parameters in the geometric construction. The
final solution can be chosen to have its ADM mass arbitrarily
close to the initial one.

The method uncovered in Corvino's thesis has been applied and
extended in a number of important ways.  The ``asymptotic
simplicity" model for isolated gravitational systems proposed
by Penrose~\cite{penrose:asymptotic} has been very influential.
This model assumes existence of smooth conformal completions to
study global properties of asymptotically flat space-times. The
question of existence of such vacuum space-times was open until
Chru\'sciel and Delay~\cite{ChDelay2}, and subsequently
Corvino~\cite{CorvinoAHP}, used this type of gluing
construction to demonstrate the existence of infinite
dimensional families of  vacuum initial data sets which evolve
to asymptotically simple space-times. The extension of the
gluing method to non-time-symmetric data was done
in~\cite{ChDelay,CorvinoSchoen2}. This allowed for the
construction of space-times which are exactly Kerr outside of a
compact set,  as well as showing that one can specify other
types of useful asymptotic behavior.

\subsubsection{Conformal gluing}
\label{IMPgluing} In~\cite{IMP1}, Isenberg, Mazzeo and Pollack
developed a gluing construction for initial data sets
satisfying certain natural non-degeneracy assumptions.  The
perspective taken there was to work within the conformal
method, and thereby establish a gluing theorem for solutions of
the determined system of PDEs given by (\ref{vacmom:noncmc})
and (\ref{vacham:noncmc}).  This was initially done only within
the setting of constant mean curvature initial data sets and in
dimension $n=3$ (the method was extended to all higher
dimensions in~\cite{IMaxP}).  The construction of~\cite{IMP1}
allowed one to combine initial data sets by taking a connected
sum of their underlying manifolds, to add wormholes (by
performing codimension $3$ surgery on the underlying,
connected, 3-manifold) to a given initial data set, and to
replace arbitrary small neighborhoods of  points in an initial
data set with asymptotically hyperbolic ends.

In~\cite{IMP2} this gluing construction was extended to only
require that the mean curvature be constant in a small
neighborhood of  the point about which one wanted to perform a
connected sum. This extension enabled the authors to show that
one can replace an arbitrary small neighborhood of a generic
point in any initial data set with an asymptotically {\em flat}
end.  Since it is easy to see that CMC solutions of the vacuum
constraint equations exist on any compact manifold~\cite{Witt},
this leads to the following result which asserts that there are
no topological obstructions to asymptotically flat solutions of
the constraint equations.

\begin{theorem}[\cite{IMP2}]
Let $\hyp$ be any closed $n$-dimensional manifold, and $p\in\hyp$. Then $\hyp\setminus\{p\}$
admits an asymptotically flat initial data set satisfying the vacuum constraint equations.
\end{theorem}

\subsubsection{Initial data engineering}
\label{IDE} The gluing constructions of~\cite{IMP1}
and~\cite{IMP2} are performed using a determined elliptic
system provided by the conformal method, which necessarily
leads to a global deformation of the initial data set, small
away from the gluing site. Now, the ability of the Corvino
gluing technique to establish compactly supported deformations
invited the question of whether these conformal gluings could
be localized.  This was answered in the affirmative
in~\cite{ChDelay} for CMC initial data under the additional,
generically satisfied~\cite{CHBeignokids}, assumption that
there are no KIDs in a neighborhood of the gluing site.

In~\cite{CIP:CMP,CIP:PRL}, this was substantially improved upon
by combining the gluing construction of~\cite{IMP1} together
with the Corvino gluing technique of~\cite{Corvino, ChDelay2},
to obtain a localized gluing construction in which the only
assumption is the absence of KIDs near points.  For a given
$n$-manifold $\hyp$ (which may or may not be connected)  and
two points $p_a\in \hyp$, $a=1,2$, we let $\tilde\hyp$ denote
the manifold obtained by replacing small geodesic balls around
these points by a neck $S^{n-1}\times I$. When $\hyp$ is
connected this corresponds to performing codimension $n$
surgery on the manifold.  When the points $p_a$ lie in
different connected components of $\hyp$, this corresponds to
taking the connected sum of those components.

\begin{Theorem}[\cite{CIP:CMP,CIP:PRL}]
\label{Tlgluingv} Let $(\hyp, \threeg, K)$ be a smooth vacuum
initial data set, with $M$ not necessarily connected, and
consider two open sets $\Omega_a\subset \hyp$, $a=1,2$, with
compact closure and smooth boundary such that
$$\mbox{ the set
of KIDs, $\mcK(\Omega_a)$, is trivial.}
$$
Then for all $p_a\in \Omega_a$, $\epsilon >0$, and $k\in \N$
there exists a smooth vacuum initial data set
$(\tilde\hyp,\threeg(\epsilon),K(\epsilon))$ on the glued
manifold $\tilde\hyp$ such that
$(\threeg(\epsilon),K(\epsilon))$ is $\epsilon$-close to
$(\threeg, K)$ in a $C^k\times C^k$ topology away from
$B(p_1,\epsilon)\cup B(p_2,\epsilon)$. Moreover
$(\threeg(\epsilon),K(\epsilon))$ coincides with $(\threeg, K)$
away from $\Omega_1\cup \Omega_2$.
\end{Theorem}

This result is sharp in the following sense: first note that,
by the positive mass theorem, initial data for Minkowski
space-time cannot locally be glued to anything which is
non-singular and vacuum. This meshes with the fact that for
Minkowskian initial data, we have $\mcK(\Omega)\ne\{0\}$ for
any open set $\Omega$. Next,  recall that by the results
in~\cite{CHBeignokids}, the no-KIDs hypothesis in Theorem
\ref{Tlgluingv} is generically satisfied. Thus, the result can
be interpreted as the statement that for generic vacuum initial
data sets the local gluing can  be performed around arbitrarily
chosen points $p_a$. In particular the collection of initial
data with generic regions $\Omega_a$ satisfying the hypotheses
of Theorem~\ref{Tlgluingv} is not empty.

The proof of Theorem \ref{Tlgluingv} is a mixture of gluing
techniques developed in~\cite{IMaxP,IMP1} and those
of~\cite{CorvinoSchoen2, Corvino, ChDelay}. In fact, the proof
proceeds initially via a generalization of the analysis
in~\cite{IMP1} to compact manifolds with boundary.  In order to
have CMC initial data near the gluing points, which the
analysis based on~\cite{IMP1} requires,  one makes use of the
work of Bartnik~\cite{bartnik:variational} on the plateau
problem for prescribed mean curvature spacelike hypersurfaces
in a Lorentzian manifold.

\bams{
Arguments in the spirit of those of the proof of
Theorem~\ref{Tlgluingv} lead to the construction of
\emph{many-body initial data}~\cite{CCI,CCI2}: starting from
initial data for $N$ gravitating isolated systems, one can
construct a new initial data set which comprises isometrically
compact subsets of each of the original systems, as large as
desired, in a distant configuration.
}

An application of the gluing techniques
concerns the question of the existence of CMC slices in
space-times with compact Cauchy surfaces.
In~\cite{bartnik:cosmological}, Bartnik showed that there exist
maximally extended, globally hyperbolic solutions of the
Einstein equations \textit{with dust} which admit no CMC
slices. Later, Eardley and Witt (unpublished) proposed a scheme
for showing that similar vacuum solutions exist, but their
argument was incomplete. It turns out that these ideas can be
implemented using Theorem \ref{Tlgluingv}, which leads to:

\begin{Corollary}\cite{CIP:CMP,CIP:PRL}
\label{noCMCslices} There exist maximal globally hyperbolic
vacuum space-times with compact Cauchy surfaces which contain
no compact spacelike hypersurfaces with constant mean
curvature.
\end{Corollary}

\bams{%
Compact Cauchy surfaces with constant mean curvature are useful
objects, as the existence of one such surface gives rise to a
unique foliation  by such surfaces~\cite{BF78}, and hence a
canonical choice of time function (often referred to as CMC or
York time). Foliations by CMC Cauchy surfaces have also been
extensively used in numerical analysis to explore the nature of
cosmological singularities. Thus the demonstration that there
exist space-times with no such surfaces has a negative impact
on such studies.

One natural question  is the extent to which space-times with
no CMC slices are common among solutions to the vacuum Einstein
equations with a fixed spatial  topology.\opp{oppmax} It is
expected that the examples constructed
in~\cite{CIP:CMP,CIP:PRL} are not isolated. In general, there
is a great deal of flexibility (in the way of free parameters)
in the local gluing construction. This can  be used to produce
one parameter families of distinct sets of  vacuum initial data
which lead to space-times as in Corollary \ref{noCMCslices}.
What is less obvious is how to prove that all members of  these
families give rise to {\em distinct} maximally extended,
globally hyperbolic vacuum space-times.

A deeper question is whether a sequence of space-times
which admit constant mean curvature Cauchy surfaces may
converge, in a strong topology, to one which admits no such
Cauchy surface. (See~\cite{Bartnik84,
bartnik:cosmological,Gerhardt83} for general criteria leading
to the existence of CMC Cauchy surfaces.)

\subsubsection{Non-zero cosmological constant}
 \label{ssSnonzeroLambda}
Gluing constructions have also been carried out  with a
non-zero cosmological constant~\cite{ChPollack,CPP,ChDelayAH}.
One aim is to construct space-times which coincide,  in the
asymptotic region, with the corresponding black hole models. In
such space-times one has complete control of the geometry in
the domain of dependence of the asymptotic region, described
there by the Kottler metrics \eq{basemet}. For time-symmetric
slices of these space-times, the constraint equations reduce to
the equation  for constant scalar curvature $R=2\Lambda$.
Gluing constructions have been previously carried out  in this
context, especially in the case of $\Lambda>0$, but
in~\cite{ChPollack,CPP,ChDelayAH} the emphasis is on gluing
with compact support, in the spirit of Corvino's thesis and its
extensions already discussed.

The  time-symmetric slices  of the $\Lambda>0$ Kottler
space-times provide ``Delaunay" metrics (see~\cite{ChPollack}
and references therein), and the main result  of
\cite{ChPollack,CPP} is the construction of large families of
metrics with exactly Delaunay ends. When $\Lambda<0$ the focus
is on asymptotically hyperbolic metrics with constant negative
scalar curvature. With hindsight, within the family of Kottler
metrics with $\Lambda \in \R$ (with $\Lambda =0$ corresponding
to the Schwarzschild metric), the gluing in the $\Lambda>0$
setting is technically easiest, while that with $\Lambda<0$ is
the most difficult. This is due to the fact that for
$\Lambda>0$ one deals with one linearized operator with a
one-dimensional kernel; in the case $\Lambda=0$ the kernel is
$(n+1)$-dimensional; while for $\Lambda<0$ one needs to
consider a one-parameter family of operators with
$(n+1)$-dimensional kernels.
}

 \section{Evolution}
 \label{SE}

In Section~\ref{SEe} we have seen that solutions of the vacuum
Einstein equations can be constructed by solving a Cauchy
problem. It is then of interest to inquire about the global
properties of the resulting space-times.
\altreverse{%
A key example to keep in mind in this context is provided
by the Taub--NUT metrics~\cite{Taub,NUT}, which exhibit
incomplete geodesics within compact sets, closed causal curves,
inequivalent extensions of maximal globally hyperbolic regions
(to be defined shortly) and inequivalent conformal boundary
completions at infinity. In particular they provide an example
of \emph{non-uniqueness} of solutions of the Cauchy problem, a
problem that we address in
the next section.%
}{
As an illustration of
what can happen, we start with a short description of a
solution of the vacuum Einstein equations with interesting
dynamical properties, the Taub -- Newman Unti Tamburino
(Taub--NUT) metric.

\subsection{Taub--NUT space-times}
 \label{ATNUT}
The Taub--NUT metrics provide interesting examples of
pathological behavior: incomplete geodesics within compact
sets, closed causal curves, inequivalent extensions of maximal
globally hyperbolic regions (to be defined shortly) and
inequivalent conformal boundary completions at infinity. For
all these reasons they are a must  in any introductory
discussion of general relativity.

The Taub--NUT metrics~\cite{Taub,NUT} are solutions of the
vacuum Einstein equations on space-time manifolds $\mcM_I$ of
the form $$ \mcM_I:=I\times S^3\;,$$ where $I$ is an interval.
They take the form~\cite{Misner}
\begin{eqnarray}
& -U^{-1}dt^2  +(2\ell )^2U\sigma^2_1 + (t^2 + \ell ^2)(\sigma^2_2 +
\sigma^2_3)\,, \label{MBianchi} & \\ & \displaystyle U(t) = -1 +
{2(mt + \ell ^2)\over
  t^2 + \ell ^2}\,. \label{UBianchi}
\end{eqnarray}
Here $\ell $ and $m$ are real numbers with $\ell  > 0$.
Further, the one-forms $\sigma_1$, $\sigma_2$ and $\sigma_3$
form a basis for the set of  left-invariant one-forms on
$SU(2)\approx S^3$: If
$$i_{S^3}:S^3\to \R^4$$ is the standard embedding of $S^3$ into
$\R^4$, then one can take
 \beal{sigmaforTN} \sigma_1 &=& 2i_{S^3}^*(x\,dw - w\,
dx + y\, dz - z\, dy)\;,
\\\nonumber
\sigma_2 &=& 2i_{S^3}^*(z\,dx - x\, dz + y\, dw - w\, dy)\;,
\\
\sigma_3 &=& 2i_{S^3}^*(x\,dy - y\, dx + z\, dw - w\, dz)\;.
 \nonumber
\eea

The metric \eq{MBianchi} is invariant under a left
$SU(2)$--action, and a further $U(1)$ action consisting of
right-rotations of $\sigma_2$ and $\sigma_3$ amongst
themselves, so that the connected component of the identity of
the group of isometries of $\fourg$ is $SU(2)\times U(1)$.

The function $U$ always has two zeros,
$$U(t)= \frac {(t_+-t)(t-t_-)}{t^2+\ell ^2}\;,$$ where $$ t_\pm :=
m\pm \sqrt{m^2+\ell ^2}\;.$$ It follows that $I$ has to be
chosen so that $t_\pm\not\in I$. The space-time
$(\mcM_{(t_-,t_+)},\fourg)$ will be referred to as \emph{the
Taub space-time}~\cite{Taub}. It is not very difficult to show
that the Taub space-times exhaust the collection of maximal
globally hyperbolic vacuum space-times evolving from
$SU(2)\times U(1)$--invariant Cauchy data on $S^3$.

The significance of these metrics to strong cosmic censorship,
which we are about to discuss, stems from the following
observation
(see~\cite{Misner,ChImaxTaubNUT,ChConformalBoundary}):

\begin{Theorem}
\label{TsccTN} Consider the Taub space-time
$(\mcM_{(t_-,t_+)},\fourg)$ defined above.
  Then
 \begin{enumerate}
 \item $(\mcM_{(t_-,t_+)},\fourg)$ is maximal globally hyperbolic.
 \item There exists an uncountable number of analytic, vacuum,
 simply connected,
 non-equivalent extensions of $(\mcM_{(t_-,t_+)},\fourg)$.
 \end{enumerate}
\end{Theorem}
}

 \subsection{Strong cosmic censorship}
 \label{sSScc}
The \emph{strong cosmic censorship (SCC) problem} concerns
{predictability}:  Indeed, a fundamental requirement of
physically relevant  equations is that solutions should be
\emph{uniquely determined by initial data}. So it is important
to inquire about predictability in general relativity.

In other words, we would like to know whether or not the
solutions provided by Theorem~\ref{Tglobinspace} are unique.
Now, it is easy to see that there can be no uniqueness unless
some restrictions on the development are imposed: consider for
example $(-\infty,1)\times \R^n$, $\R\times \R^n$ and
$(\R\times \R^n)\setminus\{(1,\vec 0)\}$ equipped with the
obvious flat metric. All three space-times contain the
spacelike surface $(\{0\}\times \R^n,\delta, 0)$, where
$\delta$ is the Euclidean metric on $\R^n$. The first two are
globally hyperbolic developments of the given initial data, but
the third is not, as it is not globally hyperbolic. And
obviously these are not isometric: e.g,.  the second is
geodesically complete, while the other two are not. So to
guarantee uniqueness some further conditions are needed.

The key existence \emph{and uniqueness} theorem in this context
is due to Choquet-Bruhat and
Geroch~\cite{ChoquetBruhatGeroch69}
(compare~\cite{CBY,HE,Chorbits}). Some terminology is needed: a
space-time $(\mcM,\fourg)$
 is said to be a \emph{development} of an initial data set
 $(\hyp,\threeg,K)$ if there exists an embedding $i:\hyp\to \mcM$
 such that $i(\hyp)$ is a Cauchy surface for $(\mcM,\fourg)$, with
$$
 i^* \fourg = \threeg\;,
$$
and with $K$ being the pull-back to $\hyp$ of the extrinsic
curvature tensor (second fundamental form) of $i(\hyp)$.
{ We will say that a development \mcmg\ is \emph{maximal globally
hyperbolic} if the following implication holds: if $\psi:\mcM\to
\mcM'$ is an isometric embedding of $\mcM$ into $(\mcM',\fourg')$,
and if $\psi(\mcM)\ne \mcM'$, then $\mcM'$ is \emph{not} globally
hyperbolic.

Note that we are not imposing any field equations on
$(\mcM',\fourg')$. One could similarly define a notion of
maximality within the class of \emph{vacuum} space-times, {but
this would lead to a weaker statement  of the Choquet-Bruhat --
Geroch theorem, which for simplicity is presented in the smooth
case:

\begin{theorem}[Existence of maximal globally hyperbolic developments~\cite{ChoquetBruhatGeroch69}]
 \label{TCBG}
For any smooth vacuum initial data $(\hyp,\threeg,K)$   there
exists a unique, up to isometric diffeomorphism, vacuum
development $(\mcM,\fourg)$,   which is inextendible in the
class of smooth globally hyperbolic  Lorentzian manifolds.
 \end{theorem}

This theorem can be thought of as the equivalent of the usual
ODE theorem of existence of maximal solutions. The
generalization is, however, highly non-trivial because while
the proof for ODEs deals with subsets of $\R$,
Theorem~\ref{TCBG} deals with manifolds which are dynamically
obtained by patching together local solutions. The main
difficulty is to prove that the patching leads to a Hausdorff
topological space. The argument makes use of Lorentzian
causality theory, which in turn relies heavily on $C^2$
differentiability of the metric. 
To obtain a version of Theorem~\ref{TCBG} with
lower differentiability, as in Theorem~\ref{Tglobinspace} or in
Remark~\ref{Rlcp}, one would need to show that the relevant
parts of causal theory can be repeated in the wider setting.
 \opp{oppmgh}

While Theorem~\ref{TCBG} is highly satisfactory, it does not
quite prove what one wants, because \emph{uniqueness is claimed
in the globally hyperbolic class only}.  But {we have  seen in
Theorem~\ref{TsccTN}} that there exist vacuum space-times with
non-unique extensions of a maximal globally hyperbolic region.
In such examples the space-time $(\mcM,\fourg)$ of
Theorem~\ref{TCBG} is unique in the class of globally
hyperbolic space-times, but it can be extended in more than one
way to strictly larger vacuum solutions. In such cases the
extension always takes places across a \emph{Cauchy horizon},
as defined in Section~\ref{SDoD}

So one cannot expect uniqueness in general. However, it has
been suggested by Penrose~\cite{PenroseSCC} that non-uniqueness
happens only in very special circumstances. The following
result of Isenberg and
Moncrief~\cite{VinceJimcompactCauchyCMP,VinceJimcompactCauchy,VinceJimHigh}
(compare~\cite{HIW}) indicates that this might indeed be the
case:

\begin{Theorem}
\label{TMI} Let $(\mcM,\fourg)$ be a vacuum analytic space-time
containing an analytic compact Cauchy horizon $\mcH$. If the
null geodesics threading $\mcH $ are closed, then the Cauchy
horizon is a \emph{Killing horizon}; in particular the isometry
group of $(\mcM,\fourg)$ is at least one-dimensional.
\end{Theorem}

The hypotheses of analyticity, compactness, and closed
generators are of course highly restrictive. In any case it is
conceivable that some kind of local {isometries need to occur
in space-times with Cauchy horizons when those conditions are
not imposed; indeed,} all known examples have this property.
But of course existence of local isometries is a highly
non-generic property, {even when vacuum equations are
imposed~\cite{CHBeignokids}, so a version of Theorem~\ref{TMI}
without those undesirable hypotheses would indeed establish
SCC.}

Whether or not Cauchy horizons require Killing vector fields, a
loose mathematical formulation of strong cosmic censorship, as
formulated in~\cite{ChrCM} following Moncrief and
Eardley~\cite{EM} and Penrose~\cite{PenroseSCC}, is the
following:
 \opp{oppkch}
\bean
 && \mbox{\em Consider the collection of initial data for, say,
vacuum or electro--vacuum}
 \\&& \nonumber
 \mbox{\em  space-times, with the initial data
surface $\hyp$ being compact, or with}
 \\&& \nonumber
 \mbox{\em  asymptotically flat initial data
$(\hyp,\threeg,K)$ . For generic such data the}
 \\&&
 \mbox{\em  maximal globally hyperbolic development is
inextendible.}
 \nonumber
 \eea

Because of the difficulty of the strong cosmic censorship
problem, a full understanding of the issues which arise in this
context seems to be completely out of reach at this stage.
There is therefore some interest in trying to understand that
question under various restrictive hypotheses, {\em e.g.},
symmetry. The simplest case, of spatially homogeneous
space-times, has turned out to be surprisingly difficult,
because of the intricacies of the dynamics of some of the
Bianchi models discussed in Section~\ref{sSBA}, and has been
settled in the affirmative in~\cite{ChRendall} {(compare
Theorem~\ref{TRingBianchi} below).}

\subsubsection{Gowdy toroidal metrics}
 \label{ssSGtm}
The next simplest case is that of \emph{Gowdy metrics} on $\To
^3:=S^1\times S^1 \times S^1$: by definition,
\begin{equation}\label{eq:gowdy}
 g=e^{(\tau-\lambda)/2}(-e^{-2\tau}d\tau^{2}+d\theta^2)
+e^{-\tau}[e^{P}d\sigma^2+2e^{P}Qd\sigma d\delta+
(e^{P}Q^2+e^{-P})d\delta^2],
\end{equation}
where $\tau\in\mathbb{R}$ and $(\theta,\sigma,\delta)$ are
coordinates on $\To ^{3}$, with the functions $P,Q$ and
$\lambda$ depending only  on $\tau$ and $\theta$.  The  metric
of a maximal globally hyperbolic $U(1)\times U(1)$--symmetric
vacuum space-time with $\To^ 3$--Cauchy surfaces can be
globally written~\cite{ChANOP} in the form \eq{eq:gowdy}
provided that the
\emph{twist constants} vanish:
\bel{twist}
 c_a:=\epsilon_{\alpha\beta\gamma\delta} X_1^\alpha
X_2^\beta \nabla^\gamma X_a^\delta=0\;, \qquad a=1,2
 \;,
\ee
where the $X_a$'s are the Killing vectors generating the
$U(1)\times U(1)$ action. The condition $c_1=c_2=0$ is
equivalent to the requirement that the family of planes
$\mathrm{span}\{X_1, X_2\}^\perp$ is integrable.

For metrics of the form \eq{eq:gowdy}, the Einstein vacuum equations
become a set of \emph{wave-map} equations
\begin{eqnarray}
P_{\tau\tau}-e^{-2\tau}P_{\theta\theta}-
e^{2P}(Q_{\tau}^2-e^{-2\tau}Q_{\theta}^2) & = & 0, \label{eq:g1}\\
Q_{\tau\tau}-e^{-2\tau}Q_{\theta\theta}
+2(P_{\tau}Q_{\tau}-e^{-2\tau}P_{\theta}Q_{\theta}) & = &
0,\label{eq:g2}
\end{eqnarray}
which are supplemented by ODE's for the function
$\lambda$:
\begin{eqnarray} \lambda_{\tau} & = &
P_{\tau}^{2}+e^{-2\tau}P_{\theta}^{2}+
e^{2P}(Q_{\tau}^{2}+e^{-2\tau}Q_{\theta}^{2}),\label{eq:gc1}\\
\lambda_{\theta} & = &
2(P_{\theta}P_{\tau}+e^{2P}Q_{\theta}Q_{\tau}). \label{eq:gc2}
\end{eqnarray}
Here we write $P_\tau$ for $\partial_\tau P$, etc.
 }

Initial data on $\To^3$ for $P$ and $Q$ have to satisfy an
integral constaint,
\begin{equation}\label{eq:constraintG}
\int_{S^{1}}(P_{\theta}P_{\tau}+e^{2P}Q_{\theta}Q_{\tau})d\theta=0
 \;,
\end{equation}
which is a consequence of \eq{eq:gc2} and of periodicity in
$\theta$.   The metric function $\lambda$ is obtained by
integrating (\ref{eq:gc1})-(\ref{eq:gc2}). Global existence of
solutions to (\ref{eq:g1})-(\ref{eq:g2}) was proved
in~\cite{Moncrief:Gowdy} when the initial data are given on a
hypersurface $\{\tau=\const\}$, and in~\cite{ChANOP} for
general $U(1)\times U(1)$--symmetric Cauchy surfaces.

The question of SCC in this class of metrics has been settled
by Ringstr\"om, who proved that the set of smooth initial data
for Gowdy models on $\To^ 3$ that do \emph{not} lead to the
formation of Cauchy horizons contains a set which is open and
dense within the set of all smooth initial data. More
precisely, Ringstr\"om's  main result
(see~\cite{RingstroemGowdy,RingstroemSCC} and references
therein) is the following:

\begin{Theorem}
 \label{TRingstroem}
Let $\tau_0\in \R$ and let
$\mathcal{S}=\{\left(Q(\tau_0),P({\tau_0}),Q_\tau{(\tau_0)},P_\tau{(\tau_0)}\right)\}$
be the set of smooth initial data  for \eq{eq:g1}-\eq{eq:g2}
satisfying (\ref{eq:constraintG}). There is a subset
$\mathcal{G}$ of $\mathcal{S}$ which is open with respect to
the $C^{2}\times C^{1}$ topology, and dense with respect to the
$C^{\infty}$ topology, such that the space-times of the form
(\ref{eq:gowdy}) corresponding to initial data in $\mathcal{G}$
are causally geodesically complete in one time direction,
incomplete in the other time direction, and the Kretschmann
scalar, $R_{\alpha\beta\gamma\delta}
R^{\alpha\beta\gamma\delta}$, becomes unbounded in the
incomplete direction of causal geodesics.
\end{Theorem}

This result does indeed establish SCC in this class of metrics:
to see that the resulting space-times are inextendible in the
category of $C^3$ manifolds with $C^2$ Lorentzian metrics, note
that the existence of any such extension would imply existence
of geodesics which are incomplete in the original space-time,
and along which every curvature scalar is bounded.

Theorem~\ref{TRingstroem} is complemented by the results
in~\cite{ChCh2,MR82b:83024,CIM}, where infinite dimensional
families of (nongeneric) solutions which \emph{are extendible}
across a Cauchy horizon are constructed.

The key to the understanding of the global structure of the
Gowdy space-times is the analysis of the behavior of the
functions $P$ and $Q$ as $\tau \to \pm \infty$. The asymptotic
behavior of those functions, established by Ringstr\"om, can
then be translated into statements about the behavior of the
space-time geometry as those limits are approached. A central
element of the proof is the existence of a \emph{velocity
function}
$$
 v(\theta):= \lim_{\tau\to\infty}\sqrt{P_\tau^2+e^{2P} Q_\tau^2
 }
 \;.
$$
Essential steps in Ringstr\"om's analysis are provided by the work
on Fuchsian PDEs of Kichenassamy and
Rendall~\cite{KichenassamyRendall,Rendall:2000ih}, as well as the
study of the action of Geroch transformations by Rendall and
Weaver~\cite{RendallWeaver} (compare~\cite{ChCh2}). See
also~\cite{ChLake} for the related problem of an exhaustive
description of Cauchy horizons in those models.

\subsubsection{Other $U(1)\times U(1)$-symmetric models}
 \label{ssSUoneUone}
The existence of two Killing vectors is also compatible with $S^3$,
$L(p,q)$ (``lens" spaces), and $S^1\times S^2$ topologies. Thus, to
achieve a complete understanding of the set of spatially compact
initial data with precisely two Killing vectors one needs to extend
Ringstr\"om's analysis to those cases.
 \opp{oppgalileo}
There is an additional difficulty that arises because of the
occurrence of axes of symmetry, where the ($1+1$)--reduced equations
have the usual singularity associated with polar coordinates.
Nevertheless, in view of the analysis by Christodoulou and
Tahvildar-Zadeh~\cite{CT93,CT293} (see also~\cite{ChANOP}), the
global geometry of \emph{generic} maximal globally hyperbolic
solutions with those topologies is reasonably well understood. This
leads one to expect that one should be able to achieve a proof of
SCC in those models using simple abstract arguments, but this
remains to be seen.

Recall, finally, that general models with two Killing vectors $X_1$
and $X_2$  on $\To^ 3$ have non-vanishing \emph{twist constants}
\eq{twist}. The Gowdy metrics are actually ``zero measure" in the
set of all $U(1)\times U(1)$ symmetric metrics on $\To^ 3$ because
$c_a\equiv 0$ for the Gowdy models. The equations for the resulting
metrics are considerably more complicated when the $c_a$'s do not
vanish, and only scant rigorous information is available on the
global properties of the associated
solutions~\cite{BCIM,IW,Rendall:1996nu}. It seems urgent to study
the dynamics of those models,
 \opp{oppgalileo2}
as they are expected to display~\cite{BIW} ``oscillatory behavior"
as the singularity is approached, in the sense of
Section~\ref{sSBKL}. Thus, they should provide the simplest  model
in which to study this behavior.

\subsubsection{Spherical symmetry}
\label{ssDafermos} One could think that the simplest possible
asymptotically flat model for studying the dynamics of the
gravitational field will be obtained by requiring spherical
symmetry, since then the equations should reduce to wave
equations in only two variables, $t$ and $r$. Unfortunately,
for vacuum space-times this turns out to be useless for this
purpose because of Birkhoff's theorem~\cite{Birkhoff23}, which
asserts that spherically symmetric vacuum metrics are static.
So, if one wishes to maintain spherical symmetry, supplementary
fields are needed. The case of a scalar field was studied in a
series of  {intricate} papers over 13 years by Christodoulou,
beginning with~\cite{ChristodoulouCMP86}, and culminating
in~\cite{ChristodoulouAnnals99}  with the verification of the
strong  cosmic censorship conjecture within the model.
Christodoulou further established ``weak cosmic censorship" in
this class, an issue to which we return in the next section,
and exhibited non-generic examples for which the conclusions of
these conjectures fail~\cite{demetrios:scalar10}.

The situation changes when electromagnetic fields are
introduced. The analysis by
Dafermos~\cite{DafermosCBH1,DafermosCBH2} of the spherically
symmetric Einstein-Maxwell-scalar field equations yields a
detailed picture of the interior of the black hole for this
model, in terms of initial data specified on the event horizon
and on an ingoing null hypersurface. When combined with the
work by Dafermos and
Rodnianski~\cite{dafermos:rodnianski:price} on Price's law, one
obtains the following global picture: initial data with a
compactly supported scalar field, and containing a trapped
surface (see Section~\ref{sstmts} below), lead to space-times
which \emph{either} contain a degenerate (extremal) black hole,
\emph{or} develop a Cauchy horizon, with a space-time metric
that can be continued past this horizon \emph{in a $C^0$, but
not $C^1$ manner}. It seems that not much is known about the
properties of the degenerate solutions, which are presumably
non-generic; it would be of interest to clarify
that.\opp{oppdafermos} {In any case, the work} shows that
strong cosmic censorship holds within the class of
nondegenerate solutions with trapped surfaces, at the $C^1$
level, leaving behind the perplexing possibility of continuous
extendability of the metric.

The reader is referred to~\cite{Andrev,ChrCM,RendallLiving} and
references therein for further reading on SCC.

\subsection{Weak cosmic censorship}
 \label{sSWcc}
The strong cosmic censorship conjecture is an attempt to
salvage predictability of Einstein's theory of gravitation.
There exists  a variant thereof which addresses the fact that
we do not seem to observe any of the singularities that are
believed to accompany gravitational collapse. The hope is then
that, generically, in asymptotically flat space-times, any
singular behavior that might form as a result of gravitational
collapse, such as causality violations,   lack of
predictability, or curvature singularities, will be
\emph{clothed by an event horizon}. For this, one introduces
the notion of \emph{future null infinity}, which is an
idealized boundary attached to space-time that represents,
loosely speaking, the end points of null geodesics escaping to
infinity. (In stationary situations this is closely related to
the region $\Mext$ of \eq{docdef}.) The \emph{black hole event
horizon} is then the boundary of the past of null infinity;
compare \eq{hdefeh} and \eq{epm}. One then wishes the part of
the space-time that lies outside the black hole region to be
well-behaved and ``sufficiently large". This is the content  of
the \emph{weak cosmic censorship} conjecture, originally due to
Penrose~\cite{PenroseSCC},  as made precise by
Christodoulou~\cite{ChristodoulouCQG99}: \emph{for generic
asymptotically flat initial data, the maximal globally
hyperbolic  development has a complete future null
infinity}.\opp{oppwcc} Heuristically this means that,
disregarding exceptional sets of initial data, no singularities
are observed at large distances, even when the observations are
continued indefinitely. One should remark that, despite the
names, the strong and weak cosmic censorship conjectures are
logically independent; neither follows from the other. Note
also that some predictability of Einstein's theory would be
salvaged if strong cosmic censorship failed with weak cosmic
censorship being verified, since then the failure of
predictability would be invisible to outside observers.

Both cosmic censorship conjectures are intimately related to
the issue of \emph{gravitational collapse},  the dynamical
formation of black holes and singularities, first observed for
a homogeneous dust model by Oppenheimer and Snyder in
1939~\cite{OppenheimerSnyder}, visualized in Figure~\ref{FOS}.
\begin{figure}[t]
\begin{center} {
\psfrag{infty}{\Huge{$\infty$}}
\psfrag{singularity}{\huge{singularity}}
\psfrag{collapse}{\huge{collapsing matter}}
\resizebox{2.5in}{!}{\includegraphics{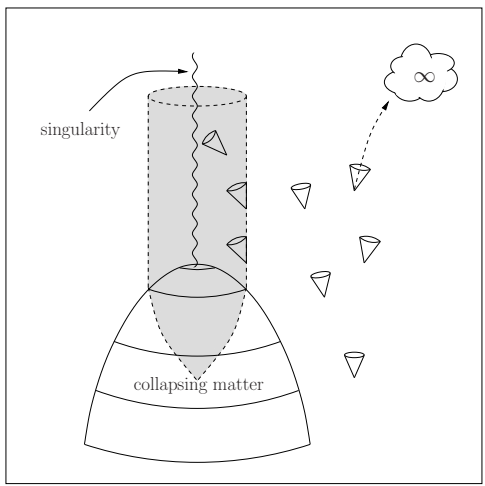}} }
\end{center}
\caption{Light-cones in the Oppenheimer-Snyder collapse.$^{\mbox{\scriptsize \ref{fCW}}}$\label{FOS}}
\end{figure}

So far the only complete analysis of weak cosmic censorship in
a field theoretical model is that of the spherically symmetric
scalar field model studied by
Christodoulou~\cite{ChristodoulouAnnals99,demetrios:scalar10},
already mentioned in Section~\ref{ssDafermos}.

\subsection{Stability of vacuum cosmological models}
 \label{sSvcm}
Not being able to understand the dynamics of all solutions, one
can ask whether some features of certain particularly important
solutions persist under small perturbations of initial data.
For example, will geodesic completeness still hold for
space-times arising from small perturbations of Minkowskian
initial data? Or, will   a global, all encompassing,
singularity persist under perturbations of Bianchi IX initial
data (see Section~\ref{sSBA}).  Such questions are the object
of stability studies.

\subsubsection{$U(1)$ symmetry}
 \label{ssSUone}
Our understanding of models with \emph{exactly one} Killing
vector is dramatically poorer than that of $U(1)\times U(1)$
symmetric space-times. Here one only has \emph{stability}
results, for small perturbations within the $U(1)$ isometry
class in the expanding direction (``away from the
singularity"):
 \opp{oppuone}
In~\cite{ChBCargese} Choquet-Bruhat considers  $U(1)$ symmetric
initial data $(\threeg,K)$ for the vacuum Einstein equations on
a manifold of the form $\hyp\times S^1$, where $\hyp$ is a
compact surface of {genus} ${ g>1}$. It is assumed that
$\tr_\threeg K$ is constant, and that $(\threeg,K)$ are
sufficiently close to $(\threeg_0,K_0)$, where $\threeg_0$ is a
product metric
$$\threeg_0=\gamma+dx^2
 \;,
$$
with $\gamma$ being a metric of constant Gauss curvature on
$\hyp$, and with $K_0$  proportional to $\threeg_0$. The sign
of the trace of $K_0$ determines an expanding time direction
and a contracting one. Under those conditions, Choquet-Bruhat
proves that the solution exists for an {infinite proper time}
in the expanding direction. The analysis builds upon previous
work by Choquet-Bruhat and Moncrief~\cite{01713418}, where a
supplementary polarization condition has been imposed. Not much
is known in the {contracting} direction in the $U(1)$-symmetric
models (see, however,~\cite{Isenberg:2002jg}), where ``mixmaster behavior"%
\footnote{See the discussion after Theorem~\ref{TRingAtt}, and
Section~\ref{sSBKL}.}
is expected~\cite{BKL,BGIMW}; compare~\cite{Berger:2000uf}.

\subsubsection{Future stability of hyperbolic models}
The proof of the above result bears some similarity to the
\emph{future stability} theorem of {Andersson and
Moncrief}~\cite{AndMon}, as generalized
in~\cite{AnderssonMoncriefStability2}, for spatially compact
hyperbolic models \emph{without any symmetries}. Those authors
consider initial data near  a negatively curved compact space
form, with the extrinsic curvature being close to a multiple of
the metric, obtaining future geodesic completeness in the
expanding direction. The control of the solution is obtained by
studying the Bel-Robinson tensor and its higher-derivatives
analogues. A striking  ingredient of the proof is  an
elliptic-hyperbolic system of equations, used to obtain local
existence in time~\cite{AnderssonMoncriefAIHP}.

\subsection{Stability of Minkowski space-time}
 \label{sSsmst}
\subsubsection{The Christodoulou-Klainerman proof}
 \label{sSCKp}
One of the flagship results in mathematical general relativity
is nonlinear stability of Minkowski space-time, first proved by
{Christodoulou and
Klainerman}~\cite{ChristodoulouKlainerman93}. One starts with
an asymptotically flat vacuum initial data set $(\threeg,K)$ on
$\R^3$. Under standard asymptotic flatness conditions, for
$(\threeg,K)$ sufficiently close to Minkowskian data,  the
maximal globally hyperbolic development $(\mcM,\fourg)$ of the
data contains a maximal hypersurface, i.e., a hypersurface
satisfying  $\mathrm{tr}_\threeg K=0$; this follows from the
results in~\cite{BCOM,christodoulou:murchadha,Bartnik84}. So
without loss of generality one can, in the small data context,
assume that the initial data set is maximal.

The precise notion of smallness needed for the
Christodoulou-Klainerman theorem is defined as follows: For
$p\in\Sigma\approx \R^3$, $a>0$,  consider the quantity
\be Q(a,p)=a^{-1}\int_\Sigma\{\sum^1_{\ell=0}
(d_p^2+a^2)^{\ell+1}|\nabla^\ell \mbox{Ric}|^2 +
\sum_{\ell=1}^2(d_p^2+a^2)^\ell |\nabla^\ell K|^2\}d\mu_g\ ,
\label{ChKlcond} \ee where $d_p$ is the geodesic distance
function from $p$, $\mbox{Ric}$ is the Ricci tensor of the
metric $g$, $d\mu_g$ is the Riemannian measure of the metric
$g$ and $\nabla$ is the Riemannian connection of $g$. Let
\[
Q_*=\inf_{a>0,\, p\in \Sigma} Q(a,p)\ .
\]
 Christodoulou and Klainerman prove causal geodesic completeness of
$(\mcM,\fourg)$ provided that $Q_*$ is sufficiently small. The
proof proceeds via an extremely involved bootstrap argument
involving a foliation by maximal hypersurfaces $\Sigma_t$
together with an analysis of the properties of an \emph{optical
function $u$}. {In the context here this is a solution of the
eikonal equation
$$
 \fourg^{\a\b} \pr_a
u\pr_\b u=0
 \;,
$$
the level sets $C_u$  of which intersect $\Sigma_t$ in spheres
which expand as $t$ increases. }
{We have:

\begin{Theorem}[{Global Stability of   Minkowski space-time}]
 \label{TKlNi}  {  Assume that
$(\hyp,\threeg,K)$ is  maximal, with%
\footnote{A function $f$ on
$\hyp$ is $o_k(r^{-\la})$ if $r^{\la+i}\nab ^i f \to 0$ as
$r\to \infty$ for all $i=0,\ldots k$. }
 \bel{Bieridecay} \threeg_{ij} = \de_{ij}+o_{3}(r^{-1/2}),\quad
K_{ij}=o_2(r^{-3/2})
 \;.
 \ee
There is an $\epsilon >0$
such that if $Q_*<\epsilon$, then the maximal globally
hyperbolic development $(\mcM,\fourg)$ of
$(\hyp,\threeg,K)$ is geodesically complete.
 }
\end{Theorem}

The above version of Theorem~\ref{TKlNi} is due to
Bieri~\cite{BieriZipser}. The original formulation
in~\cite{ChristodoulouKlainerman93} assumes moreover that
 \bel{ChKldecay} \threeg =(1+2M/r)\de+o_{4}(r^{-3/2}),\quad
K=o_3(r^{-5/2})
 \;,
 \ee
and in the definition \eq{ChKlcond} a term involving $K$ with
$\ell=0$ is added.}

By definition, asymptotically flat initial data sets approach
the Minkowskian ones as one recedes to infinity. One therefore
expects that at sufficiently large distances one should obtain
``global existence", in the sense that the maximal globally
hyperbolic development contains complete \emph{outgoing} null
geodesics. This question has been addressed by {Klainerman and
Nicol\`o}~\cite{KlainermanNicoloBook,KlainermanNicoloPeeling,klainerman:nicolo:review};
the reader is referred to those references for   precise
statements of the hypotheses made:

\begin{Theorem}
 \label{TKlN}
Consider an asymptotically flat initial data set
$(\hyp,\threeg,K)$, with maximal globally hyperbolic
development $(\mcM,
 \fourg)$. Let $\Omega_r$ denote a conditionally compact
 domain bounded by a coordinate sphere $S_r\subset \Mext$.
 There exists $R>0$ such that for all $r\ge R$ the generators
 of the boundary  $\partial J^+(\Omega_r)$ of the
domain of influence $J^+(\Omega_r)$  of $\Omega_r$ are future-complete.
\end{Theorem}

Both in~\cite{ChristodoulouKlainerman93} and in
\cite{KlainermanNicoloBook} one can find detailed information
concerning the behavior of null hypersurfaces as well as the rate at
which various components of the Riemann curvature tensor approach
zero along timelike and null geodesics.


\subsubsection{The Lindblad-Rodnianski proof}
 \label{sLoizelet}
A completely new proof of stability of Minkowski space-time has
been given by {Lindblad and
Rodnianski}~\cite{LindbladRodnianski,LindbladRodnianski2}. The
method provides less detailed asymptotic information than
\cite{ChristodoulouKlainerman93}
and~\cite{KlainermanNicoloBook} on various quantities of
interest but is much simpler. The argument is flexible enough
to allow the inclusion of a scalar field, or of a Maxwell
field~\cite{LoizeletCRAS,Loizelet:these}
(compare~\cite{BieriZipser} for an analysis along the lines of
the Christodoulou-Klainerman approach), and generalizes to
higher dimensions~\cite{CCL}. Further it allows the following,
less restrictive than that in
\cite{ChristodoulouKlainerman93,KlainermanNicoloBook},
asymptotic behavior of the initial data, for some $\alpha>0$:
\bel{dec} \threeg=(1+2m/r)\delta + O(r^{-1-\alpha})\;,\qquad K=
O(r^{-2-\alpha})
 \;.
\ee

Lindblad and Rodnianski consider the Einstein-Maxwell equations
with a neutral scalar field:
\begin{equation}
\label{em1}  R_{\mu\nu}-
\frac{R}{2}g_{\mu\nu} = T_{\mu\nu}+\hat{T}_{\mu\nu}
\;,
\end{equation}
with
$$ \hat{T}_{\mu\nu}=\pa_\mu\psi\, \pa_\nu\psi
 -\frac 12{g_{\mu\nu}} \big( g^{\alpha\beta} \pa_\alpha \psi\,
 \pa_\beta\psi \big)\;,\quad T_{\mu\nu}=2(F_{\mu\lambda}F_{\nu}^
{~\lambda}-\frac{1}{4}
g_{\mu\nu}F^{\lambda\rho}F_{\lambda\rho})
\;.
$$
The initial data are prescribed on ${\mathbb R}^{n}$, so
that the Maxwell field $F$ has a global potential $A$,
$F_{\mu\nu
}=\partial_{\mu}A_{\nu}-\partial_{\nu}A_{\mu}
$. The matter field equations read
\be
\label{em1x}
D_{\mu}F^{\mu\nu}=0\;, \quad \Box_g \psi = 0\;.
\end{equation}
The initial data, denoted by $(\mathring {\threeg}, \mathring
K, \mathring A, \mathring E, \psi_0, \psi_1)$ (where, roughly
speaking, $\mathring A$ is the initial value for the Maxwell
potential and $\mathring E$ is the initial value for the
electric field),
satisfy the following asymptotic conditions, for $r=\val{x}
\rightarrow \infty$, with some $\alpha
> 0$:
\begin{equation}\label{ci}
 \begin{array}{ll}
 \mathring{\threeg}_{ij}=\begin{cases}(1+\frac{2m}{r})\delta_{ij}+O(r^{-1-\alpha})\;,\;\text{for  }n=3\;,\\
 \delta_{ij}+O(r^{\frac{1-n}{2}-\alpha})\;,\;\text{for  }n\geq 4\;,\end{cases}\\
 \mathring{A}=O(r^{\frac{1-n}{2}-\alpha})\;, \quad
   \mathring{K}_{ij}=O(r^{-\frac{n+1}{2}-\alpha})\;, \quad \mathring{E}=O(r^{-\frac{n+1}{2}-\alpha})\;, \\
   \psi_0:=\psi\ini=O(r^{\frac{1-n}{2}-\alpha})\;,\quad
   \psi_1:=\der_t\psi\ini=O(r^{-\frac{n+1}{2}-\alpha})\;
\end{array}
\end{equation}
\bams{%
One also supposes that the Einstein-Maxwell constraint
equations hold initially:
 \begin{equation}\label{con} \left \{ \begin{array}{ll}
\mathring R=\mathring{K}^{ij}\mathring{K}_{ij}-\mathring{K}_{i}{}^{i}\mathring{K}_{j}{}^{j}+
2\mathring{E}_{i}\mathring{E}^{i} +\mathring{F}_{ij}\mathring{F}^{ij}+|\nab\psi_0|^2 + |\psi_1|^2\;,
\\     \nabla^{j}\mathring{K}_{ij}-\nabla_{i}\mathring{K}_{j}{}^{j}=\mathring{F}_{0j}\mathring{F}_i{}^{j} +\nab_i\psi_0 \,\psi_1
~~\;,
\\\nabla_i\mathring{F}^{0i}=0\;,
 \end{array} \right.\end{equation}
where $\mathring R$ is the scalar curvature of the metric $\mathring\threeg$.
}
 The strategy is to impose \emph{globally} the wave
coordinates condition
\begin{equation}\label{ch}
\partial_\mu\left(g^{\mu\nu}\sqrt{\val{\det g}}\right)=0\quad
\forall\nu=0,...,n,
\end{equation}
as well as the \emph{Lorenz gauge} for the electromagnetic
potential $A_\mu$,
\begin{align}\label{jl}
&\partial_\mu\left(\sqrt{\val{\det g}}A^\mu\right)=0\;.
\end{align}
\bams{%
Letting $\Box_g=g^{\mu\nu}\partial_\mu \partial_\nu$,
the dynamical equations take the form
\begin{equation}\label{em3}
 {\Box}_g\begin{pmatrix}h_{\mu\nu}^1\\A_\sigma\\
 \psi\end{pmatrix}
= \begin{pmatrix}S_{\mu\nu} -2\der_\mu\psi\der_\nu\psi\\S _\sigma\\0\end{pmatrix}
  -\begin{pmatrix} \Box_g h_{\mu\nu}^0\\0\\0\end{pmatrix}\;,\\
\end{equation}
where the source terms $S_{\mu\nu}$ and $S _\sigma$  are
bilinear in the derivatives of the fields, with coefficients
depending upon the metric.
}
The initial data are decomposed as
\begin{equation}\label{petith}
h^1_{\mu\nu}=h_{\mu\nu}-h^0_{\mu\nu}\;, \quad \mbox{with} \ h^0_{\mu\nu}(t)=\begin{cases}\chi (r/t)\chi (r)
\frac{2m}{r}\delta_{\mu\nu}\quad\text{for  }n=3\;,\\0\quad\text{for
}n\geq 4\;,\end{cases}
\ee
where  $\chi\in C^{\infty}$ is any function such that $\chi(s)$
equals $1$ for $s\geq 3/4$ and $0$ for $s\leq 1/2$.
The proof relies heavily on the structure of the nonlinear terms in wave coordinates.

Recall that there exists an extensive literature on wave
equations in $3+1$ dimensions with nonlinearities satisfying
the \emph{null
condition}~\cite{KlainermanGlobalCPAMTwo,Klainerman:null}, but
the nonlinearities that arise \emph{do not} satisfy that
condition. The argument works only because different components
of $h$ can be treated on a different footing. Indeed, for
solutions of the wave equation on Minkowski space-time, the
derivatives in directions tangent to the light cones decay
faster than the transverse ones. But the wave coordinates
condition \eq{ch} can be used to express the transverse
derivatives of some components of $g_{\mu\nu}$ in terms of
tangential derivatives of the remaining ones. This provides
control of the nonlinearities.

We also note the small data global existence results
of~\cite{LiChen,HormanderGlobal} on $\mathbb R^{n+1}$,
$n\geq4$,, and of~\cite{Christodoulou:global} for odd $n\ge 5$.
The structure conditions there are general enough to cover the
Einstein equations in wave coordinates, but the assumptions on
the fall-off of initial data exclude non-trivial solutions of
the vacuum
constraint equations%
\footnote{In~\cite{LiChen,HormanderGlobal} compactly supported
data are considered. In the theorem for general quasi-linear
systems given in~\cite{Christodoulou:global} the initial data
are in a Sobolev space which requires fall-off at infinity
faster than $r^{-n-3/2}$. In both cases the positive energy
theorem implies that such
initial data lead to Minkowski space-time.}%

We have:\opp{oppphg}

\begin{Theorem} \label{main} Consider smooth initial data  $(\mathring{\threeg},\mathring{K},\mathring{A},\mathring{E},\psi_0,\psi_1)$  on
$\mathbb{R}^{n}$, $n\ge 3$, satisfying \reff{ci} together with
the Einstein-Maxwell constraint equations. Let $N\in \N$,
suppose that $N_n:=N+[\frac{n+2}{2}]-2\geq 6+2[\frac{n+2}{2}]$,
and set
\begin{align}
E_{N_n,\gamma}(0)=&\sum_{0\leq\ii\leq N_n}\left(
\val{\val{(1+r)^{1/2+\gamma+\val{I}}\nabla\nabla^Ih_0^1}}_{L^2}^2 +
\val{\val{(1+r)^{1/2+\gamma+\val{I}}\nabla^I\mathring{K}}}_{L^2}^2
\right.\\\nonumber
&\left.+\val{\val{(1+r)^{1/2+\gamma+\val{I}}\nabla\nabla^I
\mathring{A}}}_{L^2}^2 +\val{\val{(1+r)^{1/2+\gamma+\val{I}}\nabla^I
\mathring{E}}}_{L^2}^2\right.\\\nonumber
&\left. +\|(1+r)^{1/2+\ga+|I|}\nabla\, \nabla^I \psi_0\|_{L^2}+
\|(1+r)^{1/2+\ga+|I|} \nabla^I \psi_1\|_{L^2}\right)\;.
\end{align}
Let $m$ be the ADM mass of $\mathring\threeg$. For every
$\gamma_0>0$ there exists $\varepsilon_0>0$, with
$\gamma_0(\varepsilon_0)\rightarrow 0 $ as
$\varepsilon_0\rightarrow 0$, such that if
\begin{equation}\label{small}
\sqrt{E_{N_n,\gamma}(0)}+m\leq\varepsilon_0,
\end{equation}
for some $\gamma>\gamma_0$,
then the maximal globally hyperbolic development of the initial
data is geodesically complete.
\end{Theorem}

\bams{%
The proof by Lindblad and Rodnianski  is an ingenious
and intricate analysis of the coupling between the
wave-coordinates gauge and the  evolution equations. One makes
a clever guess of how the fields decay in space and time,
encoded in the following weighted energy functional,
\begin{align}\label{ENkt}
{\mycal  E}^{\mbox{\scriptsize \,Matter}}_{N_n}(t)&=\sup_{0\leq\tau\leq
t}\sum_{Z\in{\mycal  Z},\val{I}\leq N_n}
\int_{\sum_{\tau}}\left(\val{\partial Z^Ih^1}^2+\val{\partial
Z^IA}^2 +\val{\partial Z^I\psi}^2\right)w(q)\,d^n x\;,
\end{align}
where
\beq\label{wq}
 w(q)=\begin{cases}
1+(1+\val{q})^{1+2\gamma},\quad q>0\;,\\
1+(1+\val{q})^{-2\mu},\quad q<0\;,
\end{cases}
\ee
with $q=r-t$ , $\mu>0\;$ and $\;0<\gamma<1\;$. Here ${\mycal
Z}$ denotes the collection of the following  generators of the
conformal Lorentz group, first used to study the decay of
solutions of the Minkowski wave equation by
Klainerman~\cite{KlainermanGlobalCPAMTwo}:
\begin{equation}\label{627}
\der_\alpha,\quad
  x_\alpha\frac{\der}{\der x^\beta}- x_\beta\frac{\der}{\der x^\alpha}\;,\quad
  x^\alpha \frac{\der}{\der x^\alpha}\;.
\end{equation}

One  argues by continuity: one chooses  $0<\delta<\frac{1}{4}$,
and one considers the maximal time  $T$ so that the inequality
\begin{equation}\label{hypo1}
\mathcal{E}_{N_n}^{\mbox{\scriptsize \,Matter}}(t)\leq
2C_{N_n}\varepsilon^2(1+t)^{2\delta}
\end{equation}
holds for $0\leq t\leq T$.   A sophisticated method, using the
Klainerman-Sobolev inequalities~\cite{KlainermanGlobalCPAMTwo},
together with a new weighted energy inequality, allows one to
show that \eq{hypo1}  holds for $0\leq t \leq T$ with a smaller
constant on the right-hand-side, contradicting maximality of
$T$, and thus proving global existence.
%
%
%

A long standing question in the study of asymptotically flat
space-times is that of the existence of an asymptotic expansion
of the metric as one recedes to infinity along outgoing null
cones,
see~\cite{Friedrich:Pune,FriedrichRadiative,penrose:scri}.
Neither the analysis
of~\cite{LindbladRodnianski,LindbladRodnianski2}, nor that
in~\cite{KlainermanNicoloBook,ChristodoulouKlainerman93},
provides sufficient information. It would be of interest to
clarify that.
}

\subsection{Towards stability of Kerr: wave equations on black hole backgrounds}
 \label{sswebhb}
\opp{oppkerr}

Since the pioneering work of  Christodoulou and Klainerman on
stability of Minkowski space-time, many researchers have been
looking into ways to address the question of stability of Kerr
black holes. The first naive guess would be to study stability
of Schwarzschild black holes, but those cannot be stable since
a generic small perturbation will introduce angular momentum.
The current strategy   is to study, as a first step, linear
wave equations on black hole backgrounds, with the hope that
sufficiently robust linear decay estimates can be bootstrapped
to produce a nonlinear stability proof. Due to limited space we
will not review those results, referring the reader to recent
important papers on the
subject~\cite{DafermosRodnianskiKerr,DafermosRodnianskiDecay,%
DafermosRodnianskiClay,TataruSchwarzschild,BlueMaxwell,%
BlueAndersson,BlueSterbenz,TataruDecay,TataruTohaneanu,DSS},
see also~\cite{FSYBAMS} and references therein.

 \subsection{Bianchi $A$ metrics}
 \label{sSBA}
Another important example of the intricate dynamical behavior
of  solutions of the Einstein equations is provided by the
\emph{``Bianchi $A$"} vacuum metrics. The key insight provided
by these space-times is the supposedly chaotic behavior of
large families of metrics in this class when a singularity is
approached. This dynamics has been conjectured to be generic;
we will return to this issue in Section~\ref{sSBKL}. As will be
seen shortly, in Bianchi $A$ space-times the Einstein evolution
equations reduce to a polynomial dynamical system on an
algebraic four-dimensional submanifold of $\R^5$. The spatial
parts of the Bianchi geometries provide a realization of six,
out of eight, homogeneous geometries in three dimensions which
form the basis of Thurston's geometrization program.

For our purposes here we define the Bianchi space-times as
\emph{maximal globally hyperbolic vacuum developments of
initial data which are invariant under a simply transitive
group of isometries}. Here the transitivity of the isometry
group is meant at the level of initial data, and \emph{not} for
the space-time. The name is a tribute to Bianchi, who gave the
classification of three dimensional Lie algebras which
underline the geometry here. These metrics split into two
classes, Bianchi $A$ and Bianchi $B$, as follows: Let $G$ be a
$3$-dimensional Lie group, and let $Z_{i}$, $i=1,2,3$ denote a
basis of left-invariant vector fields on $G$. Define the
\emph{structure constants} $\gamma_{ij}^k$ by the formula
$$
[Z_{i},Z_{j}]=\gamma_{ij}{}^{k}Z_{k}
 \;.
$$
The Lie algebra and Lie group are said to be of class $A$ if
$\gamma_{ik}{}^{k}=0$; class $B$ are the remaining ones. The
classes $A$ and $B$ correspond in mathematical terminology to
the \emph{unimodular} and \emph{non-unimodular} Lie algebras. A
convenient parameterization of the structure constants is
provided by the symmetric matrix $n^{ij}$ defined as
\begin{equation}\label{eq:ndef}
n^{ij}=\frac{1}{2}\gamma_{kl}{}^{(i}\epsilon_{}^{j)kl}.
\end{equation}
This implies $ \gamma_{ij}{}^{k}=\epsilon_{ijm}n^{km}$.
The Bianchi $A$ metrics are then divided into six classes,
according to the eigenvalues of the matrix $n^{ij}$, as
described in Table~\ref{table:bianchiA}.
\begin{table}[t]
\caption{Lie groups of Bianchi class
$A$.\label{table:bianchiA}}
\begin{tabular}{@{}ccccl}
Bianchi type & $n_{1}$ & $n_{2}$ & $n_{3}$ & Simply connected group\\
I                   & 0 & 0 & 0 & \mbox{\rm Abelian $\R^3$}\\
II                  & + & 0 & 0  & \mbox{Heisenberg } \\
V$\mathrm{I}_{0}$   & 0 & + & $-$  & \mbox{Sol (isometries of
  the Minkowski plane $\R^{1,1}$)} \\
VI$\mathrm{I}_{0}$  & 0 & + & +  & \mbox{universal cover of Euclid (isometries of $\R^2$)} \\
VIII                & $-$ & + & +  &  \mbox{universal cover of $SL(2,\R)$} \\
IX                  & + & + & +  & $SU(2)$ \\
\end{tabular}
\end{table}
For the Bianchi IX metrics, of particular interest to us here,
the group $G$ is $SU(2)$. Thus, the Taub metrics discussed in
Section~\ref{ATNUT} are members of the Bianchi IX family,
distinguished by the existence of a further $U(1)$ factor in
the isometry group.

Let $G$ be any three-dimensional Lie group, the Lie algebra of
which belongs to the Bianchi $A$ class. (The $G$'s are closely
related to the Thurston geometries,
see~Table~\ref{table:bianchiA};
compare~\cite[Table~2]{Andrev}).
%
%
Denote by $\{\sigma^i\}$ the basis dual to $\{Z_i\}$. It is not
too difficult to show that both A and B Bianchi metrics can be
globally written as
 \bel{bianchimet}
 \fourg = -dt^2 +\threeg_{ij}(t) \sigma^i \sigma^j
 \;, \qquad t\in I\;,
 \ee
with a maximal time interval $I$.

There are various ways to write the Einstein equations for a metric
of the form \eq{bianchimet}. We use the formalism introduced by
Wainwright and Hsu~\cite{WainwrightHsu}, which has proven to be
most useful for analytical
purposes~\cite{AlanMixmaster,Ringstroem1,Ringstroem2}, and we follow
the presentation in~\cite{Ringstroem1}.
Let
\[
\sigma_{ij}=K_{ij}-\frac{1}{3}\tr_\threeg K \threeg_{ij}
 \;, \qquad \theta:=\tr_\threeg K
 \;,
\]
be the trace-free part of the extrinsic curvature tensor of the
level sets of $t$.
%
Away from the (isolated) points at which $\theta$ vanishes, one
can introduce
\begin{eqnarray*}
\Sigma_{ij} & = & \sigma_{ij}/\theta
 \;,\\
N_{ij} & = & n_{ij}/\theta
 \;,\\
B_{ij} & = & 2N_{i}^{\ k}N_{kj}-N^{k}_{\ k}N_{ij}
 \;,\\
S_{ij} & = & B_{ij}-\frac{1}{3}B^{k}_{\ k}\delta_{ij}
 \;.
\end{eqnarray*}
Set
 $\Sp=\frac{3}{2}(\Sigma_{22}+\Sigma_{33})$ and
$\Sm=\sqrt{3}(\Sigma_{22}-\Sigma_{33})/2$.   If we let $N_{i}$ be
the eigenvalues of $N_{ij}$, the vacuum Einstein equations (a
detailed derivation of which can be found in~\cite{Ringstroem1})
lead to the following \emph{autonomous, polynomial} dynamical system
\begin{eqnarray}
\No' & = & (q-4\Sp)\No \;, \nonumber \\
\Nt' & = & (q+2\Sp +2\sqrt{3}\Sm)\Nt \;,  \nonumber \\
\Nth' & = & (q+2\Sp -2\sqrt{3}\Sm)\Nth  \;, \label{eq:whsu}\\
\Sp' & = & -(2-q)\Sp-3S_{+}  \nonumber\;,  \\
\Sm' & = & -(2-q)\Sm-3S_{-} \;, \nonumber
\end{eqnarray}
where a prime denotes derivation with respect to  a new time
coordinate $\tau$ defined by
\begin{equation}\label{eq:dtdtau}
\frac{d t}{d\tau}=\frac{3}{\theta}
 \;.
\end{equation}
Further,
\begin{eqnarray}
q & = & 2(\Sp^2+\Sm^2)\;,  \nonumber \\
S_{+} & = & \frac{1}{2}[(\Nt-\Nth)^2-\No(2\No-\Nt-\Nth)]\;,
\label{eq:whsudef}\\
S_{-} & = & \frac{\sqrt{3}}{2}(\Nth-\Nt)(\No-\Nt-\Nth)\;. \nonumber
\end{eqnarray}
The vacuum constraint equations reduce to one equation,
\begin{equation}
\Sp^2+\Sm^2+\frac{3}{4}[\No^2+\Nt^2+\Nth^2-2(\No\Nt+\Nt\Nth+\Nth\No)]
=1 \;. \label{eq:constraint}
\end{equation}
The points $(\No,\Nt,\Nth,\Sp,\Sm)$ can be classified according
to the values of $\No,\Nt,\Nth$ in the same way as the $n^i$'s
in Table \ref{table:bianchiA}. The sets $N_{i}>0$, $N_{i}<0$
and $N_{i}=0$ are invariant under the flow determined by
\eq{eq:whsu}, and one can therefore classify solutions to
(\ref{eq:whsu})-(\ref{eq:constraint}) accordingly. Bianchi IX
solutions correspond, up to symmetries of the system, to points
with all $N_{i}$'s positive, while for Bianchi $VIII$ solutions
one can assume that  two $N_{i}$'s are positive  and the third
is negative.

Points with $N_1=N_2=N_3=0$ correspond to Bianchi $I$ models.
The associated vacuum metrics were first derived by Kasner, and
take the form
\begin{equation}\label{eq:kasner}
d s^{2}=-d t^{2}+\sum_{i=1}^{3}t^{2p_{i}}d x^{i}\otimes d x^{i} \;,
\quad p_1+p_2+p_3=p_1^2+p_2^2+p_3^2=1
 \;.
\end{equation}
An important role in the analysis of \eq{eq:whsu} is played by
the \emph{Kasner circle}, defined as the set $\{q=2\}$. These
points belong to the configuration space, as determined by
\eq{eq:constraint}, for Bianchi I models, but the equation
$q=2$ is incompatible with \eq{eq:constraint} for Bianchi IX
metrics. Nevertheless, we shall see shortly that the Kasner
circle plays an essential role in the analysis of the Bianchi
IX dynamics.

The set $\Sm=0$, $\Nt=\Nth$, together with its permutations, is
invariant under the flow of
(\ref{eq:whsu})-(\ref{eq:constraint}). In the Bianchi IX case
these are the Taub solutions. In the Bianchi $VIII$ case the
corresponding explicit solutions, known as the NUT metrics,
have been found by Newman, Tamburino and Unti~\cite{NUT}, and
they exhibit properties similar to the Bianchi IX Taub
solutions discussed in Section~\ref{ATNUT}.

The \emph{$\omega$-limit} of  an orbit $\gamma$ of a dynamical
system is defined as the set of accumulation points of that
orbit. In~\cite{Ringstroem1,Ringstroem2}, Ringstr\"om proves
the following:

\begin{Theorem}
 \label{TRingAtt}
The $\omega$-limit set of each non-NUT Bianchi $VIII$ orbit
contains at least two distinct points  on the Kasner circle.
Similarly, non-Taub--NUT Bianchi IX orbits have at least three
distinct $\omega$-limit  points  on the Kasner circle.
\end{Theorem}

The picture which emerges from a numerical analysis of
\eq{eq:whsu} (see~\cite{BevMixmaster,CornishLevin} and
references therein) is the following: Every non-Taub--NUT
Bianchi IX orbit   approaches some point on the Kasner circle;
there it performs a ``bounce",  after which it eventually
approaches another point on the Kasner circle, and so on.
Theorem~\ref{TRingAtt} establishes the validity of this
picture. The numerical analysis further suggests that generic
orbits will have a dense $\omega$-limit set on the Kasner
circle; this is compatible with,  but
 \opp{oppdense}
does not follow from, Ringstr\"om's analysis. It has been
argued that the map which associates to each bounce the nearest
point on the Kasner circle possesses chaotic features; this is
at the origin of the ``mixmaster behavior" terminology,
sometimes used in this context.
\begin{figure}[t]
\begin{center}
\resizebox{!}{1.8in}{\includegraphics{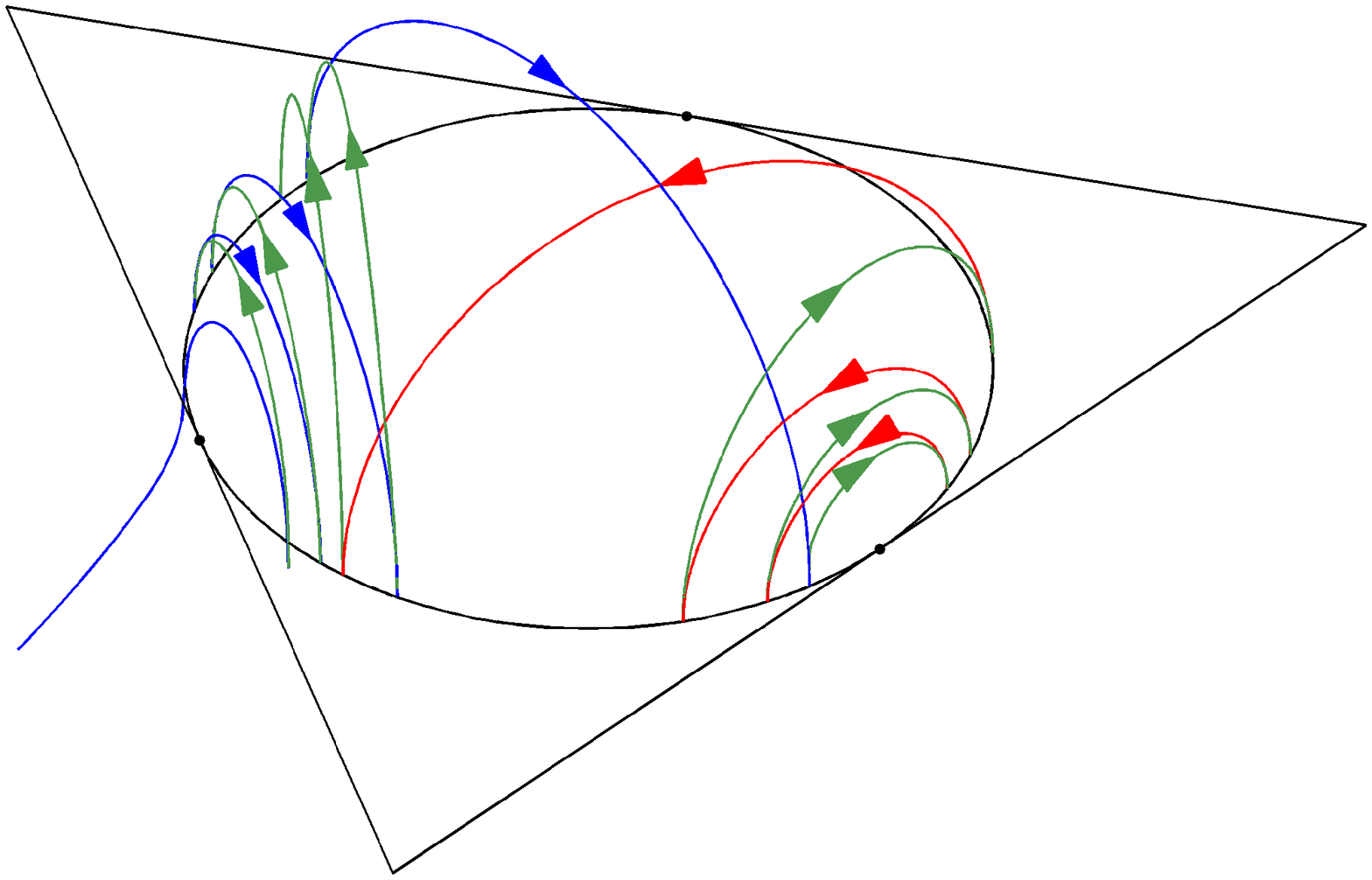}}%
%
\resizebox{!}{1.8in}{\includegraphics{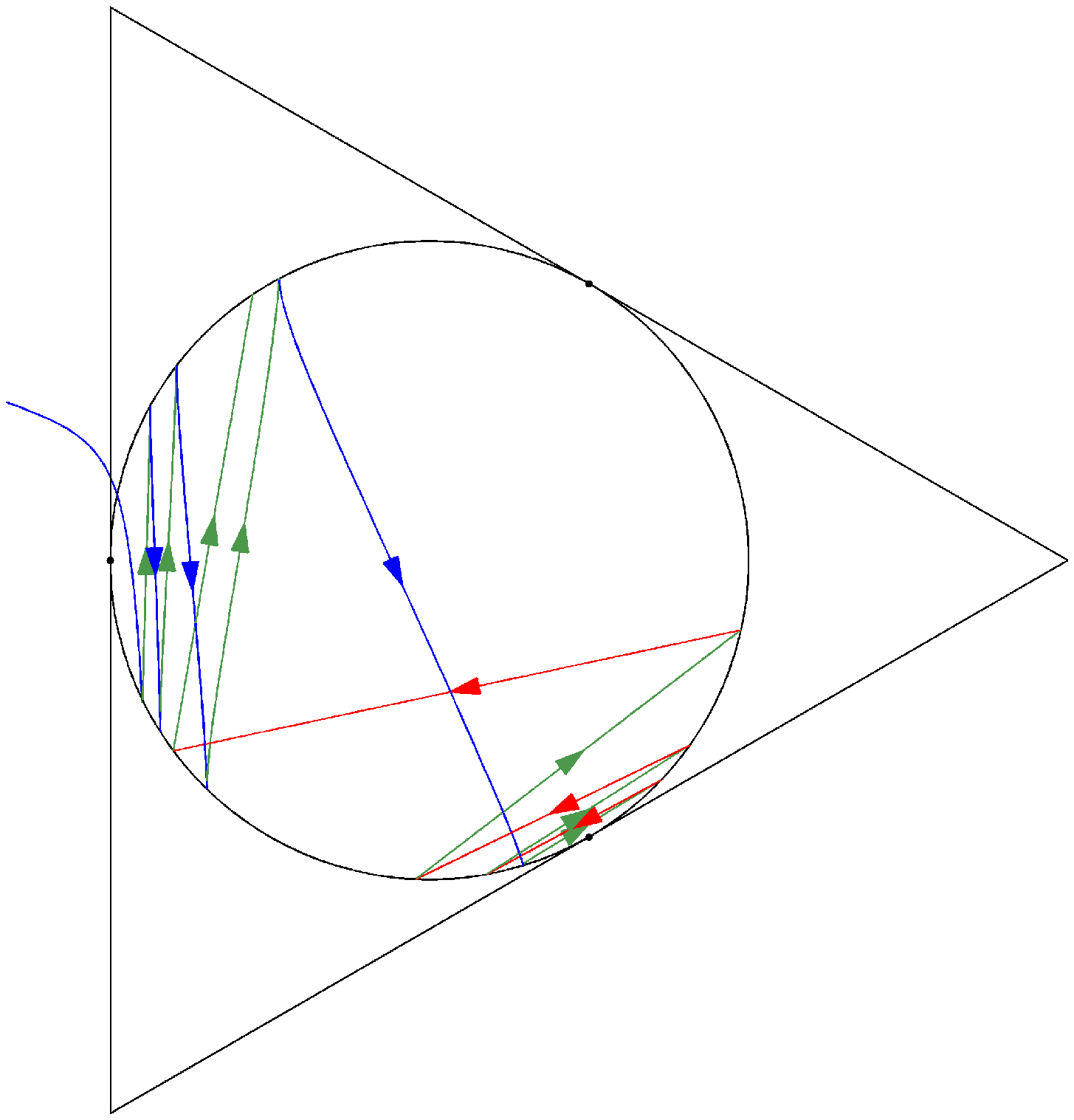}}
\caption{A few ``bounces" in a typical Bianchi IX orbit;
figures and numerics by Woei-Chet Lim. The vertical axis
represents $N_1$ (red), $N_2$ (green), $N_3$ (blue), with only
the biggest of the $N_i$'s plotted.  The Kasner circle and the
triangle for the Kasner billiard in the
$(\Sigma_+,\Sigma_-)$--plane are shown. The projected trajectories can be seen
to approach   the billiard ones.\label{fig:billiard}}
\end{center}
\end{figure}
%
Major progress concerning this issue has been achieved
in~\cite{Liebscher,Beguin,ReitererTrubowitzBianchi}, where
existence of orbits exhibiting the above behaviour has been
established, but the question of what happens for \emph{all},
or for \emph{generic} orbits remains open.

The following result of Ringstr\"om~\cite{Ringstroem1} provides
further insight into the geometry of Bianchi IX space-times:

\begin{Theorem}
\label{TRingBianchi} In all maximal globally hyperbolic
developments \mcmg\ of non-Taub--NUT Bianchi IX vacuum initial
data or of non-NUT Bianchi $VIII$ vacuum initial data the
Kretschmann scalar
\[
 R_{\alpha\beta\gamma\delta}R^{\alpha\beta\gamma\delta}
\]
is unbounded along inextendible causal geodesics.
\end{Theorem}

Note that the observation of curvature blow-up provides  a
proof, alternative to that of~\cite{ChRendall}, of the
non-existence of Cauchy horizons in generic Bianchi IX models.

We close this section by mentioning that no similar rigorous
results are known concerning the global dynamical properties of
Bianchi models of class B; compare~\cite{HHW}.

\subsection{The mixmaster conjecture}
\label{sSBKL} The \emph{most important question} in the study
of the Cauchy problem is that of the \emph{global properties of
the resulting space-times}. So far we have seen examples of
geodesically complete solutions (e.g., small perturbations of
Minkowski space-time), or all-encompassing singularities (e.g.,
generic Bianchi models), or of Cauchy horizons (e.g., Taub--NUT
metrics). The geodesically complete solutions are satisfying
but dynamically uninteresting, while the strong cosmic
censorship conjecture expresses the hope that Cauchy horizons
will almost never occur. So it appears essential to have a good
understanding of the remaining cases, presumably corresponding
to singularities. Belinski, Khalatnikov and
Lifschitz~\cite{BKL} suggested that, near singularities, at
each space point the dynamics of the gravitational field
resembles that of generic Bianchi  metrics, as described in
Section~\ref{sSBA}. Whether or not this is true, and in which
sense, remains to be seen; in any case the idea, known as
\emph{the BKL} conjecture, provided guidance
--- and still does --- to a significant body of research on general
relativistic singularities;
see~\cite{BarrowTipler,EardleyLiangSachs,DamourHenneauxNicolai}
and references therein. This then leads to the mathematical
challenge of making sense of the associated  slogan, namely
that \emph{the singularity in generic gravitational collapse is
spacelike, local, and oscillatory}. Here \emph{spacelike} is
supposed to mean that strong cosmic censorship holds. The term
\emph{local} refers to the idea that, near generic
singularities, there should exist coordinate systems in which
the metric asymptotes to a solution of equations in which
\emph{spatial} derivatives of appropriately chosen fields have
been neglected.\footnote{The resulting truncated equations
should then presumably resemble the equations satisfied by
spatially homogeneous metrics. However, different choices of
quantities which are expected to be time-independent will lead
to different choices of the associated notion of homogeneity;
for instance, in~\cite{BKL} the types Bianchi VIII and IX are
singled out; the notion of genericity of those types within the
Bianchi $A$ class is read from  Table~\ref{table:bianchiA} as
follows: ``something that can be non-zero is more generic than
something that is". On the other hand, the analysis
in~\cite{HeinzleUggla} seems to single out Bianchi VI$_{-1/9}$
metrics.}
Finally, \emph{oscillatory} is supposed to convey the idea that
the approximate solutions will actually be provided by the
Bianchi IX metrics.

The main rigorous evidence for a relatively large class of
vacuum\footnote{See, however,~\cite{AnderssonRendall,DHRW} for
a class of space-times with sources;~\cite{DHRW} also covers
vacuum in space dimensions $n\ge 10$.\label{fAR}}
space-times with singularities which are \emph{spacelike and
local} in the sense described above is  Ringstr\"om's
Theorem~\ref{TRingstroem}, describing generic Gowdy metrics,
but the resulting singularities are \emph{not} oscillatory.
This is not in contradiction with the conjecture, since the
Gowdy metrics are certainly not generic, whether in the space
of all metrics, or in the space of $U(1)\times U(1)$ symmetric
ones: As mentioned in Section~\ref{ssSUone}, generic
$U(1)\times U(1)$ metrics have non-vanishing twist constants
$c_a$ as defined by \eq{twist}. The numerical studies
of~\cite{BIW} suggest that the switching-on of the twist
constants will indeed generically lead to some kind of
oscillatory behavior.

In fact, BKL put emphasis on Bianchi IX models, while some
other authors seem to favor   Bianchi VI$_{-1/9}$, or
not-necessarily Bianchi,
oscillations~\cite{BKL,HeinzleUggla,UgglaEllis:pa,BGIMW,DamourHenneauxNicolai}.
It has moreover been suggested that the oscillatory behavior
disappears in space-time dimensions higher than
ten~\cite{DHS,DamourHenneauxNicolai}, and large families of
non-oscillatory solutions with singularities have indeed been
constructed in~\cite{DHRW}. This leads naturally to the
following, somewhat loose, conjecture:

 \begin{Conjecture}[Mixmaster conjecture]
 \label{Cmixmaster} Let $n+1\le 10$.
  \opp{oppmixmaster}
There exist open sets of vacuum metrics for which some natural
geometric variables undergo  oscillations of increasing complexity
along inextendible geodesics of unbounded curvature.
 \end{Conjecture}

The BKL conjecture would thus be a more precise version of the
above, claiming moreover genericity of the behavior, and
pointing out to the Bianchi dynamics as the right model. Those
properties are so speculative that we decided not to include
them in Conjecture~\ref{Cmixmaster}.

The only examples  so far of  oscillatory singularities
\emph{which are not spatially homogeneous}  have been
constructed by Berger and Moncrief~\cite{Berger:2000uf}. There,
a solution-generating transformation is applied to Bianchi IX
metrics, resulting in non-ho\-mogeneous solutions governed by
the ``oscillatory" functions arising from a non-Taub Bianchi IX
metric. The resulting metrics  have at least one but not more
than two Killing vectors. The analysis complements  the
numerical evidence for oscillatory behavior in $U(1)$ symmetric
models presented in~\cite{BevVince:U1chaos}.







%




%

%

%


%

%


%

%

%

%


%

%

%




\section{Marginally trapped surfaces}
\label{SMts}

There have been some interesting recent developments at the
interface of space-time geometry and the theory of black holes
associated with the notion of {\it marginally outer trapped
surfaces}. Let $\S$ be a co-dimension two  spacelike
submanifold of a space-time $\mcM$.  Under suitable orientation
assumptions, there exist two families of future directed null
geodesics issuing orthogonally from $\S$. If one of the
families has vanishing expansion along $\S$, then $\S$ is
called a marginally outer trapped surface (or an apparent
horizon).  The notion of a marginally outer trapped surface was
introduced early on in the development of the theory of black
holes, as the occurrence of the former signals the presence of
the latter. More recently, marginally outer trapped surfaces
have played a fundamental role in quasi-local descriptions of
black holes, and have been useful in numerical simulations of
black hole space-times; see e.g.~\cite{AKLivingReviews}.
Marginally outer trapped surfaces arose in a more purely
mathematical context in the work of Schoen and
Yau~\cite{SchoenYauPMT2} concerning the existence of solutions
to the Jang equation, in connection with their proof of the
positivity of mass.

Mathematically, marginally outer trapped surfaces may be viewed
as space-time analogues of minimal surfaces in Riemannian
manifolds. Despite the absence of a variational
characterization 
like that for minimal surfaces,\footnote{There seems to be no
analogue of the area functional.}  marginally trapped surfaces
have recently been shown to satisfy a number of analogous
properties, cf., in particular,
\cite{AMS1,AMS2,AndM1,AndM2,AG,Eichmair,GallowaySchoen} and the recent review \cite{AndEM}.  The aim of this section is to describe some of these  mathematical developments.

\subsection{Null hypersurfaces} \label{sSnullhyp}
Each family of null geodesics issuing orthogonally from $\S$, as described above,
forms a smooth null hypersurface near $\S$.  It would be useful at this stage
to discuss some general aspects of such hypersurfaces.
Null hypersurfaces have an interesting
geometry, and play an important role in general relativity.  In particular, as we have
seen, they represent
{\it horizons} of various sorts, such as the event horizons discussed in Section~\ref{Sbh}.

Let $(\mcM^{n+1},\fourg)$ be a space-time,
with $n \ge 2$.
 A smooth null hypersurface in $\mcM$ is a smooth
co-dimension one submanifold $\mcN = \mcN^n$ of $\mcM$ such that the
restriction of $g$ to each tangent space $T_p\mcN$ of $\mcN$ is degenerate.
This, together with the Lorentz signature, implies that there is a unique direction
of degeneracy in each tangent space $T_p\mcN$.   Thus, every null hypersurface
$\mcN$ comes equipped  with a smooth future directed {\it null} vector field $K $  ($g(K,K) = 0$) defined on, and tangent to $\mcN$, such that the normal space of $K$ at each  $p\in \mcN$ coincides
with the tangent space of $\mcN$ at $p$, i.e., $K_p^{\perp} = T_p\mcN$ for all
$p\in \mcN$.  Tangent vectors to $\mcN$, transverse to $K$, are then
necessarily spacelike.
The null vector field $K$ associated to $\mcN$  is unique up to positive
pointwise rescaling.  However, there is, in general, no canonical way to set
the scaling.

Two simple examples arise  in Minkowski space $\Min{n}$.  The
past and future cones $\d I^-(p)$ and $\d I^+(p)$ are smooth
null hypersurfaces away from the vertex $p$. Each nonzero null
vector $v \in T_p\Min{n}$ determines a null hyperplane $\Pi =
\{q \in  \Min{n} : \eta( \vv{pq}, v) = 0\}$, where $\eta  $ is
the Minkowski metric, and $\vv{pq}$ is the tangent vector at
$p$ representing the displacement from $p$ to $q$.

It is a fundamental fact that {\it the integral curves of $K$
are null geodesics}, though perhaps not affinely parameterized
- this will depend on the scaling of $K$.  Thus $\mcN$ is ruled
by null geodesics, called the null generators of $\mcN$.  For
example the future cone $\mcN =  \d I^+(p) \setminus \{p\}$ in
Minkowski space is ruled by future directed null rays emanating
from $p$.

The {\it null expansion scalar} $\th$ of $\mcN$ with respect to
$K$ is a smooth function on $\mcN$ that gives a measure of the
average expansion of the null generators of $\mcN$ towards the
future.  In essence, $\th$ is defined as the divergence of the
vector field $K$ along $\mcN$.   To be precise, given $p \in
\mcN$, let $\Pi_{n-1}$ be a co-dimension one  subspace of
$T_p\mcN^n$ transverse to $K_p$.  The metric $g$, restricted to
$\Pi_{n-1}$, will be positive definite. Let $\{e_1, e_2, ...
,e_{n-1}\}$ be an orthonormal basis for $\Pi_{n-1}$ with
respect to $g$.  Then $\th$ at $p$ is defined as,
\beq\label{thdef} \th(p) = \sum_{i=1}^{n-1} g(\D_{e_i}K,e_i)
\,. \eeq Interestingly, due to the fact that $K$ is null, this
value is independent of the choice of transverse subspace
$\Pi_{n-1}$, as well as of the choice of an orthonormal basis
for $\Pi_{n-1}$, and so the expansion scalar $\th$ is well
defined.

While $\th$  depends on the choice of $K$, it does so in a simple way.  As easily follows
from Equation \eqref{thdef}, a positive rescaling of $K$ rescales $\th$ in the same way:
If $\tilde K = fK$ then $\tilde \th = f \th$.  Thus the  {\it sign} of the null expansion
$\th$ does not depend on the scaling of $K$: $\th > 0$ means expansion on average
of the null generators, and $\th < 0$
means  contraction on average.  In Minkowski space, the future null cone
$\mcN =\d I^+(p)\setminus\{p\}$  has $\th > 0$, and the past cone,
 $\mcN =\d I^-(p) \setminus\{p\}$) has $\theta <0$.

It is useful to understand how the null expansion varies as one
moves along a null generator of $\mcN$.  Let $s \to \eta(s)$ be
a null geodesic generator of $\mcN$, and assume $K$ is scaled
so that  $\eta$ is affinely parameterized.   Then it can be
shown that the null expansion scalar $\th = \th(s)$ along
$\eta$ satisfies the propagation equation, \beq\label{ray}
\frac{d\th}{ds} = -{\rm Ric}(\eta',\eta') - \sigma^2 -
\frac1{n-1}\theta^2 \,, \eeq where $\s\ge 0$, the {\it shear
scalar}, measures the deviation from perfect isotropic
expansion.  Equation \eqref{ray} is known in the relativity
community as the Raychaudhuri equation (for a null geodesic
congruence)~\cite{HE}, and, together with a timelike version,
plays an important role in the proofs of the classical
Hawking-Penrose singularity theorems~\cite{HE}.  There are
well-known Riemannian counterparts to this equation, going back
to work of Calabi~\cite{CalabiRicci}.

Equation \eqref{ray} shows how the curvature of space-time
influences the expansion of the null generators.  We consider
here a simple application of the Raychaudhuri equation.

\begin{prop}
 \label{nonneg}
Let $\mcM$ be a space-time which obeys the null energy
condition, ${\rm Ric}\,(X,X) = R_{\a\b}X^{\a}X^{\b} \ge 0$ for
all null vectors $X$, and let $\mcN$ be a smooth null
hypersurface in $\mcM$. If the null generators of $\mcN$ are
future geodesically complete then the null generators of $\mcN$
have nonnegative expansion, $\th\ge0$.
\end{prop}

\proof Suppose $\theta < 0$ at $p\in  \mcN$. Let $\eta:
[0,\infty) \to \mcN$, $s\to \eta(s)$, be the null geodesic
generator of $\mcN$ passing through $p = \eta(0)$; by rescaling
$K$ if necessary, we can assume $\eta$ is affinely
parameterized. Let $\th = \th(s)$, $s \in [0,\infty)$, be the
null expansion of $\mcN$ along $\eta$; hence $\th(0) < 0$.
Raychaudhuri's equation and the null energy condition imply
that $\th = \th(s)$  obeys the inequality, \beq\label{ineq}
\frac{d\theta}{ds} \le - \frac1{n-1}\theta^2 \,, \eeq and hence
$\theta < 0$ for all $s> 0$. Dividing through by $\theta^2$
then gives, \beq\label{blowup}
\frac{d}{ds}\left(\frac1{\theta}\right) \ge \frac1{n-1} \, ,
\eeq which implies $1/\theta\to 0$, i.e., $\theta \to -\infty$
in finite affine parameter time, contradicting the  smoothness
of $\th$.\qed

\ss We wish to indicate the connection of Proposition
\ref{nonneg} with the theory of black holes.  In fact, this
proposition is the most rudimentary form of Hawking's  famous
area theorem~\cite{HE}.  Let $\mcM$ be  a standard black hole
space-time, as defined for example in~\cite{HE}.  It is  not
necessary to go into the technical details of the definition.
It suffices to say that in $\mcM$ there exists a region $\mcB$,
the black hole region, from which signals (future directed
causal curves) cannot ``escape to infinity" (recall the example
of the Schwarzschild solution discussed in Section~\ref{sSSm}).
The boundary of this region  is the event horizon $\mcE$,
which, in general, is a Lipschitz hypersurface  ruled by future
inextendible null geodesics, called its  null generators. If
$\mcE$ is smooth and if its generators are future complete then
Proposition \ref{nonneg} implies that $\mcE$ has nonnegative
null expansion.  This in turn implies that ``cross-sections" of
$\mcE$ are nondecreasing in area as one moves towards the
future, as asserted by the area theorem.  In the context of
black hole thermodynamics, the area theorem is referred to as
the second law of black mechanics,  and provides  a  link
between gravity and quantum physics.  As it turns out, the area
theorem remains valid without imposing any smoothness
assumptions; for a recent study of the area theorem, which
focuses on these issues of regularity, see~\cite{ChDGH}.

\subsection{Trapped and marginally trapped surfaces}
 \label{sstmts}
We begin with some definitions. Let $\S =\S^{n-1}$, $n \ge 3$,
be a spacelike submanifold of co-dimension two in a space-time
$(\mcM^{n+1}, \fourg)$.  Regardless of the dimension of
space-time, we shall refer to $\S$ as a surface, which it
actually is in the $3+1$ case. We are primarily interested in
the case where $\S$ is compact (without boundary), and so we
simply assume this from the outset.

Each normal space of $\S$, $[T_p\S]^{\perp}$, $p \in \S$, is
timelike and $2$-dimensional, and hence admits two future
directed null directions orthogonal to $\S$. Thus, if the
normal bundle is trivial, $\S$ admits two smooth nonvanishing
future directed null normal vector fields $l_+$ and $l_-$,
which are unique up to positive pointwise scaling, see
Figure~\ref{fbothnulls}. By convention, we refer to $l_+$ as
outward pointing and $l_-$ as inward pointing.\footnote{In many
situations, there is a natural choice of ``inward" and
``outward".}
\begin{figure}[ht]
\begin{center} {
 \psfrag{NM}{\huge $\phantom{\,x}l_-$}
 \psfrag{NP}{\huge $l_+$}
\resizebox{2.5in}{!}{\includegraphics{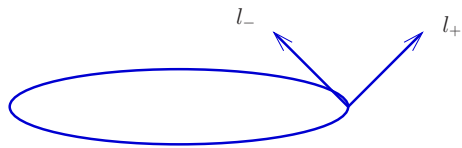}}
}
\caption{The null future normals $l_\pm$ to $\S$.
\label{fbothnulls}}
\end{center}
\end{figure}
In relativity it is standard to decompose the second fundamental form
of $\S$ into two scalar valued {\it null second forms}  $\chi_+$ and $\chi_-$, associated
to $l_+$ and $l_-$, respectively.
For each $p \in \S$, $\chi_{\pm} : T_p\S \times T_p\S \to \bbR$ is the bilinear form defined  by,
\begin{align}\label{form}
\chi_{\pm}(X, Y) = \fourg(\D_X l_{\pm}, Y) \quad \mbox{for all } X,Y \in T_p\S \,.
\end{align}
A standard argument shows  that $\chi_{\pm}$ is symmetric.
Hence, $\chi_+$ and $\chi_-$ can be traced with respect to the
induced metric $\twog$ on $\S$ to obtain the  null mean
curvatures (or null expansion scalars),
\beq \th_{\pm} ={\rm
tr}_{\twog}\, \chi_{\pm} =  {\twog}^{ij}(\chi_{\pm})_{ij}  =
\div_{\S}l_{\pm}  \,. \eeq $\th_{\pm}$ depends on the scaling
of $l_{\pm}$ in a simple way. As follows from Equation
\eqref{form}, multiplying $l_{\pm}$ by  a positive  function
$f$ simply scales $\th_{\pm}$ by the same function.
Thus, the  {\it sign} of  $\th_{\pm}$ does not depend on the scaling of $l_{\pm}$.
Physically, $\th_+$ (resp., $\th_-$) measures the divergence
of the  outgoing (resp., ingoing) light rays emanating
from $\S$.

It is useful to note the connection between
the null expansion scalars $\th_{\pm}$ and
the  expansion of the generators of a null hypersurface, as discussed
in Section~\ref{sSnullhyp}.  Let $\mcN_+$ be the null hypersurface, defined and
smooth near $\S$, generated by the null geodesics passing through
$\S$ with initial  tangents $l_+$.
Then $\th_+$ is the null expansion of $\mcN_+$ {\it restricted to} $\S$; $\th_-$ may be described similarly.

Let $\hypM$ be a spacelike hypersurface in a space-time
$(\calM,\fourg)$, with induced metric $\threeg$ and second
fundamental form $K$, and  suppose $\S$ is embedded as a
$2$-sided hypersurface in $\hypM$.  Then the null expansions
$\th_{\pm}$ can be expressed  in terms of the initial data
$(\hypM, \threeg, K)$ as follows. Since $\S$ is $2$-sided in
$\hypM$, it admits a smooth unit normal field $\nu$ in $\hypM$,
unique up to sign.  By convention, we refer to such a choice as
outward pointing.  Letting $u$ denote the future directed unit
normal to $\hypM$, $l_+= u+ \nu$ (resp., $l_- = u - \nu$) is a
future directed outward (resp., future directed inward)
pointing null normal vector field along $\S$. Let $\th_{\pm}$
be the null expansion with respect to the null normal
$l_{\pm}$.  Then, \beq\label{thid} \th_{\pm} = {\rm tr}_{\S}K
\pm H \,, \eeq where ${\rm tr}_{\S}K$ is the trace of the
projection of $K$ into $\S$ with respect to the induced metric
on $\S$, and $H$ is the mean curvature of $\S$ in $\hypM$.

For round spheres in Euclidean slices of Minkowski space, with
the obvious choice of inside and outside, one has $\th_- < 0$
and $\th_+ >0$. In fact, this is the case in general for large
``radial" spheres in {\it asymptotically flat} spacelike
hypersurfaces.  However, in regions of space-time where the
gravitational field is strong, one may have both $\th_- < 0$
and $\th_+ < 0$, in which case $\S$ is called a {\it trapped
surface}.  For example the black hole region, $0 < r < 2m$,  in
$(n+1)$-dimensional Schwarzschild space-time (see
Section~\ref{sSSm})  is foliated by spherically symmetric
$(n-1)$-spheres, all of which are trapped surfaces.  Under
appropriate energy and causality conditions, the occurrence of
a trapped surface signals  the onset of gravitational collapse.
This is the implication of the Penrose singularity
theorem~\cite{Psing}, the first of the famous singularity
theorems. 

\begin{thm}[Penrose~\cite{Psing}] \label{singthm}
Let $\mcM$ be a globally hyperbolic space-time with noncompact
Cauchy surfaces satisfying the null energy condition.  If $\mcM$ contains a trapped surface $\S$
then $\mcM$ is future null geodesically incomplete.
\end{thm}

Recall from Section~\ref{sSnullhyp} that the null energy
condition is the curvature requirement, ${\rm Ric}\,(X,X) =
R_{\a\b}X^{\a}X^{\b} \ge 0$ for all null vectors $X$. If a
space-time $\mcM$  obeys the Einstein equation \eqref{eeqn},
then one can express the null energy condition in terms of the
energy momentum tensor: $\mcM$ obeys the null energy condition
if and only if $\calT(X,X) = T_{ij}X^iX^j \ge 0$ for all null
vectors $X$.

In studying an {\it isolated} gravitating system, such as the
gravitational collapse  of a star, it is customary to model the
situation by a space-time which is  asymptotically flat.  In
this context, the assumption of Theorem \ref{singthm}  that the
space-time admits a noncompact Cauchy surface is natural. The
conclusion in the theorem of future null geodesic
incompleteness is an indication that  space-time  ``comes to an
end" or develops  a singularity somewhere in the causal future.
However,  the theorem gives no information about the nature of
the singularity.

Existence of \emph{vacuum} asymptotically flat initial data
sets, with one asymptotic region and containing compact trapped
surfaces, has been established by Beig and
\'O~Murchadha~\cite{Beigneill:trapped}.

An intriguing question that arises in this context is whether a
trapped surface can develop dynamically from initial data that
did not contain any.  This has been addressed by Christodoulou
in~\cite{ChristodoulouBHFormation}, where a formidable analysis
of the focusing effect of sufficiently strong incoming
gravitational waves is presented. Some further developments can
be found in~\cite{RodnianskiKlainerman:scarred}.
\bams{%
\proof[Remark on the proof of Theorem~\ref{singthm}] Under the
assumption of future null completeness, one uses the trapped
condition to show that the achronal boundary $\partial
I^+(\S)$, which is generated by null geodesic segments issuing
orthogonally from $\S$,  is compact.  Then, flowing along the
integral curves of a timelike vector field establishes a
homemorphism from $\partial I^+(\S)$ to any Cauchy surface $S$,
contradicting the noncompactness assumption.
%
%
\qed}

\ss To continue our discussion, consider again the general
setting of a spacelike surface $\S^{n-1}$ in a space-time
$\mcM^{n+1}$, with future directed null normal fields $l_{\pm}$
and associated null expansion scalars $\th_{\pm}$. Focusing
attention on just the outward null normal $l_+$, we say that
$\S$ is an {\it outer trapped surface} (resp., {\it weakly
outer trapped surface}) if $\th_+ < 0$ (resp., $\th_+ \le 0$).
If $\th_+$ vanishes, we say that $\S$ is a {\it marginally
outer trapped surface}, or MOTS for short. In what follows we
will be primarily concerned with properties of MOTSs.

MOTSs arise naturally in a number of situations.   As an
outgrowth of their work on the positive energy theorem, Schoen
and Yau~\cite{SchoenYau83} showed that suitable conditions on
the energy density and momentum density of an asymptotically
flat initial data set insure the presence of a MOTS; see
also~\cite{YauBH}. Next, as follows from our comments about the
area theorem in Section~\ref{sSnullhyp}, cross-sections of the
event horizon in black hole space-times have nonnegative
expansion $\th \ge 0$. (By  a cross-section, we mean a smooth
intersection of the event horizon with a spacelike
hypersurface.) In the steady state limit this expansion goes to
zero.  Thus, it is a basic fact that cross-sections of the
event horizon in stationary black hole space-times are MOTSs.
For dynamical black hole space-times, MOTSs typically occur in
the black hole region, i.e., the region inside the event
horizon.  While there are heuristic arguments for the existence
of MOTSs in this situation, based on looking at the boundary of
the `trapped region'~\cite{HE,Wald:book} within a given
spacelike slice, using an approach put forth by
Schoen~\cite{schoen:miamiwaves}, their existence has been rigorously
established under natural physical conditions, first by Andersson and
Metzger~\cite{AndM2} in dimension three, and then by
Eichmair~\cite{Eichmair} up to dimension seven; see
Section~\ref{ssEMOTSs}.

As noted earlier,  MOTSs may be viewed as  space-time analogues of minimal
surfaces in Riemannian geometry.  In fact, as follows from Equation
\eqref{thid}, in the time-symmetric case ($K=0$)
a MOTS   is simply a minimal surface in $\hypM$.   Of importance for certain applications
is the fact, first discussed by Andersson, Mars and Simon~\cite{AMS1, AMS2}, that MOTS admit a notion of stability analogous
to that for minimal surfaces.

\subsection{Stability of MOTSs}
\label{stabmots} In Riemannian geometry, a minimal surface
(surface with vanishing mean curvature) is stable provided, for
a suitable class of variations, the second variation of area is
nonnegative, $\delta^2A \ge 0$.   Stability of minimal surfaces
can also be characterized in terms of the associated stability
operator.  This latter approach extends to MOTSs, as we now
describe.

Let $\S$ be a MOTS in $\hypM$ with outward unit normal $\nu$.  Consider a normal variation of $\S$ in $\hypM$, i.e. a map $F: (-\e,\e) \times \S \to \hypM$,
such that  (i) $F(0,\cdot) = {\rm id_{\S}}$  and (ii)~$\left . \frac{\d F}{\d t} \right |_{t=0} = \phi\nu$,
$\phi \in C^{\infty}(\S)$.
Let $\th(t)$ denote the null expansion of $\S_t := F(t,\S)$
with respect to $l_t = u + \nu_t$, where $u$ is the future
directed timelike unit normal to $\hypM$ and $\nu_t$ is the
outer unit normal  to $\S_t$ in $\hypM$.   A computation shows,
\beq\label{thder} \left . \frac{\d\th}{\d t} \right |_{t=0}   =
L(\f) \;, \eeq where $L : C^{\infty}(\S) \to C^{\infty}(\S)$ is
the operator~\cite{GallowaySchoen,AMS2}, \beq\label{stabop}
L(\phi)  = -\triangle \phi + 2\<X,\D\phi\>  + \left( \frac12 S
- (\rho + J(\nu)) - \frac12 |\chi|^2+{\rm div}\, X - |X|^2
\right)\phi \,. \eeq In the above, $\triangle$, $\D$ and ${\rm
div}$ are the Laplacian, gradient and divergence operator,
respectively, on $\S$, $S$ is the scalar curvature of $\S$,
$\rho$ and $J$ are the energy density and momentum density,
respectively,  as defined in Equations \eqref{eq:c1},
\eqref{eq:c2}, $X$ is the vector field  on $\S$  defined by
taking the tangential part of $\D_{\nu}u$ along $\S$, and
$\<\,,\,\>$ denotes the induced metric  on $\S$.

In the time-symmetric case,  $\th$ in \eqref{thder} becomes the
mean curvature $H$, the vector field $X$ vanishes and $L$
reduces to the classical stability operator (linearization of
the mean curvature operator)  of minimal surface theory. In
analogy with the minimal surface case, we refer to $L$ in
\eqref{stabop} as the stability operator associated with
variations in the null expansion $\th$.  Although in general
$L$ is not self-adjoint,  its principal eigenvalue (eigenvalue
with smallest real part) $\la_1(L)$ is real.   Moreover there
exists an associated eigenfunction $\phi$ which is positive on
$\S$. Continuing the analogy with the minimal surface case, we
say that a MOTS is stable provided $\la_1(L) \ge 0$.  (In the
minimal surface case this is equivalent to the second variation
of area being nonnegative.)  It follows from basic properties
of $L$ that a MOTS $\S$ is stable if and only if there exists a
normal variation of $\S$, with $\phi > 0$, such that $\left .
\frac{\d\th}{\d t} \right |_{t=0} \ge 0$.

Stable MOTSs arise naturally in physical situations, for
example, as {\it  outermost} MOTSs. We say $\S$ is an {\it
outermost} MOTS in $\hypM$ provided there are no weakly outer
trapped ($\th_+ \le 0$) surfaces outside of, and homologous
to,~$\S$.  We say $\S$ is a {\it weakly outermost} MOTS in
$\hypM$ provided there are no outer trapped  ($\th_+ < 0$)
surfaces outside of, and homologous to, $\S$.   Clearly,
``outermost" implies ``weakly outermost". Moreover we have the
following~\cite{AMS1,AMS2}:

\begin{prop} Weakly outermost MOTSs are stable.
\end{prop}

To see this consider a variation $\{\S_t\}$ of a weakly
outermost MOTS $\S$ with variation vector field $\calV =
\phi\nu$, where $\phi$ is a positive eigenfunction associated
to the principal eigenvalue $\la_1 = \la_1(L)$.   If $\la_1<
0$, then Equation \eqref{thder} implies $\left .
\frac{\d\th}{\d t} \right |_{t=0} = \la_1 \phi < 0$. Since
$\th(0) = 0$, this implies that, for small $t > 0$, $\S_t$ is
outer trapped, contrary to $\S$ being weakly outermost.

A standard fact in the theory of black holes is that, for black
hole space-times obeying the null energy condition,  there can
be  no weakly outer trapped surfaces contained in the domain of
outer communications (the region outside of all black holes and
white holes).   It follows that compact cross-sections of the
event horizon in stationary black hole space-times obeying the
null energy condition are stable MOTSs.  Moreover, results of
Andersson and Metzger~\cite{AndM1, AndM2} provide natural
criteria for the existence of outermost MOTSs in  initial data
sets containing trapped regions; see Section~\ref{ssEMOTSs}.

Stable MOTSs share a number of properties in common with  stable minimal surfaces.
This sometimes  depends on the following fact.   Consider the
``symmetrized" operator
$L_0: C^{\infty}(\S) \to C^{\infty}(\S)$,
\beq\label{symop}
L_0(\phi)  = -\triangle \phi  + \left( \frac12 S - (\rho + J(\nu)) - \frac12 |\chi|^2\right)\phi \,.
\eeq
formally obtained  by  setting $X= 0$ in \eqref{stabop}.   The key argument in~\cite{GallowaySchoen}
shows the following (see also~\cite{AMS2,Galloway:mots}).

\begin{prop}\label{eigen}
$\la_1(L_0) \ge \la_1(L)$.
\end{prop}

In the next subsection we consider an application
of stable MOTS to the topology of black holes.

\subsection{On the topology of black holes}   A useful step in the proof of
black hole uniqueness  (see Section~\ref{sSubh}) is Hawking's
theorem on the topology of black holes~\cite{HE} which asserts
that compact cross-sections of the event horizon in
$3+1$-dimensional, appropriately regular, asymptotically flat
stationary black hole space-times obeying the dominant energy
condition are topologically 2-spheres. As shown by
Hawking~\cite{HawkingLesHouchesI}, this conclusion also holds
for outermost MOTSs in space-times that are not necessarily
stationary.   The proof in both cases is variational in nature,
and relies on the classical Gauss-Bonnet theorem. Developments
in physics related to string theory have lead to an increased
interest in the study of  gravity, and in particular black
holes, in higher dimensions; see e.g.~\cite{EmparanReallReview}
for a recent review. The remarkable example of Emparan and
Reall~\cite{EmparanReall} of a $4+1$ asymptotically flat
stationary vacuum black space-time with horizon topology $S^1
\times S^2$, the so-called ``black ring", shows that horizon
topology need not be spherical in higher dimensions.  This
example naturally led to the question of what are the allowable
horizon topologies in higher dimensional black hole
space-times. This question was addressed in the papers
of~\cite{GallowaySchoen,Galloway:mots}, resulting in a natural
generalization of Hawking's topology theorem to higher
dimensions, which we now discuss.

Consider a space-time $(\calM^{n+1}, \fourg)$, $n \ge 3$,
satisfying the Einstein equations (not necessarily vacuum), and
for simplicity assume that the cosmological constant vanishes,
$\Lambda = 0$. Let $\hypM$ be a spacelike hypersurface in
$\calM$, which gives rise to the initial data set $(\hypM,
\threeg, K)$, as in Section~\ref{stabmots}.  Recall from
Equation \eq{DEC}  that the dominant energy condition holds
with respect to this initial data set provided $\rho \ge |J|$
along $\hypM$, where $\rho$ and $J$ are the energy density and
momentum density, respectively,  as defined in Section \ref{CE}
(but with $\Lambda$ set to zero).

The following result, obtained in~\cite{GallowaySchoen}, gives a natural extension of Hawking's
black hole topology theorem to higher dimensions.  Recall, a Riemannian manifold $\S$ is of
{\it positive Yamabe type} if it admits a metric of positive scalar curvature.

\begin{thm}\label{posyam}
Let  $\S^{n-1}$ be a stable MOTS in an initial data set $(\hypM^n,\threeg,K)$, $n \ge~3$.
\ben
\item If $\rho > |J|$ along $\S$ then $\S$ is of positive Yamabe type.
\item If $\rho \ge |J|$ along $\S$ then $\S$ is of positive Yamabe type {\rm unless}
$\S$ is Ricci flat (flat if $n=2,3$), $\chi = 0$ and $\rho + J(\nu) = 0$ along $\S$.
\een
\end{thm}

In the time-symmetric case, Theorem \ref{posyam} reduces to the
classical result of Schoen and Yau~\cite{SchoenYauManuscripta},
critical to their study of manifolds of positive scalar
curvature, that a compact stable minimal hypersurface in a
Riemannian manifold of positive scalar curvature is of positive
Yamabe type.

The key to the proof of Theorem \ref{posyam} is Proposition
\ref{eigen}, which, since $\S$ is assumed stable, implies that
$\la_1(L_0) \ge 0$, where $L_0$ is the operator given in
\eqref{symop}.   Now, in effect, the proof has been reduced to
the Riemannian case.   Consider $\S$ in the conformally related
metric, $\tilde\twog = \f^{\frac2{n-2}}\twog$, where $\twog$ is
the induced metric on $\S$ and $\phi$ is a positive
eigenfunction corresponding to $\la_1(L_0)$. The scalar
curvatures $\tilde S$ and $S$ of the metrics $\tilde \gamma$
and $\gamma$, respectively, are related by (compare
\eq{ConfReq})
 \begin{align*}
\tilde S & =  \phi^{-\frac{n}{n-2}}\left(-2\triangle\f + S \f +\frac{n-1}{n-2}\frac{|\D\f|^2}{\f}\right) \nonumber \\\
& =  \phi^{-\frac{2}{n-2}}\left(2\la_1(L_0) +  2(\rho + J(\nu)) +  |\chi|^2
+\frac{n-1}{n-2}\frac{|\D\f|^2}{\f^2}\right) \,,
\end{align*}
where for the second equation we have used  \eqref{symop}.
Since $\rho + J(\nu)  \ge \rho - |J| \ge 0$, we have that
$\tilde S \ge 0$.   By further standard metric deformations,
the scalar curvature of $\S$ can be made strictly positive,
unless various quantities vanish identically.

According to Theorem \ref{posyam}, apart from certain
exceptional circumstances, a stable marginally outer trapped
surface $\S$ in a space-time $\calM$ obeying the dominant
energy condition is of positive Yamabe type.  Assume for the
following discussion that $\S$ is orientable.  Then, in the
standard case: $\dim \calM =3+1$ (and hence $\dim \S = 2$),
Gauss-Bonnet tells us that if $\S$ is positive Yamabe then $\S$
is topologically a two-sphere, and we recover Hawking's
theorem. In higher dimensions,  much is now known about
topological obstructions to the existence of  metrics of
positive scalar curvature. While the first major result along
these lines is the famous theorem of
Lichnerowicz~\cite{Lichnerowicz63} concerning the vanishing of
the $\hat A$ genus, a key advance in our understanding was made
in the late 1970s and early 1980s by Schoen and
Yau~\cite{SchoenYauAnnMath79,SchoenYauManuscripta}, and Gromov
and Lawson~\cite{GL1,GL2}. Let us focus on the case: dim $\calM
= 4+1$, and hence dim $\S = 3$.  Then by results of Schoen-Yau
and Gromov-Lawson, in light of the  resolution of the
Poincar\'e conjecture, $\S$ must be diffeomorphic to a finite
connected sum of  spherical spaces (spaces which are covered by
the $3$-sphere) and  $S^2 \times S^1$'s.  Indeed, by the prime
decomposition theorem, $\S$ can be expressed as a connected sum
of spherical spaces, $S^2 \times S^1$'s, and $K(\pi,1)$
manifolds (manifolds whose universal covers are contractible).
But as $\S$ admits a metric of positive scalar curvature, it
cannot have any $K(\pi,1)$'s in its prime decomposition.
  Hence the basic horizon topologies in dim $\calM=4+1$ are $S^3$ and
$S^2 \times S^1$ (in the sense that $\S$ is ``built up" from
such spaces), both of which are realized by nontrivial black
hole space-times. \opp{oppgs} It remains an interesting open
question which topologies of positive Yamabe type can be
realized as outermost MOTSs; see~\cite{SchwartzF} for examples
involving products of spheres.

A drawback of Theorem \ref{posyam} is that it allows certain possibilities
that one would like to rule out: for example, the theorem does not rule out the possibility of a vacuum
black hole space-time with toroidal horizon topology.  (This  borderline case
also arises in the proof of Hawking's theorem).
In fact, one can construct
examples of stable toroidal MOTSs in space-times obeying the dominant
energy condition.  Such MOTSs cannot, however, be outermost, as the
following theorem asserts.

\begin{thm}[\cite{Galloway:mots}]\label{posyam2}
Let $\S$ be an outermost MOTS in the spacelike hypersurface
$\hyp$, and assume the dominant energy condition \eqref{DECsts}
holds in a space-time neighborhood of $\S$.\footnote{Note that
since we are assuming $\Lambda =0$ here, the dominant energy
condition
 is equivalent to
the condition, $G_{\rho\nu}X^{\rho}Y^{\nu} \ge 0$ for all
future directed causal vectors $X,Y$, where $G_{\mu\nu}$  is
the Einstein tensor.} Then $\S$ is of positive Yamabe type.
\end{thm}

As an immediate corollary, we have that {\it compact cross-sections of event horizons
in regular
stationary black hole space-times obeying the dominant energy condition are
of positive Yamabe type.}  In particular, there can be no toroidal horizons.

Theorem \ref{posyam2} is an immediate consequence of the following rigidity result.

\begin{thm}[\cite{Galloway:mots}]\label{rigid}
Let $\S$ be a weakly outermost MOTS in the spacelike
hypersurface $\hypM$, and assume the dominant energy condition
holds in a space-time neighborhood of $\S$.  If $\S$ is not of
positive Yamabe type then there exists an outer neighborhood $U
\approx [0,\e) \times \S$ of $\S$ in $\hyp$ such that each
slice $\S_t = \{t\} \times \S$, $t \in [0,\e)$ is a MOTS.
\end{thm}

Theorem \ref{rigid} is proved in two stages. The first stage,
and the main effort, is to establish Theorem \ref{rigid}
subject to the additional assumption that $\hypM$ has {\it
nonpositive} mean curvature, $\tau \le 0$.  This is a purely
initial data result, and the proof is carried out in two steps.
The first step involves an inverse function theorem argument to
show that an outer neighborhood of $\S$ can be foliated by
surfaces $\S_t$ of constant null expansion, $\th(t) = c_t$.
This uses the stability of $\S$ in a critical way.  In the
second step it is shown that all of these constants are zero,
$c_t = 0$.   It is here where the sign of the mean curvature of
$\hypM$ needs to be controlled.  Once having proved Theorem
\ref{rigid}, subject to the condition, $\tau \le 0$, the next
stage, which is actually easy, is   a deformation argument
(specifically, a deformation of $\hypM$ near $\S$) that reduces
the problem to the case  $\tau \le 0$.

We remark that Riemannian versions of Theorem \ref{rigid} had previously been
considered in~\cite{CaiGallowayRigidity,CaiProceedings}.

\subsection{Existence of MOTSs}
 \label{ssEMOTSs}

As mentioned earlier, compact cross-sections of the event
horizon in regular stationary black hole space-times are
necessarily MOTSs.   In dynamical black hole space-times, it is
typical for trapped or outer trapped surfaces to form in the
black hole region. But the occurrence of an outer trapped
surface in a spacelike hypersurface that obeys a mild
asymptotic flatness condition leads to the existence of a MOTS,
as follows from  the existence result alluded to near the end
of Section 7.2, and which we now
discuss~\cite{AndM2,Eichmair,schoen:miamiwaves}.

Let $\Omega$ be a relatively compact domain in a spacelike
hypersurface $\hyp$, with smooth boundary $\d \Omega$. We
assume that $\d\Omega$ decomposes as a disjoint union of
components, $\d\Omega  = \SI  \cup \SO$,  where we think of
$\SI $ as the ``inner" boundary and $\SO$ as the ``outer"
boundary of $\Omega$ in $\hyp$.  We choose  the normal along
$\SI $ that points into $\Omega$, and the normal along $\SO$
that points out of $\Omega$, so that both normals point towards
the region exterior to $\SO$. Thus, $\SI $ is outer trapped if
$\th < 0$ with respect to the future directed null normal field
along $\SI $  that projects into $\Omega$.  We say that $\SO$
is {\it outer untrapped} if $\th > 0$ with respect to the
future directed null normal along $\SO$ that projects out of
$\Omega$.  Heuristically,  if $\SO$ is lying in a region where
$\hyp$ is ``flattening out" then we expect $\SO$ to be outer
untrapped.

With this notation and terminology we have the following existence result for MOTS.

\begin{thm}[\cite{AndM1,Eichmair}]\label{motsexist}
Let $\hyp^n$ be a spacelike hypersurface in a space-time
$(\calM^{n+1},\fourg)$, with $n \le 7$.  Let $\Omega$ be a relatively compact
domain in $\hyp$, with smooth boundary $\d\Omega = \SI  \cup \SO$, such that
the inner boundary $\SI $ is outer trapped and the outer boundary $\SO$ is outer untrapped, as described above.  Then there exists a smooth  compact
MOTS in $\Omega$ homologous to $\SI $.
\end{thm}

Moreover, it has been shown that the MOTS constructed in
Theorem~\ref{motsexist} is stable \cite{AndM1, AndEM}.

In the time-symmetric case, Theorem \ref{motsexist} reduces to a well-known
existence result for stable minimal surfaces.  In the time-symmetric case the barrier conditions
in Theorem \ref{motsexist} simply say that $\Omega$ is a mean convex domain.
One can then minimize area in the homology class of a surface in $\Omega$
parallel to $\SI $ and apply standard compactness and regularity results of geometric measure theory~\cite{FedererMeasureTheory} to obtain a smooth (provided $\dim \hyp \le 7$) stable minimal surface in $\Omega$ homologous to $\SI $.  However,  since  MOTS do not arise as stationary points of some elliptic functional, such a procedure
does not work for general initial data.  A completely different approach must be taken.

In fact, the proof of Theorem \ref{motsexist}  is based on  Jang's equation
\cite{Jang78} which is closely related to the MOTS condition $\th^+ = 0$.
Given an initial data set $(\hyp, \fourg,K)$, consider the
Riemannian product manifold, $\hat \hyp = \bbR \times M$,
$\hat \fourg = dt^2 + g$, and extend $K$ to $\hat \hyp$ by taking it to be
constant along the $t$-lines.   Given a function $f$ on $\hyp$, consider
its graph, $N_f = {\rm graph}\, f = \{(t,x)\in \hat \hyp: t= f(x), x \in \hyp\}$, equipped with the induced metric.  Then Jang's equation is the equation,
\beq\label{jang}
(H+P)(f) = H(f) + P(f) = 0 \,,
\eeq
where $H(f) =$ the mean curvature of $N_f$ and
$P = {\rm tr}\,_{N_f} K$ (compare with Equation \eqref{thid}).

Schoen and Yau~\cite{SchoenYauPMT2} established existence and
regularity for Jang's equation with respect to asymptotically
flat initial sets as part of their approach to proving the
positive mass theorem for general, nonmaximal, initial data
sets.   In the process they discovered an obstruction to global
existence: Solutions to Jang's equation  tend to blow-up in the
presence of MOTS in the initial data $(\hyp,g,K)$.   This
problematic blow-up behavior that Schoen and Yau had to contend
with has now been turned on its head to become a feature of
Jang's equation:  In order to establish the existence of MOTS,
one induces blow-up of the Jang equation.   This is the
approach taken in~\cite{Eichmair,AndM2}  in a somewhat
different situation.

In order to obtain solutions to Jang's equation one considers the regularized
equation,
\beq\label{jangreg}
(H+P)(f_t) = tf_t    \,.
\eeq
In~\cite{Eichmair} Eichmair uses a Perron method to obtain solutions $f_t$
to \eqref{jangreg} for $t$ sufficiently small, with  values that tend to infinity 
in a small collared neighborhood of $\SO$, and that tend to
minus infinity
in a small collared neighborhood of $\SI $, as $t\to 0$.   The construction of Perron sub and super
solutions makes use of the barrier conditions.   Using the ``almost minimizing"
property~\cite{Eichmair} of the graphs $N_t := {\rm graph}\, f_t$,  one is able to pass to a smooth
subsequential limit manifold $N$, bounded away from $\d\Omega$, each component
of which is either a cylinder or a graph that asymptotes to a cylinder.   The projection into
$\hyp$ of such a  cylinder produces  the desired MOTS.

Under the barrier conditions of Theorem \ref{motsexist},
Andersson and Metzger~\cite{AndM2} were able to make use of
this basic existence result,  to establish for $3$-dimensional
initial data sets, the existence of an {\it outermost} MOTS,
thereby providing a rigorous proof of a long held ``folk
belief" in the theory of black holes that the boundary of the
so-called outer trapped region in a time slice of space-time
is a smooth MOTS.   The key to proving the existence of an
outermost MOTS is a compactness result for stable MOTS which
follows from the extrinsic curvature estimates obtained by
Andersson and Metzger in~\cite{AndM1}, together with an area
bound obtained in~\cite{AndM2}.  This area bound follows from
an outer injectivity radius estimate for a certain class of
MOTS, established  by an interesting surgery procedure.   The
powerful methods developed by Eichmair
in~\cite{Eichmair,Eichmair2} can be used to extend this result
to $n$-dimensional initial data sets, $3 \le n \le 7$.

Consider, now, a foliation $\{\hyp_t\}$, $a < t < b$ of a
region of a space-time $(\mcM^{n+1},\fourg)$ by spacelike
hypersurfaces $\hyp_t$.  If each slice $M_t$ admits an
outermost MOTS $\Si_t$, then the family of MOTSs $\{\Si_t\}$
may form a hypersurface in space-time. A smooth spacelike
hypersurface $H$ in a space-time $(\mcM^{n+1},\fourg)$ foliated
by MOTS is called a {\it marginally outer trapped tube} (MOTT).
In \cite{AMS1,AMS2}, Andersson, Mars and Simon have obtained a
rigorous existence result for MOTTs.  Consider a spacelike
foliation $\{\hyp_t\}$, $a < t < b$, and suppose $\Si_{t_0}$ is
a MOTS in $\hyp_{t_0}$.   They prove that if $\Si_{t_0}$ is
{\it strictly} stable, i.e. if the principal eigenvalue of the
associated stability operator \eqref{stabop} is strictly
positive, then there exists a MOTT $H$ such  that for $t$ close
to $t_0$, $\S_t := H \cap \hyp_t$ is a MOTS.  For further related
results see \cite{AndMMS}.

Many challenging questions concerning the global existence and
behavior of MOTTs remain open.  Building on the analysis of
Dafermos~\cite{DafermosCBH2}, Williams~\cite{WilliamsMTT} gave
sufficient conditions on spherically symmetric black hole
space-times satisfying the dominant energy condition, insuring
that a spherically symmetric MOTT exists, is achronal, and is
asymptotic to the event horizon. \opp{oppmtt} {Understanding
the generic asymptotic behavior of MOTTs in the nonspherically
symmetric case remains an interesting open problem.}

A MOTT satisfying certain supplementary conditions gives rise
to the notion of a {\it dynamical horizon}, which provides an
alternative, quasi-local description of a black hole. For an
extensive review of dynamical horizons and related concepts,
including physical applications, see~\cite{AKLivingReviews}.


\appendix

{\small \section{Open problems}
 \label{ssopp}
We compile here a list of interesting open problems discussed
in the paper. They all appear to be difficult, of varying
degrees of difficulty, with some {most likely} intractable in
the foreseeable future.

\begin{enumerate}
\item
\bams{%
 Classify all vacuum near-horizon geometries with compact
    cross-sections, i.e., Riemannian metrics on compact
    manifolds satisfying \eq{vEe},
    p.~\pageref{nearHorgeom},  in low dimensions.
\item
    }
    Remove the hypotheses of analyticity, non-degeneracy,
    and connectedness in the black-hole uniqueness
    Theorem~\ref{Tubh}, p.~\pageref{bhanalyticity}.
\item Construct a five-dimensional, stationary, \regular{}
    vacuum black hole with two-dimensional group of
    isometries, or show that there are no such black holes.
    More generally, classify
    such black holes. Compare p.~\pageref{bhhigher}.
\item Prove that the positive energy theorem holds in all
    dimensions without the spin assumption. Compare
\bams{%
p.~\pageref{opppet} and
}
p.~\pageref{opppet2}.
\item Find the optimal differentiability conditions for
    Theorem~\ref{Tlcp}. More generally, construct a
    coherent local well posedness theory for the evolution
    and constraint equations for metrics with low
    differentiability; compare Remark~\ref{Rlcp},
    p.~\pageref{oppdiff}.
\item Describe in a constructive way the set of solutions
    of the vacuum constraint equations on compact,
    asymptotically flat, and asymptotically hyperbolic
    manifolds, with arbitrary smooth initial data. Compare
    p.~\pageref{nonCMC}.
    \bams{%
    \item  Find a well posed
    initial
boundary value problem for the vacuum Einstein equations
    which is well suited for numerical treatment, and prove
    numerical convergence. See Section~\ref{sSIbvp},
    p.~\pageref{oppibvp}.
    }
\item Formulate, and prove, a precise version of
    Conjecture~\ref{Cmixmaster}, p.~\pageref{oppmixmaster}.
    More generally, formulate and prove a precise version
    of the BKL conjecture, or find an open set of metrics
    developing a singularity which do not exhibit a
    BKL-type behavior.
\item Show that generic Bianchi IX orbits have a dense
    $\omega$--limit set on the Kasner circle; compare
    p.~\pageref{oppdense}.
\item Show that uniqueness and existence of maximal
    globally hyperbolic developments, Theorem~\ref{TCBG},
    p.~\pageref{TCBG}, holds in an optimal weak
    differentiability class. This requires revisiting the
    whole causality theory for metrics of low
    differentiability.
\item Remove the condition of closed generators in
    Theorem~\ref{TMI}, p.~\pageref{oppkch}. More generally,
    show that existence of Cauchy horizons, not necessarily
    compact or analytic, implies existence of local
    isometries, or construct a counterexample.
\item Generalize Ringstr\"om's Theorem~\ref{TRingstroem},
    p.~\pageref{TRingstroem}, to Gowdy models on $S^2\times
    S^1$, on $S^3$, and on lens spaces $L(p,q)$.
\item Show that some twisting  $U(1)\times U(1)$ symmetric
    vacuum models on $\T^3$ have mixmaster behavior. More
    generally, find an open set of such models with
    mixmaster behavior. Even better, analyze exhaustively
    the asymptotic behavior of those models; see
    Section~\ref{ssSUoneUone}, p.~\pageref{oppgalileo}.
\item  Show that degenerate asymptotically flat spherically
    symmetric Einstein-Maxwell-scalar field solutions are
    non-generic; compare Section~\ref{ssDafermos},
    p.~\pageref{oppdafermos}. Analyze what happens in
    solutions in which trapped surfaces do not form. More
    generally, prove weak and strong cosmic censorship
    within this class of space-times.
\item Find an open set of $U(1)$ symmetric metrics where
    the dynamics can be analyzed in the contracting
    direction. More generally, analyze exhaustively the
    dynamics of those models, see Section~\ref{ssSUone},
    p.~\pageref{oppuone}.
\item Show that polyhomogeneous initial data in
    Theorem~\ref{main}, p.~\pageref{main}, lead to
    solutions with polyhomogeneous behavior at null
    infinity.
\item Show that the Kerr solution is stable against small
    vacuum perturbations, compare Section~\ref{sswebhb},
    p.~\pageref{sswebhb}.
\item  Determine what topologies of positive Yamabe type
    can be realized as outermost MOTSs in space-times
    satisfying the dominant energy condition, compare
    p.~\pageref{oppgs}.
\item Prove initial data equivalents of
    Theorems~\ref{posyam2} and \ref{rigid},
    p.~\pageref{rigid}.
\item   Describe the generic behavior near the event
    horizon of marginally trapped tubes in black hole
    space-times, see p.~\pageref{oppmtt}.
\end{enumerate}
}

\bigskip

{\sc Acknowledgements:}  We are grateful to the Banff
International Research Station and to the Mittag-Leffler
Institute for hospitality at various stages of work on this
project. PTC and DP acknowledge the friendly hospitality of the
University of Miami during part of the work on this paper. We
are indebted to Jim Isenberg and Catherine Williams for many
useful comments on a previous version of the manuscript.
Further thanks are due to Catherine Williams for providing some
figures, and for helping with  others; and to Woei-Chet Lim for
providing Figure~\ref{fig:billiard}.

\bibliographystyle{amsplain}
\bibliography{./../references/newbiblio,%
./../references/newbib,%
./../references/reffile,%
./../references/bibl,%
./../references/Energy,%
./../references/hip_bib,%
./../references/dp-BAMS,%
./../references/prop2,%
./../references/besse2,%
./../references/netbiblio,%
./../references/bartnik,%
./../references/myGR}

\def\polhk#1{\setbox0=\hbox{#1}{\ooalign{\hidewidth
  \lower1.5ex\hbox{`}\hidewidth\crcr\unhbox0}}}
  \def\polhk#1{\setbox0=\hbox{#1}{\ooalign{\hidewidth
  \lower1.5ex\hbox{`}\hidewidth\crcr\unhbox0}}} \def\cprime{$'$}
  \def\cprime{$'$} \def\cprime{$'$} \def\cprime{$'$}
\providecommand{\bysame}{\leavevmode\hbox to3em{\hrulefill}\thinspace}
\providecommand{\MR}{\relax\ifhmode\unskip\space\fi MR }
\providecommand{\MRhref}[2]{%
  \href{http://www.ams.org/mathscinet-getitem?mr=#1}{#2}
}
\providecommand{\href}[2]{#2}
\begin{thebibliography}{100}

\bibitem{AIK}
S.~Alexakis, A.D. Ionescu, and S.~Klainerman, \emph{{Hawking's local rigidity
  theorem without analyticity}},  (2009), arXiv:0902.1173.

\bibitem{ICA07}
P.~Allen, A.~Clausen, and J.~Isenberg, \emph{Near-constant mean curvature
  solutions of the {E}instein constraint equations with non-negative {Y}amabe
  metrics}, Class.\ Quantum Grav. \textbf{25} (2008), 075009, 15,
  arxiv:0710.0725v1[gr-qc]. \MR{MR2404418}

\bibitem{Andrev}
L.~Andersson, \emph{The global existence problem in general relativity}, The
  Einstein Equations and the Large Scale Behavior of Gravitational Fields (P.T.
  Chru\'{s}ciel and H.~Friedrich, eds.), Birkh{\"a}user, Basel, 2004,
  arXiv:gr-qc/9911032v4, pp.~71--120.

\bibitem{BlueAndersson}
L.~Andersson and P.~Blue, \emph{{Hidden symmetries and decay for the wave
  equation on the Kerr spacetime}},  (2009), arXiv:0908.2265 [math.AP].

\bibitem{AndEM}
L.~Andersson, M.~Eichmair, and J.~Metzger, \emph{{Jang's equation and its
  applications to marginally trapped surfaces}},  (2010), arXiv:1006.4601
  [gr-qc].

\bibitem{AndMMS}
L.~Andersson, M.~Mars, J.~Metzger, and W.~Simon, \emph{The time evolution of
  marginally trapped surfaces}, Classical Quantum Gravity \textbf{26} (2009),
  085018, 14. \MR{MR2524562 (2010i:83005)}

\bibitem{AMS1}
L.~Andersson, M.~Mars, and W.~Simon, \emph{{Local existence of dynamical and
  trapping horizons}}, Phys.\ Rev.\ Lett. \textbf{95} (2005), 111102.

\bibitem{AMS2}
\bysame, \emph{{Stability of marginally outer trapped surfaces and existence of
  marginally outer trapped tubes}}, Adv. Theor. Math. Phys. \textbf{12} (2008),
  853--888. \MR{MR2420905 (2010a:83005)}

\bibitem{AndM1}
L.~Andersson and J.~Metzger, \emph{{Curvature estimates for stable marginally
  trapped surfaces}},  (2005), arXiv:gr-qc/0512106.

\bibitem{AndM2}
\bysame, \emph{{The area of horizons and the trapped region}}, Commun.\ Math.\
  Phys. \textbf{290} (2009), 941--972, arXiv:0708.4252 [gr-qc]. \MR{MR2525646
  (2010f:53118)}

\bibitem{AnderssonMoncriefStability2}
L.~Andersson and V.~Moncrief, \emph{Einstein spaces as attractors for the
  {Einstein} flow}, in preparation.

\bibitem{AnderssonMoncriefAIHP}
\bysame, \emph{Elliptic-hyperbolic systems and the {E}instein equations}, Ann.\
  Henri Poincar\'e \textbf{4} (2003), 1--34, arXiv:gr-qc/0110111.

\bibitem{AndMon}
\bysame, \emph{Future complete vacuum space-times}, The Einstein Equations and
  the Large Scale Behavior of Gravitational Fields (P.T. Chru\'{s}ciel and
  H.~Friedrich, eds.), Birkh{\"a}user, Basel, 2004, pp.~299--330,
  arXiv:gr--qc/0303045. \MR{MR2098919 (2006c:83004)}

\bibitem{AnderssonRendall}
L.~Andersson and A.D. Rendall, \emph{Quiescent cosmological singularities},
  Commun.\ Math.\ Phys. \textbf{218} (2001), 479--511, arXiv:gr-qc/0001047.
  \MR{2002h:83072}

\bibitem{ADM}
R.~Arnowitt, S.~Deser, and C.W. Misner, \emph{The dynamics of general
  relativity}, Gravitation: An introduction to current research, Wiley, New
  York, 1962, pp.~227--265. \MR{MR0143629 (26 \#1182)}

\bibitem{AG}
A.~Ashtekar and G.J. Galloway, \emph{Some uniqueness results for dynamical
  horizons}, Adv. Theor. Math.\ Phys. \textbf{9} (2005), 1--30. \MR{MR2193368
  (2006k:83101)}

\bibitem{AKLivingReviews}
A.~Ashtekar and B.~Krishnan, \emph{{Isolated and dynamical horizons and their
  applications}}, Living Rev. Rel. \textbf{7} (2004), gr-qc/0407042.

\bibitem{AshtekarLewandowski}
A.~Ashtekar and J.~Lewandowski, \emph{{Background independent quantum gravity:
  A status report}}, Class.\ Quantum Grav. \textbf{21} (2004), R53--R152,
  arXiv:gr-qc/0404018.

\bibitem{barcelo-2005-8}
C.~Barcel\'o, S.~Liberati, and M.~Visser, \emph{Analogue gravity}, Living Rev.\
  Rel. \textbf{8}, 12, arXiv:gr-qc/0505065.

\bibitem{Barcelo:2001ah}
\bysame, \emph{Analog gravity from field theory normal modes?}, Class.\ Quantum
  Grav. \textbf{18} (2001), 3595--3610, arXiv:gr-qc/0104001.

\bibitem{BarrowTipler}
J.D. Barrow and F.J. Tipler, \emph{Analysis of the generic singularity studies
  by {B}elinskii, {K}halatnikov, and {L}ifschitz}, Phys.\ Rep. \textbf{56}
  (1979), 371--402. \MR{MR555355 (83c:83005)}

\bibitem{Bartnik84}
R.~Bartnik, \emph{The existence of maximal hypersurfaces in asymptotically flat
  space-times}, Commun.\ Math.\ Phys. \textbf{94} (1984), 155--175.

\bibitem{Bartnik86}
\bysame, \emph{The mass of an asymptotically flat manifold}, Commun.\ Pure
  Appl.\ Math. \textbf{39} (1986), 661--693.

\bibitem{bartnik:variational}
\bysame, \emph{Regularity of variational maximal surfaces}, Acta Math.
  \textbf{161} (1988), 145--181.

\bibitem{bartnik:cosmological}
\bysame, \emph{Remarks on cosmological spacetimes and constant mean curvature
  surfaces}, Commun.\ Math.\ Phys. \textbf{117} (1988), 615--624.

\bibitem{Bartnikopen}
\bysame, \emph{Some open problems in mathematical relativity}, Conference on
  Mathematical Relativity (Canberra, 1988), Proc.\ Centre Math.\ Anal.\
  Austral.\ Nat.\ Univ., vol.~19, Austral.\ Nat.\ Univ., Canberra, 1989,
  pp.~244--268. \MR{MR1020805 (90g:83001)}

\bibitem{bartnik:phase}
\bysame, \emph{Phase space for the {E}instein equations}, Commun.\ Anal.\ Geom.
  \textbf{13} (2005), 845--885, arXiv:gr-qc/0402070. \MR{MR2216143
  (2007d:83012)}

\bibitem{BCOM}
R.~Bartnik, P.T. Chru\'{s}ciel, and N.~\'O Murchadha, \emph{{On maximal
  surfaces in asymptotically flat space-times.}}, Commun.\ Math.\ Phys.
  \textbf{130} (1990), 95--109.

\bibitem{BartnikIsenberg}
R.~Bartnik and J.~Isenberg, \emph{The constraint equations}, The Einstein
  equations and the large scale behavior of gravitational fields, Birkh\"auser,
  Basel, 2004, pp.~1--38. \MR{MR2098912 (2005j:83007)}

\bibitem{BaumgarteShapiro}
T.W. Baumgarte and S.L. Shapiro, \emph{Numerical integration of {E}instein's
  field equations}, Phys.\ Rev.\ D (3) \textbf{59} (1999), 024007, 7,
  arXiv:gr-qc/9810065. \MR{MR1692065 (2000b:83001)}

\bibitem{BeemEhrlichEasley}
J.K. Beem, P.E. Ehrlich, and K.L. Easley, \emph{Global {L}orentzian geometry},
  {Second} ed., Marcel Dekker Inc., New York, 1996. \MR{MR1384756 (97f:53100)}

\bibitem{Beguin}
F.~B\'eguin, \emph{{Aperiodic oscillatory asymptotic behavior for some
  {Bianchi} spacetimes}},  (2010), arXiv:1004.2984 [gr-qc].

\bibitem{BeigTT}
R.~Beig, \emph{T{T}-tensors and conformally flat structures on
  {$3$}-manifolds}, Mathematics of gravitation, Part I (Warsaw, 1996), Banach
  Center Publ., vol.~41, Polish Acad. Sci., Warsaw, 1997, pp.~109--118.
  \MR{MR1466511 (98k:53040)}

\bibitem{ChBeig1}
R.~Beig and P.T. Chru\'{s}ciel, \emph{Killing vectors in asymptotically flat
  space-times: {I. A}symptotically translational {K}illing vectors and the
  rigid positive energy theorem}, Jour.\ Math.\ Phys. \textbf{37} (1996),
  1939--1961, arXiv:gr-qc/9510015.

\bibitem{ChBeigKIDs}
\bysame, \emph{Killing {I}nitial {D}ata}, Class.\ Quantum. Grav. \textbf{14}
  (1997), A83--A92, A special issue in honour of Andrzej Trautman on the
  occasion of his 64th Birthday, J.Tafel, editor. \MR{MR1691888 (2000c:83011)}

\bibitem{ChBeig3}
\bysame, \emph{The asymptotics of stationary electro-vacuum metrics in odd
  space-time dimensions}, Class.\ Quantum Grav. \textbf{24} (2007), 867--874.
  \MR{MR2297271}

\bibitem{CHBeignokids}
R.~Beig, P.T. Chru\'{s}ciel, and R.~Schoen, \emph{{KIDs} are non-generic},
  Ann.\ H.\ Poincar\'e \textbf{6} (2005), 155--194, arXiv:gr-qc/0403042.
  \MR{MR2121280 (2005m:83013)}

\bibitem{Beigneill:trapped}
R.~Beig and N.{\'O} Murchadha, \emph{Vacuum space-times with future trapped
  surfaces}, Class.\ Quantum Grav. \textbf{13} (1996), 739--751. \MR{MR1383704
  (97e:83005)}

\bibitem{BKL}
V.A. Belinskii, I.M. Khalatnikov, and E.M. Lifshtitz, \emph{Oscillatory
  approach to a singular point in the relativistic cosmology}, Adv.\ Phys.
  \textbf{19} (1970), 525--573.

\bibitem{BCIM}
B.~Berger, P.T. Chru\'{s}ciel, J.~Isenberg, and V.~Moncrief, \emph{Global
  foliations of vacuum space-times with {{$T^2$}} isometry}, Ann.\ Phys.\ (NY)
  \textbf{260} (1997), 117--148.

\bibitem{BIW}
B.~Berger, J.~Isenberg, and M.~Weaver, \emph{Oscillatory approach to the
  singularity in vacuum space-times with {$T^2$} isometry}, Phys.\ Rev.
  \textbf{D64} (2001), 084006, arXiv:gr-qc/0104048, erratum-ibid. D67, 129901
  (2003).

\bibitem{BevMixmaster}
B.K. Berger, \emph{Hunting local mixmaster dynamics in spatially inhomogeneous
  cosmologies}, Class.\ Quantum Grav. \textbf{21} (2004), S81--S95, A
  space-time safari: Essays in honour of Vincent Moncrief. \MR{MR2053000
  (2005f:83066)}

\bibitem{BGIMW}
B.K. Berger, D.~Garfinkle, J.~Isenberg, V.~Moncrief, and M.~Weaver, \emph{The
  singularity in generic gravitational collapse is spacelike, local, and
  oscillatory}, Mod. Phys. Lett. \textbf{A13} (1998), 1565--1574,
  arXiv:gr-qc/9805063.

\bibitem{Berger:2000uf}
B.K. Berger and V.~Moncrief, \emph{Exact {U(1) symmetric cosmologies with local
  M}ixmaster dynamics}, Phys.\ Rev. \textbf{D62} (2000), 023509,
  arXiv:gr-qc/0001083.

\bibitem{BevVince:U1chaos}
\bysame, \emph{Signature for local mixmaster dynamics in {$\rm U(1)$} symmetric
  cosmologies}, Phys.\ Rev.\ D \textbf{62} (2000), 123501, 9,
  arXiv:gr-qc/0006071. \MR{MR1813870 (2001k:83027)}

\bibitem{BernalSanchez}
A.N. Bernal and M.~S{\'a}nchez, \emph{Smoothness of time functions and the
  metric splitting of globally hyperbolic space-times}, Commun.\ Math.\ Phys.
  \textbf{257} (2005), 43--50. \MR{MR2163568 (2006g:53105)}

\bibitem{Besse}
A.L. Besse, \emph{{Einstein manifolds}}, Ergebnisse der Mathematik und ihrer
  Grenzgebiete. 3. Folge, vol.~10, Springer Verlag, Berlin, New York,
  Heidelberg, 1987.

\bibitem{BieriZipser}
L.~Bieri and N.~Zipser, \emph{Extensions of the stability theorem of the
  {M}inkowski space in general relativity}, AMS/IP Studies in Advanced
  Mathematics, vol.~45, American Mathematical Society, Providence, RI, 2009,
  pp.~xxiv+491. \MR{MR2531716}

\bibitem{Birkhoff23}
G.~D. Birkhoff, \emph{Relativity and modern physics}, Harvard UP, 1923.

\bibitem{Birmingham}
D.~Birmingham, \emph{Topological black holes in anti-de {Sitter} space},
  Class.\ Quantum Grav. \textbf{16} (1999), 1197--1205, arXiv:hep-th/9808032.
  \MR{MR1696149 (2000c:83062)}

\bibitem{BishopWinicour}
N.T. Bishop, R.~G{\'o}mez, L.~Lehner, M.~Maharaj, and J.~Winicour,
  \emph{Characteristic initial data for a star orbiting a black hole}, Phys.\
  Rev.\ D \textbf{72} (2005), 024002, 16. \MR{MR2171948 (2006d:83051)}

\bibitem{BlueMaxwell}
P.~Blue, \emph{Decay of the {Maxwell field on the S}chwarzschild manifold},
  Jour.\ Hyp.\ Differ. Equ. \textbf{5} (2008), 807--856, arXiv:0710.4102
  [math.AP]. \MR{MR2475482 (2010d:35371)}

\bibitem{BlueSterbenz}
P.~Blue and J.~Sterbenz, \emph{Uniform decay of local energy and the
  semi-linear wave equation on schwarzchild space}, Commun.\ Math.\ Phys.
  \textbf{268} (2005), 481--504, arXiv:math.AP/0510315. \MR{MR2259204
  (2007i:58037)}

\bibitem{BoothFairhurst}
I.~Booth and S.~Fairhurst, \emph{Extremality conditions for isolated and
  dynamical horizons}, Phys.\ Rev.\ D \textbf{77} (2008), 084005, 14 pp.
  \MR{MR2443517 (2009k:83046)}

\bibitem{BrayPenroseIneq}
H.L. Bray, \emph{Proof of the {R}iemannian {P}enrose inequality using the
  positive mass theorem}, Jour.\ Diff.\ Geom. \textbf{59} (2001), 177--267.
  \MR{MR1908823 (2004j:53046)}

\bibitem{BrayICM}
\bysame, \emph{Black holes and the {P}enrose inequality in general relativity},
  Proceedings of the International Congress of Mathematicians, Vol. II
  (Beijing, 2002) (Beijing), Higher Ed. Press, 2002, pp.~257--271. \MR{1 957
  038}

\bibitem{BrayNotices}
\bysame, \emph{Black holes, geometric flows, and the {P}enrose inequality in
  general relativity}, Notices Amer. Math. Soc. \textbf{49} (2002), 1372--1381.
  \MR{MR1936643 (2003j:83052)}

\bibitem{ChBray}
H.L. Bray and P.T. Chru\'{s}ciel, \emph{The {P}enrose inequality}, The Einstein
  Equations and the Large Scale Behavior of Gravitational Fields (P.T.
  Chru\'{s}ciel and H.~Friedrich, eds.), Birkh{\"a}user, Basel, 2004,
  pp.~39--70, arXiv:gr--qc/0312047.

\bibitem{BrayKhuri2}
H.L. Bray and M.A. Khuri, \emph{{A Jang Equation Approach to the Penrose
  Inequality}},  (2009), arXiv:0910.4785 [math.DG].

\bibitem{BrayKhuri}
\bysame, \emph{{PDE's which imply the Penrose conjecture}},  (2009),
  arXiv:0905.2622 [math.DG].

\bibitem{BrayLee}
H.L. Bray and D.A. Lee, \emph{{On the Riemannian Penrose inequality in
  dimensions less than eight}}, Duke Math.\ Jour. \textbf{148} (2009), 81--106,
  arXiv:0705.1128. \MR{MR2515101}

\bibitem{BrayMiao}
H.L. Bray and P.~Miao, \emph{On the capacity of surfaces in manifolds with
  nonnegative scalar curvature}, Invent. Math. \textbf{172} (2008), 459--475.
  \MR{MR2393076}

\bibitem{BraySchoen}
H.L. Bray and R.~Schoen, \emph{Recent proofs of the {R}iemannian {P}enrose
  conjecture}, Current developments in mathematics, 1999 (Cambridge, MA), Int.
  Press, Somerville, MA, 1999, pp.~1--36. \MR{1 990 246}

\bibitem{BHMS}
Hubert Bray, Sean Hayward, Marc Mars, and Walter Simon, \emph{Generalized
  inverse mean curvature flows in spacetime}, Commun.\ Math.\ Phys.
  \textbf{272} (2007), 119--138, arXiv:gr-qc/0603014. \MR{MR2291804
  (2008c:53065)}

\bibitem{BMG}
P.~Breitenlohner, D.~Maison, and G.~Gibbons, \emph{{$4$}-dimensional black
  holes from {K}aluza-{K}lein theories}, Commun.\ Math.\ Phys. \textbf{120}
  (1988), 295--333. \MR{MR973537 (89j:83018)}

\bibitem{BF78}
D.~Brill and F.~Flaherty, \emph{Maximizing properties of extremal surfaces in
  general relativity}, Ann.\ Inst.\ H.\ Poincar\'e Sect. A (N.S.) \textbf{28}
  (1978), 335--347. \MR{MR0479299 (57 \#18741)}

\bibitem{BLP}
D.~Brill, J.~Louko, and P.~Peldan, \emph{Thermodynamics of (3+1)-dimensional
  black holes with toroidal or higher genus horizons}, Phys.\ Rev. \textbf{D56}
  (1997), 3600--3610, arXiv:gr-qc/9705012.

\bibitem{Brout:1995rd}
R.~Brout, S.~Massar, R.~Parentani, and P.~Spindel, \emph{A primer for black
  hole quantum physics}, Phys.\ Rept. \textbf{260} (1995), 329--454.

\bibitem{BILY}
R.~Budic, J.~Isenberg, L.~Lindblom, and P.~Yasskin, \emph{On the determination
  of the {C}auchy surfaces from intrinsic properties}, Commun.\ Math.\ Phys.
  \textbf{61} (1978), 87--95.

\bibitem{bunting:masood}
G.~Bunting and A.K.M. Masood{--ul--A}lam, \emph{Nonexistence of multiple black
  holes in asymptotically {E}uclidean static vacuum space-time}, Gen.\ Rel.\
  Grav. \textbf{19} (1987), 147--154.

\bibitem{CaciottaNicolo}
G.~Caciotta and F.~Nicol{\`o}, \emph{Global characteristic problem for
  {E}instein vacuum equations with small initial data {I}: {T}he initial data
  constraints}, Jour.\ Hyperbolic Diff.\ Equ. \textbf{2} (2005), 201--277,
  arXiv:gr-qc/0409028. \MR{MR2134959 (2006i:58042)}

\bibitem{CaciottaNicolo2}
\bysame, \emph{Global characteristic problem for the {Einstein} vacuum
  equations with small initial data {II}: The existence proof},  (2006),
  arXiv:gr-qc/0608038.

\bibitem{CadeauWoolgar}
C.~Cadeau and E.~Woolgar, \emph{New five dimensional black holes classified by
  horizon geometry, and a {Bianchi VI} braneworld}, Class.\ Quantum Grav.
  (2001), 527--542, arXiv:gr-qc/0011029.

\bibitem{CaiProceedings}
M.~Cai, \emph{Volume minimizing hypersurfaces in manifolds of nonnegative
  scalar curvature}, Minimal surfaces, geometric analysis and symplectic
  geometry (Baltimore, MD, 1999), Adv. Stud. Pure Math., vol.~34, Math. Soc.
  Japan, Tokyo, 2002, pp.~1--7. \MR{MR1925731 (2003f:53104)}

\bibitem{CaiGallowayRigidity}
M.~Cai and G.J. Galloway, \emph{Rigidity of area minimizing tori in 3-manifolds
  of nonnegative scalar curvature}, Commun.\ Anal. Geom. \textbf{8} (2000),
  565--573. \MR{2001j:53051}

\bibitem{CalabiRicci}
E.~Calabi, \emph{On {R}icci curvature and geodesics}, Duke Math. J. \textbf{34}
  (1967), 667--676. \MR{MR0216429 (35 \#7262)}

\bibitem{Campanelli:2005dd}
M.~Campanelli, C.~O. Lousto, P.~Marronetti, and Y.~Zlochower, \emph{{Accurate
  evolutions of orbiting black-hole binaries without excision}}, Phys.\ Rev.\
  Lett. \textbf{96} (2006), 111101.

\bibitem{Cantor77}
M.~Cantor, \emph{The existence of non-trivial asymptotically flat initial data
  for vacuum spacetimes}, Commun.\ Math.\ Phys. \textbf{57} (1977), 83--96.
  \MR{MR0462440 (57 \#2414)}

\bibitem{CarrascoMars}
A.~Carrasco and M.~Mars, \emph{{A counter-example to a recent version of the
  Penrose conjecture}},  (2009), arXiv:0911.0883 [gr-qc].

\bibitem{CarterKerr}
B.~Carter, \emph{Global structure of the {Kerr} family of gravitational
  fields}, Phys.\ Rev. \textbf{174} (1968), 1559--1571.

\bibitem{CarterlesHouches}
\bysame, \emph{Black hole equilibrium states}, Black Holes (C.\ de~Witt and B.\
  de~Witt, eds.), Gordon \& Breach, New York, London, Paris, 1973, Proceedings
  of the Les Houches Summer School.

\bibitem{Carter:1997im}
\bysame, \emph{Has the black hole equilibrium problem been solved?}, in Proc.
  of the 8th Marcel Grossmann Meeting on Relativistic Astrophysics - MG 8,
  Jerusalem, Israel, 22 - 27 June 1997, T.~Piran, Ed., World Scientific, 1999,
  arXiv:gr-qc/9712038.

\bibitem{CMSciama}
A.~Celotti, J.C. Miller, and D.W. Sciama, \emph{{Astrophysical evidence for the
  existence of black holes}}, Class.\ Quantum Grav. \textbf{16} (1999),
  A3--A21, arXiv:astro-ph/9912186.

\bibitem{ChCh2}
M.~Chae and P.T. Chru\'{s}ciel, \emph{On the dynamics of {Gowdy} space times},
  Commun.\ Pure Appl.\ Math. \textbf{57} (2004), 1015--1074,
  arXiv:gr-qc/0305029.

\bibitem{ChBlesHouches}
Y.~Choquet-Bruhat, \emph{Positive-energy theorems}, Relativity, groups and
  topology, II (Les Houches, 1983) (B.S. deWitt and R.~Stora, eds.),
  North-Holland, Amsterdam, 1984, pp.~739--785.

\bibitem{Choquet-Bruhat:safari}
\bysame, \emph{Einstein constraints on $n$ dimensional compact manifolds},
  Class.\ Quantum Grav. \textbf{21} (2004), S127--S152, arXiv:gr-qc/0311029.

\bibitem{ChBCargese}
\bysame, \emph{Future complete {U(1)} symmetric {E}insteinian space-times, the
  unpolarized case}, The Einstein Equations and the Large Scale Behavior of
  Gravitational Fields (P.T. Chru\'{s}ciel and H.~Friedrich, eds.),
  Birkh{\"a}user, Basel, 2004, pp.~251--298.

\bibitem{YCB:GRbook}
\bysame, \emph{General relativity and the {E}instein equations}, Oxford
  Mathematical Monographs, Oxford University Press, Oxford, UK, 2009.
  \MR{MR2473363 (2010f:83001)}

\bibitem{CBChristodoulou}
Y.~Choquet-Bruhat and D.~Christodoulou, \emph{Elliptic systems in
  {$H\sb{s,\delta }$} spaces on manifolds which are {E}uclidean at infinity},
  Acta Math. \textbf{146} (1981), 129--150. \MR{MR594629 (82c:58060)}

\bibitem{CCL}
Y.~Choquet-Bruhat, P.T. Chru\'{s}ciel, and J.~Loizelet, \emph{{Global solutions
  of the Einstein--Maxwell equations in higher dimension}}, Class.\ Quantum
  Grav. (2006), 7383--7394, arXiv:gr-qc/0608108.

\bibitem{CCG}
Y.~Choquet-Bruhat, P.T. Chru\'{s}ciel, and J.M. Mart\'in-Garc\'ia, \emph{{The
  light-cone theorem}}, Class.\ Quantum Grav. \textbf{26} (2009), 135011 (22
  pp), arXiv:0905.2133 [gr-qc].

\bibitem{CCM2}
\bysame, \emph{{The Cauchy problem on a characteristic cone for the Einstein
  equations in arbitrary dimensions}},  (2010), arXiv:1006.4467 [gr-qc].

\bibitem{ChoquetBruhatGeroch69}
Y.~Choquet-Bruhat and R.~Geroch, \emph{Global aspects of the {C}auchy problem
  in general relativity}, Commun.\ Math.\ Phys. \textbf{14} (1969), 329--335.
  \MR{MR0250640 (40 \#3872)}

\bibitem{CBIM91}
Y.~Choquet-Bruhat, J.~Isenberg, and V.~Moncrief, \emph{Solutions of constraints
  for {E}instein equations}, C.\ R.\ Acad.\ Sci.\ Paris S\'er.\ I Math.
  \textbf{315} (1992), 349--355. \MR{MR1179734 (93h:58151)}

\bibitem{CBIP1}
Y.~Choquet-Bruhat, J.~Isenberg, and D.~Pollack, \emph{The {E}instein-scalar
  field constraints on asymptotically {E}uclidean manifolds}, Chinese Ann.\
  Math.\ Ser.\ B \textbf{27} (2006), 31--52, arXiv:gr-qc/0506101. \MR{MR2209950
  (2007h:58049)}

\bibitem{CBIP-Leray}
\bysame, \emph{Applications of theorems of {J}ean {L}eray to the
  {E}instein-scalar field equations}, Jour.\ Fixed Point Theory Appl.
  \textbf{1} (2007), 31--46, arXiv:gr-qc/0611009. \MR{MR2282342}

\bibitem{CBIP2}
\bysame, \emph{The constraint equations for the {E}instein-scalar field system
  on compact manifolds}, Class.\ Quantum Grav. \textbf{24} (2007), 809--828,
  arXiv:gr-qc/0610045. \MR{MR2297268}

\bibitem{ChIY}
Y.~Choquet-Bruhat, J.~Isenberg, and J.W. {York$,~$Jr.}, \emph{Einstein
  constraints on asymptotically {E}uclidean manifolds}, Phys.\ Rev.\ D
  \textbf{61} (2000), 084034 (20 pp.), arXiv:gr-qc/9906095.

\bibitem{01713418}
Y.~Choquet-Bruhat and V.~Moncrief, \emph{{Future global in time Einsteinian
  space-times with {U(1)} isometry group.}}, Ann.\ Henri Poincar\'e \textbf{2}
  (2001), 1007--1064.

\bibitem{CBY}
Y.~Choquet-Bruhat and J.~York, \emph{The {C}auchy problem}, General Relativity
  (A.~Held, ed.), Plenum Press, New York, 1980, pp.~99--172.

\bibitem{Christodoulou:global}
D.~Christodoulou, \emph{Global solutions of nonlinear hyperbolic equations for
  small initial data}, Commun.\ Pure Appl.\ Math. \textbf{39} (1986), 267--282.
  \MR{MR820070 (87c:35111)}

\bibitem{ChristodoulouCMP86}
\bysame, \emph{The problem of a self-gravitating scalar field}, Commun.\ Math.\
  Phys. \textbf{105} (1986), 337--361. \MR{MR848643 (87i:83009)}

\bibitem{demetrios:scalar10}
\bysame, \emph{Examples of naked singularity formation in the gravitational
  collapse of a scalar field}, Ann.\ Math. \textbf{140} (1994), 607--653.

\bibitem{ChristodoulouAnnals99}
\bysame, \emph{The instability of naked singularities in the gravitational
  collapse of a scalar field}, Ann.\ of Math.\ (2) \textbf{149} (1999),
  183--217. \MR{MR1680551 (2000a:83086)}

\bibitem{ChristodoulouCQG99}
\bysame, \emph{On the global initial value problem and the issue of
  singularities}, Class.\ Quantum Grav. \textbf{16} (1999), A23--A35.
  \MR{MR1728432 (2001a:83010)}

\bibitem{ChristodoulouBHFormation}
\bysame, \emph{{The Formation of Black Holes in General Relativity}},  (2008),
  x+589, arXiv:0805.3880 [gr-qc]. \MR{MR2488976 (2009k:83010)}

\bibitem{ChristodoulouKlainerman93}
D.~Christodoulou and S.~Klainerman, \emph{The global nonlinear stability of the
  {M}inkowski space}, Princeton Mathematical Series, vol.~41, Princeton
  University Press, Princeton, NJ, 1993. \MR{MR1316662 (95k:83006)}

\bibitem{christodoulou:murchadha}
D.~Christodoulou and N.{\'O} Murchadha, \emph{The boost problem in general
  relativity}, Commun.\ Math.\ Phys. \textbf{80} (1981), 271--300.

\bibitem{CT293}
D.~Christodoulou and A.~Shadi Tahvildar-Zadeh, \emph{On the asymptotic behavior
  of spherically symmetric wave maps}, Duke Math.\ Jour. \textbf{71} (1993),
  31--69. \MR{94j:58044}

\bibitem{CT93}
\bysame, \emph{On the regularity of spherically symmetric wave maps}, Commun.\
  Pure Appl. Math \textbf{46} (1993), 1041--1091.

\bibitem{Chrusciel:2002mi}
P.T. Chru\'{s}ciel, \emph{Black holes}, Proceedings of the T{\"u}bingen
  Workshop on the Conformal Structure of Space-times, H.~Friedrich and J.
  Frauendiener, Eds., Springer Lecture Notes in Physics {\bf 604}, 61--102
  (2002), arXiv:gr-qc/0201053.

\bibitem{ChErice}
\bysame, \emph{Boundary conditions at spatial infinity from a {H}amiltonian
  point of view}, Topological Properties and Global Structure of Space--Time
  (P.\ Bergmann and V.\ de~Sabbata, eds.), Plenum Press, New York, 1986, pp.
  49--59, {URL} \url{http://www.phys.univ-tours.fr/~piotr/scans}.

\bibitem{ChANOP}
\bysame, \emph{On space-times with {${\rm U}(1)\times {\rm U}(1)$} symmetric
  compact {C}auchy surfaces}, Ann.\ Phys. \textbf{202} (1990), 100--150.
  \MR{MR1067565 (91h:83007)}

\bibitem{ChrCM}
\bysame, \emph{On uniqueness in the large of solutions of {E}instein equations
  (``{S}trong {C}osmic {C}ensorship'')}, Cont. Math. \textbf{132} (1992),
  235--273.

\bibitem{Chorbits}
\bysame, \emph{On completeness of orbits of {K}illing vector fields}, Class.\
  Quantum Grav. \textbf{10} (1993), 2091--2101, arXiv:gr-qc/9304029.

\bibitem{Ch:rigidity}
\bysame, \emph{On rigidity of analytic black holes}, Commun.\ Math.\ Phys.
  \textbf{189} (1997), 1--7, arXiv:gr-qc/9610011.

\bibitem{Chstatic}
\bysame, \emph{The classification of static vacuum space-times containing an
  asymptotically flat spacelike hypersurface with compact interior}, Class.\
  Quantum Grav. \textbf{16} (1999), 661--687, \emph{Corrigendum} in
  arXiv:gr-qc/9809088v2.

\bibitem{ChstaticelvacarxivErr}
\bysame, \emph{Towards the classification of static electro--vacuum space-times
  containing an asymptotically flat spacelike hypersurface with compact
  interior}, Class.\ Quantum Grav. \textbf{16} (1999), 689--704, Corrigendum in
  arXiv:gr-qc/9810022v3.

\bibitem{ChBeijing}
\bysame, \emph{Beijing lecture notes on mathematical relativity},
  \url{www.phys.univ-tours.fr/~piotr/papers/BeijingAll.pdf}, 2006.

\bibitem{ChConformalBoundary}
\bysame, \emph{Conformal boundary extensions of {Lorentzian} manifolds},
  (2006), preprint AEI-2006-039, arXiv:gr-qc/0606101.

\bibitem{ChUone}
\bysame, \emph{Mass and angular-momentum inequalities for axi-symmetric initial
  data sets. {I. Positivity of mass}}, Annals Phys. \textbf{323} (2008),
  2566--2590, doi:10.1016/j.aop.2007.12.010, arXiv:0710.3680 [gr-qc].

\bibitem{CC}
P.T. Chru\'{s}ciel and J.~Cortier, \emph{{On the geometry of {Emparan-Reall}
  black rings}},  (2008), arXiv:0807.2309 [gr-qc].

\bibitem{CCGP}
P.T. Chru\'{s}ciel, J.~Cortier, and A.~Garcia-Parrado, \emph{{On the global
  structure of the Pomeransky-Senkov black holes}},  (2009), arXiv:0911.0802
  [gr-qc].

\bibitem{CCI}
P.T. Chru\'{s}ciel, J.~Corvino, and J.~Isenberg, \emph{{Construction of
  $N$-body time-symmetric initial data sets in general relativity}},  (2009),
  arXiv:0909.1101 [gr-qc].

\bibitem{CCI2}
\bysame, \emph{{Construction of $N$-body initial data sets in general
  relativity}},  (2010), in preparation.

\bibitem{ChCo}
P.T. Chru\'{s}ciel and J.~Lopes Costa, \emph{On uniqueness of stationary black
  holes}, Ast\'erisque (2008), 195--265, arXiv:0806.0016v2 [gr-qc].

\bibitem{ChDelay2}
P.T. Chru\'{s}ciel and E.~Delay, \emph{Existence of non-trivial asymptotically
  simple vacuum space-times}, Class.\ Quantum Grav. \textbf{19} (2002),
  L71--L79, arXiv:gr-qc/0203053, erratum-ibid, 3389. \MR{MR1902228
  (2003e:83024a)}

\bibitem{ChDelay}
\bysame, \emph{On mapping properties of the general relativistic constraints
  operator in weighted function spaces, with applications}, M\'em.\ Soc.\
  Math.\ de France. \textbf{94} (2003), vi+103, arXiv:gr-qc/0301073v2.
  \MR{MR2031583 (2005f:83008)}

\bibitem{ChDelayHilbert}
\bysame, \emph{Manifold structures for sets of solutions of the general
  relativistic constraint equations}, Jour.\ Geom\ Phys. (2004), 442--472,
  arXiv:gr-qc/0309001v2. \MR{MR2085346 (2005i:83008)}

\bibitem{ChDelayAH}
\bysame, \emph{Gluing constructions for asymptotically hyperbolic manifolds
  with constant scalar curvature},  (2007), arXiv:0710.xxx [gr-qc].

\bibitem{ChDGH}
P.T. Chru\'{s}ciel, E.~Delay, G.~Galloway, and R.~Howard, \emph{Regularity of
  horizons and the area theorem}, Annales Henri Poincar\'e \textbf{2} (2001),
  109--178, arXiv:gr-qc/0001003. \MR{MR1823836 (2002e:83045)}

\bibitem{ChGstatic}
P.T. Chru\'{s}ciel and G.~Galloway, \emph{Uniqueness of static black-holes
  without analyticity},  (2010), arXiv:1004.0513 [gr-qc].

\bibitem{ChImaxTaubNUT}
P.T. Chru\'{s}ciel and J.~Isenberg, \emph{Non--isometric vacuum extensions of
  vacuum maximal globally hyperbolic space--times}, Phys. Rev. D (3)
  \textbf{48} (1993), 1616--1628. \MR{MR1236815 (94f:83007)}

\bibitem{CIM}
P.T. Chru\'{s}ciel, J.~Isenberg, and V.~Moncrief, \emph{Strong cosmic
  censorship in polarised {G}owdy space--times}, Class.\ Quantum Grav.
  \textbf{7} (1990), 1671--1680.

\bibitem{CIP:PRL}
P.T. Chru{\'s}ciel, J.~Isenberg, and D.~Pollack, \emph{Gluing initial data sets
  for general relativity}, Phys.\ Rev.\ Lett. \textbf{93} (2004), 081101,
  arXiv:gr-qc/0409047.

\bibitem{CIP:CMP}
\bysame, \emph{Initial data engineering}, Commun.\ Math.\ Phys. \textbf{257}
  (2005), 29--42, arXiv:gr-qc/0403066. \MR{MR2163567 (2007d:83013)}

\bibitem{ChLake}
P.T. Chru\'{s}ciel and K.~Lake, \emph{Cauchy horizons in {G}owdy space times},
  Class.\ Quantum Grav. \textbf{21} (2004), S153--S170, arXiv:gr-qc/0307088.

\bibitem{ChMaerten}
P.T. Chru\'{s}ciel and D.~Maerten, \emph{Killing vectors in asymptotically flat
  space-times: {II}. {A}symptotically translational {K}illing vectors and the
  rigid positive energy theorem in higher dimensions}, Jour.\ Math.\ Phys.
  \textbf{47} (2006), 022502, 10 pp., arXiv:gr-qc/0512042. \MR{MR2208148
  (2007b:83054)}

\bibitem{ChNguyen}
P.T. Chru\'{s}ciel and L.~Nguyen, \emph{{A uniqueness theorem for degenerate
  Kerr-Newman black holes}},  (2010), arXiv:1002.1737 [gr-qc].

\bibitem{CPP}
P.T. Chru\'{s}ciel, F.~Pacard, and D.~Pollack, \emph{{Singular {Yamabe metrics
  and initial data with exactly Kottler-Schwarzschild-de Sitter ends II.
  Generic} metrics}},  (2008), arXiv:0803.1817 [gr-qc].

\bibitem{ChPollack}
P.T. Chru\'{s}ciel and D.~Pollack, \emph{{Singular Yamabe metrics and initial
  data with \emph{exactly} Kottler--Schwarzschild--de Sitter ends}}, Ann.\
  Henri Poincar\'e (2008), 639--654, arXiv:0710.3365 [gr-qc].

\bibitem{CRT}
P.T. Chru\'{s}ciel, H.S. Reall, and K.P. Tod, \emph{On non-existence of static
  vacuum black holes with degenerate components of the event horizon}, Class.\
  Quantum Grav. \textbf{23} (2006), 549--554, arXiv:gr-qc/0512041.
  \MR{MR2196372 (2007b:83090)}

\bibitem{ChRendall}
P.T. Chru\'{s}ciel and A.~Rendall, \emph{Strong cosmic censorship in vacuum
  space-times with compact, locally homogeneous {C}auchy surfaces}, Ann.\
  Physics \textbf{242} (1995), 349--385. \MR{MR1349391 (96f:83085)}

\bibitem{CT}
P.T. Chru\'{s}ciel and K.P. Tod, \emph{The classification of static
  electro-vacuum space-times containing an asymptotically flat spacelike
  hypersurface with compact interior}, Commun.\ Math.\ Phys. \textbf{271}
  (2007), 577--589. \MR{MR2291788}

\bibitem{ChWald1}
P.T. Chru\'{s}ciel and R.M. Wald, \emph{Maximal hypersurfaces in stationary
  asymptotically flat space--times}, Commun.\ Math.\ Phys. \textbf{163} (1994),
  561--604, arXiv:gr--qc/9304009. \MR{MR1284797 (95f:53113)}

\bibitem{ChWald}
\bysame, \emph{On the topology of stationary black holes}, Class.\ Quantum
  Grav. \textbf{11} (1994), no.~12, L147--152, arXiv:gr--qc/9410004.
  \MR{MR1307013 (95j:83080)}

\bibitem{CornishLevin}
N.J. Cornish and J.J. Levin, \emph{The mixmaster universe: {A chaotic Farey
  tale}}, Phys.\ Rev. \textbf{D55} (1997), 7489--7510, arXiv:gr-qc/9612066.

\bibitem{Corvino}
J.~Corvino, \emph{Scalar curvature deformation and a gluing construction for
  the {E}instein constraint equations}, Commun.\ Math.\ Phys. \textbf{214}
  (2000), 137--189. \MR{MR1794269 (2002b:53050)}

\bibitem{CorvinoAHP}
\bysame, \emph{On the existence and stability of the {Penrose}
  compactification}, Annales H.\ Poincar\'e \textbf{8} (2007), 597--620.
  \MR{MR2329363}

\bibitem{CorvinoSchoen2}
J.~Corvino and R.M. Schoen, \emph{On the asymptotics for the vacuum {E}instein
  constraint equations}, Jour.\ Diff.\ Geom. \textbf{73} (2006), 185--217,
  arXiv:gr-qc/0301071. \MR{MR2225517 (2007e:58044)}

\bibitem{DafermosCBH1}
M.~Dafermos, \emph{Stability and instability of the {C}auchy horizon for the
  spherically symmetric {E}instein-{M}axwell-scalar field equations}, Ann.\ of
  Math. (2) \textbf{158} (2003), 875--928. \MR{MR2031855 (2005f:83009)}

\bibitem{DafermosCBH2}
\bysame, \emph{{The interior of charged black holes and the problem of
  uniqueness in general relativity}}, Commun.\ Pure Appl.\ Math. \textbf{58}
  (2005), 445--504, arXiv:gr-qc/0307013. \MR{MR2119866 (2006e:83087)}

\bibitem{dafermos:rodnianski:price}
M.~Dafermos and I.~Rodnianski, \emph{A proof of {P}rice's law for the collapse
  of a self-gravitating scalar field}, Invent.\ Math. \textbf{162} (2005),
  381--457, arXiv:gr-qc/0309115. \MR{MR2199010 (2006i:83016)}

\bibitem{DafermosRodnianskiKerr}
\bysame, \emph{{A proof of the uniform boundedness of solutions to the wave
  equation on slowly rotating Kerr backgrounds}},  (2008), arXiv:0805.4309
  [gr-qc].

\bibitem{DafermosRodnianskiClay}
\bysame, \emph{{Lectures on black holes and linear waves}},  (2008),
  arXiv:0811.0354 [gr-qc].

\bibitem{DafermosRodnianskiDecay}
\bysame, \emph{The red-shift effect and radiation decay on black hole
  spacetimes}, Commun.\ Pure Appl.\ Math. \textbf{62} (2009), 859--919.
  \MR{MR2527808}

\bibitem{Dain:AH}
S.~Dain, \emph{Trapped surfaces as boundaries for the constraint equations},
  Class.\ Quantum Grav. \textbf{21} (2004), 555--574, Corrigendum ib. p.~769,
  arXiv:gr-qc/0308009.

\bibitem{DainFriedrich}
S.~Dain and H.~Friedrich, \emph{Asymptotically flat initial data with
  prescribed regularity at infinity}, Commun.\ Math.\ Phys. \textbf{222}
  (2001), no.~3, 569--609. \MR{MR1888089 (2003f:58057)}

\bibitem{DamourHenneauxNicolai}
T.~Damour, M.~Henneaux, and H.~Nicolai, \emph{Cosmological billiards}, Class.\
  Quantum Grav. \textbf{20} (2003), R145--R200. \MR{MR1981434 (2004e:83124)}

\bibitem{DHRW}
T.~Damour, M.~Henneaux, A.~D. Rendall, and M.~Weaver, \emph{Kasner-like
  behaviour for subcritical {E}instein-matter systems}, Ann.\ Henri Poincar\'e
  \textbf{3} (2002), 1049--1111. \MR{MR1957378 (2004g:83092)}

\bibitem{DHS}
J.~Demaret, J.-L. Hanquin, M.~Henneaux, and P.~Spindel, \emph{Nonoscillatory
  behaviour in vacuum {K}aluza-{K}lein cosmologies}, Phys.\ Lett.\ B
  \textbf{164} (1985), 27--30. \MR{MR815631 (87e:83060)}

\bibitem{DonaldsonJDG83}
S.K. Donaldson, \emph{An application of gauge theory to four-dimensional
  topology}, Jour.\ Diff.\ Geom. \textbf{18} (1983), 279--315. \MR{MR710056
  (85c:57015)}

\bibitem{DSS}
R.~Donninger, W.~Schlag, and A.~Soffer, \emph{{A proof of Price's Law on
  Schwarzschild black hole manifolds for all angular momenta}},  (2009),
  arXiv:0908.4292 [gr-qc].

\bibitem{DossaAHP}
M.~Dossa, \emph{Probl\`emes de {C}auchy sur un cono\"\i de caract\'eristique
  pour les \'equations d'{E}instein (conformes) du vide et pour les \'equations
  de {Y}ang-{M}ills-{H}iggs}, Ann.\ Henri Poincar\'e \textbf{4} (2003),
  385--411. \MR{MR1985778 (2004h:58041)}

\bibitem{EardleyLiangSachs}
D.~Eardley, E.~Liang, and R.~Sachs, \emph{Velocity-dominated singularities in
  irrotational dust cosmologies}, Jour.\ Math.\ Phys. \textbf{13} (1972),
  99--106.

\bibitem{Eichmair}
M.~Eichmair, \emph{The plateau problem for apparent horizons}, 2007,
  arXiv:0711.4139 [math.DG].

\bibitem{Eichmair2}
\bysame, \emph{Existence, regularity, and properties of generalized apparent
  horizons}, 2010, arXiv:0805.4454 [math.DG], pp.~745--760. \MR{MR2585986}

\bibitem{Emparan:2004wy}
R.~Emparan, \emph{Rotating circular strings, and infinite non-uniqueness of
  black rings}, JHEP \textbf{03} (2004), 064.

\bibitem{EmparanReall}
R.~Emparan and H.S. Reall, \emph{A rotating black ring in five dimensions},
  Phys.\ Rev.\ Lett. \textbf{88} (2002), 101101, arXiv:hep-th/0110260.

\bibitem{EmparanReallReview}
\bysame, \emph{Black rings}, Class.\ Quantum Grav. \textbf{23} (2006),
  R169--R197, arXiv:hep-th/0608012.

\bibitem{EmparanReallLR}
\bysame, \emph{{Black Holes in Higher Dimensions}}, Living Rev.\ Rel.
  \textbf{11} (2008), 6, arXiv:0801.3471 [hep-th].

\bibitem{Peterson}
B.M.~Peterson et~al., \emph{{Central Masses and Broad-Line Region Sizes of
  Active Galactic Nuclei. II. A Homogeneous Analysis of a Large
  Reverberation-Mapping Database}}, Astrophys.\ Jour. \textbf{613} (2004),
  682--699, arXiv:astro-ph/0407299.

\bibitem{Evans:book}
L.C. Evans, \emph{Partial differential equations}, Graduate Studies in
  Mathematics, vol.~19, American Mathematical Society, Providence, RI, 1998.
  \MR{MR1625845 (99e:35001)}

\bibitem{FedererMeasureTheory}
H.~Federer, \emph{Geometric measure theory}, Springer Verlag, New York, 1969,
  ({D}ie {G}rundlehren der mathematischen {W}issenschaften, Vol. 153).

\bibitem{FiguerasLucietti}
P.~Figueras and J.~Lucietti, \emph{{On the uniqueness of extremal vacuum black
  holes}}, Class.\ Quantum Grav. \textbf{27} (2010), 095001, arXiv:0906.5565
  [hep-th].

\bibitem{FSYBAMS}
F.~Finster, N.~Kamran, J.~Smoller, and S.-T. Yau, \emph{Linear waves in the
  {K}err geometry: a mathematical voyage to black hole physics}, Bull.\ Amer.\
  Math.\ Soc. (N.S.) \textbf{46} (2009), 635--659, arXiv:0801.1423 [math-ph].
  \MR{MR2525736}

\bibitem{FischerMarsdenHI}
A.E. Fischer and J.E. Marsden, \emph{The initial value problem and the
  dynamical formulation of general relativity}, Einstein Centenary Volume
  (Hawking and Israel, eds.), Cambridge University Press, Cambridge, 1979,
  pp.~138--211.

\bibitem{ChBActa}
Y.~Four{\`e}s-Bruhat, \emph{Th\'eor\`eme d'existence pour certains syst\`emes
  d'\'equations aux d\'eriv\'ees partielles non lin\'eaires}, Acta Math.
  \textbf{88} (1952), 141--225.

\bibitem{Frauendiener:Penrose}
J.~Frauendiener, \emph{On the {P}enrose inequality}, Phys.\ Rev.\ Lett.
  \textbf{87} (2001), 101101, 4 pp., arXiv:gr-qc/0105093. \MR{MR1854297
  (2002f:83054)}

\bibitem{FriedrichRadiative}
H.~Friedrich, \emph{On purely radiative space-times}, Commun.\ Math.\ Phys.
  \textbf{103} (1986), 35--65. \MR{MR826857 (87e:83029)}

\bibitem{Friedrich:hyperbolicreview}
\bysame, \emph{Hyperbolic reductions for {E}instein's equations}, Class.\
  Quantum Grav. \textbf{13} (1996), 1451--1469.

\bibitem{Friedrich:Pune}
\bysame, \emph{Einstein's equation and geometric asymptotics}, Gravitation and
  Relativity: At the turn of the Millenium (Pune) (N.~Dadhich and J.~Narlikar,
  eds.), IUCAA, 1998, Proceedings of GR15, pp.~153--176.

\bibitem{FriedrichNagy}
H.~Friedrich and G.~Nagy, \emph{The initial boundary value problem for
  {E}instein's vacuum field equation}, Commun.\ Math.\ Phys. \textbf{201}
  (1998), 619--655.

\bibitem{FRW}
H.~Friedrich, I.~R{\'a}cz, and R.M. Wald, \emph{On the rigidity theorem for
  space-times with a stationary event horizon or a compact {C}auchy horizon},
  Commun.\ Math.\ Phys. \textbf{204} (1999), 691--707, arXiv:gr-qc/9811021.

\bibitem{FriedrichRendall00}
H.~Friedrich and A.D Rendall, \emph{The {C}auchy problem for the {E}instein
  equations}, Einstein's field equations and their physical implications,
  Lecture Notes in Phys., vol. 540, Springer, Berlin, 2000,
  arXiv:gr-qc/0002074, pp.~127--223. \MR{MR1765130 (2001m:83009)}

\bibitem{Fronsdal}
C.~Fronsdal, \emph{Completion and embedding of the {S}chwarzschild solution},
  Phys. Rev. (2) \textbf{116} (1959), 778--781. \MR{MR0110524 (22 \#1402)}

\bibitem{Galloway:cauchy}
G.J. Galloway, \emph{{Some results on Cauchy surface criteria in Lorentzian
  geometry}}, Illinois Jour.\ Math. \textbf{29} (1985), 1--10.

\bibitem{Galloway:fitopology}
\bysame, \emph{A ``finite infinity'' version of the {FSW} topological
  censorship}, Class.\ Quantum Grav. \textbf{13} (1996), 1471--1478.
  \MR{MR1397128 (97h:83065)}

\bibitem{Galloway:mots}
\bysame, \emph{{Rigidity of marginally trapped surfaces and the topology of
  black holes}}, Commun.\ Anal.\ Geom. \textbf{16} (2008), 217--229,
  gr-qc/0608118.

\bibitem{GallowaySchoen}
G.J. Galloway and R.~Schoen, \emph{A generalization of {Hawking's} black hole
  topology theorem to higher dimensions}, Commun.\ Math.\ Phys. \textbf{266}
  (2005), 571--576, arXiv:gr-qc/0509107. \MR{MR2238889 (2007i:53078)}

\bibitem{Gerhardt83}
C.~Gerhardt, \emph{{$H$}-surfaces in {L}orentzian manifolds}, Commun.\ Math.\
  Phys. \textbf{89} (1983), 523--553. \MR{MR713684 (85h:53049)}

\bibitem{GerochDoD}
R.~Geroch, \emph{Domain of dependence}, Jour.\ Math. Phys. \textbf{11} (1970),
  437--449.

\bibitem{GibbonsHawkingCEH}
G.W. Gibbons and S.W. Hawking, \emph{Cosmological event horizons,
  thermodynamics, and particle creation}, Phys.\ Rev. \textbf{D15} (1977),
  2738--2751.

\bibitem{Gillessen:2010ty}
S.~Gillessen et~al., \emph{{The power of monitoring stellar orbits}},  (2010),
  arXiv:1002.1224 [astro-ph.GA].

\bibitem{GL1}
M.~Gromov and {H.B.~Lawson, Jr.}, \emph{Spin and scalar curvature in the
  presence of a fundamental group. {I}}, Ann. of Math. (2) \textbf{111} (1980),
  209--230. \MR{MR569070 (81g:53022)}

\bibitem{GL2}
\bysame, \emph{Positive scalar curvature and the {D}irac operator on complete
  {R}iemannian manifolds}, Inst. Hautes \'Etudes Sci. Publ. Math. (1983),
  83--196. \MR{MR720933 (85g:58082)}

\bibitem{Hajicek3Remarks}
P.~H{\'a}j{\'{\i}}{\v{c}}ek, \emph{Three remarks on axisymmetric stationary
  horizons}, Commun.\ Math.\ Phys. \textbf{36} (1974), 305--320. \MR{MR0418816
  (54 \#6852)}

\bibitem{Harmark}
T.~Harmark, \emph{Stationary and axisymmetric solutions of higher-dimensional
  general relativity}, Phys.\ Rev.\ D (3) \textbf{70} (2004), 124002, 25,
  arXiv:hep-th/0408141. \MR{MR2124693 (2005k:83136)}

\bibitem{HarmarkOlesen}
T.~Harmark and P.~Olesen, \emph{Structure of stationary and axisymmetric
  metrics}, Phys.\ Rev.\ D (3) \textbf{72} (2005), 124017, 12, hep-th/0408141.
  \MR{MR2198031 (2007i:83042)}

\bibitem{HartleHawking}
J.B. Hartle and S.W. Hawking, \emph{Solutions of the {E}instein--{M}axwell
  equations with many black holes}, Commun.\ Math.\ Phys. \textbf{26} (1972),
  87--101.

\bibitem{HawkingPenrose}
S.~W. Hawking and R.~Penrose, \emph{The singularities of gravitational collapse
  and cosmology}, Proc.\ Roy.\ Soc.\ London Ser.\ A \textbf{314} (1970),
  529--548. \MR{MR0264959 (41 \#9548)}

\bibitem{Ha1}
S.W. Hawking, \emph{Black holes in general relativity}, Commun.\ Math.\ Phys.
  \textbf{25} (1972), 152--166.

\bibitem{HawkingLesHouchesI}
\bysame, \emph{The event horizon}, Black holes/{L}es astres occlus (B.S.
  DeWitt, ed.), Gordon and Breach Science Publishers, 1973, pp.~xii+552+176.
  \MR{MR0408678 (53 \#12441)}

\bibitem{HE}
S.W. Hawking and G.F.R. Ellis, \emph{The large scale structure of space-time},
  Cambridge University Press, Cambridge, 1973, Cambridge Monographs on
  Mathematical Physics, No. 1. \MR{MR0424186 (54 \#12154)}

\bibitem{HPP}
E.~Hebey, F.~Pacard, and D.~Pollack, \emph{A variational analysis of
  {Einstein-scalar field Lichnerowicz equations on compact Riemannian}
  manifolds}, Commun.\ Math.\ Phys. \textbf{278} (2008), no.~1, 117--132,
  arXiv:gr-qc/0702031. \MR{MR2367200}

\bibitem{HeinzleUggla}
J.M. Heinzle, C.~Uggla, and N.~Rohr, \emph{The cosmological billiard
  attractor}, Adv. Theor. Math. Phys. \textbf{13} (2009), 293--407,
  arXiv:gr-qc/0702141. \MR{MR2481269 (2010i:83117)}

\bibitem{HennigAnsorgCederbaum}
J.~Hennig, M.~Ansorg, and C.~Cederbaum, \emph{A universal inequality between
  the angular momentum and horizon area for axisymmetric and stationary black
  holes with surrounding matter}, Class.\ Quantum Grav. \textbf{25} (2008),
  162002, 8. \MR{MR2429717 (2009e:83092)}

\bibitem{mh-inegalite-penrose}
M.~Herzlich, \emph{A {Penrose-}like inequality for the mass of {R}iemannian
  asymptotically flat manifolds}, Commun.\ Math.\ Phys. \textbf{188} (1997),
  121--133.

\bibitem{Heusler:book}
M.~Heusler, \emph{Black hole uniqueness theorems}, Cambridge University Press,
  Cambridge, 1996.

\bibitem{HHW}
C.G. Hewitt, J.T. Horwood, and J.~Wainwright, \emph{Asymptotic dynamics of the
  exceptional {B}ianchi cosmologies}, Class.\ Quantum Grav. \textbf{20} (2003),
  1743--1756. \MR{MR1981447 (2004g:83027)}

\bibitem{HollandsDegenerate}
S.~Hollands and A.~Ishibashi, \emph{{All vacuum near horizon geometries in
  arbitrary dimensions}},  (2009), arXiv:0909.3462 [gr-qc].

\bibitem{HICMP}
\bysame, \emph{On the `stationary implies axisymmetric' theorem for extremal
  black holes in higher dimensions}, Commun.\ Math.\ Phys. \textbf{291} (2009),
  no.~2, 443--471. \MR{MR2530167}

\bibitem{HIW}
S.~Hollands, A.~Ishibashi, and R.M. Wald, \emph{A higher dimensional stationary
  rotating black hole must be axisymmetric}, Commun.\ Math. Phys. \textbf{271}
  (2007), 699--722, arXiv:gr-qc/0605106.

\bibitem{HY2}
S.~Hollands and S.~Yazadjiev, \emph{{A Uniqueness theorem for 5-dimensional
  {Einstein-Maxwell} black holes}}, Class.\ Quantum Grav. \textbf{25} (2008),
  095010, arXiv:0711.1722 [gr-qc]. \MR{MR2417776 (2009c:83061)}

\bibitem{HY3}
\bysame, \emph{{A uniqueness theorem for stationary Kaluza-Klein black holes}},
   (2008), arXiv:0812.3036 [gr-qc].

\bibitem{HY}
\bysame, \emph{Uniqueness theorem for 5-dimensional black holes with two axial
  {K}illing fields}, Commun.\ Math. Phys. \textbf{283} (2008), 749--768,
  arXiv:0707.2775 [gr-qc]. \MR{MR2434746}

\bibitem{HolstScheelLindblom}
M.~Holst, L.~Lindblom, R.~Owen, H.P. Pfeiffer, M.A. Scheel, and L.E. Kidder,
  \emph{Optimal constraint projection for hyperbolic evolution systems}, Phys.
  Rev. D (3) \textbf{70} (2004), 084017, 17. \MR{MR2117121 (2005j:83008)}

\bibitem{HNT07}
M.~Holst, G.~Nagy, and G.~Tsogtgerel, \emph{Rough solutions of the {Einstein
  constraint equations on closed manifolds without near-CMC} conditions},
  Commun.\ Math.\ Phys., arXiV:0712.0798 [gr-qc]. \MR{MR2500992}

\bibitem{HNT08}
\bysame, \emph{Far-from-constant mean curvature solutions of {E}instein's
  constraint equations with positive {Y}amabe metrics}, Phys.\ Rev.\ Lett.
  \textbf{100} (2008), 161101, 4. \MR{MR2403263}

\bibitem{HormanderGlobal}
L.~H\"ormander, \emph{On the fully nonlinear {C}auchy problem with small data.
  {II}}, Microlocal analysis and nonlinear waves (Minneapolis, MN, 1988--1989),
  IMA Vol. Math. Appl., vol.~30, Springer, New York, 1991, pp.~51--81.
  \MR{MR1120284 (94c:35127)}

\bibitem{Horowitz:1996rn}
G.T. Horowitz, \emph{Quantum states of black holes}, {R.M.~Wald (ed.), Black
  holes and relativistic stars. Symposium held in Chicago, IL, USA, December
  14-15, 1996. Univ. of Chicago Press, Chicago, Ill., pp.~241--266},
  arXiv:gr-qc/9704072.

\bibitem{HuiskenIlmanen}
G.~Huisken and T.~Ilmanen, \emph{The inverse mean curvature flow and the
  {R}iemannian {P}enrose inequality}, Jour.\ Diff.\ Geom. \textbf{59} (2001),
  353--437. \MR{MR1916951 (2003h:53091)}

\bibitem{IonescuKlainerman2}
A.D. Ionescu and S.~Klainerman, \emph{{Uniqueness results for ill posed
  characteristic problems in curved space-times}},  (2007), arXiv:0711.0042
  [gr-qc].

\bibitem{IonescuKlainerman1}
\bysame, \emph{{On the uniqueness of smooth, stationary black holes in
  vacuum}}, Invent. Math. \textbf{175} (2009), 35--102, arXiv:0711.0040
  [gr-qc]. \MR{MR2461426 (2009j:83053)}

\bibitem{Jimconstraints}
J.~Isenberg, \emph{Constant mean curvature solutions of the {Einstein}
  constraint equations on closed manifolds}, Class.\ Quantum Grav. \textbf{12}
  (1995), 2249--2274. \MR{MR1353772 (97a:83013)}

\bibitem{IMaxP}
J.~Isenberg, D.~Maxwell, and D.~Pollack, \emph{Gluing of non-vacuum solutions
  of the {Einstein} constraint equations}, Adv.\ Theor.\ Math.\ Phys.
  \textbf{9} (2005), 129--172. \MR{MR2193370 (2006j:83004)}

\bibitem{IMP1}
J.~Isenberg, R.~Mazzeo, and D.~Pollack, \emph{Gluing and wormholes for the
  {E}instein constraint equations}, Commun.\ Math.\ Phys. \textbf{231} (2002),
  529--568, arXiv:gr-qc/0109045. \MR{MR1946448 (2004a:83006)}

\bibitem{IMP2}
\bysame, \emph{On the topology of vacuum spacetimes}, Ann.\ Henri Poincar\'e
  \textbf{4} (2003), 369--383. \MR{MR1985777 (2004h:53053)}

\bibitem{VinceJimcompactCauchy}
J.~Isenberg and V.~Moncrief, \emph{Symmetries of cosmological {C}auchy horizons
  with exceptional orbits}, Jour.\ Math. Phys. \textbf{26} (1985), 1024--1027.

\bibitem{IM}
\bysame, \emph{Some results on non--constant mean curvature solutions of the
  {E}instein constraint equations}, Physics on Manifolds (M.~Flato, R.~Kerner,
  and A.~Lichnerowicz, eds.), Kluwer Academic Publishers, Dordrecht, 1994, Y.
  Choquet--Bruhat Festschrift, pp.~295--302.

\bibitem{VinceJim:noncmc}
\bysame, \emph{A set of nonconstant mean curvature solutions of the {E}instein
  constraint equations on closed manifolds}, Class.\ Quantum Gravity
  \textbf{13} (1996), 1819--1847. \MR{MR1400943 (97h:83010)}

\bibitem{Isenberg:2002jg}
\bysame, \emph{Asymptotic behavior of polarized and half-polarized {U(1)}
  symmetric vacuum spacetimes}, Class.\ Quantum Grav. \textbf{19} (2002),
  5361--5386, arXiv:gr-qc/0203042.

\bibitem{VinceJimHigh}
\bysame, \emph{{Symmetries of Higher Dimensional Black Holes}}, Class.\ Quantum
  Grav. \textbf{25} (2008), 195015, arXiv:0805.1451.

\bibitem{IsenbergOMurchadha}
J.~Isenberg and N.~{\'O} Murchadha, \emph{Non-{CMC} conformal data sets which
  do not produce solutions of the {E}instein constraint equations}, Class.\
  Quantum Grav. \textbf{21} (2004), S233--S241, A space-time safari: Essays in
  honour of Vincent Moncrief; arXiv:gr-qc/0311057. \MR{MR2053007 (2005c:83003)}

\bibitem{IW}
J.~Isenberg and M.~Weaver, \emph{On the area of the symmetry orbits in {$T^2$}
  symmetric space-times}, Class.\ Quantum Grav. \textbf{20} (2003), 3783--3796,
  arXiv:gr-qc/0304019.

\bibitem{Israel:vacuum}
W.~Israel, \emph{Event horizons in static vacuum space-times}, Phys. Rev.
  \textbf{164} (1967), 1776--1779.

\bibitem{Israel:bhreview}
\bysame, \emph{{Dark stars: The evolution of an idea}}, {S.W.~Hawking and
  W.~Israel (eds.), Three hundred years of gravitation. Cambridge: Cambridge
  University Press, pp.~199-276}, 1987.

\bibitem{Jang78}
P.~S. Jang, \emph{On the positivity of energy in general relativity}, Jour.\
  Math.\ Phys. \textbf{19} (1978), 1152--1155.

\bibitem{Johnston}
W.R. Johnston, \emph{List of black holes candidates},
  \url{http://www.johnstonsarchive.net/relativity/bhctable.html}.

\bibitem{KayWald}
B.S. Kay and R.M. Wald, \emph{Theorems on the uniqueness and thermal properties
  of stationary, nonsingular, quasi-free states on space-times with a bifurcate
  horizon}, Phys.\ Rep. \textbf{207} (1991), 49--136.

\bibitem{KichenassamyRendall}
S.~Kichenassamy and A.~Rendall, \emph{Analytic description of singularities in
  {G}owdy space-times}, Class.\ Quantum Grav. \textbf{15} (1998), 1339--1355.

\bibitem{KlainermanGlobalCPAMTwo}
S.~Klainerman, \emph{Uniform decay estimates and the {L}orentz invariance of
  the classical wave equation}, Commun.\ Pure Appl.\ Math. \textbf{38} (1985),
  321--332. \MR{MR784477 (86i:35091)}

\bibitem{Klainerman:null}
\bysame, \emph{The null condition and global existence to nonlinear wave
  equations}, Nonlinear systems of partial differential equations in applied
  mathematics, Part 1 (Santa Fe, N.M., 1984), Lectures in Appl.\ Math.,
  vol.~23, Amer. Math. Soc., Providence, RI, 1986, pp.~293--326. \MR{MR837683
  (87h:35217)}

\bibitem{klainerman:nicolo:review}
S.~Klainerman and F.~Nicol{\`o}, \emph{On local and global aspects of the
  {Cauchy} problem in general relativity}, Class.\ Quantum Grav. \textbf{16}
  (1999), R73--R157. \MR{MR1709123 (2000h:83006)}

\bibitem{KlainermanNicoloBook}
\bysame, \emph{The evolution problem in general relativity}, Progress in
  Mathematical Physics, vol.~25, Birkh{\"a}user, Boston, MA, 2003. \MR{1 946
  854}

\bibitem{KlainermanNicoloPeeling}
\bysame, \emph{Peeling properties of asymptotically flat solutions to the
  {E}instein vacuum equations}, Class.\ Quantum Grav. \textbf{20} (2003),
  3215--3257. \MR{1 992 002}

\bibitem{KlainermanRodnianski:r3}
S.~Klainerman and I.~Rodnianski, \emph{Ricci defects of microlocalized
  {E}instein metrics}, Jour.\ Hyperbolic Differ.\ Equ. \textbf{1} (2004),
  85--113. \MR{MR2052472 (2005f:58048)}

\bibitem{KlainermanRodnianski:r2}
\bysame, \emph{The causal structure of microlocalized rough {E}instein
  metrics}, Ann.\ of Math.\ (2) \textbf{161} (2005), 1195--1243. \MR{MR2180401
  (2007d:58052)}

\bibitem{KlainermanRodnianski:r1}
\bysame, \emph{Rough solutions of the {E}instein-vacuum equations}, Ann.\ of
  Math.\ (2) \textbf{161} (2005), 1143--1193. \MR{MR2180400 (2007d:58051)}

\bibitem{RodnianskiKlainerman:scarred}
\bysame, \emph{{On emerging scarred surfaces for the Einstein vacuum
  equations}},  (2010), arXiv:1002.2656 [gr-qc].

\bibitem{Kottler}
F.~Kottler, \emph{{\"Uber die physikalischen Grundlagen der Einsteinschen
  Gravitationstheorie}}, Annalen der Physik \textbf{56} (1918), 401--462.

\bibitem{KramerNeugebauerDoubleKerr}
D.~Kramer and G.~Neugebauer, \emph{The superposition of two {K}err solutions},
  Phys. Lett. A \textbf{75} (1979/80), 259--261. \MR{MR594394 (81m:83014)}

\bibitem{KRSW}
H.O. Kreiss, O.~Reula, O.~Sarbach, and J.~Winicour, \emph{{Well-posed
  initial-boundary value problem for the harmonic {E}instein equations using
  energy estimates}}, Class. Quant. Grav. \textbf{24} (2007), 5973--5984,
  arXiv:0707.4188 [gr-qc].

\bibitem{Kruskal}
M.D. Kruskal, \emph{Maximal extension of {S}chwarzschild metric}, Phys. Rev.
  (2) \textbf{119} (1960), 1743--1745. \MR{MR0115757 (22 \#6555)}

\bibitem{KunduriLucietti2}
H.K. Kunduri and J.~Lucietti, \emph{{A classification of near-horizon
  geometries of extremal vacuum black holes}}, Jour.\ Math.\ Phys. \textbf{50}
  (2009), 082502, arXiv:0806.2051 [hep-th].

\bibitem{KunduriLucietti}
\bysame, \emph{{Static near-horizon geometries in five dimensions}}, Class.\
  Quantum Grav. \textbf{26} (2009), 245010, arXiv:0907.0410 [hep-th].

\bibitem{LeeParker}
J.M. Lee and T.H. Parker, \emph{The {Y}amabe problem}, Bull.\ Amer.\ Math.\
  Soc. (N.S.) \textbf{17} (1987), 37--91. \MR{MR888880 (88f:53001)}

\bibitem{Leonhardt:Piwnicki}
U.~Leonhardt and P.~Piwnicki, \emph{Relativistic effects of light in moving
  media with extremely low group velocity}, Phys.\ Rev.\ Lett. \textbf{84}
  (2000), 822--825.

\bibitem{Leray}
J.~Leray, \emph{Hyperbolic differential equations}, mimeographed notes, 1953,
  Princeton.

\bibitem{LP1}
J.~Lewandowski and T.~{Paw\l owski}, \emph{Extremal isolated horizons: A local
  uniqueness theorem}, Class.\ Quantum Grav. \textbf{20} (2003), 587--606,
  arXiv:gr-qc/0208032.

\bibitem{LP2}
\bysame, \emph{Quasi-local rotating black holes in higher dimension: geometry},
  Class.\ Quantum Grav. \textbf{22} (2005), 1573--1598, arXiv:gr-qc/0410146.

\bibitem{LiChen}
T.-T. Li and Y.M Chen, \emph{Global classical solutions for nonlinear evolution
  equations}, Pitman Monographs and Surveys in Pure and Applied Mathematics,
  vol.~45, Longman Scientific \& Technical, Harlow, 1992. \MR{MR1172318
  (93g:35002)}

\bibitem{Li:Tian2}
Y.~Li and G.~Tian, \emph{Nonexistence of axially symmetric, stationary solution
  of {E}instein vacuum equation with disconnected symmetric event horizon},
  Manuscripta Math. \textbf{73} (1991), 83--89.

\bibitem{Lich44}
A.~Lichnerowicz, \emph{L'int\'egration des \'equations de la gravitation
  relativiste et le probl\`eme des {$n$} corps}, Jour.\ Math.\ Pures Appl.\ (9)
  \textbf{23} (1944), 37--63. \MR{MR0014298 (7,266d)}

\bibitem{Lichnerowicz63}
A.~Lichnerowicz, \emph{Spineurs harmoniques}, C.R. Acad. Sci. Paris S\'er. A-B
  \textbf{257} (1963), 7--9.

\bibitem{Liebscher}
S.~Liebscher, J.~Harterich, K.~Webster, and M.~Georgi, \emph{{Ancient Dynamics
  in {Bianchi} Models: Approach to Periodic Cycles}},  (2010), arXiv:1004.1989
  [gr-qc].

\bibitem{LindbladRodnianski2}
H.~Lindblad and I.~Rodnianski, \emph{The global stability of the {Minkowski}
  space-time in harmonic gauge},  (2004), arXiv:math.ap/0411109.

\bibitem{LindbladRodnianski}
\bysame, \emph{Global existence for the {E}instein vacuum equations in wave
  coordinates}, Commun.\ Math.\ Phys. \textbf{256} (2005), 43--110,
  arXiv:math.ap/0312479. \MR{MR2134337 (2006b:83020)}

\bibitem{Lockhart}
R.B. Lockhart, \emph{Fredholm properties of a class of elliptic operators on
  noncompact manifolds}, Duke Math.\ J. \textbf{48} (1981), 289--312.
  \MR{MR610188 (82j:35050)}

\bibitem{LockhartMcOwenActa}
R.B. Lockhart and R.C. McOwen, \emph{On elliptic systems in {${\bf R}\sp{n}$}},
  Acta Math. \textbf{150} (1983), 125--135. \MR{MR697610 (84d:35048)}

\bibitem{LockhartMcOwenActa-cor}
\bysame, \emph{Correction to: ``{O}n elliptic systems in {${\bf R}\sp n$}''
  [{A}cta {M}ath. {\bf 150} (1983), no. 1-2, 125--135]}, Acta Math.
  \textbf{153} (1984), 303--304. \MR{MR766267 (86a:35049)}

\bibitem{LockhartMcOwenPisa}
\bysame, \emph{Elliptic differential operators on noncompact manifolds}, Ann.\
  Scuola Norm.\ Sup.\ Pisa Cl.\ Sci.\ (4) \textbf{12} (1985), 409--447.
  \MR{MR837256 (87k:58266)}

\bibitem{LoizeletCRAS}
J.~Loizelet, \emph{Solutions globales des \'equations d'{Einstein-Maxwell} en
  jauge harmonique et jauge de {Lorenz}}, Comptes Rendus Acad.\ Sci.\ S\'er.\ I
  \textbf{342} (2006), 479--482.

\bibitem{Loizelet:these}
\bysame, \emph{Probl\`emes globaux en relativit\'e g\'en\'erale}, Ph.D. thesis,
  Universit\'e de Tours, 2008,
  \url{www.phys.univ-tours.fr/~piotr/papers/TheseTitreComplet.pdf}.

\bibitem{Majumdar}
S.D. Majumdar, \emph{A class of exact solutions of {E}instein's field
  equations}, Phys. Rev. \textbf{72} (1947), 390--398.

\bibitem{MalecPenrose}
E.~Malec, \emph{Isoperimetric inequalities in the physics of black holes}, Acta
  Phys.\ Polon. B \textbf{22} (1991), 829--858. \MR{MR1151689 (93e:83046)}

\bibitem{MMS}
E.~Malec, M.~Mars, and W.~Simon, \emph{On the {P}enrose inequality for general
  horizons}, Phys.\ Rev.\ Lett. \textbf{88} (2002), 121102,
  arXiv:gr-qc/0201024.

\bibitem{TataruSchwarzschild}
J.~Marzuola, J.~Metcalfe, D.~Tataru, and M.~Tohaneanu, \emph{Strichartz
  estimates on {Schwarzschild} black hole backgrounds}, Comm. Math. Phys.
  \textbf{293} (2010), 37--83, arXiv:0802.3942 [math.AP]. \MR{MR2563798}

\bibitem{Masood}
A.K.M. Masood{--ul--A}lam, \emph{Uniqueness proof of static charged black holes
  revisited}, Class.\ Quantum Grav. \textbf{9} (1992), L53--L55.

\bibitem{Maxwell:compact}
D.~Maxwell, \emph{Rough solutions of the {E}instein constraint equations on
  compact manifolds}, Jour.\ Hyperbolic Diff.\ Equ. \textbf{2} (2005),
  521--546, arXiv:gr-qc/0506085. \MR{MR2151120 (2006d:58027)}

\bibitem{Maxwell:AH}
\bysame, \emph{Solutions of the {E}instein constraint equations with apparent
  horizon boundaries}, Commun.\ Math.\ Phys. \textbf{253} (2005), 561--583,
  arXiv:gr-qc/0307117. \MR{MR2116728 (2006c:83008)}

\bibitem{Maxwell:rough}
\bysame, \emph{Rough solutions of the {E}instein constraint equations}, J.
  Reine Angew. Math. \textbf{590} (2006), 1--29, arXiv:gr-qc/0405088.
  \MR{MR2208126 (2006j:58044)}

\bibitem{MaxwellNonCMC}
\bysame, \emph{{A model problem for conformal parameterizations of the Einstein
  constraint equations}},  (2009), arXiv:0909.5674 [gr-qc].

\bibitem{Maxwell08}
\bysame, \emph{A class of solutions of the vacuum {Einstein} constraint
  equations with freely specified mean curvature}, Math. Res. Lett. \textbf{16}
  (2009), 627--645, arXiv:0804.0874v1 [gr-qc]. \MR{MR2525029}

\bibitem{McOwen79}
R.C. McOwen, \emph{The behavior of the {L}aplacian on weighted {S}obolev
  spaces}, Commun.\ Pure Appl.\ Math. \textbf{32} (1979), 783--795.
  \MR{MR539158 (81m:47069)}

\bibitem{McOwen80fred}
\bysame, \emph{Fredholm theory of partial differential equations on complete
  {R}iemannian manifolds}, Pacific J.\ Math. \textbf{87} (1980), 169--185.
  \MR{MR590874 (82g:58082)}

\bibitem{McOwen80}
\bysame, \emph{On elliptic operators in {${\bf R}\sp{n}$}}, Commun.\ Partial
  Differential Equations \textbf{5} (1980), 913--933. \MR{MR584101 (81k:35058)}

\bibitem{MST}
P.~Miao, Y.~Shi, and L.-F. Tam, \emph{On geometric problems related to
  {Brown-York and Liu-Yau} quasilocal mass}, 2009, arXiv.org:0906.5451
  [math.DG].

\bibitem{Miller:2003sc}
M.C. Miller and E.J.M. Colbert, \emph{Intermediate-mass black holes}, Int.\
  Jour.\ Mod.\ Phys. \textbf{D13} (2004), 1--64.

\bibitem{Misner}
C.W. Misner, \emph{Taub--{N}{U}{T} space as a counterexample to almost
  anything}, Relativity Theory and Astrophysics, AMS, Providence, Rhode Island,
  1967, Lectures in Appl. Math., vol. 8, pp.~160--169.

\bibitem{Moncrief75}
V.~Moncrief, \emph{Spacetime symmetries and linearization stability of the
  {E}instein equations. {I}}, Jour.\ Mathematical Phys. \textbf{16} (1975),
  493--498. \MR{MR0363398 (50 \#15836)}

\bibitem{Moncrief:Gowdy}
\bysame, \emph{Global properties of {G}owdy spacetimes with {$T^3\times{\Bbb
  R}$} topology}, Ann. Phys. \textbf{132} (1981), 87--107.

\bibitem{MR82b:83024}
\bysame, \emph{Infinite-dimensional family of vacuum cosmological models with
  {T}aub-{NUT} ({N}ewman-{U}nti-{T}amburino)-type extensions}, Phys.\ Rev.\ D
  \textbf{23} (1981), 312--315. \MR{82b:83024}

\bibitem{EM}
V.~Moncrief and D.~Eardley, \emph{The global existence problem and cosmic
  censorship in general relativity}, Gen. Rel. Grav. \textbf{13} (1981),
  887--892.

\bibitem{VinceJimcompactCauchyCMP}
V.~Moncrief and J.~Isenberg, \emph{Symmetries of cosmological {C}auchy
  horizons}, Commun.\ Math.\ Phys. \textbf{89} (1983), 387--413.

\bibitem{IdaMorisawa}
Y.~Morisawa and D.~Ida, \emph{A boundary value problem for the five-dimensional
  stationary rotating black holes}, Phys.\ Rev. \textbf{D69} (2004), 124005,
  arXiv:gr-qc/0401100.

\bibitem{MyersPerry}
R.C. Myers and M.J. Perry, \emph{{Black holes in higher dimensional
  space-times}}, Ann.\ Phys. \textbf{172} (1986), 304--347.

\bibitem{HennigNeugebauer}
G.~Neugebauer and J.~Hennig, \emph{Non-existence of stationary two-black-hole
  configurations}, Gen. Relativity Gravitation \textbf{41} (2009), no.~9,
  2113--2130. \MR{MR2534657}

\bibitem{neugebauer:meinel}
G.~Neugebauer and R.~Meinel, \emph{General relativistic gravitational field of
  a rigidly rotating disc of dust: Axis potential, disk metric and surface mass
  density}, Phys. Rev. Lett. \textbf{73} (1994), 2166--2168.

\bibitem{Neugebauer:2003qe}
G.~Neugebauer and R.~Meinel, \emph{Progress in relativistic gravitational
  theory using the inverse scattering method}, Jour.\ Math.\ Phys. \textbf{44}
  (2003), 3407--3429, arXiv:gr-qc/0304086.

\bibitem{NUT}
E.~Newman, L.~Tamburino, and T.~Unti, \emph{Empty-space generalization of the
  {S}chwarzschild metric}, Jour.\ Math.\ Phys. \textbf{4} (1963), 915--923.
  \MR{MR0152345 (27 \#2325)}

\bibitem{NicolasDissMath}
J.-P. Nicolas, \emph{Dirac fields on asymptotically flat space-times},
  Dissertationes Math. (Rozprawy Mat.) \textbf{408} (2002), 1--85. \MR{1 952
  742}

\bibitem{NirenbergWalker}
L.~Nirenberg and H.F. Walker, \emph{The null spaces of elliptic partial
  differential operators in {${\bf R}\sp{n}$}}, J.\ Math.\ Anal.\ Appl.
  \textbf{42} (1973), 271--301, Collection of articles dedicated to Salomon
  Bochner. \MR{MR0320821 (47 \#9354)}

\bibitem{Novello:2001fv}
M.~Novello, J.M. Salim, V.A.~De Lorenci, and E.~Elbaz, \emph{Nonlinear
  electrodynamics can generate a closed spacelike path for photons}, Phys.\
  Rev. \textbf{D63} (2001), 103516 (5 pp.).

\bibitem{Novello:2002qg}
M.~Novello, M.~Visser, and G.~Volovik, \emph{Artificial black holes}, River
  Edge, USA: World Scientific (2002) 391 p.

\bibitem{OMYork73}
N.~O'Murchadha and J.W. {York$~$Jr.}, \emph{Existence and uniqueness of
  solutions of the {H}amiltonian constraint of general relativity on compact
  manifolds}, Jour.\ Mathematical Phys. \textbf{14} (1973), 1551--1557.
  \MR{MR0332094 (48 \#10421)}

\bibitem{ONeill}
B.~O'Neill, \emph{Semi--{R}iemannian geometry}, Academic Press, New York, 1983.

\bibitem{BONeillKerr}
\bysame, \emph{The geometry of {K}err black holes}, A.K.~Peters, Wellesley,
  Mass., 1995.

\bibitem{OppenheimerSnyder}
J.~R. Oppenheimer and H.~Snyder, \emph{On continued gravitational contraction},
  Phys.\ Rev. \textbf{56} (1939), 455--459.

\bibitem{Orlik}
P.~Orlik, \emph{Seifert manifolds}, Springer-Verlag, Berlin, 1972, Lecture
  Notes in Mathematics, Vol. 291. \MR{MR0426001 (54 \#13950)}

\bibitem{PacardRiviere}
F.~Pacard and T.~Rivi{\`e}re, \emph{Linear and nonlinear aspects of vortices},
  Progress in Nonlinear Differential Equations and their Applications, 39,
  Birkh\"auser Boston Inc., Boston, MA, 2000, The Ginzburg-Landau model.
  \MR{MR1763040 (2001k:35066)}

\bibitem{Padmanabhan:2003gd}
T.~Padmanabhan, \emph{{Gravity and the thermodynamics of horizons}}, Phys.
  Rept. \textbf{406} (2005), 49--125, arXiv:gr-qc/0311036.

\bibitem{Papapetrou:mp}
A.~Papapetrou, \emph{A static solution of the equations of the gravitational
  field for an arbitrary charge distribution}, Proc. Roy. Irish Acad.
  \textbf{A51} (1947), 191--204.

\bibitem{ParkerTaubes}
T.~Parker and C.H. Taubes, \emph{On {W}itten's proof of the positive energy
  theorem}, Commun.\ Math.\ Phys. \textbf{84} (1982), 223--238. \MR{MR661134
  (83m:83020)}

\bibitem{Paschalidis}
V.~Paschalidis, \emph{{Mixed Hyperbolic-Second-Order Parabolic Formulations of
  General Relativity}},  (2007), arXiv:0704.2861 [gr-qc].

\bibitem{penrose:asymptotic}
R.~Penrose, \emph{Asymptotic properties of fields and space--times}, Phys.\
  Rev.\ Lett. \textbf{10} (1963), 66--68.

\bibitem{Psing}
\bysame, \emph{Gravitational collapse and space-time singularities}, Phys.\
  Rev.\ Lett. \textbf{14} (1965), 57--59. \MR{MR0172678 (30 \#2897)}

\bibitem{penrose:scri}
\bysame, \emph{Zero rest-mass fields including gravitation: {A}symptotic
  behaviour}, Proc.\ Roy.\ Soc.\ London Ser. A \textbf{284} (1965), 159--203.
  \MR{MR0175590 (30 \#5774)}

\bibitem{PenroseSCC}
\bysame, \emph{Gravitational collapse --- the role of general relativity},
  Riv.\ del Nuovo Cim.\ (numero speziale) \textbf{1} (1969), 252--276.

\bibitem{PlanchonRodnianski}
F.~Planchon and I.~Rodnianski, \emph{On uniqueness for the {Cauchy} problem in
  general relativity},  (2007), in preparation.

\bibitem{PS}
A.A. Pomeransky and R.A. Sen'kov, \emph{{Black ring with two angular momenta}},
   (2006), hep-th/0612005.

\bibitem{Pretorius}
F.~Pretorius, \emph{Evolution of binary black hole space-times}, Phys.\ Rev.\
  Lett. \textbf{95} (2005), 121101, arXiv:gr-qc/0507014.

\bibitem{RaczWald2}
I.~R{\'a}cz and R.M. Wald, \emph{Global extensions of space-times describing
  asymptotic final states of black holes}, Class.\ Quantum Grav. \textbf{13}
  (1996), 539--552, arXiv:gr-qc/9507055. \MR{MR1385315 (97a:83071)}

\bibitem{Raymond}
F.~Raymond, \emph{Classification of the actions of the circle on
  {$3$}-manifolds}, Trans. Amer. Math.\ Soc. \textbf{131} (1968), 51--78.
  \MR{MR0219086 (36 \#2169)}

\bibitem{ReitererTrubowitzBianchi}
M.~Reiterer and E.~Trubowitz, \emph{{The {BKL} Conjectures for Spatially
  Homogeneous Spacetimes}},  (2010), arXiv:1005.4908 [gr-qc].

\bibitem{RendallCIVP}
A.D. Rendall, \emph{Reduction of the characteristic initial value problem to
  the {C}auchy problem and its applications to the {E}instein equations},
  Proc.\ Roy.\ Soc.\ London A \textbf{427} (1990), 221--239. \MR{MR1032984
  (91a:83004)}

\bibitem{Rendall:1996nu}
\bysame, \emph{Existence of constant mean curvature foliations in space-times
  with two-dimensional local symmetry}, Comm.\ Math.\ Phys. \textbf{189}
  (1997), 145--164, arXiv:gr-qc/9605022.

\bibitem{AlanMixmaster}
\bysame, \emph{Global dynamics of the mixmaster model}, Class.\ Quantum Grav.
  \textbf{14} (1997), 2341--2356. \MR{MR1468587 (98j:83024)}

\bibitem{RendallLiving}
\bysame, \emph{Local and global existence theorems for the {E}instein
  equations}, Living Reviews in Relativity \textbf{1} (1998), {URL}
  \url{http://www.livingreviews.org}.

\bibitem{Rendall:2000ih}
\bysame, \emph{Fuchsian analysis of singularities in {G}owdy space-times beyond
  analyticity}, Class.\ Quantum Grav. \textbf{17} (2000), 3305--3316,
  arXiv:gr-qc/0004044.

\bibitem{RendallWeaver}
A.D. Rendall and M.~Weaver, \emph{Manufacture of {G}owdy space-times with
  spikes}, Class.\ Quantum Grav. \textbf{18} (2001), 2959--2975,
  arXiv:gr-qc/0103102.

\bibitem{Ringstroem1}
H.~Ringstr{\"o}m, \emph{Curvature blow up in {B}ianchi {VIII} and {IX} vacuum
  space-times}, Class.\ Quantum Grav. \textbf{17} (2000), 713--731,
  arXiv:gr-qc/9911115. \MR{MR1744051 (2001d:83070)}

\bibitem{Ringstroem2}
\bysame, \emph{The {B}ianchi {IX} attractor}, Ann. Henri Poincar\'e \textbf{2}
  (2001), 405--500, arXiv:gr-qc/0006035. \MR{MR1846852 (2002h:83075)}

\bibitem{RingstroemGowdy}
\bysame, \emph{Existence of an asymptotic velocity and implications for the
  asymptotic behavior in the direction of the singularity in {$T\sp
  3$}-{G}owdy}, Commun.\ Pure Appl.\ Math. \textbf{59} (2006), 977--1041.
  \MR{MR2222842}

\bibitem{RingstroemSCC}
\bysame, \emph{Strong cosmic censorship in {$T^{3}$-{G}owdy} space-times}, Ann.
  of Math. (2) \textbf{170} (2008), 1181--1240. \MR{MR2600872}

\bibitem{RobinsonKerr}
D.C. Robinson, \emph{Uniqueness of the {Kerr} black hole}, Phys.\ Rev.\ Lett.
  \textbf{34} (1975), 905--906.

\bibitem{RobinsonSP}
\bysame, \emph{A simple proof of the generalization of {I}srael's theorem},
  Gen.\ Rel.\ Grav. \textbf{8} (1977), 695--698.

\bibitem{Ruback}
P.~Ruback, \emph{A new uniqueness theorem for charged black holes}, Class.\
  Quantum Grav. \textbf{5} (1988), L155--L159.

\bibitem{Sarbachetalt}
O.~Sarbach, G.~Calabrese, J.~Pullin, and M.~Tiglio, \emph{Hyperbolicity of the
  {BSSN system of Einstein} evolution equations}, Phys.\ Rev. \textbf{D66}
  (2002), 064002, arXiv:gr-qc/0205064.

\bibitem{SchoenCatini}
R.~Schoen, \emph{Variational theory for the total scalar curvature functional
  for {R}iemannian metrics and related topics}, Topics in calculus of
  variations (Montecatini Terme, 1987), Lecture Notes in Math., vol. 1365,
  Springer, Berlin, 1989, pp.~120--154. \MR{MR994021 (90g:58023)}

\bibitem{schoen:miamiwaves}
\bysame, \emph{{\rm Lecture at the {Miami Waves Conference}}}, January 2004.

\bibitem{SchoenClay}
\bysame, \emph{Mean curvature in {R}iemannian geometry and general relativity},
  Global theory of minimal surfaces, Clay Math.\ Proc., vol.~2, Amer.\ Math.\
  Soc., Providence, RI, 2005, pp.~113--136. \MR{MR2167257 (2006f:53044)}

\bibitem{SchoenYauAnnMath79}
R.~Schoen and S.-T. Yau, \emph{Existence of incompressible minimal surfaces and
  the topology of three-dimensional manifolds with nonnegative scalar
  curvature}, Ann.\ of Math.\ (2) \textbf{110} (1979), 127--142. \MR{MR541332
  (81k:58029)}

\bibitem{SchoenYauPMT1}
\bysame, \emph{On the proof of the positive mass conjecture in general
  relativity}, Commun.\ Math.\ Phys. \textbf{65} (1979), 45--76. \MR{MR526976
  (80j:83024)}

\bibitem{SchoenYauManuscripta}
\bysame, \emph{On the structure of manifolds with positive scalar curvature},
  Manuscripta Math. \textbf{28} (1979), 159--183. \MR{MR535700 (80k:53064)}

\bibitem{SchoenYauELMGR}
\bysame, \emph{The energy and the linear momentum of space-times in general
  relativity}, Commun.\ Math.\ Phys. \textbf{79} (1981), 47--51. \MR{MR609227
  (82j:83045)}

\bibitem{SchoenYauPMT2}
\bysame, \emph{Proof of the positive mass theorem. {II}}, Commun.\ Math.\ Phys.
  \textbf{79} (1981), 231--260. \MR{MR612249 (83i:83045)}

\bibitem{SchoenYau83}
\bysame, \emph{The existence of a black hole due to the condensation of
  matter}, Commun.\ Math.\ Phys. \textbf{90} (1983), 575--579.

\bibitem{SchoenYauKlienian}
\bysame, \emph{Conformally flat manifolds, {K}leinian groups and scalar
  curvature}, Invent.\ Math. \textbf{92} (1988), 47--71. \MR{MR931204
  (89c:58139)}

\bibitem{SchoenYauLectures}
\bysame, \emph{Lectures on differential geometry}, Conference Proceedings and
  Lecture Notes in Geometry and Topology, I, International Press, Cambridge,
  MA, 1994. \MR{MR1333601 (97d:53001)}

\bibitem{SchwartzF}
F.~Schwartz, \emph{{Existence of outermost apparent horizons with product of
  spheres topology}}, Commun.\ Anal.\ Geom. \textbf{16} (2008), 799--817,
  arXiv:0704.2403 [gr-qc]. \MR{MR2471370 (2009m:53100)}

\bibitem{Seifert}
H.J. Seifert, \emph{Smoothing and extending cosmic time functions}, Gen.\ Rel.\
  Grav. \textbf{8} (1977), 815--831. \MR{MR0484260 (58 \#4185)}

\bibitem{ShiTam3}
Y.~Shi and L.-F. Tam, \emph{Quasi-local mass and the existence of horizons},
  Commun.\ Math.\ Phys. \textbf{274} (2007), 277--295, arXiv:math.DG/0511398.
  \MR{MR2322904 (2008g:53095)}

\bibitem{ShibataNakamura}
M.~Shibata and T.~Nakamura, \emph{Evolution of three-dimensional gravitational
  waves: harmonic slicing case}, Phys.\ Rev.\ D (3) \textbf{52} (1995),
  5428--5444. \MR{MR1360417 (96g:83050)}

\bibitem{Simon:elvac}
W.~Simon, \emph{Radiative {E}instein-{M}axwell spacetimes and `no--hair'
  theorems}, Class.\ Quantum Grav. \textbf{9} (1992), 241--256.

\bibitem{SmithTataru:sharp}
H.F. Smith and D.~Tataru, \emph{Sharp local well-posedness results for the
  nonlinear wave equation}, Ann.\ of Math.\ (2) \textbf{162} (2005), 291--366.
  \MR{MR2178963 (2006k:35193)}

\bibitem{Sudarsky:wald}
D.~Sudarsky and R.M. Wald, \emph{Extrema of mass, stationarity and staticity,
  and solutions to the {E}instein--{Y}ang--{M}ills equations}, Phys.\ Rev.
  \textbf{D46} (1993), 1453--1474.

\bibitem{SyngeSchwarzschild}
J.L. Synge, \emph{The gravitational field of a particle}, Proc.\ Roy.\ Irish
  Acad.\ Sect.\ A. \textbf{53} (1950), 83--114. \MR{MR0039426 (12,546g)}

\bibitem{SzabadosLR}
L.~Szabados, \emph{Quasi-local energy-momentum and angular momentum in {GR: A}
  review article}, Living Rev. \textbf{4} (2004), URL
  \url{http://relativity.livingreviews.org/Articles/lrr-2004-4}.

\bibitem{Szekeres}
Gy. Szekeres, \emph{On the singularities of a {R}iemannian manifold}, Gen.\
  Rel.\ Grav. \textbf{34} (2002), 2001--2016, Reprinted from Publ. Math.
  Debrecen {{\bf{7}}} (1960), 285--301 [ MR0125541 (23 \#A2842)]. \MR{MR1945497
  (2003j:83073a)}

\bibitem{TataruDecay}
D.~Tataru, \emph{Local decay of waves on asymptotically flat stationary
  space-times},  (2008), arXiv:0910.5290 [math.AP].

\bibitem{TataruTohaneanu}
D.~Tataru and M.~Tohaneanu, \emph{Local energy estimate on kerr black hole
  backgrounds},  (2008), arXiv:0810.5766 [math.AP].

\bibitem{Taub}
A.H. Taub, \emph{Empty space-times admitting a three parameter group of
  motions}, Ann.\ of Math.\ (2) \textbf{53} (1951), 472--490. \MR{MR0041565
  (12,865b)}

\bibitem{Taubes82}
C.M. Taubes, \emph{Self-dual {Y}ang-{M}ills connections on non-self-dual
  {$4$}-manifolds}, Jour.\ Diff.\ Geom. \textbf{17} (1982), 139--170.
  \MR{MR658473 (83i:53055)}

\bibitem{JP}
T.Johannsen and D.~Psaltis, \emph{{Testing the No-Hair Theorem with
  Observations in the Electromagnetic Spectrum: II. Black-Hole Images}},
  (2010).

\bibitem{UgglaEllis:pa}
C.~Uggla, H.~van Elst, J.~Wainwright, and G.F.R. Ellis, \emph{The past
  attractor in inhomogeneous cosmology}, Phys.\ Rev. \textbf{D68} (2003),
  103502 (22~pp.), arXiv:gr-qc/0304002.

\bibitem{Unruh}
W.G. Unruh, \emph{Dumb holes and the effects of high frequencies on black hole
  evaporation}, Phys.\ Rev. \textbf{D51} (1995), 2827--2838,
  arXiv:gr-qc/9409008.

\bibitem{WainwrightHsu}
J.~Wainwright and L.~Hsu, \emph{A dynamical systems approach to {B}ianchi
  cosmologies: orthogonal models of class {$A$}}, Class.\ Quantum Grav.
  \textbf{6} (1989), 1409--1431. \MR{MR1014971 (90h:83033)}

\bibitem{Wald:book}
R.M. Wald, \emph{General relativity}, University of Chicago Press, Chicago,
  1984.

\bibitem{Wald:LR}
\bysame, \emph{The thermodynamics of black holes}, Living Reviews \textbf{4}
  (2001), arXiv:gr-qc/9912119, URL \url{http://www.livingreviews.org/}.

\bibitem{WangYau}
M.-T. Wang and S.-T. Yau, \emph{A generalization of {L}iu-{Y}au's quasi-local
  mass}, Commun.\ Anal.\ Geom. \textbf{15} (2007), 249--282. \MR{MR2344323
  (2008h:53046)}

\bibitem{WangYau5}
\bysame, \emph{Isometric embeddings into the {M}inkowski space and new
  quasi-local mass}, Commun.\ Math.\ Phys. \textbf{288} (2009), no.~3,
  919--942. \MR{MR2504860 (2010d:53077)}

\bibitem{WangYau6}
\bysame, \emph{Quasilocal mass in general relativity}, Phys.\ Rev.\ Lett.
  \textbf{102} (2009), no.~2, no. 021101, 4. \MR{MR2475769 (2010b:83015)}

\bibitem{Weinstein1}
G.~Weinstein, \emph{On rotating black--holes in equilibrium in general
  relativity}, Commun.\ Pure Appl. Math. \textbf{XLIII} (1990), 903--948.

\bibitem{Weinstein:trans}
\bysame, \emph{On the force between rotating coaxial black holes}, Trans. of
  the Amer. Math. Soc. \textbf{343} (1994), 899--906.

\bibitem{WilliamsMTT}
C.~Williams, \emph{Asymptotic behavior of spherically symmetric marginally
  trapped tubes}, Ann.\ Henri Poincar\'e \textbf{9} (2008), 1029--1067,
  arXiv:gr-qc/0702101. \MR{MR2453255 (2009h:83095)}

\bibitem{Witt}
D.M. Witt, \emph{Vacuum space-times that admit no maximal slice}, Phys.\ Rev.\
  Lett. \textbf{57} (1986), 1386--1389. \MR{MR857257 (87h:83008)}

\bibitem{WittenPMT}
E.~Witten, \emph{A new proof of the positive energy theorem}, Commun.\ Math.\
  Phys. \textbf{80} (1981), 381--402. \MR{MR626707 (83e:83035)}

\bibitem{YauBH}
S.-T. Yau, \emph{Geometry of three manifolds and existence of black hole due to
  boundary effect}, Adv. Theor. Math. Phys. \textbf{5} (2001), 755--767.
  \MR{MR1926294 (2003j:53052)}

\bibitem{York74}
J.W. {York$~$Jr.}, \emph{Covariant decompositions of symmetric tensors in the
  theory of gravitation}, Ann.\ Inst.\ H.\ Poincar\'e Sect.\ A (N.S.)
  \textbf{21} (1974), 319--332. \MR{MR0373548 (51 \#9748)}

\bibitem{Ziolkowski}
J.~{Zi\'o\l kowski}, \emph{Black hole candidates},  (2003),
  arXiv:astro-ph/0307307.

\end{thebibliography}

\end{document}